\documentclass[rmp,aps,nofootinbib,amsmath,floatfix,twocolumn, showopacs,10pt]{revtex4-1}
\usepackage{natbib}
\usepackage[mathscr]{eucal}
\usepackage{graphicx}
\usepackage{epsfig}
\usepackage{bm}
\usepackage{amssymb}
\usepackage{amsmath}
\usepackage[normalem]{ulem}
\usepackage{color}
\usepackage{subfigure,epsfig,epstopdf}
\usepackage{psfrag}
\usepackage[normalem]{ulem}  
\usepackage{xcolor}
\usepackage{pdfcomment}

\newcommand{\red}{\textcolor{black}}

\newcommand{\be}{\begin{equation}}
\newcommand{\dd}{\displaystyle}
\newcommand{\ee}{\end{equation}}
\newcommand{\bea}{\begin{eqnarray}}
\newcommand{\eea}{\end{eqnarray}}

\newcommand{\de}{\partial}
\newcommand{\A}{\alpha}
\newcommand{\B}{\beta}

\newcommand{\fourier}[1]{\int \frac{d^4#1}{(2\pi)^4}}
\newcommand{\intspace}[1]{\int d^4 {#1}\,}
\newcommand{\ha}{\frac{1}{2}}
\newcommand{\rr}{{\bm x}}
\newcommand{\q}[2]{ {\bm q}_{#1}^{#2}}

\newcommand{\setq}[2]{\{{\bm q}_{#1}^{#2}\}}

\newcommand{\fiak}[3]{\phi_{#1}^{#2}(#3)}
\newcommand{\dm}[1]{\delta\mu_{#1}}
\newcommand{\coleps}{\varepsilon^{\alpha\beta I}}
\newcommand{\flaeps}{\epsilon_{ij I}}
\newcommand{\colepsg}{\varepsilon^{\alpha\beta \gamma}}
\newcommand{\flaepsg}{\epsilon_{ij k}}
\newcommand{\colepst}{\varepsilon^{\alpha\beta 3}}
\newcommand{\flaepst}{\epsilon_{ij 3}}
\newcommand{\vu}{{\bm u}}
\newcommand{\cross}[1]{#1\!\!\!/}
\newcommand{\piak}[3]{{\cal{P}}_{#1}^{#2}(#3)}
\newcommand{\eps}{\varepsilon}
\newcommand{\al}{\alpha}
\newcommand{\apjl}{{\rm ApJ Lett.}}	
\newcommand{\mnras}{{Mon.Not.Roy.Astron.Soc.}}
\newcommand{\iaucirc}{{IAU Circ.}}

\def \Tr {{\rm Tr}}

\def\be{\begin{eqnarray}}
\def\ee{\end{eqnarray}}

\begin{document}

\preprint{}

\title{Crystalline color superconductors}

\author{Roberto Anglani}
\affiliation{Institute of Intelligent Systems for Automation, National Research Council,
CNR-ISSIA, Via Amendola 122/D-O, I-70126 Bari}
\author{Roberto Casalbuoni}
\affiliation{Department of Physics, University of Florence and INFN Via G. Sansone 1, 50019 Sesto Fiorentino (FI), Italy}
\author{Marco Ciminale}
\affiliation{Ministero dell'Istruzione, dell'Universit\`a e della Ricerca (MIUR),
Viale Trastevere 76/a, 00153 Roma, Italy }
\author{Raoul Gatto}
\affiliation{Departement de Physique Theorique, Universite de Geneve, CH-1211 Geneve 4, Switzerland}
\author{Nicola Ippolito}
\affiliation{INFN, Sezione di Bari, Via E. Orabona 4, 70126 Bari, Italy}
\author{Massimo Mannarelli}
\email{massimo@lngs.infn.it}
\affiliation{INFN, Laboratori Nazionali del Gran Sasso, Via G. Acitelli 22, 67100 Assergi (AQ), Italy}
\author{Marco Ruggieri}
\affiliation{Department of Physics and Astronomy, University of Catania, Via S. Sofia 64, I-95125 Catania}

\begin{abstract}
Inhomogeneous superconductors and inhomogeneous superfluids appear in a variety of contexts including quark matter at extreme densities, fermionic systems of cold atoms, type-II cuprates, and organic superconductors. In the present review the focus is on properties of quark matter at high baryonic density, which may exist in the interior of compact stars. The conditions realized in these stellar objects tend to disfavor standard symmetric BCS pairing and may favor an inhomogeneous color superconducting phase. The properties of inhomogeneous color superconductors are discussed in detail and in particular of crystalline color superconductors. The possible astrophysical signatures associated with the presence of crystalline color superconducting phases within the core of compact stars are also reviewed. \\ \\
\textit{We dedicate this paper to the memory of Giuseppe Nardulli for his important collaboration to the general subject of this review.}
\end{abstract}

\maketitle

\tableofcontents

\section{Introduction}
\label{sec:intro}

Ideas about color superconducting (CSC)  matter date back to more than 30 years ago \cite{Collins,Barrois:1977xd,Frautschi:1978rz,Bailin:1983bm},
but  this phenomenon has only recently received a great deal of  consideration
(for recent reviews see \textcite{Rajagopal:2000wf,Hsu:2000sy,Hong:2000ck,Alford:2001dt,Nardulli:2002ma,
Schafer:2003vz, Rischke:2003mt, Alford:2007xm}). Color superconductivity is the quark matter analog of the standard electromagnetic superconductivity and is believed to be the ground state of hadronic matter at sufficiently large baryonic densities.
At very high density the naive expectation, due to  asymptotic freedom, is that quarks  form a Fermi
sphere of almost free fermions. However,   Bardeen,
Cooper and Schrieffer (BCS) \cite{Cooper:1956, Bardeen:1957-1, Bardeen:1957-2} have shown that  the Fermi surfaces of free
fermions are unstable in presence of an attractive, arbitrary
small, interaction between fermions. In quantum chromodynamics (QCD) the attractive interaction between quarks can be due to instanton exchange \cite{Schafer:1996wv}, at intermediate densities, or to gluon exchange in the $\bar
3$ color channel, at  higher densities. Therefore,   one expects that at  high densities quarks  form a coherent state of Cooper pairs.

It should be noted that  the mentioned old papers  \cite{Collins,Barrois:1977xd,Frautschi:1978rz,Bailin:1983bm}
were  based on the existence of the attractive $\bar{3}$ color channel
and on analogies with ordinary superconductors. The main  result of these analyses was that  quarks form Cooper pairs with a gap of order a few MeV.
In more recent times  two papers by \textcite{Rapp:1997zu} and \textcite{Alford:1997zt}, have brought this result  to question.  These authors
considered diquark condensation arising from instanton-mediated interactions and although their approximations are not under rigorous quantitative control, the result was that  gaps can be as large as $100$ MeV.  

Color superconductivity offers a clue to the behavior of strong interactions at very
 high baryonic densities, an issue of paramount relevance for
 the understanding of the physics of
 compact stars and of heavy ion collisions. In the asymptotic regime it is  possible to understand the structure of the quark condensate from basic considerations. Consider the matrix element
\begin{equation}
\langle 0|\psi_{is}^\alpha\psi_{jt}^\beta|0\rangle\, , \label{condensate-general}
\end{equation}
where $\psi_{is}^\alpha$, $\psi_{jt}^\beta$ represent the quark fields, and $\alpha,\beta=1,2,3$, $s,t=1,2$,  $i,j=1,\cdots, N_f$ are color, spin and flavor indices, respectively.
For sufficiently large quark chemical potential, $\mu$,
 \red{assuming the orbital angular momentum state  be in a s-wave}, 
the color spin and flavor structure can be completely fixed by the following arguments:
\begin{itemize}
\item Antisymmetry in color indices $(\alpha,\beta)$ in order to
have attraction. \item Antisymmetry in spin indices $(s,t)$ in
order to  have a spin zero condensate. 
\item Given the
structure in color and spin, Fermi statistics requires
antisymmetry in flavor indices.
\end{itemize}
\red{The isotropic structure of
the  condensate with vanishing total angular momentum is favored with respect to higher spin or higher orbital angular momentum condensates because a larger portion of the phase space around the Fermi surface is available for pairing.}
Since the quark   spin and momenta  in the pair are opposite, it
follows that the left(right)-handed quarks can pair only with
left(right)-handed quarks.  Considering three-flavor quark matter at large baryonic density, the  so-called color-flavor locked (CFL) phase \cite{Alford:1998mk} turns out to be  thermodynamically favored, with condensate 
\begin{equation}
\langle 0|\psi_{iL}^\alpha\psi_{jL}^\beta|0\rangle=-\langle
0|\psi_{iR}^\alpha\psi_{jR}^\beta|0\rangle \propto
\Delta_{\rm CFL}\sum_{I=1}^3\coleps\flaeps\,,
\label{condensate-CFL}\end{equation}
\red{where $\Delta_{\rm CFL}$ is the pairing gap and $\varepsilon^{\alpha\beta\gamma}$ and $\epsilon_{ijk}$ are the completely antisymmetric Levi-Civita symbols in color and flavor space, respectively. We have suppressed spinorial  indices and neglected pairing in the color sextet channel. Pairing in the color sextet channel is automatically  induced by the quark color structure \cite{Alford:1998mk, Alford:1999pa}, but the condensate in this channel is much smaller than in the color antitriplet channel \cite{Schafer:1999fe, Shovkovy:1999mr} and in most cases it can be  neglected \cite{Rajagopal:2000wf}.}

\red{The reason of the name ``color-flavor locked"  is that only simultaneous transformations in color and in flavor spaces
leave the condensate invariant.  The corresponding  symmetry breaking
pattern is, indeed, the following}
\be\begin{split}
  &SU(3)_c\otimes SU(3)_L \otimes SU(3)_R\otimes U(1)_B \\ & \to SU(3)_{c+L+R}
  \otimes Z_2\,,
\label{eq:breakingCFL}
\end{split}
\ee
where $SU(3)_{c+L+R}$ is the diagonal global subgroup of the three
$SU(3)$ groups and the $Z_2$ group means that the quark fields
can still be multiplied by $-1$. According to the symmetry breaking pattern, the $17$ generators of   chiral  symmetry,  color symmetry and $U(1)_B$ symmetry are spontaneously broken.  The 8 broken generators of the color gauge group correspond to the 8 longitudinal degrees of freedom of the
gluons and according to the Higgs-Anderson mechanism these gauge bosons acquire a Meissner mass. The  diquark condensation induces a Majorana-like mass term in the fermionic sector which is not  diagonal in color and flavor indices. Thus, the fermionic excitations consist of  gapped  modes with mass proportional to $\Delta_{\rm CFL}$\footnote{This is a feature of all homogenous superconducting phases in weak coupling: the fermionic excitations which are charged with respect to the condensate acquire a Majorana-like mass term proportional to the pairing gap.}.  The low-energy  spectrum consists of  9 Nambu-Goldstone bosons (NGB) organized in an  octet, associated with
the breaking of the flavor group, and in a singlet, associated
 with the breaking of the baryonic number. For nonvanishing  quark masses the octet of NGBs becomes massive,  but the singlet NGB is protected by the  $U(1)_B$ symmetry; it remains massless and determines the superfluid properties of the CFL phase.
 The effective theory describing the NGBs for the CFL phase
 has been studied by \textcite{Casalbuoni:1999wu, Son:1999cm, Son:2002zn}.
The CFL condensate also breaks  the axial $U(1)_A$ symmetry; given that at very high densities the explicit axial symmetry breaking is weak, one has to include the corresponding  pseudo-NGB  in the low-energy spectrum.

After the first attempts with instanton-induced interaction many authors tried various approaches  for calculating the magnitude of the gap parameters in the CSC phases (for references see the review by \textcite{Rajagopal:2000wf}).
 Dealing with QCD the ideal situation would be if these kind of calculations could fall within the scope of lattice gauge theories. Unfortunately, lattice methods rely on Monte Carlo sampling techniques that are unfeasible  at finite density because  the fermion determinant becomes complex. Although various approximation schemes have been developed, for instance, Taylor expansion in the chemical potential \cite{Allton:2003vx},
reweighing techniques \cite{Fodor:2001au}, analytical continuation of calculation employing imaginary  baryonic chemical potential \cite{Roberge:1986mm, Alford:1998sd} or heavy Wilson quarks \cite{Fromm:2011qi}, no definite results have been obtained so far for large values of the baryonic chemical potential and physical quark masses.

In the absence of suitable lattice methods, quantitative analyses of color superconductivity have followed two distinct paths.
The first path is  semiphenomenological, and based on
simplified models.  The main feature of these models is that they should incorporate the most important physical effects while being at the same time tractable within present mathematical techniques.
All these models have free parameters that  are adjusted to give rise to a reasonable vacuum physics.

Examples of these kind of techniques include Nambu-Jona Lasinio (NJL) models in which the interaction between quarks is replaced by a four-fermion
interaction originating from  instanton exchange \cite{Alford:1997zt, Rapp:1997zu, Berges:1998rc} or where the  four-fermion interaction is modeled by that induced by single-gluon exchange \cite{Alford:1997zt, Alford:1999pa}.
Random matrix models have been studied by \textcite{Vanderheyden:2000ti} and instanton liquid models have been investigated by \textcite{Carter:1998ji, Rapp:1999qa, Rapp:2000zd}. 
Although none of these methods has a firm theoretical basis, all of them yield results all in fairly qualitative agreement.
This is probably due to the fact that what really matters is the existence of an attractive interaction between quarks and  that the parameters of the various models are chosen in such a way to reproduce the chirally broken ground state.
The gap parameter evaluated within these models varies between
 tens of MeV up to
$100$~MeV. The critical temperature is typically the same found in normal superconductivity, that is about one half of the gap.

The second path starts from first principles and relies on the property of asymptotic freedom of QCD. Various results have been obtained starting from the QCD action, employing   renormalization group techniques or   through the Schwinger-Dyson equation by \textcite{Evans:1998nf, Schafer:1998na, Son:1998uk, Schafer:1999jg, Pisarski:1999tv, Hong:1999fh, Brown:1999aq, Evans:1999at}. In particular,  \textcite{Son:1998uk}, using the renormalization group near the Fermi surface has obtained  the asymptotic form of the gap and corrections have been evaluated  by \textcite{Brown:1999aq}.

The result  of the above-mentioned methods is that the  CFL phase  is  the thermodynamically favored state of matter at asymptotic densities. Qualitatively one can understand this result considering that in the CFL phase  quarks of all three flavors  participate coherently in pairing.  Since superconductivity is a cooperative phenomenon, the larger  the number of fermions that participate in pairing, more energetically favored is the superconducting phase.

In the description of color superconductivity  one has to deal with various scales, the chemical potential $\mu$, the gap parameter, which we shall generically indicate with $\Delta$,  the constituent strange quark mass $M_s$ and the screening/damping scale $g_s \mu$, where $g_s$ is the QCD coupling. One typically has that $\mu\gg g_s \mu \gg \Delta$, whereas the strange quark mass can be considered as a free parameter, although in some models it can be computed self-consistently.

\begin{figure}[t]
\begin{center}
\includegraphics[width=8cm]{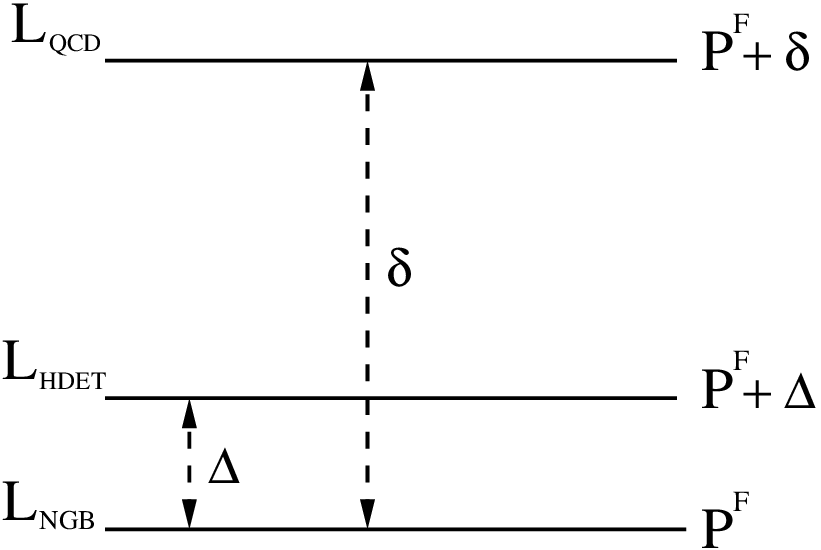}
\end{center}
\caption{Schematic representation of the hierarchy of effective Lagrangians characteristic 
 of high-density QCD. \label{hierarchy}}
\end{figure}

Quantum chromodynamics at high density is conveniently studied through a hierarchy of
effective field theories, schematically depicted in Fig.~\ref{hierarchy}.
The starting point is the fundamental
QCD Lagrangian, then  one can obtain the low-energy effective Lagrangian
 through different methods. One way  is to integrate out high-energy degrees of freedom as shown by \textcite{Polchinski:1992ed}.   The physics is particularly simple for energies close to the Fermi
energy where all the interactions are irrelevant
except for a four-fermion interaction coupling pair of fermions with
opposite momenta. This is nothing but the interaction giving rise
to  BCS condensation, which can be  described using the
high-density effective theory (HDET) \cite{Nardulli:2002ma, Hong:1998tn,Hong:1999ru,Beane:2000ms,Casalbuoni:2000na, Schafer:2003jn}. 
The HDET  is based on the fact that at vanishing temperature and large chemical potentials antiparticle fields decouple
and the only relevant fermionic degrees of freedom are  quasiparticles  and quasiholes close to the Fermi surface.
 In the  HDET Lagrangian  the great advantage is that  the effective fermionic fields have no spin structure  and therefore the theory is particularly simple to handle.  

This description is supposed to hold up to a cutoff
$P^F+\delta$, with $\delta$  smaller than the Fermi momentum, $P^F$, but
bigger than the gap parameter, \textit{i.e.} $\Delta\ll\delta\ll P^F$. Considering 
momenta much smaller than $\Delta$ all the gapped
particles decouple and one is left with the low-energy modes as
NGBs, ungapped fermions and holes and massless gauge
fields according to the symmetry breaking scheme. 
In the case of CFL and other CSC phases, such
effective Lagrangians have been derived by \textcite{Casalbuoni:1999wu, Casalbuoni:2000cn, Rischke:2000cn}. The parameters of the effective Lagrangian can be evaluated at each step of the
hierarchy by matching the Green's functions with the ones evaluated
at the upper level. For the CFL phase the effective Lagrangian of the superfluid mode  associated with the breaking of $U(1)_B$   may also be determined by symmetry arguments  alone as shown by \textcite{Son:2002zn}.

\red{In the high-density limit one  neglects the quark masses, and the CFL is believed to be the favored phase. On the other hand, considering a quark chemical potential of the same order of magnitude of the strange quark mass tends to disfavor the CFL pairing. The reason is that 
the typical effect of   quark masses is to produce a mismatch between Fermi surfaces.}  Neglecting light quark masses and assuming  (for simplicity) that quarks have all the same chemical potentials, the  Fermi spheres have now different radii
 \be
 P_s^F=\sqrt{\mu^2-M_s^2},~~~P_u^F=P_d^F=\mu\,.\label{simple-mismatch}\ee
Thus, increasing $M_s$ for a fixed value of $\mu$, increases the mismatch between the Fermi surface of strange quarks and the Fermi surfaces of up and down quarks (which in this simplistic case are equal).

The standard  BCS mechanism assumes that the Fermi momenta of the fermionic
species that form Cooper pairs are equal. When there is a mismatch it is not guaranteed that BCS pairing  takes place, because the condensation of fermions with different Fermi momenta has a free energy cost. \red{As first shown for weakly interacting  two-level systems by \textcite{Chandrasekhar, Clogston},  for mismatches below the  Chandrasekhar-Clogston (CC) limit} there is still condensation  and in the case at hand it means that the CFL phase is favored. However, for large values of the strange quark mass the assumptions leading to prove that the favored phase is CFL should be reconsidered. According to Eq.~\eqref{simple-mismatch} if the  strange quark mass is  about the quark chemical potential, then strange quarks decouple, and the corresponding favored
condensate should consist of only up and down quarks. With only two flavors of quarks, and due to the antisymmetry in color, the condensate must
necessarily choose a direction in color space and one possible pairing pattern is 
\begin{equation}
\begin{split}
&\langle 0|\psi_{iL}^\alpha\psi_{jL}^\beta|0\rangle \red{=-\langle
0|\psi_{iR}^\alpha\psi_{jR}^\beta|0\rangle} \propto
\Delta_{\rm 2SC}\, \varepsilon^{\alpha\beta 3}\epsilon_{ij3}\,, \\ &\alpha,\beta\in SU_c(2)~~~i,j\in SU(2)_L\,.\label{condensate-2SC}
\end{split}
\end{equation}
This phase of matter is known as two-flavor color superconductor (2SC)  and  $\Delta_{\rm 2SC}$ is the corresponding gap parameter. This phase is characterized by the presence of 2 ungapped quarks, $q_{ub}, q_{db}$ and 4  gapped quasiparticles  given by the combinations  $q_{dr}-q_{ug}$ and  $q_{ur}-q_{dg}$ of the quark fields, where the color indices of the fundamental representation $1,2,3$ have been identified with $r,g,b$ (red, green and blue).
In case massive strange quarks are present the corresponding phase is named
2SC+s and eventually strange quarks may by themselves form a spin-1 condensate \cite{Pisarski:1999bf}.

In the 2SC phase the symmetry breaking pattern is completely different from the
three-flavor case and it turns out to be
\be
\begin{split}
 & SU(3)_c\otimes SU(2)_L\otimes SU(2)_R\otimes U(1)_B\\ & \to  SU(2)_c
  \otimes SU(2)_L\otimes SU(2)_R\otimes U(1)_{\tilde B}\otimes Z_2\,.
\label{breaking-pattern-2SC}
\end{split}
\ee
The chiral group remains unbroken, meaning that there are no NGBs. The original color
symmetry group is broken to $SU(2)_c$ and since three color generators are unbroken, only five gluons acquire a Meissner mass. Even though  $U(1)_B$
is spontaneously broken there is an unbroken $U(1)_{\tilde B}$ global symmetry, with  $\tilde B$ given by a combination of  $B$ and of the eighth color generator, playing the same role of the original baryonic number symmetry.  In particular, this means that unlike CFL matter, 2SC matter  is not superfluid.  One can construct an effective
theory describing the emergence of the unbroken subgroup
$SU(2)_c$ and the low-energy excitations, much in the same way as
one builds  chiral effective Lagrangian with effective fields
at zero density. This development can be
found in the work by  \textcite{Casalbuoni:2000cn,Rischke:2000cn}.\\

\red{The CFL phase and the 2SC phase have been the first phases to be proposed and have been extensively studied. These are homogeneous phases, meaning that the condensate is not space dependent. The arising of inhomogeneous condensates for  imbalanced Fermi momenta in quark matter has only lately attracted the interest of the high-energy community. Actually, this  is  quite a general  problem arising not only in high-density QCD but also in condensed matter systems and in ultracold atom systems. There are several similarities between these systems, see \textcite{Casalbuoni:2003wh} for a review and for a discussion of  two-flavor QCD with imbalanced  Fermi momenta.  In the present Review we focus on high-density QCD   giving  a detailed and self-contained presentation of the various properties of two- and three-flavor quark matter with mismatched  Fermi spheres.} 

Studying the  pairing mechanisms of quark matter in systems with mismatched Fermi spheres is relevant when considering  realistic conditions, \textit{i.e.} conditions that can be realized in a compact stellar object (CSO). This is a real possibility since the central densities for these stars is very large, conceivably reaching $10^{15}$
g/cm$^{3}$, whereas the temperature is of the order of tens of keV, much less than the critical temperature for color superconductivity. The various processes taking place in CSOs produce a more complicated mismatch than the one presented in Eq.~\eqref{simple-mismatch}.  \red{The reason is that matter
inside a CSO should be electrically neutral, in $\beta$ equilibrium and in a color singlet state. 
 If electrons are present
(as generally required by electrical neutrality) the  $\beta$-equilibrium condition forces the chemical potentials
of quarks with different electric charges to be different. As far as color is concerned, it is possible to impose a simple
condition, that is color neutrality, because  it has been shown by \textcite{Amore:2001uf} (in the two-flavor case) that there is a small free energy cost in projecting color singlet states out of color neutral ones. Since the condensate is in general not diagonal in color indices, the requirement of color neutrality determines a mismatch between the chemical potentials of quarks with different colors.
 Thus, the  effect of the mass of the strange quark, $\beta$ equilibrium
and color and electric neutrality, is to produce a stress on the Fermi 
spheres of quarks with different flavor and color, trying to pull them apart. If the stress is sufficiently large the CFL phase cannot be realized, but condensation can still take place in different channels, depending on the parameters of the system.} Besides the above-mentioned standard 2SC and 2SC+s phases, the two-flavor superconducting phase 2SCus, with pairing between up and strange quarks can be favored; see \textit{e.g.} \textcite{Iida:2003cc, Ruester:2006aj} for different pairing patterns. For very large mismatches among the three flavors of quarks only the interspecies single-flavor spin-1 pairing may take place \cite{Bailin:1979nh, Alford:2002rz, Alford:1997zt, Buballa:2002wy, Schafer:2000tw, Schmitt:2004et, Schmitt:2002sc}, see \textit{e.g.} \textcite{Alford:2007xm} for an extended discussion on these topics.

In the  2SC phase, $\beta$ equilibrium and neutrality conditions tend to induce a chemical potential mismatch, $\delta\mu$,  between  up and down quarks. A remarkable property is that for  $\vert\delta\mu\vert=\Delta_{\rm 2SC}$
gapless fermionic modes appear and therefore the corresponding phase has been named g2SC by  \textcite{Shovkovy:2003uu,Huang:2003xd}, with ``g" standing for gapless.  The g2SC phase  has the same condensate of the 2SC phase reported in Eq.~\eqref{condensate-2SC}, and consequently the ground states of the 2SC and of the g2SC phases  share the same symmetry. However,  these two phases have a different low-energy spectrum,  due to  the fact that in the g2SC phase  only two  fermionic modes are gapped.
 The  g2SC phase  is energetically favored with respect to the 2SC phase and unpaired quark matter  in a certain range of values of the four-fermion interaction strength when one considers $\beta$ equilibrium, color
 and electrical neutrality \cite{Shovkovy:2003uu}. 

Pinning down the correct ground state of  neutral quark matter in $\beta$ equilibrium  is not simple because another difficulty emerges. This problem,  already present in  simple two-level systems, see for example \textcite{Gubankova:2008ya}, has a rather general character \cite{Alford:2005qw}, and  is due to
an instability connected to the Meissner mass. For sufficiently large chemical potential differences, the system becomes \textit{magnetically} unstable, meaning that the Meissner mass becomes imaginary. In the 2SC phase the color group is broken to $SU(2)_c$ and 5 out of 8 gluons acquire a mass. Four of these masses turn out to be imaginary in the 2SC phase for $\Delta_{\rm 2SC}/\sqrt{2} < \delta\mu< \Delta_{\rm 2SC}$, thus in this range of $\delta\mu$  the 2SC phase is \textit{chromomagnetically} unstable \cite{Huang:2004am, Huang:2004bg}. Increasing  the chemical potential difference the instability gets worse, because   at the phase transition  from the 2SC phase to the g2SC phase   all the five gluon masses become pure imaginary. 

\red{An analogous phenomenon arises in three-flavor quark matter   in the  gapless CFL (gCFL) phase \cite{Alford:2003fq, Alford:2004hz, Alford:2004zr, Fukushima:2004zq}. The gapless color-flavor locked phase has 
been proposed as the favored ground state for sufficiently large mismatch between up, down and  strange quarks and  occurs in  color and electrically neutral quark matter in $\beta$ equilibrium  for $
{M_s^2}/{2\mu}\gtrsim \Delta_{\rm CFL}$.   However, this phase  turns out to be chromomagnetically unstable \cite{Casalbuoni:2004tb, Fukushima:2005cm}, because when gapless fermionic modes appear the Meissner masses of some gluons become imaginary.}

Quite generally, the imaginary value of the Meissner mass can be understood as a tendency of the system toward an inhomogeneous phase \cite{Iida:2006df, Gubankova:2008ya, Hong:2005jv}. This can be easily seen in a  toy model system for the case of a  $U(1)$ symmetry, in which one can show that the coefficient of the gradient term of the low-energy fluctuations around the ground state  of the effective action is proportional to the Meissner mass squared \cite{Gubankova:2008ya}.

There is a variety  of solutions  that have been proposed for the chromomagnetic instability and that can be realized depending on the particular conditions considered. As we have already discussed,  the chromomagnetic instability is a serious problem
not only for the gapless phases (g2SC and gCFL) but also for the 2SC phase.
In the latter case, the vector condensates of gluons  with a value of about $10$ MeV can cure  the
instability \red{\cite{Gorbar:2005rx, Gorbar:2006up, Gorbar:2007vx, Fukushima:2006su, Kiriyama:2006ui}. The corresponding phase has been named gluonic phase and is characterized by the nonvanishing value of some gluon condensates and the spontaneous breakdown of the color,  electromagnetic  and  rotational symmetries down to the $SO(2)$ rotational symmetry. As discussed by \textcite{Gorbar:2006up, Gorbar:2007vx}, the chromomagnetic instabilities of the 4-7th gluons and of the 8th gluon might be related to two different phenomena. 
 The 4-7th instability seems to indicate the Bose-Einstein condensation of plasmons. Thus at the CC limit should correspond a second order phase transition to a state with uniform plasmon \textit{chromoelectric} condensate, which can be taken to be   $\langle {\bm A}^6\rangle$, inducing a nonvanishing value of $\langle {\bm A}^3\rangle$ and of $\langle {A}^3_0\rangle$.
On the other hand, the existence of gapless fermionic modes in the g2SC phase may  indicate the existence of a   $\langle {\bm A}^8\rangle$ condensate.} The chromomagnetic instability of the gapped 2SC phase can also be removed by the formation of an inhomogeneous condensate of charged gluons \cite{Ferrer:2007uw}. \red{The finite temperature case has been discussed by  \textcite{Kiriyama:2006jp, He:2006vr}, finding that
in the weak and intermediate coupling regime, the 2SC and g2SC phases are stabilized by temperatures of the order of tens of MeV.}

For the cases in which the chromomagnetic instability is related to the presence of gapless modes,   \textcite{Hong:2005jv} studied the possibility that  a secondary gap opens at the Fermi surface. The solution of the instability is due to a mechanism that stabilizes the system  preventing the appearance of gapless modes. This solution represents one of the few cases in which the instability may be cured by means of a different homogeneous condensate. However, the secondary gap turns out to be extremely small and at  temperatures typical of CSOs it is not  able to fix the chromomagnetic instability \cite{Alford:2005kj}.

For three-flavor quark matter two inhomogeneous superconducting phases have been proposed. If kaon condensation takes place in the CFL phase \cite{Bedaque:2001je, Kaplan:2001qk}, the chromomagnetic instability might drive the system toward an inhomogeneous state in which a kaon condensate current is generated,  balanced by a counterpropagating current in  the opposite direction carried by gapless quark quasiparticles. This phase of matter, named curCFL-$K^0$,  has been studied by \textcite{Kryjevski:2005qq} and turns out to be chromomagnetically stable.

The second possibility is the crystalline color
superconducting (CCSC) phase \cite{Alford:2000ze, Bowers:2001ip, Kundu:2001tt, Leibovich:2001xr, Bowers:2002xr, Casalbuoni:2001gt, Casalbuoni:2002pa, Casalbuoni:2002my, Casalbuoni:2003sa, Casalbuoni:2004wm, Casalbuoni:2005zp, Mannarelli:2006fy}, which is  the QCD analogue of a form of non-BCS pairing
first proposed by Larkin, Ovchinnikov, Fulde and Ferrell
(LOFF) \cite{LO, FF}.  The condensate characteristic of this phase is given by
\begin{equation}
\begin{split}
\langle
0|\psi_{iL}^\alpha\psi_{jL}^\beta|0\rangle = &-\langle
0|\psi_{iR}^\alpha\psi_{jR}^\beta|0\rangle\\ &\propto
 \sum_{I=1}^3\
 \Delta_I \coleps\flaeps \hspace{-.3cm}
\sum_{\q{I}{m}\in\setq{I}{}} e^{2i\q{I}{m}\cdot\bm x},
 \label{condensate-crystal}
\end{split}
\end{equation}
which is similar to the  condensate reported in Eq.~\eqref{condensate-CFL} but now there are three gap parameters, each having a  periodic modulation in space. The modulation of the $I$'th
condensate is defined by the vectors $\q{I}{m}$, where $m$ is the index which identifies the elements of the  set  $\setq{I}{}$.  In position space, this corresponds to condensates that vary
 like $\sum_m \exp(2 i {\bm q^m}\cdot {\bm x})$, meaning that
the ${\bm q}^m$'s are the reciprocal vectors which define the crystal
structure of the condensate. 

\red{The case of two-flavor CCSC, first  proposed by \textcite{Alford:2000ze}, corresponds to the vanishing of all but one gap parameter in Eq.~\eqref{condensate-crystal} and represents a candidate phase for curing the chromomagnetic instability in the two-flavor case.} Indeed, the chromomagnetic stability of a simple two-flavor  periodic structure with a gap parameter modulated by a single plane wave with wave vector $\bm q$ (hereafter we shall refer to this phase as Fulde-Ferrell (FF) structure \cite{FF})  has been considered by \textcite{Giannakis:2004pf, Giannakis:2005vw, Giannakis:2005sa},  where it  has been  shown that
\begin{itemize}
  \item The presence of the chromomagnetic instability in g2SC is
  exactly what one needs in order that the FF phase is
  energetically favored \cite{Giannakis:2004pf}.
  \item  The FF phase in the two-flavor case has no chromomagnetic
  instability (though it has gapless modes) at least in the weak
  coupling limit \cite{Giannakis:2005vw,Giannakis:2005sa}.
\end{itemize}
The stability of the FF phase in the strong coupling case  has been studied  by \textcite{Gorbar:2005tx}, in which  it is shown that for large values of the gap parameter the FF phase cannot cure the chromomagnetic instability.
\textcite{Nickel:2008ng}   have questioned whether among the possible one-dimensional
periodic modulations the LOFF solution is the favored one.   According to  \textcite{Nickel:2008ng},  for two-flavor quark matter a solitonic ground state is  favored with respect to  FF  in the range of values $0.7 \Delta \lesssim \delta\mu \lesssim 0.78\, \Delta$. However, at least in weak coupling,  the FF phase is not the  crystalline structure one should compare to. The  FF phase is slightly energetically favored with respect to unpaired quark matter and 2SC quark matter for $ \Delta/\sqrt 2< \delta\mu < 0.754\, \Delta$, but more complicated crystalline structures  have larger condensation energies in a larger range of values of $\delta\mu$ \cite{Bowers:2002xr}.

The stability analysis  of the three-flavor CCSC phase \red{has only been}  performed  for  a simple structure made of  two plane waves   by a Ginzburg-Landau (GL)  expansion \cite{Ciminale:2006sm}. This particular  three-flavor CCSC phase turns out to be  chromomagnetically stable, but the stability of more complicated crystalline structures has not been studied, although by general arguments  they are expected to be stable, at least in the weak coupling limit.  

Whether or not the  crystalline color superconducting phase is the correct ground state for quark systems with  mismatched  Fermi surfaces  has not been proven yet. In any case   it represents an appealing candidate because in this phase   quark pairing has no energy cost proportional to $\delta\mu$. The reason is that    pairing occurs between quarks living on their own Fermi surfaces. However, this kind of pairing can take place  only if  Cooper pairs   have nonzero total
momentum $2{\bm q}$ and therefore it has an  energy cost corresponding  to the kinetic energy needed for the creation of  quark currents.   Moreover, pairing can take place only in restricted phase space regions, meaning that the condensation energy is smaller than in the homogeneous phase.    The vector ${\bm q}$ has a  magnitude proportional to the chemical potential  splitting between Fermi surfaces, whereas  its direction  is spontaneously chosen by the system. In case one considers structures composed by a set of vectors  $\{{\bm q}_I\}$, one has to find the arrangement that minimizes the free energy of the system \cite{Bowers:2002xr, Rajagopal:2006ig}. This is a rather complicated task that is achieved  by analyzing some ansatz structures and comparing the corresponding free energy. \\

The  presence  of CCSC matter within CSOs may lead to a number of observable signatures associated with

\begin{enumerate}
\item \label{gwaves} Gravitational wave emission.
\item \label{glitches} Anomalies in the rotation frequency (known as glitches).
\item \label{cool} Cooling processes.
\item \label{mass-radius} Mass-radius relation.
\end{enumerate}

Point \ref{gwaves}.  relies on the observation that pulsars can be continuous sources of  gravitational waves if their mass distribution is not axis-symmetric.  The large shear modulus characteristic of the CCSC phase  allows the presence of big deformations of the  star, usually called ``mountains", making CSOs with a CCSC core strong sources of  gravitational waves. A different source of gravitational waves are the unstable oscillations of CSOs with a crystalline crust. 

Regarding point \ref{glitches}.,  the large rigidity of the CCSC phase  makes the crystalline phases of quark matter unique among all forms of matter proposed as candidates for explaining stellar glitches. 

Regarding point~\ref{cool}., one of the interesting properties of the CCSC phase is that  some quarks at their respective Fermi surfaces are unpaired. For this reason their neutrino emissivity
and heat capacity are only quantitatively smaller than
those of unpaired quark matter, not parametrically suppressed.
This suggests that neutron stars with crystalline quark matter
cores will cool down by the direct Urca reactions,
\textit{i.e.} more rapidly than in standard cooling scenarios.

Point   \ref{mass-radius}. is related to the fact that  recent observations of very massive compact stars seem to challenge the  possibility that CSOs have a CCSC core. \\

\begin{figure}[th!]
\begin{center}
\includegraphics[width=2.8in,angle=0]{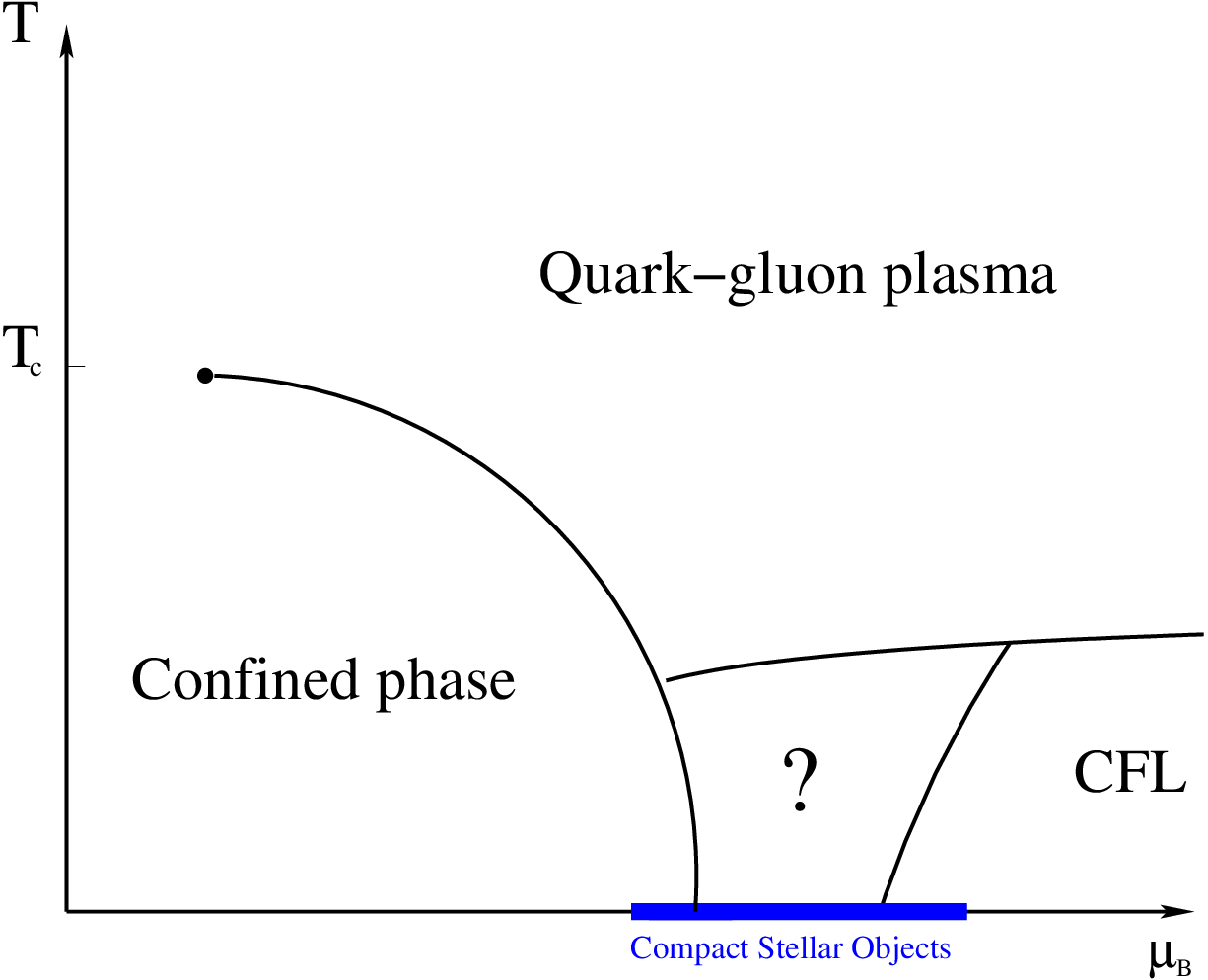}
\end{center}
\caption{(color online). Schematic phase diagram of strongly interacting matter as a function of the  baryonic chemical potential and  temperature.  At low temperatures and low densities matter consists of confined  hadrons. At high temperatures  quark and gluons degrees of freedom are liberated  forming the quark-gluon plasma.  At low temperatures and  very high densities  the CFL phase is favored. At   densities and  temperatures relevant for compact stellar objects, the CFL phase may be superseded by some different color superconducting phase or by some other phase of matter. The thick blue segment represents the possible range of baryonic chemical potential reachable in compact stars.}\label{fig-phase1}
\end{figure}

Summarizing, we can say that the state of matter at asymptotic densities is well defined and should correspond to the CFL condensate. At intermediate and more realistic densities it is not clear which is the ground state of matter. Our knowledge of the phases of matter can be  represented in the so-called QCD phase diagram,  schematically depicted  in Fig.~\ref{fig-phase1}. At low density and low temperature quarks are confined in hadrons but increasing the energy scale quarks and gluons  degrees of freedom are liberated. At high temperature this leads to the formation of a plasma of quarks and gluons, while at large densities matter should be in a color superconducting phase.  \red{If the conditions realized in CSOs favor the presence of inhomogeneous CSC phase, there might be distinctive astrophysical signatures of its presence.} 

Apart from the phases we have discussed other possibilities may be realized  in  the density regime relevant for CSOs.  Here we only mention that  it has been recently proposed  one more  candidate phase, the so-called  quarkyonic phase \cite{McLerran:2007qj}, which is characterized by a nonvanishing baryon number density and found to be a  candidate phase at least for a  large number of colors. \red{The possibility that also the quarkyonic phase shows a crystalline structure, in the so-called quarkyonic chiral spiral  state, has been discussed by  \textcite{Kojo:2010fe}.}
Another possibility is that the constituent value of the strange quark mass is so small that the  CFL phase is the dominant one down to the phase transition to the hadronic phase. In this case, a rather interesting possibility is that there is no phase transition between the CFL phase and the hadronic phase (hypernuclear matter), in the so-called quark-hadron continuity scenario \cite{Schafer:1998ef}.

\red{This review is organized as follows. Section~\ref{sec:Chapter2}  is devoted to the study of the two-flavor inhomogeneous phases. Here we  go from mismatched Fermi spheres, discussed in Sec.~\ref{sec:mismatched}, to the analysis of the gapless 2SC phases of QCD in Sec.~\ref{sec:g2SC}, and a detailed study of the crystalline phase in Sec.~\ref{sec:2flavor-crystals}. An approximate method based on the Ginzburg-Landau expansion is introduced in Sec.~\ref{sec:GL-2flavor}. A different approximation based on an  expansion around the gapless modes is discussed in 
Sec~\ref{sec:disp2flavor} and applied to the analysis of the dispersion laws of fermionic quasiparticles and of the corresponding specific heats. A third approximation method based on a smearing procedure is discussed in Sec.~\ref{sec:2flavor-smearing}.   Low-energy phonon excitations and their contributions to the specific heat are discussed in Sec.~\ref{phonons}. The stability analysis of the CCSC phase is considered in Sec.~\ref{sec:2stability}. A discussion of the solitonic ground state is presented in Sec.~\ref{sec:solitonic}. In Sec.~\ref{sec:cond-cold} we  briefly report on relevant results obtained in condensed matter and ultracold fermionic systems. In Sec.~\ref{sec:Chapter3} we  turn to  the three-flavor case. In particular, in Sec.~\ref{sec:gCFL} we  discuss the gapless CFL phase and  its instability. In Sec.~\ref{sec:3flavor_1PW} various aspects of the three-flavor CCSC phase made of two plane waves are discussed. Section \ref{sec:3flavor-crystals} is dedicated to the Ginzburg-Landau analysis of three-flavor  crystalline structures.  In Sec.~\ref{sec:shear}  the Nambu-Goldstone and the phonon modes are studied and we report  an analysis of the shear modulus of the two energetically favored crystalline phases. In Sec.~\ref{sec:Astrophysics} we  discuss whether the presence of  an inhomogeneous color superconducting phase  within the core of a compact star may lead to  observable effects. Gravitational wave emission is discussed in Sec.~\ref{sec:gwaves}; glitches are discusses in Sec.~\ref{sec:glitches} and  the cooling of toy model compact stars with a CCSC core is discussed in Sec.~\ref{sec:cooling}; the mass-radius relation for some models of hybrid CSO with a CCSC core  is discussed in Sec.~\ref{sec:mass-radius}. In Sec.~\ref{sec:Conclusion} we draw our conclusions and outlook.}

\section{The two-flavor inhomogeneous phases }
\label{sec:Chapter2}
The inhomogeneous two-flavor crystalline color superconducting (CCSC) phase is an extension to QCD of the phase proposed in condensed matter systems  by Fulde and Ferrell \cite{FF} and by Larkin and Ovchinnikov \cite{LO} (LOFF). 
Some aspects of this phase  have been
previously reviewed by  \textcite{Casalbuoni:2003wh}, discussing  the analogy between high-density QCD and condensed matter systems, as well.
Therefore, we will focus here on recent results, and in particular
 we discuss one of the main properties of this phase, namely its \textit{chromomagnetic} stability. This important property is not shared with homogeneous gapless
 color superconducting (CSC) phases \red{(at least in weak coupling)}, and therefore  strongly motivates its study.

\subsection{Mismatched Fermi spheres}\label{sec:mismatched}

Before discussing the case of two-flavor quark matter, we show how gapless superconductivity  may arise
considering the simpler case of a nonrelativistic two-level fermionic gas, thus avoiding the formal complications due to flavor and color degrees of freedom. For definiteness, we review the  system discussed by \textcite{Gubankova:2008ya}  consisting of two unbalanced populations of fermionic species $\psi_1$ and $\psi_2$, with opposite spin, at vanishing temperature, having the hamiltonian density 
\begin{equation}
{\cal H} = \sum_{s=1,2}\psi^\dagger_s\left(-\frac{\nabla^2}{2m} - \mu_s\right)\psi_s 
-g\,\psi_1^\dagger\psi^\dagger_2 \psi_2\psi_1~, \label{eq:H1}
\end{equation}
where $g > 0$ is the four-fermion coupling constant. The chemical potentials of the two species can be written as
$\mu_1 = \mu + \delta\mu$ and $\mu_2 = \mu - \delta\mu$, so that $\mu$ is the average of the two chemical potentials 
and $2\delta\mu$ their difference. The effect of the attractive interaction between fermions 
is to induce the difermion condensate 
\begin{equation}
\langle\psi_s(x)\psi_t(x) \rangle = \frac{\Delta(x)}{g}i (\sigma_2)_{st}~,
\label{eq:bilinear}
\end{equation}
which spontaneously breaks the global symmetry corresponding to particle number conservation. 
As a result  the fermionic excitation spectrum consists of two Bogolyubov modes with dispersion laws
\begin{equation}
E_a = \delta\mu + \sqrt{\xi^2 + \Delta_0^2}~,~~~~~E_b = -\delta\mu + \sqrt{\xi^2 + \Delta_0^2}~,
\label{eq:H2}
\end{equation}
with $\xi = -\mu + p^2/2m$ and $\Delta_0$ is the homogenous mean field solution.  Without loss of generality we take $\delta\mu> 0$, then from Eq.~\eqref{eq:H2} we infer that tuning the chemical potential difference to  values $\delta\mu \geq\Delta_0$, the mode $b$ becomes  gapless.
This phase corresponds to a superconductor with one gapped and one gapless fermionic mode and is named gapless homogeneous superfluid. 

\begin{figure*}[t]
\includegraphics[width=12cm]{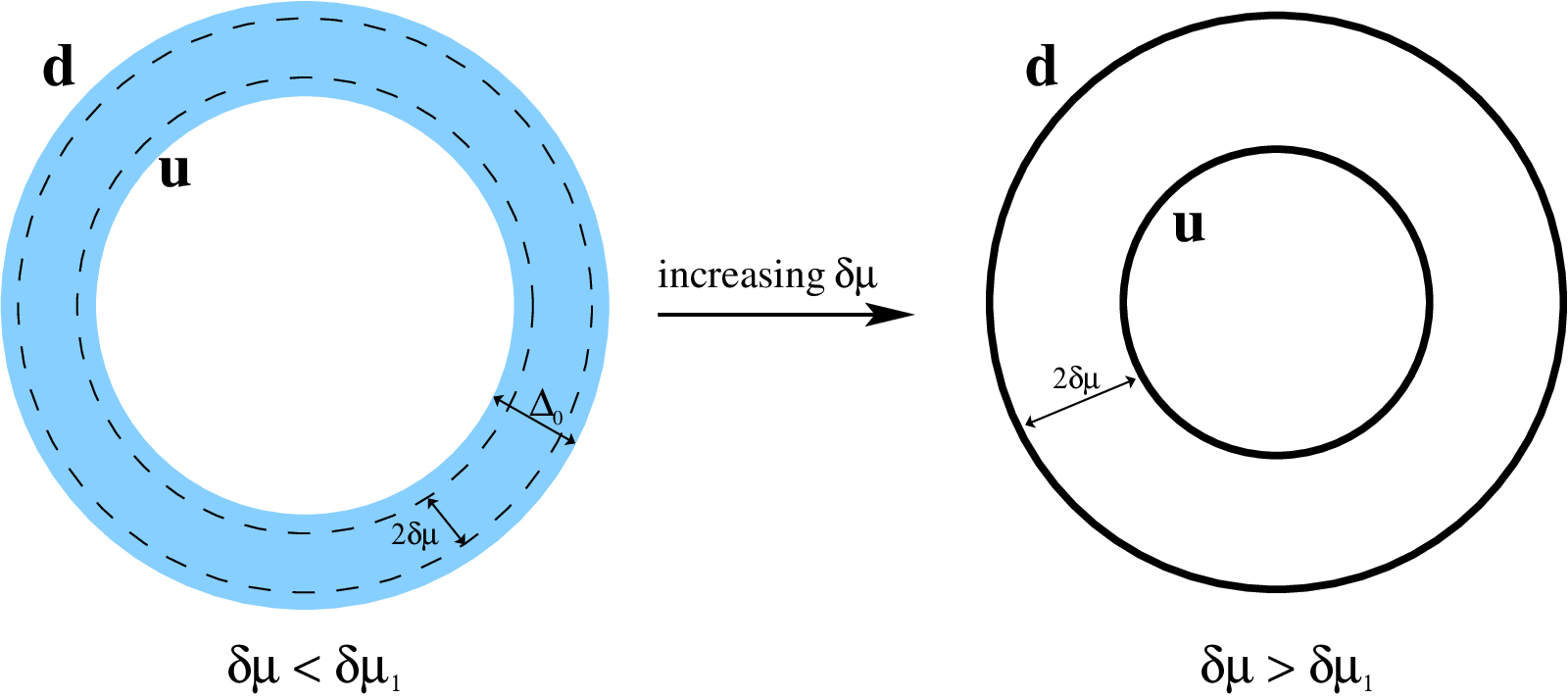}
\caption{\label{Fig:Clogston} (color online). Pictorial description of the behavior of the Fermi spheres of two different populations of fermions, with up ($\bf u$) and down ($\bf d$) spins, 
with increasing $\delta\mu$. \red{Left panel: the dashed black lines would correspond to the Fermi spheres of the two populations in the noninteracting case.  In the presence of a weak attractive interaction the BCS pairing produces a smearing of the Fermi spheres corresponding to the gray (light blue online) region.  Right panel: for $\delta\mu  > \delta\mu_1=\Delta_0/\sqrt{2}$ the Fermi spheres of the two populations (solid black lines) are widely separated and the  BCS homogenous phase is no more energetically favored.}}
\end{figure*}

In the above naive discussion we did not take into account that  increasing  $\delta\mu$  the difference between the  free energy of the superfluid phase, $\Omega_s$, and of the normal phase, $\Omega_n$ decreases;  eventually the normal phase becomes energetically favored for sufficiently large $\delta\mu$.
In weak coupling it is possible to show that  the two free energies  become equal at $\delta\mu_1 = \Delta_0/\sqrt{2}$ (corresponding to the so-called Chandrasekhar-Clogston (CC) limit \cite{Chandrasekhar, Clogston}), that is before the fermionic excitation spectrum becomes gapless.  At this critical value of $\delta\mu$ a first order transition to the normal phase takes place and the superfluid phase becomes metastable. The reason for this behavior can be qualitatively understood as follows. Pairing results in an energy gain of the order of $\Delta_0$, however BCS pairing takes place between fermions with equal and opposite momenta. When a mismatch between the Fermi spheres is present it tends to disfavor the BCS pairing, because for having  equal momenta fermions must pay an energy cost of the order of $\delta\mu$. Therefore, when $\delta\mu > c \Delta_0$, where $c$ is some number,  pairing cannot take place. In the weak coupling limit, one finds that $c = 1/\sqrt{2}$.  This behavior is pictorially depicted in
Fig. \ref{Fig:Clogston}.

Considering homogeneous phases,  a metastable superconducting phase exists for 
$\delta\mu  \geq \delta\mu_1$, but still the system cannot develop fermionic massless modes, because at $\delta\mu = \Delta_0$  various instabilities appear   \cite{Pao, Sheehy:2006qc, Mannarelli:2006hr, Gubankova:2006gj,  Gubankova:2008ya, Wu:2003zzh}. To explain what happens, let us consider the low-energy spectrum of the system, which can be described considering the fluctuations of $\Delta(x)$ around the mean field solution $\Delta_0$. The oscillations in the magnitude of the condensate are described by the Higgs mode, $\lambda(x)$, while the phase fluctuations are described by the Nambu-Goldstone (or Andersson-Bogolyubov) mode $\phi(x)$. Integrating out the fermionic degrees of freedom results in the Lagrangian density \cite{Gubankova:2008ya}

\begin{figure}[b!]
\begin{center}
\includegraphics[width=8cm]{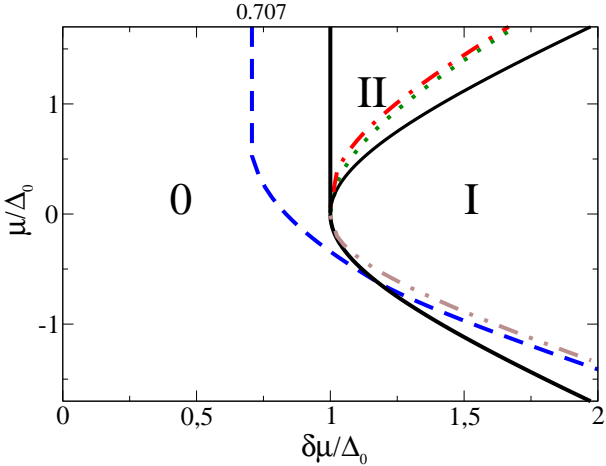}
\end{center}
\caption{\label{Fig:xreview} (color online). Exclusion plot in the ($\delta\mu/\Delta_0$, $\mu/\Delta_0$)
plane according to conditions 1, 2 and 3 (see the text). Condition 1 excludes the region above the dot-dot-dashed (brown) line; condition 2 excludes the region above the dot-dashed (red) line; condition 3 excludes the region above the dotted (green) line. The region above the dashed (blue) line corresponds to  $\Omega_s > \Omega_n$ and is therefore metastable or unstable. On
the top of the figure the Chandrasekhar-Clogston limit $\delta\mu/\Delta_0 =
1/\sqrt{2}\simeq 0.707$ is indicated.  Regions  featuring  zero ({\bf 0}), one ({\bf I}) or two ({\bf II}) gapless surfaces in momentum space are separated by solid lines. These lines are determined by finding the zeros of  $E_b$, given in Eq.~\eqref{eq:H2}. Adapted from \textcite{Gubankova:2008ya}.}
\end{figure}

\be\begin{split}
{\cal L}_{\lambda,\phi} =&  A (\partial_t \phi)^2 - \frac{B}{3} (\bm \nabla\phi)^2  -C \lambda^2 \\  &+D (\partial_t \lambda)^2 - \frac{E}{3} (\bm\nabla\lambda)^2  \label{L-lambda-phi} \label{L-lambda-phi}\,.
\end{split}
\ee



The stability of the system is guaranteed when all the coefficients $A$, $B$, $C$, $D$, $E$ are positive. $A$ and $D$ turn out to be always positive, then we define the three stability conditions  

\begin{enumerate}
\item The Higgs has a positive  squared mass: $C > 0$ \label{list1}
\item The space derivative of the Higgs must be positive: $E > 0$ \label{list2}
\item The space derivative of the NGB must be positive: $B >0$ \label{list3}
\end{enumerate}
Condition \ref{list1} is not satisfied if we are expanding around a maximum of the free energy, because the Higgs mass is proportional to the curvature of the free energy at the stationary point. The fact that the condition \ref{list2} is not satisfied signals that the system is unstable toward space fluctuations of the absolute value of the condensate, while condition  \ref{list3} is not fulfilled when the system is unstable toward space fluctuations  of the phase of the condensate. Clearly the conditions \ref{list2} and \ref{list3} are related and tell us that when a large mismatch between the Fermi sphere is present,  the system prefers to move to a phase in which the translation symmetry is spontaneously broken. In other words, the fact that the homogeneous phase is unstable towards space fluctuations of the condensate means that the energetically favored condensate is the one having a spacial modulation,  that is an  inhomogeneous condensate. For this aspect one may actually think of $\delta\mu$ as the control parameter for the transition from a homogenous phase to an inhomogeneous phase. We shall further elaborate on this point in Sec.~\ref{sec:momentum-susceptibility}, when discussing the momentum susceptibility. 
\red{The results of the analysis concerning  the stability conditions of the considered two-level model are reported in Fig.~\ref{Fig:xreview}.   The three conditions above are simultaneously violated in weak coupling for  $\delta\mu/ \Delta_0 \ge 1$, but they are violated at different values of this ratio in the strong coupling regime.  The most stringent is the condition \ref{list1}, excluding the region above the dot-dot-dashed  line. Condition \ref{list2} and condition \ref{list3} exclude the region above the dot-dashed  and dotted  lines, respectively. The region above the dashed line corresponds to $\Omega_s>\Omega_n$. Therefore in weak coupling (that is for $\Delta_0 \ll \mu$) the homogeneous phase is metastable for $\Delta_0/\sqrt{2} \le \delta\mu \le \Delta_0$.    With  increasing  coupling strength it is possible to force the system into a stable homogeneous gapless phase (corresponding to the region between the solid line and the dashed line at the bottom of Fig.~\ref{Fig:xreview}), but this happens when $\mu \sim - \Delta_0$, deep in the Bose-Einstein condensate (BEC) limit. }

In weak coupling  the  homogenous BCS  phase can be  energetically favored  for $ \delta\mu > \delta\mu_1$ if there is a way of reducing $\Omega_s - \Omega_n$, and this is indeed what happens in \red{some CSC phases}  in which  the color and electrical neutrality conditions may disfavor the normal phase.    On the other hand, this does not imply that the system has gapless modes. Indeed, in weak coupling  the stability of the region with  gapless modes  is  controlled by  conditions  \ref{list2} and \ref{list3}.  Decreasing  $\Omega_s-\Omega_n$ does not \textit{per se} guarantee that these conditions are satisfied. In fact, 
it turns out that  the gapless homogeneous phase is in general not accessible in weak coupling, because when $\delta\mu > \Delta_0$ both conditions  \ref{list2} and \ref{list3} are not satisfied, meaning that the solution with $\Delta_0 \neq 0$ is unstable. 

Gauging the $U(1)$ global symmetry, it is possible to show that the condition $B<0$ is equivalent to the condition that the Meissner mass squared of the gauge field becomes negative, which corresponds to a \textit{magnetic} instability.  Therefore the magnetic instability is related to the fact that we are expanding the free energy around a local maximum. This statement  is rather general and indeed in Sec.~\ref{sec:2stability} we shall see that an analogous conclusion can be drawn for the  2SC phase. Notice that   increasing the temperature of the system does not help to recover from this instability \cite{Alford:2005qw}. Indeed, the effect of the temperature is to produce a smoothing of the dispersion law, which has the effect of increasing the instability region to values $\delta\mu < \Delta_0$. \\

Summarizing, we have seen  that for the simplest case of a weakly interacting two-flavor system, for $\delta\mu > \delta\mu_1$  the superfluid homogenous phase is metastable, while  for $\delta\mu > \Delta_0$, $\Omega_s$  does not have a local minimum in $\Delta_0 \neq 0$ and it is unstable toward fluctuations of the condensate. In general, the three conditions above should be simultaneously satisfied for having a stable (or metastable) vacuum. The gapless phase is only accessible for homogenous superfluids deep in the strong coupling regime, for negative values of the chemical potential.   \\

A different possibility is that gapless modes arise at weak coupling in an inhomogeneous superfluid.  As we shall see in the following sections,   the inhomogeneous LOFF  phase is energetically favored in a certain range of values of $\delta\mu$ larger than the CC limit, see  Sec.~\ref{sec:2flavor-crystals},   it has  gapless fermionic excitations, see  Sec.~\ref{sec:disp2flavor}, and  it is (chromo)magnetically stable, see Sec. \ref{sec:2stability}. It is  important to remark that the presence of a gapless fermionic spectrum is not in contrast with the existence of superconductivity  \cite{gen66}, \textit{e.g.} type II superconductors have gapless fermionic excitations for sufficiently large magnetic fields \cite{gen66, Sarma-book}.  

\subsection{Gapless 2SC phase of QCD}\label{sec:g2SC}
The gapless 2SC phase (g2SC in the following) of QCD was proposed by \textcite{Shovkovy:2003uu} (see also \textcite{Huang:2003xd})
as a CSC phase which may sustain large Fermi surface mismatches. 
However, it was soon realized by the same authors that this phase is  chromomagnetic unstable  \cite{Huang:2004am,Huang:2004bg}, 
meaning that the masses of some  gauge fields become imaginary.  
In the following we briefly discuss the properties of the  g2SC phase at vanishing
temperature, and then we  deal with the problem of the chromomagnetic instability.

We consider neutral two-flavor quark matter at finite chemical
potential described by the following Lagrangian density:
\begin{equation}
{\cal L} = \bar\psi\left(i\gamma_\mu\partial^\mu - m
+\bm\mu\gamma_0\right)\psi + {\cal L}_{int}~, \label{eq:L2sc1}
\end{equation}
where  $\psi\equiv\psi^\alpha_{ i}$, $i=1,2$ $\alpha =1, 2, 3$ corresponds to
a quark spinor of flavor $i$ and color $\alpha$. The current quark mass is denoted by
$m$ (we take the isospin symmetric limit $m_u = m_d = m$), and
${\cal L}_{int}$ is an interaction Lagrangian that will be
specified later.

In Eq.~\eqref{eq:L2sc1}, $\bm\mu$ is the quark chemical potential
matrix with color and flavor indices, given by
\begin{equation}
\bm\mu \equiv\mu_{i j, \alpha\beta} = \left(\mu~\delta_{ij}-\mu_e~
Q_{ij} \right)\delta_{\alpha\beta} + \frac{2}{\sqrt{3}}\mu_8
(T_8)_{\alpha\beta}\delta_{ij}~, \label{eq:muijab}
\end{equation}
where   the quark electric charge matrix
and \red{the $SU(3)$ color generators are respectively   \be Q_{ij}=\text{diag}(Q_u,Q_d)\,, \qquad T_a= \frac{\lambda_a}{2} \label{eq:TA}\,,\ee  with $\lambda_a$ the Gell-Mann matrices  for $a=1,\dots,8$.} 
With $\mu_e$ and $\mu_8$ we denote respectively the
electron and the color chemical potential. Since $\bm\mu$ is diagonal in color and flavor spaces, we can indicate its element with $\mu_{i\alpha}$, \textit{e.g.}  $\mu_{ub}$ is the chemical potential of up blue  quarks.
A chemical
potential along the third direction of color, $\mu_3$, can be
introduced besides $\mu_8$, but, for all the
cases that we discuss in this section, we require the ground state
to be invariant under the $SU(2)_c$ color subgroup; this makes the
introduction of $\mu_3$ unnecessary.

As interaction Lagrangian density we consider  the NJL-like model 
\begin{widetext}
\begin{equation}
{\cal L}_{int} = G_S\left[(\bar\psi\psi)^2 + (\bar\psi i
\gamma_5\bm\tau\psi)^2\right]
+G_D\left[(\bar\psi^C\epsilon\varepsilon i\gamma_5\psi)_{k\gamma}
   (i\bar\psi\epsilon\varepsilon i\gamma_5\psi^C)_{k\gamma}\right]~,   \label{eq:Lint}
\end{equation}
\end{widetext}
where $\psi^C = C\bar\psi^T$ denotes the
charge-conjugate spinor, with $C=i\gamma_2\gamma_0$ the charge
conjugation matrix. The matrices $\varepsilon$ and $\epsilon$
denote the antisymmetric tensors in flavor and color space,
respectively;  we used  in the second term on the right hand side of Eq.~\eqref{eq:Lint}  the shorthand notation 
\be
(\bar\psi^C\epsilon\varepsilon i\gamma_5\psi)_{k\gamma} \equiv (\bar\psi^{\alpha C}_{i}\flaepsg\colepsg i\gamma_5\psi^\beta_{j})\,,
\ee
and an analogous expression for the other bilinear.
\red{For the 2SC phase considered in this 
section we assume condensation takes place only in the $k=\gamma=3$
channel.} 
In Eq.~\eqref{eq:Lint} two coupling constants are
introduced in the scalar-pseudoscalar quark-antiquark channel,
denoted by $G_S$, and in the scalar diquark channel, denoted by
$G_D$. \textcite{Huang:2003xd} chose the parameters of the model  to reproduce the pion decay constant in the vacuum,
$f_\pi=93$ MeV, and the vacuum chiral condensate $\langle\bar u
u\rangle^{1/3}$ = $\langle\bar d d\rangle^{1/3}$ = $-250$ MeV.
Moreover, an ultraviolet cutoff $\Lambda$ is introduced to
regularize the divergent momentum integrals.  The parameter set
of \textcite{Huang:2003xd} is given by
\begin{equation}
\Lambda=653.3~\text{MeV}~,~~~G_S = 5.0163~\text{GeV}^{-2}~.
\end{equation}
The relative strength between the couplings in the quark-antiquark and
quark-quark channels could be fixed by a Fierz rearranging of the quark-antiquark
interaction, see for example \textcite{Buballa:2003qv}. 
\red{For example,
considering  interactions with  the quantum numbers of the one gluon exchange,
 the Fierz transformation gives $G_D/ G_S =0.75$.} However, non perturbative
in-medium effects might change this value. Therefore, in the work by \textcite{Huang:2003xd}  the ratio of $G_D$ to $ G_S$ is considered as a free parameter.

\red{At high density  \textcite{Huang:2003xd}  consider only the case $m
= 0$ and   vanishing  chiral condensate}.  When
$m\neq0$ the chiral condensate in the ground state does not vanish,
but its effects are presumably negligible,  giving  a small shift of the quark
Fermi momenta. This shift  might change the numerical
value of the electron chemical potential only of some few percent.
Hence, the main results by \textcite{Huang:2003xd} should not change
much if a nonvanishing value of the current quark mass is considered.

Once the Lagrangian density is specified, the goal is to compute
the thermodynamic potential. In the mean field (and one
loop) approximation, this can be done easily using standard
 techniques. The mean
field Lagrangian density can be written  within the Nambu-Gorkov formalism,  in the compact form 
\begin{equation}
{\cal L} = \chi^\dagger S^{-1}\chi -\frac{\Delta_{\rm 2SC}^2}{4 G_D}~,
\label{eq:Lagrangian2SC}\end{equation}
where 
\be
\chi = \left(\begin{array}{l} \psi \\
\psi^C
\end{array}\right)\ ,
\ee 
is the Nambu-Gorkov spinor and the gap parameter, $\bm\Delta \equiv \Delta_{\rm 2SC}
\varepsilon^{\alpha\beta 3}\varepsilon_{ij 3} C\gamma_5$, is  included in the inverse propagator 
\begin{equation}
S^{-1} =  \left(\begin{array}{cc}
  i\gamma_\mu\partial^\mu + \bm\mu \gamma_0 & \bm\Delta \\
  \bm\Delta^\dagger & i\gamma_\mu\partial^\mu - \bm\mu \gamma_0 \\
\end{array}\right) ~,
\label{eq:Sinvert}\end{equation}
as an off-diagonal term in the ``Nambu-Gorkov space".

We shall focus on the zero temperature regime (for a discussion of the rather uncommon temperature  behavior of the g2SC phase  see  \textcite{Huang:2003xd}), which is  relevant for astrophysical
applications.  The  one loop
expression of the thermodynamic potential can be determined from the inverse propagator in Eq.\eqref{eq:Sinvert}; for vanishing temperature it is given by
\begin{equation}
\Omega= \frac{\Delta_{\rm 2SC}^2}{4 G_D}-\frac{\mu_e^4}{12\pi^2}
 -\sum_n\int\frac{d\bm p}{(2\pi)^3}|E_n|~,
\label{eq:rambo2}
\end{equation}
 for a derivation see for example \textcite{Buballa:2003qv}.
 The second addendum corresponds to the  electron free energy
(electron masses have been neglected). The last addendum
is the contribution due to the quark determinant. The sum
runs over the twelve fermion propagator poles, six of them
corresponding to quarks and the other six corresponding to
antiquarks:
\begin{eqnarray}
E_{1,2} &=& |\bm p| \mp \mu_{ub}~,\\
E_{3,4} &=& |\bm p| \mp \mu_{db}~,\\
E_{5,6} &=& \delta\mu + \sqrt{(|\bm p| \mp \bar\mu)^2 + \Delta_{\rm 2SC}^2}~,\\
E_{7,8} &=& -\delta\mu + \sqrt{(|\bm p| \mp \bar\mu)^2 +
\Delta_{\rm 2SC}^2}~,
       \label{eq:DL3}
\end{eqnarray}
and $E_9 = E_5$, $E_{10} = E_6$, $E_{11} = E_7$, $E_{12} = E_8$.
Here we have introduced the shorthand notation
\be
\bar\mu &=&\mu -\frac{\mu_e}{6} +
\frac{\mu_8}{3}~, \qquad \delta\mu = \frac{\mu_e}{2}~.
\ee
Using the explicit form of the dispersion laws, the free energy can be written as
\begin{eqnarray}
\Omega &=& -\frac{\mu_e^4}{12\pi^2}
-\frac{\mu_{ub}^4}{12\pi^2}
 -\frac{\mu_{db}^4}{12\pi^2} -\frac{\Lambda^4}{2\pi^2} -2\int_0^\Lambda\frac{p^2 dp}{\pi^2} \nonumber \\
 && \times
    \left(\sqrt{(p-\bar\mu)^2 + \Delta_{\rm 2SC}^2} + \sqrt{(p+\bar\mu)^2 + \Delta_{\rm 2SC}^2}\right)
       \nonumber\\
 &&-2\theta(\delta\mu-\Delta_{\rm 2SC})\int_{\mu_-}^{\mu_+}\frac{p^2 dp}{\pi^2}
    \left(\delta\mu - \sqrt{(p-\bar\mu)^2 + \Delta_{\rm 2SC}^2}\right)\nonumber\\
        \label{eq:THP1}
\end{eqnarray}
where $\mu_\pm = \bar\mu \pm \sqrt{\delta\mu^2 - \Delta_{\rm 2SC}^2}$.

The value of $\Delta_{\rm 2SC}$ is determined by the solution of the
equation
\begin{equation}
\frac{\partial\Omega}{\partial\Delta_{\rm 2SC}} = 0~,\label{eq:GE}
\end{equation}
  with the neutrality constraints,
\begin{equation}
n_8 = -\frac{\partial\Omega}{\partial\mu_8}=0~,
  ~~~n_Q =-\frac{\partial\Omega}{\partial\mu_e}=0~,
   \label{eq:NC}
\end{equation}
which fix the values of $\mu_e$ and $\mu_8$.

The numerical analysis by \textcite{Huang:2003xd} shows that
$\mu_8$ is much smaller than $\mu_e$ and $\Delta_{\rm 2SC}$, both for
$\Delta_{\rm 2SC}\geq\delta\mu$ and for $\Delta_{\rm 2SC} < \delta\mu$. As a
consequence, it is possible to simplify the equations for the gap
parameter and the electron chemical potential, Eq.~\eqref{eq:GE}
and Eq.~\eqref{eq:NC} respectively, by putting $\mu_8 = 0$.  Therefore,
the properties of the system depend only on the values  $\Delta_{\rm 2SC}$ and $\mu_e$ and on the couplings $G_D$ and $G_S$. The result by \textcite{Huang:2003xd} can be summarized as follows:

\begin{itemize}
\item For $G_D/G_S \gtrsim 0.8$, \textit{strong coupling}, the 2SC phase is the only \red{homogeneous} stable phase
\item For $0.7 \lesssim G_D/G_S \lesssim 0.8$, \textit{intermediate coupling}, the g2SC phase is allowed for $\delta\mu > \Delta_{\rm 2SC}$
\item For $G_D/G_S \lesssim 0.7$, \textit{weak coupling}, only unpaired quark matter is favored.
\end{itemize}

In the g2SC phase   the quasiparticle fermionic spectrum consists of four gapless modes and two gapped modes, whereas in the 2SC phase there are two gapless fermionic modes  and four gapped fermionic modes. In the latter case the only gapless modes correspond to the up and down blue  quarks that do not participate in pairing.

\subsubsection{Meissner masses of gluons in the g2SC phase}\label{sec:meissner_g2SC}

The diquark condensate of the 2SC phase induces the symmetry breaking pattern reported in Eq.~\eqref{breaking-pattern-2SC}; in  particular the group $SU(3)_c\otimes U(1)_{em}$ is broken down to $ SU(2)_c\otimes \tilde U(1)_{em}$, where  $\tilde U(1)_{em}$ is the gauge group corresponding to the rotated massless photon associated with the unbroken generator
\be \tilde Q = Q \cos \theta-\frac{g_s}eT_8 \sin\theta \,,\ee 
where $g_s$ and $e$ denote the strong and the electromagnetic
couplings respectively, and $Q$ and $T_8$  are defined in Eq.~\eqref{eq:TA}. The mixing coefficients have been determined by \textcite{Alford:1999pb}  (see also \textcite{Gorbar:2000ms}) and are given by
\be \cos \theta = \frac{\sqrt 3 g_s}{\sqrt{3 g_s^2 + e^2}} \qquad \sin \theta = \frac{e}{\sqrt{3 g_s^2 + e^2}} \,. \ee
The linear combination 
\be  T_{\bar 8} = T_8 \cos \theta+\frac{g_s}e Q \sin\theta \,, \label{eq:eightbar}\ee
is orthogonal to $\tilde Q$ and gives the broken generator;  the corresponding gauge field, which we shall refer to as the $\bar 8$ mode, acquires a Meissner mass. Actually, the NJL-like Lagrangian  in Eqs.\eqref{eq:L2sc1} and \eqref{eq:Lint}, has only global symmetries, but gauging the $SU(3)_c$ group and the $U(1)$ subgroup of $SU(2)_L \times SU(2)_R$, one has that the spontaneous symmetry breaking leads to the generation of Meissner masses for the  five gluons associated  with the broken generators. To compute these masses, we define the gauge boson polarization
tensor, see for example \textcite{bellac2000thermal},
\begin{equation}
\Pi_{ab}^{\mu\nu}(p) = -\frac{i}{2}\int\frac{d^4 q}{(2\pi)^4}
   \text{Tr}\left[\Gamma_a^\mu S(q) \Gamma_b^\nu S(q-p)\right]~,
\label{eq:GBPT}
\end{equation}
where $S(p)$ is the quark propagator in momentum space, which can be obtained from Eq.~\eqref{eq:Sinvert}, and  \begin{eqnarray}
\Gamma_a^\mu &=& g_s\,\gamma^\mu~\text{diag}\left(T_a,-T_a^T\right)~,
\label{eq:VMcolor}\\
\Gamma_9^\mu &=& e~\gamma^\mu~ \text{diag}\left(Q, -Q\right)~,
\label{eq:VMem}
\end{eqnarray}
are the interaction vertex matrices.  The trace in Eq.~\eqref{eq:GBPT} is taken over  Dirac, Nambu-Gorkov, color and flavor indices;   $a,b = 1, \dots, 8$  indicate the adjoint color and we use the convention that the  component with $a,b=9$ corresponds to the photon. 

The screening masses of the gauge bosons are defined
in terms of the eigenvalues of the polarization
tensor and in the basis in which $\Pi_{ab}^{\mu\nu}$ is diagonal
the Debye masses and the Meissner masses
are  respectively defined as
\begin{eqnarray}
 {\cal M}_{D,a}^2 &=& -\lim_{\bm p\rightarrow 0}\Pi_{aa}^{00}(0, \bm p )~,
 \label{eq:debye-generic}\\
{\cal M}_{M,a}^2 &=& -\frac{1}{2}\lim_{\bm p\rightarrow 0}
 \left(g_{ij}+\frac{\bm p_i \bm p_j}{\vert \bm p\vert^2}\right)\Pi_{aa}^{ij}(0,\bm p )~.
\label{eq:meissner-generic}\end{eqnarray}
Both masses have been evaluated in the 2SC phase by  \textcite{Rischke:2000qz,Rischke:2002rz,Schmitt:2003aa, Rischke:2003mt}.
The Debye masses of all gluons are related to the chromoelectric screening and are always real, therefore do
not affect the stability of the 2SC and g2SC phases. The Meissner
masses of the gluons with adjoint color $a=1,2,3$ are always
zero, because they are associated with the unbroken color
subgroup $SU(2)_c$. 
\red{It is interesting to note that in the 2SC phase  $\Pi^{00}_{ab}$ is diagonal in the $a,b$ indices and therefore there is no need to diagonalize it \cite{Schmitt:2003aa, Rischke:2003mt}. This happens because of a cancellation between the
contribution of the blue ungapped quarks with that of the gapped excitations.  The  magnetic components of the  $8$th gluon and of the photon do instead mix and the polarization tensor has to be diagonalized for extracting the Meissner mass. In contrast, in the CFL phase both the electric and magnetic sectors mix   because of the absence of ungapped excitations \cite{Schmitt:2003aa}.}

For nonvanishing values of $\delta\mu$ the Meissner  masses have been 
evaluated by \textcite{Huang:2004am,Huang:2004bg}. Gluons with adjoint color $a=4,5,6,7$, are degenerate and in
the limit $\mu_8 = 0$ their  Meissner  masses are given by 
\begin{equation}\begin{split}
{\cal M}_{M,4}^2 = &\frac{4\alpha_s\bar\mu^2}{\pi}
               \left[
                     \frac{\Delta_{\rm 2SC}^2 - 2\delta\mu^2}{2\Delta_{\rm 2SC}^2} \right. \\ &\left.
               + \frac{\delta\mu}{\Delta_{\rm 2SC}^2}\sqrt{\delta\mu^2 - \Delta_{\rm 2SC}^2}\,
                      \theta(\delta\mu-\Delta_{\rm 2SC})
                 \right]~.
\label{eq:Pi}\end{split}
\end{equation}
The squared Meissner mass turns out to be negative not only in the gapless
phase, $\Delta_{\rm 2SC}/\delta\mu < 1$, but also in the gapped phase,
when $\Delta_{\rm 2SC}/\delta\mu < \sqrt{2}$. This result seems in contrast with the result of the previous section, where an imaginary Meissner mass was related to the existence of a local maximum of the free energy arising at $\delta\mu = \Delta$.
However, from the analysis of the 2SC free energy of the system, one can see that when $\Delta_{\rm 2SC}/\delta\mu = \sqrt{2}$ the state with $\Delta_{\rm 2SC} \neq 0$ corresponds to a saddle point in the $\Delta-\delta\mu$ plane. The neutrality condition transforms this saddle point into a local minimum. However,  as explained in the previous section, the  gauge fields can be related to the fluctuations of the gap parameter. These fluctuations can probe all the directions in the  $\Delta-\delta\mu$ plane around the stationary point and would result in a low-energy Lagrangian with  dispersion laws akin to  those discussed in Eqs.\eqref{L-lambda-phi}  with the coefficients $E$ and $B$ negative. 

Finally, let us consider  the  $\bar 8$ mode, which is associated with the broken generator  defined in Eq.~\eqref{eq:eightbar}.  The corresponding Meissner mass can be obtained diagonalizing the polarization tensor in Eq.~\eqref{eq:GBPT}  in the subspace $a,b \in \{8,9\}$, or more directly, by substituting $T_a \rightarrow T_{\bar 8}$ in the vertex factors of  the polarization tensor.  The squared Meissner mass of the $\bar 8$ mode turns out to be
\begin{equation}
{\cal M}_{M,\bar 8}^2 = \frac{4(3\alpha_s + \alpha) \bar\mu^2}{27\pi}
    \left(1 - \delta\mu\frac{\theta(\delta\mu-\Delta_{\rm 2SC})}{\sqrt{\delta\mu^2 - \Delta_{\rm 2SC}^2}}
    \right),
\label{eq:8til}
\end{equation}
and becomes negative for $\Delta_{\rm 2SC}/\delta\mu < 1$.  As shown by \textcite{Gatto:2007ja}, the instability in this  sector is transmitted to a gradient instability of the pseudo-Goldstone boson related to the $U(1)_A$
symmetry which is broken by the diquark condensate. Although in vacuum
the $U(1)_A$ symmetry is explicitly broken by instantons,
 at finite chemical potential instantons are Debye
screened and  $U(1)_A$ can be considered as an approximate
symmetry, which is then spontaneously broken by the diquark
condensate.

\subsection{The two-flavor crystalline color superconducting phase}\label{sec:2flavor-crystals}
Since the homogeneous g2SC phase is chromomagnetically unstable, the question arises of the possible existence of a different superconducting phase for large mismatch between the Fermi spheres. There are several candidate phases, which include the gluonic phase \cite{Gorbar:2005rx, Gorbar:2006up}, the solitonic phase (see Sec. \ref{sec:solitonic}) and the CCSC phase.

In this section we  review some of the main results about
the two-flavor CCSC  phase.  First, we describe the one plane wave ansatz in the framework of the simple model discussed in Sec.~\ref{sec:mismatched}; then, we turn to the CCSC phase and report on the Ginzburg-Landau (GL) analysis of various
crystalline structures. We relax the constraint of
electrical and color neutrality, and treat the difference of
chemical potentials between $u$ and $d$ quarks, 2$\delta\mu$, as a free
parameter. 

\subsubsection{From one plane wave to the crystalline phase}\label{sec:1pwans}
In  Sec.~\ref{sec:mismatched} we have shown that in weak coupling  the CC limit signals that the standard BCS phase becomes metastable, but does not forbid the existence of different forms of superconductivity. In particular, it does not forbid the existence of Cooper pairs with nonvanishing total momentum.   It was shown by Larkin and Ovchinnikov \cite{LO} and Fulde and Ferrel \cite{FF},  in the context of electromagnetic superconductors, that in a certain range of values of $\delta\mu$ it might be energetically favored the realization of Cooper pairs with nonzero total momentum.  For the simple two-level system discussed in Sec. \ref{sec:mismatched},  Cooper pairs with momentum $2 \bm q$ can be described by considering the difermion condensate in Eq.~\eqref{eq:bilinear} given by 
\begin{equation}
\langle\psi_s(x)\psi_t(x) \rangle = \frac{\Delta}{g}\, i (\sigma_2)_{st}e^{2 i {\bm q} \cdot {\bm x}}~,
\label{FF-ansatz}
\end{equation}
and we shall call this state of matter the FF phase.

\begin{figure*}[t]
\begin{center}
\psfrag{\theta}{$\vartheta_q$}
\subfigure{\includegraphics[width=7cm]{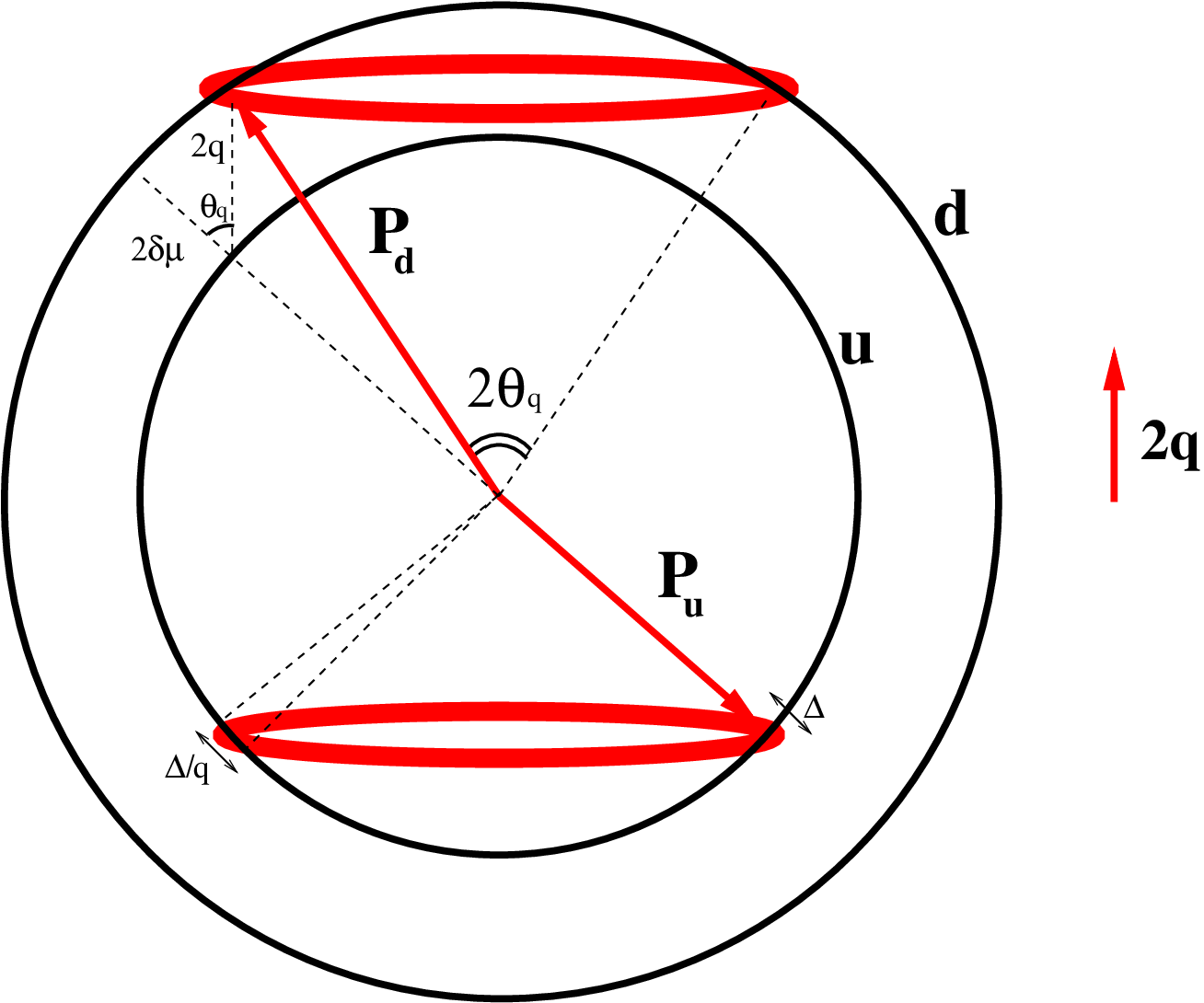}}\hspace{1.5cm}
\subfigure{\includegraphics[width=7cm]{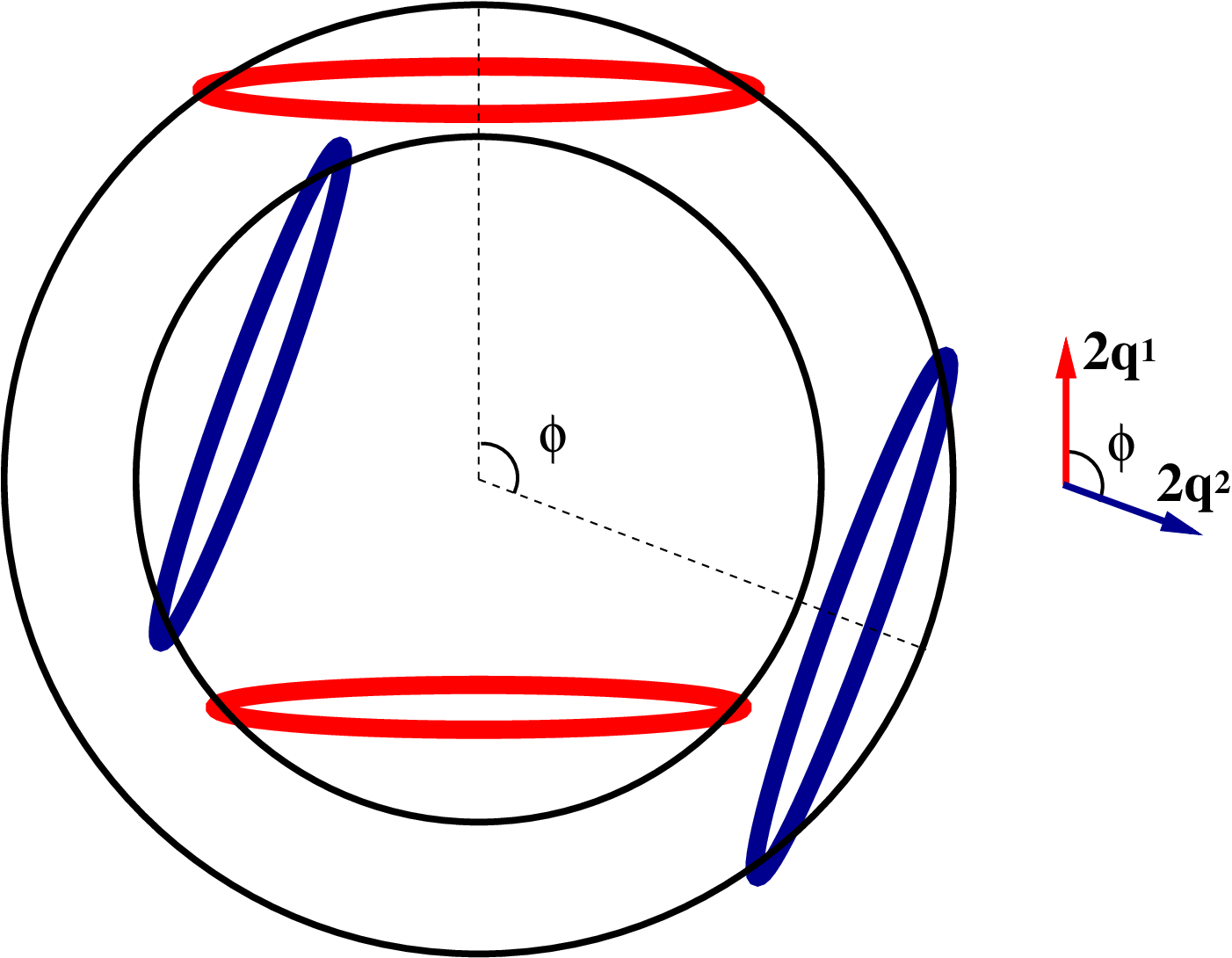}}
\end{center}
\caption{\label{Fig:FF} (color online). Pictorial description of the LOFF pairing in the weak coupling approximation. When $\delta\mu  > \Delta_0/\sqrt{2}$ the BCS homogenous pairing is not  energetically allowed, but pairing between fermions with total nonvanishing momentum can be realized.  Left panel: In the FF phase  pairing takes place in two ribbons  on the top  the Fermi spheres of up and down fermions, such that $\bm P_u + \bm P_d = 2 \bm q$, having opening angle $2 \vartheta_q \simeq 2 \arccos( \delta\mu/q) \simeq 67^{\circ}$, thickness $\Delta$ and angular width $\Delta/q$, see Sec.\ref{sec:gap}.  Right panel: Structure obtained with two plane waves corresponding to two vectors $2 \bm q^1$ (light gray, red online) and $2 \bm q^2$ (dark gray, blue online) with relative angle $\phi$. The size and the opening angle of each ribbon is as in the FF phase.   The structure with $\phi = 180^{\circ}$ is called the ``strip". For illustrative purposes, we have greatly exaggerated the splitting between the Fermi surfaces, relative to the values used in the calculations reported in Sec.~\ref{sec:2flavor-crystals}.}
\end{figure*}

Notice that this ansatz breaks rotational symmetry because there is a privileged direction corresponding to $\bm q$. In the left panel of  Fig.~\ref{Fig:FF}    the two Fermi spheres of fermions are pictorially shown and the gray (red online) ribbons correspond to the  regions in momentum space where pairing occurs. Pairing between fermions of different spin can only take place in a restricted region of momentum space and this implies that $\Delta < \Delta_0$. The reason why the FF phase is energetically favored with respect to the normal phase is that no energy cost proportional to $\delta\mu$ has to be payed for allowing the formation of Cooper pairs. The only energetic price to pay is due to the kinetic energy associated with Cooper pairs:  there is a spontaneous generation of   a supercurrent in the direction of $ {\bm q}$, which is balanced by a current of normal fermions in the opposite direction \cite{FF}. 
The gap
parameter in Eq.~\eqref{FF-ansatz} can be determined 
solving a gap equation under the constraint that the modulus of the Cooper momentum, $q$,  
minimizes the free energy. The result is that at $\delta\mu\approx\delta\mu_1 = \Delta_0/\sqrt{2}$ there is a first
order phase transition from the homogeneous BCS phase to the FF phase.
 Increasing further $\delta\mu$ results in a smooth
decreasing of the gap function of the FF phase, until at a critical
value  $\delta\mu_2$ a second order
phase transition to the normal phase takes place. In the weak
coupling limit $\delta\mu_2 = 0.754\Delta_0$; the range
$[\delta\mu_1,\delta\mu_2]$ is called the  LOFF {\em window}. In
the LOFF window, the optimal value of $q$ turns out to be
approximately constant, $q\approx 1.2\delta\mu$.

\textcite{Alford:2000ze} presented the two-flavor QCD  analog  of the FF phase.  In this case the condensate has the same color, spin and flavor structure of the 2SC condensate, but with the plane wave space dependence characteristic of the FF phase, that is 

\be \langle  \psi^\alpha_{i}(x) C\gamma_5 \psi^\beta_{j}(x) \rangle \propto  \Delta_{i j}^{\alpha \beta} = \Delta e^{i 2 \bm q \cdot  \bm x}\colepst\flaepst ~.
\label{FF-2flavor}
\ee

As noted by \textcite{Alford:2000ze}, the FF condensate induces a spin-1
condensate as well; however, its effect is found to be numerically
small, and it will be  neglected here. As in the simple  two-level system, in the two-flavor FF phase it is possible to determine the free energy using a NJL-like model with the condensate given
 in \eqref{FF-2flavor}.   \red{To this end it is convenient using the Nambu-Gorkov formalism discussed in Sec~\ref{sec:g2SC} in a slightly different way. 
In the FF case  we are considering pairing between  $u$
quarks with momentum ${\bm p}+{\bm q}$ and $d$ quarks with momentum
$-{\bm p}+{\bm q}$ so that the total momentum is $2 \bm q$ as in Eq.~\eqref{FF-2flavor}.  In momentum space, the standard two-flavor Nambu-Gorkov spinor reads
\be
\chi({\bm p}) = \left(\begin{array}{l} \psi_u({\bm p})\\ \psi_d({\bm p}) \\
\psi_u^C(-{\bm p})\\ \psi_d^C(-{\bm p})
\end{array}\right)\,, \label{NGspinor-standard}
\ee 
and thus the corresponding propagator has
terms in it that are not only off-diagonal in the Nambu-Gorkov space, but also 
off-diagonal in {\it momentum} space. Thus any calculation looks complicated in this basis, but it  simplifies  changing the Nambu-Gorkov basis as follows \cite{Bowers:2001ip} \be
\chi({\bm p}) = \left(\begin{array}{l} \psi_u({\bm p}+{\bm q})\\ \psi_d({\bm p}-{\bm q}) \\
 \psi_u^C(-{\bm p}-{\bm q})\\ \psi_d^C(-{\bm p}+{\bm q})
\end{array}\right)\ .
\label{NGspinor2a} \ee The effect of this momentum shift is to eliminate the dependence on $\bm q$  in the  off-diagonal terms in the Nambu-Gorkov propagator. Indeed, the pair condensate  described by
terms in the fermion propagator occurring in the $\psi_u$-$\psi^C_d$ and
$\psi_d$-$\psi_u^C$ entries  are now independent of $\bm q$, 
 making the propagator diagonal in ${\bm p}$-space and the calculation tractable.
Of course the above choice of the momentum shift is not unique: being ${\bm p}$   an integration variable we can shift it for example by $\bm p \rightarrow \bm p - \bm q$ thus rewriting the Nambu-Gorkov spinor in the two-flavor phase as
\be
\chi({\bm p}) = \left(\begin{array}{l} \psi_u({\bm p})\\ \psi_d({\bm p}-2{\bm q}) \\
\psi_u^C(-{\bm p})\\ \psi_d^C(-{\bm p}+2{\bm q})
\end{array}\right)\,,
\label{NGspinor2b} \ee
is always possible.  }
 
The inverse propagator in the shifted basis has  formally the same expression given in Eq.~\eqref{eq:Sinvert}, with space-independent off-diagonal terms, but now the derivatives act on spinors with shifted momenta, see \textcite{Bowers:2001ip} for more details. The  gap parameter   can then be determined solving the corresponding  gap equation  under the constraint that the favored value of  $q$ minimizes the free energy. The  results are the same obtained in the two-level system; in particular  the LOFF window and $q$ have the same expressions reported above (but now $2 \delta\mu$ is the difference of
chemical potentials between $u$ and $d$ quarks).

From Fig.~\ref{Fig:FF}, it should be clear that an immediate generalization of the FF phase can be obtained adding more ribbons on the top of the Fermi spheres, corresponding to different vectors ${\bm q}_m$, with ${\bm q}_m \in \{{\bm q} \}$, where $\{{\bm q} \}$ is some set of vectors to be determined by minimizing the free energy of the system and $m$ is a label that identifies the vectors of the set. This in turn, corresponds to  considering  inhomogeneous CSC phases with a more general
ansatz than in Eq.~\eqref{FF-2flavor}, where  the single plane wave  is replaced by a superposition of plane waves,  that is
\begin{equation}
\Delta e^{2i \bm q \cdot \bm x} \rightarrow \sum_{{\bm q}^m \in \{{\bm q} \}}\, \Delta_{{\bm q}^m} e^{2i\bm q^m \cdot\bm
x}~. \label{1cs}
\end{equation}
To simplify the analysis, a set of assumptions are used. The vectors ${\bm q^m}$ are taken with equal length, thus  we can write ${\bm q^m} = q  \,{\bm n^m} $ and the set of vectors characterizing each condensate  can be indicated with $\{{\bm n} \}$. The  set of vectors $\{{\bm n} \}$ identifies the vertices of a crystalline structure, thus  at each set $\{{\bm n} \}$ corresponds a particular crystalline  phase.  As a further simplifying assumption, the coefficients $\Delta_{{\bm q}^m}$ are taken independent of ${\bm q}^m $ and we shall indicate their common value with $\Delta$. In other words, we  consider condensates with 
\begin{equation}
\Delta ({\bm x}) =  \Delta \sum_{m=1}^P \,  e^{2i q \bm n^m \cdot\bm
x}\,, \label{crystals}
\end{equation}
where $P$ is the number of vectors $\bm n^m$. The simplest example is clearly the FF condensate, depicted in the left panel of Fig.~\ref{Fig:FF}, characterized by a  single plane wave, thus corresponding to $P=1$. The case with $P=2$ is reported in the right panel of Fig.\ref{Fig:FF}, in this case the ``crystalline" structure is completely determined by $\phi$, the relative angle between $\bm n^1$ and $\bm n^2$; more complicated structures can be pictorially represented in a similar way.

It is important to stress that the crystalline structure  is determined by the modulation of the condensate, but
the underlying fermions are not arranged in an ordered pattern, indeed fermions are superconducting, that is they  form a superfluid of charged carriers.  

\red{The computation of the spectrum and of the free energy of a system with a general crystalline condensate cannot be obtained by the momentum shift technique discussed above. The reason is that by a momentum shift we can  eliminate the dependence on only one of the vector $\bm q^m$ in the off-diagonal term of the Nambu-Gorkov propagator.  As a consequence, the use of
some approximation is necessary.  In the next section we discuss the GL approximation for the evaluation of the free energy and in Sec.~\ref{sec:disp2flavor} we present a method for determining  the low-energy fermionic spectrum. A different approximation method is discussed in Sec.~\ref{sec:2flavor-smearing}.} 

\subsubsection{Ginzburg-Landau analysis}
\label{sec:GL-2flavor}

A viable method for the evaluation of the free energy of some crystalline structures is  the
GL expansion, which is obtained expanding  $\Omega$  in powers of
$\Delta$:
\be
\Omega = \Omega_n+ P\alpha\Delta^2 + \frac{\beta}{2}\Delta^4 +
\frac{\gamma}{3}\Delta^6 + {\cal O}(\Delta^8)~,\label{eq:OGL}
\ee 
where  $\Omega_n$ is the free energy of the normal phase and the coefficients $\alpha$, $\beta$ and $\gamma$ have been computed, in the one-loop approximation, using as   microscopic theory  a NJL model by \textcite{Bowers:2002xr}. 
The GL expansion is well suited for studying second order phase transitions but might give reasonable results for \textit{soft} first order phase transitions as well. In the present  case the expansion is under control  for $\Delta/q \ll 1$ and if the coefficient $\gamma$ is positive, meaning  that the free energy is bounded from below. 

For a given crystalline
structure, the coefficients in Eq.~\eqref{eq:OGL} depend on
$\delta\mu$ and on $q$; the latter is fixed,
in the calculation by \textcite{Bowers:2002xr}, to the weak coupling
value $q \simeq 1.2\delta\mu$. For any value of $\delta\mu$ the
thermodynamic potential of a given structure is computed by
minimization with respect to $\Delta$ and then the optimal crystalline structure
is  identified with that with the lowest free energy. It is
possible to compute analytically the GL coefficients only for few
structures. \red{One example is the FF phase, in which
\begin{align}
 \alpha_{\rm FF}(q,\delta\mu) &= \frac{2\mu^2}{\pi^2} \left(-1
+
\frac{\delta\mu}{2q}\ln\left|\frac{q+\delta\mu}{q-\delta\mu}\right|
\right.\nonumber\\ &+ \left. \frac{1}{2}\ln\left|\frac{4(q^2
-\delta\mu^2)}{\Delta_{\rm 2SC}^2}\right|\right)\,, 
\label{alfaFF} \\
\beta_{\rm FF}(q,\delta\mu) &=\frac{\mu^2}{2\pi^2}\frac{1}{q^2-\delta\mu^2}
\label{betaFF}\,,\\
\gamma_{\rm FF}(q,\delta\mu)& =\frac{\mu^2}{16 \pi^2}\frac{q^2+3 \delta\mu^2}{(q^2-\delta\mu^2)^3} \,,
\label{gammaFF}
\end{align}
where $\Delta_{\rm 2SC}$ is the  2SC gap parameter.} In general,  the GL coefficients  of more complicated structures have to be computed numerically; we refer to the appendix of the work by \textcite{Bowers:2002xr} for  details. In that paper  twenty-three crystalline structures
have been studied and among them those
 with $P>9$ turn out to be largely disfavored. 
 This has been nicely explained in the weak
coupling regime: in this case, as shown in the left panel of Fig.  \ref{Fig:FF} for the FF phase, the pairing regions of the inhomogeneous superconductor can be approximated as rings on the top of the Fermi surfaces;
one ring per wave vector $\bm n$ in the set $\{\bm n\}$. The
computation of the lowest order GL coefficients shows that the
intersection of two rings is energetically
disfavored \cite{Bowers:2002xr}. As a consequence, it is natural
to expect that in the most favored structure no intersecting
rings appear. Since each ring has an opening angle of approximately $67$
degrees,  a maximum of nine rings can be accommodated on a spherical surface.

The fact that configurations with overlapping rings are disfavored can be quantitatively understood  as follows. 
For the case of  two plane waves  (right panel of Fig.  \ref{Fig:FF}), in the weak coupling approximation
there is one pairing ring for each of the two wave vectors. The quartic coefficient $\beta$  depends on the angle $\phi$  between the two wave vectors and  it diverges at \be
 \phi_0 \approx 2 \vartheta_q \approx 2\arccos\frac{\delta\mu}{q}\approx 67 ^{\circ} \label{eq:phi0}\,,\ee
corresponding to the angle at which the two pairing rings
are contiguous, meaning that for $\phi < \phi_0$ the two rings  overlap. The latter case is energetically disfavored because, being $\beta$ large and positive, the free energy \red{gain} would be smaller.

The  divergence of the coefficient $\beta(\phi)$ at $\phi = \phi_0$
is due to the two limits that have been taken to compute the free energy,
namely the GL and weak coupling limits.  A detailed explanation of what happens will be given in Sec. \ref{sec:3flavor_1PW} when  discussing a simple three-flavor crystalline structure. In any case it is clear that the divergency of a GL coefficient means that the expansion is not under control, or more precisely, that the radius of convergence of the series in Eq.~\eqref{eq:OGL} tends to zero.

Among the crystals with no intersecting rings, seven are good
candidates to be the most favored structure. Within these seven
structures, the octahedron, which corresponds to a crystal with
$P=6$ and whose wave vectors point into the direction of a
body-centered cube (bcc), is the only one with effective potential
bounded from below (that is, with $\gamma>0$). The remaining six
structures, given by different configurations with $P=7,8$ and $9$, are characterized by a potential
which is unbounded from below, at least at the considered order $\Delta^6$.
Even if in this case the free energy cannot be computed,
qualitative arguments given by \textcite{Bowers:2002xr} suggest that
the favored structure is the one with $P=8$ having wave vectors pointing
towards the vertices of a face-centered cube (fcc).

Of course, as these authors admit, their study cannot be trusted quantitatively, because of the several
limitations of the GL analysis.  First of all, the GL expansion formally
corresponds to an expansion in powers of $\Delta/q$ and therefore it is well suited for the study of second order phase transitions, but
 the condition that $\Delta/q\ll 1$ is not satisfied by all the 
crystalline structures considered by \textcite{Bowers:2002xr}. In some cases 
 the GL analysis predicts a strong first-order phase
transition to the normal state,  with
a large value of the gap at the transition point. 
Moreover, it may happen that the local minimum for small values of $\Delta/q$ is not a global minimum of the system, as discussed in Sec. \ref{sec:solitonic}. In this case the  GL expansion in   Eq.~\eqref{eq:OGL} underestimates the free energy \red{gain} 
of the system and is not able to reproduce the correct order of the phase transition. For a more reliable determination of the ground state one should consider terms of higher power in $\Delta$, which are difficult to evaluate. Finally,  the claimed favored crystalline structure, namely the fcc, has $\gamma<0$ and a global minimum cannot be found unless the coefficient ${\cal O}(\Delta^8)$ (or of higher order) is computed and found to be positive. 
 
 Because of these reasons, the
quantitative predictions of the GL analysis should be taken with a
grain of salt. One  should not trust the order of the phase transition obtained by the GL expansion and   the comparison among various crystalline structures may be partially incorrect, because it is not guaranteed that  the corresponding free energies have been accurately determined. 

On the other hand, the qualitative picture that we
can draw from it, namely the existence of crystalline structures
with lower free energy than the single plane wave, is quite
reasonable: crystalline structures benefit of more phase space
available for pairing, thus lowering the free energy. The symmetry argument is quite solid too, because it is based on the fact that configurations with overlapping pairing regions are disfavored, and as we shall see for one particular configuration in the three-flavor case in Sec. \ref{sec:3flavor_1PW}, one can prove that this statement is correct without relying on the GL expansion. We shall also show an interesting point, that the GL expansion underestimates the free energy \red{gain} of the crystalline structures. And this happens not only in the presence of a global minimum different from the local minimum around which the GL expansion is performed, see Sec. \ref{sec:solitonic},  but also comparing the GL free energy with the free energy evaluated without the $\Delta/q$ expansion.

\subsection{Fermionic dispersion laws and specific heats}\label{sec:disp2flavor}
The thermal coefficients (specific heat, thermal conductivity etc.) of quark matter at very low temperature   are of  fundamental importance for the  transport properties and  cooling mechanisms of compact stars. The largest contribution to the thermal coefficients comes from the low-energy degrees of freedom and it is therefore of the utmost importance to understand whether fermionic modes are gapped or gapless. Indeed,
the absence of a gap in the spectrum of fermions implies that quasiquarks can be excited even at low temperature and therefore the corresponding thermal coefficients are not suppressed by a factor $\approx e^{-\Delta/T}$  (which is distinctive of homogeneous BCS superconductors). 

In this section we discuss the fermion dispersion laws in the two-flavor CCSC phase for low values of momenta. Then, we use the obtained  dispersion laws for the
computation of the specific heat. A different contribution to the specific heat, due to phonons,  will be presented in Sec.~\ref{phonons}. The results discussed below do not rely on the GL approximation but are obtained by an expansion around the zeros of the full inverse propagator \cite{LO, Casalbuoni:2003sa}.

\subsubsection{Fermi quasiparticle dispersion law: general settings}\label{sec:fermi-quasiparticles}
We consider  a general difermion condensate $\Delta({\bm x})$, and determine the quasiparticle dispersion laws looking at the zero modes of the inverse propagator of the system.  Arranging the fields in the 
 Nambu-Gorkov spinor as follows, 
\begin{equation}
\chi_i^\alpha=
\left(
\begin{array}{c}
\tilde{G}^\alpha_i \\
-i(\sigma_2)_{\alpha\beta}\tilde{F}^{\beta}_i
\end{array}
\right)~,
\end{equation}
the inverse propagator is given by
\begin{widetext}
\be (S^{-1})^{\A\B}_{ij}=
\left(
\begin{array}{cc} \delta^{\A\B} [  \delta_{ij} (E + i {\bm v}
\cdot {\bm \nabla}) + \delta\mu(\sigma_3)_{ij}] & -\eps^{\A\B 3}\varepsilon_{ij3}\Delta({\bm x}) \cr -\varepsilon^{\A\B 3}\eps_{ij3}\Delta({\bm x})^{*} &
\delta^{\A\B} [ \delta_{ij} (E - i {\bm v} \cdot {\bm \nabla}) + \delta\mu(\sigma_3)_{ij}]
\end{array} \right)\,,\ee
\end{widetext}
where $E$ is the quasiparticle energy and $\bm v$ is the Fermi velocity, that for massless quarks satisfies $v=|{\bm v}|=1$. The quasiparticle spectrum can obtained  by solving the eigenvalue equation
 \be (S^{-1})^{\A\B}_{ij} \chi_{j}^{\B}=0 \label{eig1} \, .\ee
Performing the unitary transformation 
\be \bar G_i^\alpha=\left(e^{i\delta\mu\, \sigma_3\, {\bm v\cdot
\bm x}/v^2}\right)_{ij} G_j^\alpha\,,~\bar F_i^\alpha=\left(e^{-i\delta\mu\, \sigma_3\,{\bm v\cdot \bm x}/v^2}\,\sigma_2\right)_{ij} F_j^\alpha\, \nonumber \\
\label{transformation}\ee 
it is possible to eliminate the dependence on $\delta\mu$ in the eigenvalue problem and this corresponds to measuring the energy of each flavor from its Fermi energy. The resulting equations for $F_i^\alpha$ and $G_i^\alpha$ are
independent of color and flavor indices, and therefore these   indices will be omitted below. The eigenvalue problem reduces to solve the coupled differential equations:  
\bea (E + i {\bm v}\cdot {\bm\nabla}) G - i \Delta({\bm x})  F
 &=& 0\ ,
\cr (E  - i {\bm v}\cdot {\bm\nabla}) F+ i \Delta({\bm x})^{*}  G &=& 0\ .\label{eigfg}
 \eea

These equations can be used to find the dispersion laws for any inhomogeneous $\Delta({\bm x})$, and we shall consider here the periodic structures of the form given in Eq \eqref{crystals}  determining whether gapless fermionic excitations are present. We shall prove that for any crystalline structure with real-valued periodic functions $\Delta({\bm x})$ there exists  a gapless mode  iff  the set $\{\bm n \}$ does not contain the vector  $\bm n = \bm 0$. 

The proof is given below. Here we note that this theorem does not apply to the case in which $\Delta(x)$ is not real. As representative case of complex $\Delta(x)$ we consider the single plane wave (FF structure), explaining in which circumstances gapless modes arise. The theorem implies that any antipodal structure has a gapless mode, in particular the  ``strip" (corresponding to the structure depicted in  the right panel of Fig.~\ref{Fig:FF} for $\phi = 180^{\circ}$) and the cube have gapless modes.  On the other hand, the set of vectors which identify a triedral prism or a hexahedral prisms have a vector with $\bm n =0$, and the corresponding dispersion laws are gapped.

Proof:\\
A periodic solution of the system in \eqref{eigfg}  is given by the Bloch functions 
\be\label{bloch}
G({\bm x}) =  u({\bm x}) e^{i \bm {k \cdot x}} \qquad F({\bm x}) =  w({\bm x}) e^{i \bm {k \cdot x}} \,,
\ee
where $u({\bm x}) $ and $w({\bm x}) $ are periodic functions. These solutions  are periodic  if $\bm k$ is real. However, we can look for a generic solution of Eqs.~\eqref{eigfg} with $k$ complex. 
If  the  solution with a complex $k$ exists,  it means that  a gap is present in the excitation spectrum and the eigenfunctions decrease exponentially with $x$.
\red{In fact, suppose that the dispersion law has a  gap $\Delta\neq 0$, then it can be written as   $E=\sqrt{k^2 + \Delta^2}$. Since we are expanding around
$E=0$, then $k\approx\sqrt{-\Delta^2}+\textit{corrections}$, making $k$ complex. Clearly, the physical momentum is always real; the fact that we find a complex momentum only means that   the spectrum is gapped.}

Taking $E=0$ in  \eqref{eigfg}, the  two solutions of the system of equations are given by  $F_+=G_+$ or by $F_-=-G_-$, where
\be
G_{\pm}({\bm x}) = \exp\left( \pm \int_0^{\bm x_\parallel} \Delta({\bm x^\prime}) \frac{d x^\prime_{\parallel}}{v^2}   \right)\,,
\label{solutionE=0}
\ee
with $x_{\parallel} = {\bm x } \cdot {\bm v}$. 

Comparing this expression with Eq.~\eqref{bloch}, it is clear that these solutions corresponds to  Bloch functions with ${\rm Re} (k)=0$. If  $\Delta({\bm x})$ has a term with $\bm n=0$, it means that $\Delta({\bm x}) = A + f({\bm x})$, where $A$  is a constant and   $f({\bm x})$ is a periodic nonconstant function.  In this case Eq. \eqref{solutionE=0} has an exponential behavior of the type $G(\vert \bm x\vert )  \propto \exp{(\pm A\vert \bm x\vert)}$ and therefore the spectrum is gapped. On the other hand, if in the expansion of $\Delta({\bm x})$ no term with $\bm n=0$ is present then the imaginary part of $k$ vanishes and the spectrum is gapless Q.E.D.

The fermion dispersion law for the gapless modes can be determined using degenerate perturbation theory 
for $\xi = p -\mu \ll q$, \textit{i.e.} close to the Fermi sphere. At the lowest order in $\xi/q$  one finds that
\be E({\bm v}, \xi)=
\frac{\xi}{\sqrt{ A_+({\bm v})A_-({\bm v})}}  = c({\bm v}) \xi\, , \label{eps1}
\ee 
where $c({\bm v})$ is the velocity of the excitations,   and  
\be A_\pm ({\bm v})=
\frac{1}{V_c} \int_{\mathrm{cell}} d \bm x \exp\left[{\pm 2 \int
\Delta({\bm x}^{\,\prime})\frac{d  x^\prime_{\parallel}}{v^2} }\right] \, , \label{DEFA}\ee where $V_c$ is the volume
of a unit cell of the lattice. The energy of the fermionic excitations depends linearly on the residual momentum
$\xi$, but the velocity of the excitations is not isotropic. 

\begin{figure}[t]
\psfrag{ct}[][]{$ \cos\vartheta $}
{\includegraphics[width=8cm]{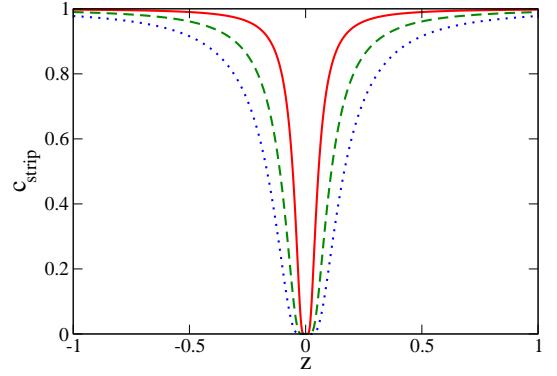}}
\caption{ (color online). Velocity of the fermionic quasiparticles as a function of $z=\cos \vartheta$ for $2\Delta/(q v)=0.1$ (solid red line), $2\Delta/(q v)=0.2$ (dashed green line), $2\Delta/(q v)=0.3$ (dotted blue line) \red{for
 the strip structure given in Eq.~\eqref{eq:strip}. }\label{dispersion-strip}}
\end{figure}

Let us now specialize Eq.~\eqref{eps1} to the case of the strip, 
\be\label{eq:strip}
\Delta({\bm x}) = 2 \Delta \cos( 2 {\bm q} \cdot {\bm x})\,, 
\ee
which corresponds to the condensate in Eq.\eqref{crystals} with $m=2$ and $\bm n^2= -\bm n^1 = \bm n$ and strictly speaking does not describe a crystal, but a condensate that is modulated in the ${\bm n}-$direction.    The coefficients in the dispersion law are given by
\begin{equation}
A_\pm^{(\mathrm{strip})}\equiv A^{(\mathrm{strip})} =I_0\left(\frac{2 \Delta}{\bm q \cdot \bm v } \right) \, ,\label{Astrip}
\end{equation}
where $I_0(z)$ is the modified Bessel function of the zeroth order.  Therefore, the velocity of the fermionic quasiparticles has the analytic expression 
\be
c_{\rm strip}(\bm v) = \frac{1}{I_0\left(\frac{2\Delta}{\bm q \cdot \bm v}\right)}\,,
\ee
which has the important property to vanish when $\bm v$ is orthogonal to $\bm n$. The reason is that in the direction orthogonal to $\bm n$ the gap is constant and  its effect is equivalent to a potential barrier.  
Taking  ${\bm n} = (0,0,1)$, the dispersion law is symmetric for rotations around the $z-$axis, for inversions with respect to the plane $z =0$ and depends only on the polar angle $\vartheta$,  between $\bm v$ and the  $z-$axis.
In Fig. \ref{dispersion-strip} we report a plot of the  velocity of fermionic quasiparticles  as a function of $\cos\vartheta $,  for three different values of the ratio $\Delta/(q v)$. 
In the ultrarelativistic case, $v=1$, the relevant case is $\Delta/q<1$ and we see that the dispersion law of fermionic quasiparticles is not much affected by the condensate for   $\cos{\vartheta} \gtrsim 0.2 \Delta/q$ and it is the same of relativistic fermions. On the other hand, for small values of $\cos\vartheta$ the fermionic velocity is exponentially suppressed and vanishes for $\vartheta=\pi/2$, meaning that fermionic quasiparticles cannot propagate in the $x-y$ plane, as discussed above.

The fact that the dispersion law is linear in $\xi$ for small values of the momentum does not assure that it is linear for any value of the momentum. Considering  $v_z \ll 1$, it is possible to solve the Eq.~\eqref{eigfg} for $\bm k$ in \eqref{bloch} along the $z-$direction, without restricting to low momenta \cite{LO};  the result is that 
\be
E^2 =  \frac{4 \Delta |v_z q|}{\pi} e^{-\frac{4 \Delta}{|v_z q|}}\left(1 -\cos \frac{\pi k}{q}\right)\,,
\ee
thus the dispersion law is linear in the residual  momentum,  only for $k/q \ll 1$.

\begin{figure*}[t]
\begin{center}
\subfigure{\includegraphics[width=8.cm]{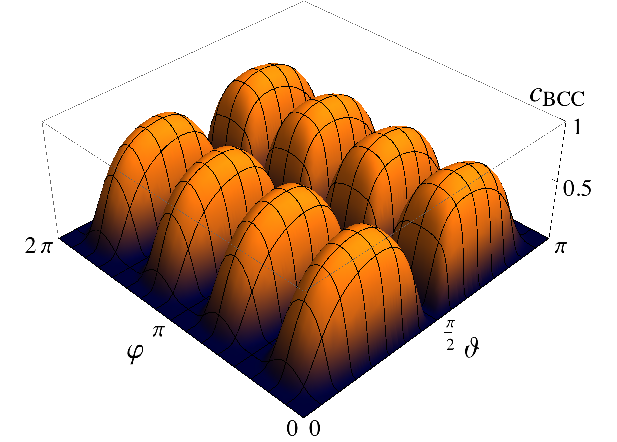}} \subfigure{\includegraphics[width=8.cm]{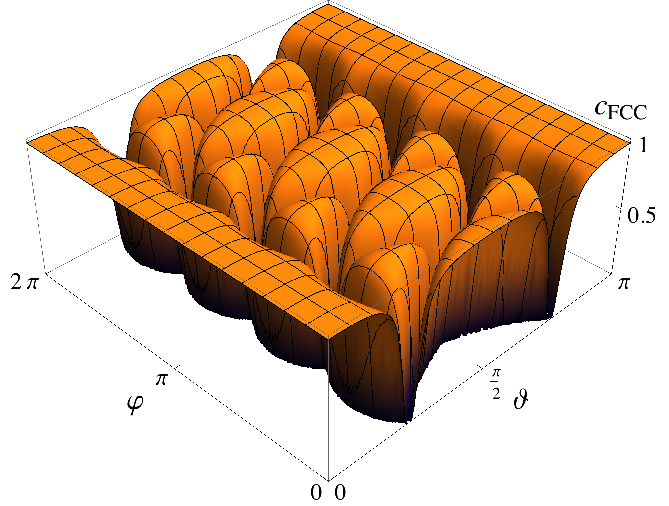}} 
\end{center}
\caption{(color online). Velocity of the fermionic quasiparticles in the bcc crystalline structure (left panel)  and in the fcc crystalline structure (right panel)  as a function of  the polar angles $\vartheta$ and $\varphi$, for $2\Delta/(q v)=0.3$. \label{fig:dispersion-bcc}}
\end{figure*}

For the octahedron, whose six wave vectors point into the direction of a
bcc structure, the corresponding gap parameter can be written as
\be
\Delta(\bm x) = 2 \Delta [ \cos (2 q x) +\cos (2 q y) +\cos (2 q z) ]\,.
\ee
It is easy to show that in this case the integral in Eq.~\eqref{DEFA} factorizes, and the dispersion law is gapless with velocity
\be
c_{\rm bcc}(\bm v) = \frac{1}{I_0\left(\frac{2\Delta}{q v_x }\right) I_0\left(\frac{2\Delta}{q v_y }\right) I_0\left(\frac{2\Delta}{q v_z }\right)  }\,. \label{eq:dispersion-bcc}
\ee
The corresponding plot is reported in Fig.\ref{fig:dispersion-bcc}, left panel, where the unit vector $\bm v$  has been expressed by the polar angles $\vartheta$ and $\varphi$. The plot has been obtained for $2\Delta/(q v)=0.3$ and considering  $\vartheta \in [0,\pi]$ and $\varphi \in [0,2\pi]$.  Note that according to Eq. \eqref{eq:dispersion-bcc},
the  velocity of the fermionic quasiparticles vanishes along the planes  $v_x=0$,  $v_y=0$ and $v_z=0$.

Unfortunately, for more complicated crystalline structures it s not possible to have an analytic expression of the fermionic velocity. One notable example is the  fcc structure which is defined by the condensate
\be
\Delta(\bm x) = 2\Delta \sum_{m=1}^4 \cos (2 q \, \bm n^m \cdot {\bm x}) \,,
\ee
where $\bm n^1 = \sqrt\frac{1}3(1,1,1)$,  $\bm n^2 = \sqrt\frac{1}3(1,1,-1)$, $\bm n^3 = \sqrt\frac{1}3(1,-1,1)$
and $\bm n^4 = \sqrt\frac{1}3(-1,1,1)$. Upon plugging this expression in Eq.\eqref{DEFA} we obtain that
\begin{equation}
A^{(\mathrm{fcc})}_\pm=\left(\frac{q}{\pi}\right)^3 \int_{cell}dV\, \exp\left\{\pm\frac{2\Delta}{q v}B\right\} \, ,\label{AformulaCUBE}
\end{equation}where the integration is over the elementary cell of
volume $(\pi/q)^3$ and
\begin{widetext}
\bea B = \sqrt{3} v \left(
\frac{\sin\,2q(x+y+z)}{v_x+v_y+v_z}+\frac{\sin\,2q(x+y-z)}{v_x+v_y-v_z}+ \frac{\sin\,2q(x-y+z)}{v_x-v_y+v_z}
+\frac{\sin\,2q(-x+y+z)}{-v_x+v_y+v_z} \right) \,. \label{defF} \eea   \end{widetext}
It is then easy to show that $A^{(\mathrm{fcc})}_+=A^{(\mathrm{fcc})}_- \equiv A^{(\mathrm{fcc})}$ and
expressing the components of the unit vector $\bm v$ in $B$ as  functions of the polar angles $\vartheta$ and $\varphi$, and upon substituting  in  $A^{(\mathrm{fcc})}$, one has that the quasiparticle velocity 
\be c_{\rm fcc}(\vartheta,\varphi) = \frac{1}{A^{(\mathrm{fcc})}(\vartheta,\varphi)}
\,,\ee 

has the behavior reported in Fig.\ref{fig:dispersion-bcc}, right panel. The plot of  $c_{\rm fcc}(\vartheta,\varphi)$ has been obtained  for $2 \Delta/(q v) = 0.3$ and considering  $\vartheta \in [0,\pi]$ and $\varphi \in [0,2 \pi]$. The velocity of the fermionic quasiparticles is equal to $1$ almost everywhere with the exclusion of a restricted region which corresponds to the zeros of $B(\vartheta,\varphi)$,  given by the solutions of   $\bm v \cdot \bm n^m =0$. 

In conclusion, we have shown that for various crystalline structures the  fermionic spectrum is gapless. For $k/q \ll 1 $ the dispersion law is linear in momentum and  the fermionic velocity  vanishes along the planes orthogonal to the direction of the vertices of the reciprocal lattice. This result remains valid also for massive quarks, since the effect of the quark mass can be accounted for by reducing $v$ \cite{Casalbuoni:2002st}.

Regarding  the FF condensate we take the direction of the
 Cooper pair total momentum $2\bm q$ along the $z-$axis. In this case, the quasiparticle spectrum has the analytical expression
 \be E_\pm = qv_z \pm
\sqrt{\xi^2+ \Delta^2} \, ,\label{disp-1wave}\ee where the quasiparticle energies are computed from the corresponding Fermi
energies $\mu_{u,d}$. Eq.(\ref{disp-1wave}) is the dispersion law of quasiparticle ($E_\pm\ge 0$) or hole states ($E_\pm<0$).
As for the case of the strip and of the fcc, the dispersion law depends on the considered direction in coordinate space. However, contrary to what happens for real crystalline structures, in the FF phase  there are directions along which the dispersion laws are gapped and directions along which  the dispersion laws are gapless. 

The zeros of the dispersion laws are located, for a given value of $\cos\vartheta$, at
\begin{equation}
\xi_0 =\pm \sqrt{(q v \cos\vartheta)^2-\Delta^2}~,
\end{equation}
and   therefore  for  $\Delta/q v >1 $ both gapless modes disappear.

\subsubsection{Specific heat of the Fermi quasiparticles}
The contribution of the Fermi quasiparticles to the specific heat
per unit volume is given by 
\begin{equation} c_V =
2 \sum_j\!\int\!\frac{d^3\!p}{(2\pi)^3}~E_j \frac{\partial
n(E_j,T)}{\partial T}~,\label{eq:CvDef} 
\end{equation}
$n(E_j,T)$ is the Fermi distribution function and the sum is over all the fermionic modes. Considering  the low temperature range, $T\ll\Delta$, which is  relevant for astrophysical applications, and using Eq.~\eqref{eps1} one obtains for a generic two-flavor  crystalline structure
 \be c_V =\, \frac{4 \mu^2 T }{3}\, \int
\frac{d \Omega}{4 \pi}\frac{1}{c(\bm v)} + \frac{2\mu^2}3\,T\ ,\label{CA}\ee 
where the first and the second addenda correspond to the contribution of the paired and the unpaired quarks respectively. Note that both contributions are linear in $T$, because all degrees of freedom have a linear dispersion law at small momenta.
This expression can be evaluated in closed form for the strip by \textcite{Casalbuoni:2003sa} obtaining \be c_V^{(\mathrm{strip})}=\frac{4 \mu^2 T}{3}~ _1F_2\left(-{1}/{2};\,{1}/{2},\,1;\,\left({\Delta}/q v
\right)^2\right) + \frac{2\mu^2}3\,T\,,\nonumber\\ \label{Cstrip}
\ee
where $_1 F_2$ denotes the generalized hypergeometric function
\cite{Gradshteyn}. Differently from the analysis by \textcite{LO}, here $v$
is not small and we can take $\Delta/q v\to 0$ near the second
order phase transition. Since for small $\Delta/q v$ one has $_1
F_2(-1/2;\,1/2,\,1; ({\Delta}/q v
)^2)\simeq 1-({\Delta}/q v)^2$, it is easily seen that the
normal Fermi liquid result is obtained for $\Delta=0$. On the other
hand, for nonvanishing  $\Delta$, the specific heat turns out to be
smaller.

In the case of the FF state, the dispersion law
of the quasiparticles is given by (\ref{disp-1wave}) and  using Eq.~(\ref{eq:CvDef})  one  obtains that in the small temperature limit ($T \ll \Delta$) and for $\Delta < q $
 \be
c_V^{(\mathrm{FF})} \simeq \frac{4 \mu^2 T }3 \sqrt{1- \frac{\Delta^2}{(q v)^2}} + \frac{2\mu^2}3\,T\,. \label{C1w}\ee  The paired quarks
contribution depends linearly on temperature because the quasiparticle dispersion law (\ref{disp-1wave}) gives rise to gapless modes when $\Delta/q v<1$. There is also an additional
contribution to the specific heat  due to gapped modes, but this contribution is exponentially suppressed with the temperature and has not been reported in Eq.~\eqref{C1w}.

\subsection{Smearing procedure and HDET approximation for two-flavor QCD}\label{sec:2flavor-smearing}
In Sec.~\ref{sec:GL-2flavor} we have discussed the  GL 
expansion of the free energy for various CCSC phases. Since the GL expansion has several limitations and is 
reliable when  $\Delta/\delta\mu$ is small, it is useful to derive a different approximation scheme. In Sec.~\ref{sec:fermi-quasiparticles}  we have presented an approximation allowing us to deal with the dispersion law of fermionic quasiparticles close to the gapless momentum. However, this method cannot be used for evaluating the free energy of the system or the low-energy Lagrangian. For these purposes a different approximation named the \textit{smearing procedure} can be used. This approximation, developed  by \textcite{Casalbuoni:2004wm} within the HDET framework, is valid when $\Delta/\delta\mu$ is large, and is thus complementary to the GL expansion.

\subsubsection{Gap equations \label{sec:gap}}
\red{The mean field Lagrangian term describing  condensation in any CCSC phase can be written as follows
\begin{equation}
{\cal L}_{\Delta}=-\frac 1
2(\psi_{i}^\alpha C \gamma_5 \psi_{j}^\beta\,\Delta_{ij}^{\alpha\beta}(\bm x)+\,{\rm h.c.})\,-\frac{\Delta^*({\bm
x})\Delta({\bm
 x})}G\,,
\label{8}
\end{equation}
where $\Delta_{ij}^{\alpha\beta}(\bm x)$ is the pertinent gap parameter and hereafter we shall indicate the NJL coupling constant in the diquark channel with $G$.   
We define the smearing procedure considering in the first place  the FF phase, with $\Delta_{ij}^{\alpha\beta}(\bm x)$ given in Eq.~\eqref{FF-2flavor}. Although this case can be solved exactly, it is useful to consider it  to fix the notation and to introduce some definitions to be used later on. We shall consider Cooper pairing of the massless  up
and down quarks,  chemical potential $\mu_u$, $\mu_d$, and we define  $\mu=(\mu_u+\mu_d)/2$ and $\delta\mu=|\mu_u-\mu_d|/2\ll\mu$.}

The calculation can be simplified using the HDET approximation~\cite{Nardulli:2002ma, Hong:1998tn,Hong:1999ru,Beane:2000ms,Casalbuoni:2000na, Schafer:2003jn}.  
We Fourier decompose
the fermionic fields as follows:
\begin{equation} \psi_i^{\alpha}(x)= \int\frac{d \Omega}{4\pi}e^{-i\mu{\bm v}\cdot{\bm x}}\,\left(\psi^\alpha_{{i},\bm
v}(x)+\psi^{\alpha-}_{{i},\bm v}(x)\right)\label{decomp-2flavor} \, ,
\end{equation}
where ${\bm v}$ is a unit three-dimensional vector whose direction is integrated
over; 
$\psi^\alpha_{{i},\bm v}(x)$ (resp. $\psi^{\alpha-}_{{i},\bm v}(x)$)
are positive (resp. negative) energy projections of the fermionic
fields with flavor $i$ and color $\alpha$ indices, as defined by
\textcite{Casalbuoni:2001gt,Nardulli:2002ma}. Note that these fields
 are written in a mixed notation, meaning that they depend both on the space coordinates and on the unit vector $\bm v$, which points to a particular direction in momentum space. Since only quasiparticles and quasiholes live close to the Fermi surface, they are the only relevant degrees of freedom. Antiparticles decouple and their contribution is suppressed by powers of $1/\mu$.  The  three-dimensional momentum
of a fermion is written as $(\mu+\xi){\bm v}$, with  $\xi$  representing the ``residual'' momentum
component.  The
integration over momentum space is separated into an angular
integration over ${\bm v}$ and a radial integration over $-\delta \le
\xi \le \delta$. The cutoff $\delta$ must be taken  smaller
than $\mu$ but   much larger than the gap in the homogeneous phase and $\delta\mu$, see Fig.\ref{hierarchy}. 
 At the leading order in $1/\mu$ the \red{free} Lagrangian can be written as
\begin{equation} 
{\cal L}= \int\frac{d{\Omega}}{4\pi}\, \left[
\psi^{\alpha \dagger}_{{i},\bm
v}\left(i V \cdot \de
+ \delta\mu_{i}\right)
\psi^{\alpha}_{{i},\bm
v} \right]~, \label{L110} 
\end{equation}
where  $V^\nu=(1,{\bm v})$, and we also define for later use $\tilde{
V}^\nu=(1,-{\bm v})$. 
In the HDET approximation the Lagrangian term in Eq.~\eqref{8} for the FF phase turns into
\begin{equation} 
{\cal L}_{\Delta}= -\frac{1}{2}\left(\int\frac{d\Omega}{4\pi}\,\Delta_{ij}^{\alpha \beta}
\psi^{\alpha T}_{{i},
-\bm v} \,C\,\gamma_5\, \psi^\beta_{j, {\bm v}}~+ {\rm h.c.}\right)\,-\frac{\Delta^2}{G}\,,
\label{eq:LD-2FL}
\end{equation}
and the  zero-temperature gap equation can be written as
 \be 1=\frac{G\rho}2\int\frac{d \Omega
}{4\pi}\int_0^{\delta}\frac{d\xi}{\sqrt{\xi^2+\Delta^2}}\, \left(1-\theta(-E_u)-\theta(-E_d)\right)\,,\nonumber\\ \label{eq:1}\ee where 
$ \rho=\frac{4\mu^2}{\pi^2} $  is the density of states  in two-flavor QCD. The quasiparticle dispersion laws have been obtained in Eq.~\eqref{disp-1wave}, but energies are now  measured from the common energy level $\mu$, and therefore
\be
E_{u,d}=\pm\delta\mu\mp {\bm q} \cdot {\bm v}+\sqrt{\xi^2+\Delta^2} \,.
\label{dispersionFF}\ee 
In Eq.~\eqref{eq:1} the contribution of hole excitations   is taken into account simply multiplying the contribution of quasiparticles times two. We observe that in general
\be
1-\theta(-x)-\theta(-y)=\theta(x)\theta(y) - \theta(-x)\theta(-y) \,,
\ee
and since $E_u$ and $E_d$ cannot  be  simultaneously negative, then
\be
1-\theta(-E_u)-\theta(-E_d)=\theta(E_u)\theta(E_d)\,,\ee 
so the integration in Eq.~\eqref{eq:1} is over a restricted region named the pairing region (PR), defined by
\be\label{PRFF} \mathrm{PR} = \left\{ (\xi,{\bm v \cdot \bm n}) \,|\, E_u>0 \,\mathrm{and}\, E_d>0; \xi \le \delta \right\} \, .\ee  More explicitly, the pairing region  is defined
by the condition  \be\begin{split}
\text{max}\left\{-1,\, z_q-\frac{\sqrt{\xi^2+\Delta^2}}q\right\} <\,{\bm v\cdot\bm n}& \\ <\text{min}\left\{1,\,z_q+\frac{\sqrt{\xi^2+\Delta^2}}q\right\}& \label{eq:2}\,,\end{split}\ee 
with 
\be z_q=\frac{\delta\mu}q\,.\label{eq:zetaq}\ee
From the above definition it follows that for small values of $\Delta$ \red{and $\xi$} the pairing region  is centered at  $ \vartheta_q = \arccos z_q$, has an angular width of order  $\Delta/q $ and a thickness of order $\Delta$, see Fig.~\ref{Fig:FF}. 

Thus, Eq.~\eqref{eq:1} can be written in a different way: 
\be \begin{split}\Delta&=\frac{G\rho}{2}\int\int_\mathrm{PR} \frac{d{\Omega}}{4\pi}  d \xi \frac{\Delta}{\sqrt{\xi^2+\Delta^2}}  \\ &= \frac{G\rho}{2}\int
 \frac{d\,{\Omega}}{4\pi}
 \int_0^\delta\,d\xi\,\frac{\Delta_{\rm eff}}{\sqrt{\xi^2+\Delta^2_{\rm eff}}}\end{split}
\label{GAPeff} \, , \ee 
where $\Delta_{\rm eff}\equiv \Delta_{\rm eff}({\bm v\cdot \bm n}, \xi)$ is defined as\bea
\Delta_{\rm eff}\,=\Delta\,\theta(E_u)\theta(E_d)\,=\, \left\{ 
\begin{array}{cc}\Delta & \text{for  }
(\xi,{\bm v \cdot \bm n}) \in \mathrm{PR}\cr&\cr 0 &
\text{elsewhere.} \\ 
\end{array} \right.
\label{GAP8.1} \eea

The above procedure defines the smearing procedure for the FF phase; it can be extended to  the case of $P$ plane waves, Eq.~\eqref{crystals}, generalizing the results of the previous equations, assuming that  in the mean field Lagrangian one can substitute $\Delta$ with $\Delta_{\rm E}\left({\bm v},p_0\right)$, where \be\Delta_{E}({\bm
v},p_0)=\sum_{m=1}^P\Delta_{\rm eff}\left({\bm v\cdot\bm
n^m},p_0\right)\label{eq13}\, ,\ee
meaning that $\Delta_{E} = n \Delta$, where $n= 1, \dots, P$. We can thus generalize the pairing region to
\be {\cal P}_n=\{({\bm v},\xi)\,|\,\Delta_E({\bm
v},\epsilon)=n\Delta\}\,. \ee 
Note that   in this equation we have
made explicit the dependence on the energy $p_0$ instead of  that on the residual momentum, because 
the pole position is in general in 
 \be \epsilon_n=\sqrt{\xi^2+n^2 \Delta^2}\ .\ee 

Correspondingly, the gap equation is \be P
\Delta=i\frac{G\rho}{2}\int\frac{d{\Omega}}{4\pi}\int
\frac{d p_0d\xi}{2\pi} \, \frac{\Delta_{E}({\bm
v},p_0)}{p_0^2-\xi^2-\Delta_{E}^2({\bm
v},p_0)}\label{29}, \ee which generalizes Eq. \eqref{GAPeff}.
The origin of the factor $P$ on the l.h.s. of this equation is as
follows. The Lagrangian contains the term \be\frac{\Delta^*({\bm
x})\Delta({\bm
 x})}G\,,
\ee which, when averaged over the cell, gets non vanishing contribution only from the diagonal terms in the double sum over the plane waves and each plane wave gives a separate contribution.

 The energy integration is performed by the residue theorem and
 the phase space is divided into different regions according to
 the pole positions. Therefore the gap equation turns out to be given by \be
 P
\Delta\ln\frac{2\delta}{\Delta_{\rm 2SC}} =\sum_{n=1}^P\int\int_{{\cal P}_n}\frac{d\Omega}{4\pi} d\xi \,
\frac{n \Delta}{\sqrt{\xi^2+n^2\Delta^2}}\label{290} \,, \ee where 
 we have used  the NJL gap equation
  \be
  \Delta_{\rm 2SC} = 2 \delta \exp\left\{-\frac{\pi^2}{2 G \mu^2 }\right\} \,,
\label{eq:2SC-gap}
 \ee 
for relating  the NJL coupling $G$ to the 2SC gap parameter and the momentum cutoff.  The first term in the
sum, corresponding to the region ${\cal P}_1$, has $P$ equal contributions with a dispersion law equal to the FF  case. This can be
interpreted as a contribution from $P$ non interacting plane waves. In the other regions the different plane waves have an overlap. Since the
definition of the regions ${\cal P}_n$ depends on the value of $\Delta$, their determination is part of the problem of solving the gap equation.

Said in a different way, in the smearing procedure the dispersion relation of the quasiparticles has several branches corresponding to the values $n\Delta$, $n=1,\cdots,P$. Therefore, the following  interpretation of the gap equation \eqref{290} can be given. Each
term in the sum corresponds to one branch of the dispersion law, \textit{i.e.} to the propagation of a gapped quasiparticle with gap $n\Delta$, which is defined in the  region  ${\cal P}_n$. However, the regions ${\cal P}_n$ do not represent a partition of the phase space since it is possible to have at the same point quasiparticles with different gaps.

\subsubsection{Numerical results: Free energy computation\label{sub}}
 \begin{table}[b]\begin{center}
 \begin{tabular}{|c|c|c|c|}
   \hline
  $ P $&$ z_q $& $\frac \Delta{ \Delta_{\rm 2SC}}
  $&$\frac{\Omega-\Omega_n}{\rho\Delta_{\rm 2SC}^2}$
\\   \hline
 1& 0.78 & 0.24 & $-3.6\times 10^{-3}$ \\
 2& 1.0 & 0.75 & -0.16 \\
  6& 0.9 & 0.28 & -0.22 \\
  8&0.9& 0.21& -0.18\\ \hline
 \end{tabular}
\end{center} 
\caption{\label{tabledeltamu1} The values of $z_q=\delta\mu /q$, of the 
  gap,  and of the free energy at  $\delta\mu = \delta\mu_1=\Delta_{\rm 2SC}/\sqrt 2$
  for different crystalline structures. See the text for more details. Adapted from \textcite{Casalbuoni:2004wm}.}\end{table}
 
 In  Table \ref{tabledeltamu1} we report the results obtained in the CC limit ($\delta\mu=\delta\mu_1=\Delta_{\rm 2SC}/\sqrt{2}$) for four crystalline structures, respectively the FF ($P=1$), the strip ($P=2$), the octahedron (bcc) ($P=6$), and the fcc ($P=8$). The  table shows that, among the four considered structures, the favored one at  $\delta\mu=\delta\mu_1$ is the  octahedron, which however does not have the largest gap $\Delta$.  The gap parameter determines the extension of the pairing regions, see  Eq.\eqref{eq:2}, but a free energy gain may result from having many small nonoverlapping pairing regions, as well.  Indeed, the strip
has the largest gap parameter, but not the largest free energy, presumably because the pairing occurs only in four large ribbons (two for each vectors $\bm q^m$, see the right panel of  Fig.\ref{Fig:FF}). Indeed, as we have shown in the previous section, the width and the thickness of the ribbons are proportional to $\Delta$. For example, in the octahedron  pairing occurs in smaller ribbons, but there are  twelve pairing regions, which give a large contribution to the free energy.

\begin{figure}[t]
\includegraphics[width=8cm]{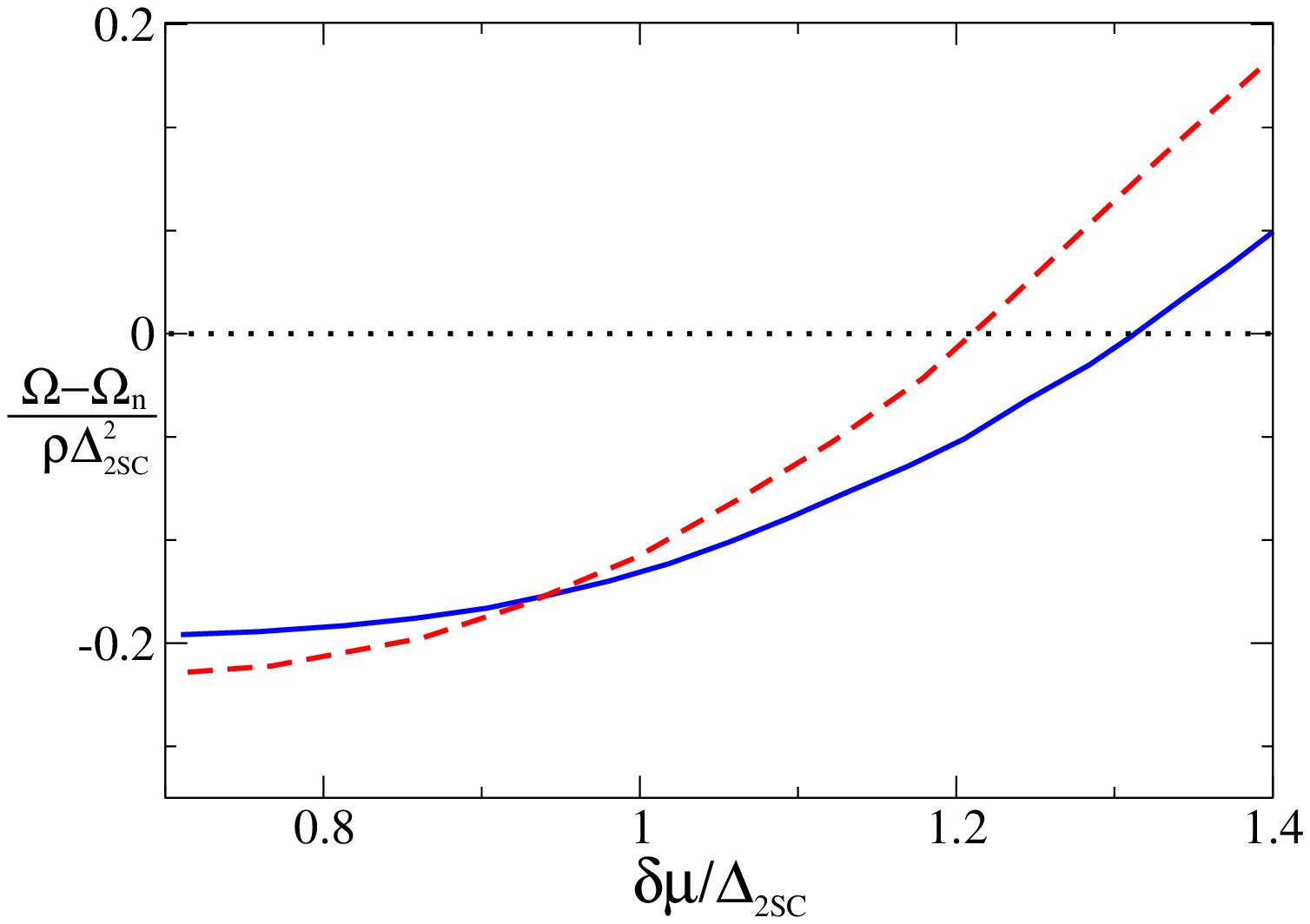}
\caption{(color online). Free energy  as a function of
$\delta\mu/\Delta_{\rm 2SC}$, for the octahedron (dashed red line) and  the  fcc (solid blue line). Free energies are measured respect to the normal phase. The octahedron is the favored structure up to $\delta\mu \simeq  0.95 \Delta_{\rm 2SC}$. In the range 
 $0.95 \Delta_{\rm 2SC}\lesssim \delta\mu \lesssim 1.32 \Delta_{\rm 2SC}$ the fcc is  favored. For larger values of  $\delta\mu$ the normal phase is energetically favored. In this plot the values of $\delta\mu$,  $z_q$ and  $\Delta$ are those that minimize the free energy. Adapted from \textcite{Casalbuoni:2004wm}.\label{omegaplot} }
\end{figure}

For the case $\delta\mu\neq\delta\mu_1$, we report in  Fig.~\ref{omegaplot} the plot of the free energy of the octahedron (dashed line) and of the fcc (full line)  structures as a function of $\delta\mu/\Delta_{\rm 2SC}$.
The octahedron is the favored structure up to $\delta\mu\simeq 0.95\, \Delta_{\rm 2SC}$; the   fcc structure is favored in the range 
$0.95 \Delta_{\rm 2SC}\lesssim \delta\mu \lesssim 1.32 \Delta_{\rm 2SC}$; for larger values of $\delta\mu$  the
normal phase becomes favored. The fcc gap parameter  is smaller than  the octahedron gap parameter for any value of $\delta\mu$, but  the fcc benefits of more pairing regions than the octahedron.

In Table \ref{tabledeltamu2} we report various  numerical results for each crystalline structure at the transition point from the CCSC phase to the normal phase: the value of $\delta\mu_2$, the computed order of the phase transition between the crystalline phase and the
 normal phase, the value of $z_q$,   see Eq.\eqref{eq:zetaq},  and of the discontinuity in $\Delta/\Delta_{\rm 2SC}$ at $\delta\mu=\delta\mu_2-0^+$. 
\begin{table}[h!]
 \begin{tabular}{|c|c|c|c|c|}\hline
$P $ &  $\delta\mu_2/ \Delta_{\rm 2SC}$  &  Order & $   z_q$ & $ \Delta/ \Delta_{\rm 2SC}$
\\
\hline
 1& 0.754 &  II&0.83 & 0  \\
 2& 0.83 &I& 1.0 & 0.81  \\
  6& 1.22 & I&0.95 & 0.43 \\
  8& 1.32 &I&0.90& 0.35  \\ \hline
\end{tabular} 
\caption{\label{tabledeltamu2} Values of $\delta\mu_2$, of the order of the phase transition between the CCSC phase and the normal phase, of
$z_q=\delta\mu /q$ and of the discontinuity of $\Delta/\Delta_{\rm 2SC}$ at the phase transition point for different crystalline structures. See the text for more details. Adapted from \textcite{Casalbuoni:2004wm}.}\end{table}

In the smearing  approximation  both the order of the transition and  the point where the transition occurs are different from  those obtained within the GL approximation. However, the difference in $\delta\mu_2$ is $\sim 10\%$ and in $z_q$ is $\lesssim 17\% $ (in GL  for any structure  $z_q \approx 0.83$). On the other hand, in agreement with the GL results, the structure with $6$ plane waves is energetically favored over the structure with $2$ plane waves or with $1$ pane wave. Therefore, increasing the pairing region leads to an increase of the free energy gain making the configuration more stable. An interesting result of the  smearing procedure is that with increasing $\delta\mu$ the configuration with $6$ plane waves  is superseded by the structure with $8$ plane waves, which has a lower $\Delta$   but a larger number of pairing regions.   This has to be contrasted with the result of the GL approximation, see Sec.\ref{sec:GL-2flavor}, where the octahedron is always disfavored with respect to the fcc (but one should consider that the GL  free energy of the fcc is not bounded from below).

\subsection{Effective Lagrangian of phonons and contribution to the specific heat}\label{phonons}
The low-energy spectrum of a periodically modulated condensate, besides Fermi quasiparticles consists  also  of massless NGBs which originate from the spontaneous breaking of translational invariance. For two-flavor quark matter in any crystalline phase, these are the  only NGBs, because no global symmetry of the system is spontaneously broken. Since the modulation of the condensate is associated with a  crystalline structure, these NGBs describe the vibrations  of the crystal and are for this reason called phonons. \red{These phonons are not the standard  pressure oscillations of a fluid, rather they are akin to second sound in standard superfluids, because  they are related to  chemical potential oscillations, see  for example \textcite{Anglani:2011cw}. } More in detail, phonons  are small position and time dependent displacements of the condensate: in the presence of  phonons, then,
\begin{equation}
\Delta(\bm x)\rightarrow
\Delta^u(x)=\Delta(\bm x-\vu(x))~,\label{phon2sc}
\end{equation}
and one may define a set of three dimension-one scalar field   $\phi^{(i)}$, by \be \frac{\phi^{(i)}}{f_\phi} = 2q\,\bm u_{i} \label{def-phi}\,,\ee with $i=1,2,3$, where $f_\phi$ is the corresponding decay constant.  In the following we shall refer to both $\bm u$ and $\bm \phi$ as the phonons.
For any crystalline structure one can deduce the general expression of  the phonon Lagrangian  from the symmetries of the  system, but the various coefficients have to be evaluated by a microscopic theory. The calculation of the low-energy coefficients  can be  done using a NJL-like model with a smearing procedure,  as in the work by \textcite{Casalbuoni:2002my, Casalbuoni:2001gt}, or by a GL expansion, as done by \textcite{Mannarelli:2007bs}. 
These  coefficients are related to the elastic properties of the crystalline structure. For the two-flavor CCSC phases  the elastic properties do not seem to be  of  particularly interest, because not all quarks condense and thus the system should not behave as a crystal under an external stress. However, as we shall discuss in  Sec.~\ref{sec:shear},  the shear modulus of various three-flavor  crystalline structures is actually associated with a form of rigidity of quark matter. 
 
The GL expansion by \textcite{Mannarelli:2007bs} provides an expression of the leading order (LO) Lagrangian density in the derivative expansion valid for a generic set $\{ \bm n \}$ of unit vectors
\begin{widetext}\be
{\cal{L}}^{\Delta^2}[{\bm u}]=\ha \frac{2\mu^2|\Delta|^2}{\pi^2(1-z_q^2)}
 \sum_{m}\left[
 \partial_0(\bm n^m \cdot \vu)\partial_0(\bm n^m \cdot\vu) -
 (\bm n^m\cdot{\bm \partial})(\bm n^m\cdot\vu) (\bm n^m\cdot{\bm\partial})(\bm n^m\cdot\vu)
 \right]\label{eq:lagrangian_GL_ph}\;,
\ee\end{widetext}
with $z_q $ given in Eq.~\eqref{eq:zetaq}. This ${\cal O}(\Delta^2)$ Lagrangian  includes  the displacement fields at  the second order and therefore the interaction terms are missing. We shall see in Sec.~\ref{sec:shear1} how this expression can be derived from a NJL-like model in the more complicated case of the three-flavor CCSC phase.

In the FF phase there is one single phonon field, $\phi$, and one privileged direction corresponding to $\bm q$. Given the space symmetries of the system,   the LO Lagrangian density in the momentum expansion is given by
\be
{\cal L}=\frac{C}{2 f_\phi^2}\left(
{\dot\phi}^2-v_\parallel^2(\nabla_\parallel\phi)^2-v_\perp^2|\nabla_\perp\phi|^2
\right)\label{lag-FF} \, ,\ee where  
$\nabla_\parallel={\bm n}\cdot\bm\nabla$,
${\bm\nabla}_\perp={\bm\nabla}-\bm n\nabla_\parallel$,  ${\bm
n}$ is the unit vector parallel to  $\bm q$.   The breaking of the rotational symmetry in  the underlying microscopic theory implies that the  velocity of propagation in the direction parallel to $\bm n$,  $v_\parallel$, can be different from the velocity of propagation in the direction orthogonal to $\bm n$, $v_\perp$. The dispersion
law, relating the phonon quasimomentum $\bm k$ and energy
$\omega$,  is given by \be \omega ({\bm k})=\sqrt{
v_\perp^2(k_x^2+k_y^2)+v_\parallel^2k_z^2 }\,,\ee 
and the contribution of phonons to the specific heat at small temperatures turns out to be\be
 c_V^{(\mathrm{FF})}=\frac{ 8\pi^2
}{15v_\perp^2
  v_\parallel}\,T^3\,~~~~~~{\rm (phonons)}\ .\label{cv4}\ee

\textcite{Casalbuoni:2003sa}, using the smearing procedure,  have obtained that  $C=\mu^2 k_r^2$, with  $k_r \propto \delta\mu^2/q^2$  a coefficient arising within the smearing procedure. The obtained values of the velocities are
\be v_\parallel^2=
\cos^2\vartheta_q\simeq 0.7\,,~\ v_\perp^2=\frac 1 2\,\sin^2\vartheta_q\simeq 0.15
\, , \label{eq:velocities}\ee with $\vartheta_q$ defined in Eq.~\eqref{eq:phi0}.   

Since the strip has the same space symmetries of the FF phase and one phonon field, the Lagrangian density has the same formal expression reported in Eq.~\eqref{lag-FF},   but the values of the longitudinal and transverse velocities can be different.  Thus, for the strip the expression of the specific heat is formally the same reported above. 
From Eq.~\eqref{eq:lagrangian_GL_ph}  we can see that the ${\cal O}(\Delta^2)$ GL expansion gives  $v_\parallel=1$ and $v_\perp=0$ for both the FF and the strip. Unfortunately, higher order  corrections in $\Delta$  of the phonon velocities have not been computed within the GL expansion, hindering the comparison with  Eq.~\eqref{eq:velocities}.
Note, however, that both methods  give a  FF  low-energy Lagrangian  with the same $O(2)$  symmetry of the microscopic system. 

\red{The low-energy oscillations of more complicated crystals  can in principle be described in a similar way. As an example, for the fcc structure the oscillations are described by  three  phonon fields,  $\phi^{(i)}$, and the LO Lagrangian density \red{compatible with the fcc symmetry} is given by \begin{widetext}
 \be {\cal L}=\frac{C}{
2f_\phi^2}\left(\sum_{i=1,2,3}({\dot\phi}^{(i)})^2-a  \sum_{i=1,2,3}|{\bm\nabla}\phi^{(i)}|^2- b  \sum_{i=1,2,3}\de_i\phi^{(i)}\de_i\phi^{(i)}- 2 c\sum_{i,j=1,2,3}\de_i\phi^{(i)}\de_j\phi^{(j)} \right)~,\label{fcc-phonon-lag}\ee \end{widetext}
where $a,b$ and $c$ are three coefficients to be determined by the microscopic theory\footnote{\red{In principle the additional term $\partial_i \phi^{(j)}\partial_j \phi^{(i)} $ should be included, see for example \textcite{Leutwyler:1996er}, but here it is assumed that it can be recast in the 
term proportional to $c$ by integration by parts.}}.  The smearing procedure gives
\begin{equation}
a=1/12\simeq 0.08~,~~~~~b=0~,~~~~~ c=(3\cos^2\vartheta_q-1)/12\simeq 0.09\,.
\label{1bis}
\end{equation}
Since for vanishing $b$ the Lagrangian in Eq.~\eqref{fcc-phonon-lag} is rotationally invariant,  it has a larger symmetry  than the underlying microscopic theory, which has the fcc symmetry. A possible explanation is that being phonons  long wavelength fluctuations of the crystal, the low-energy parameters are given by an average over the cubic  structure and thus are not sensitive to the local modulation of the condensate. On the other hand,  substituting the appropriate unit vectors in  the   ${\cal O}(\Delta^2)$ GL in Eq.~\eqref{eq:lagrangian_GL_ph} one obtains   $a=c=-b/2=1/3$. Thus, according to the GL analysis the low-energy theory and the microscopic theory have the same fcc symmetry. This might be an artifact of the GL expansion, indeed  in the GL analysis the  low-energy parameters are obtained first expanding the action in $\Delta$ and then in $\phi$, but  it  is not obvious that the two expansions commute. Nevertheless, note that  the  smearing procedure gives $a\simeq c$, compatible with the GL result $a=c$.}  

The specific heat contribution of the phonons in the fcc structure has been evaluated numerically by \textcite{Casalbuoni:2003sa}, and   the result is 
 \be
 c_V^{(\mathrm{fcc})}\approx 88~\pi^2~T^3\,~~~~~~~~~~~~{\rm (phonons)}\,.\label{cv4315}\ee
\subsection{Chromomagnetic stability of the two-flavor crystalline phase}\label{sec:2stability}
As discussed in Section \ref{sec:meissner_g2SC}, the 2SC phase is  chromomagnetically unstable for $\delta\mu > \delta\mu_1$. This instability could be interpreted as the tendency of the system to generate a net momentum of the quark pair, as  shown by \textcite{Giannakis:2004pf}.  Therefore, the chromomagnetic instability can be interpreted as as  a tendency to develop quark currents,  which in turn is equivalent  to the FF phase, where diquark carry momentum $2 \bm q$.
 
In this section, we review the results by \textcite{Giannakis:2004pf},
which relate the Meissner mass of the  $\bar 8$ mode  of the 2SC to
the momentum susceptibility and  also the computation of the
Meissner tensor in the crystalline phases \cite{Giannakis:2004pf, Giannakis:2005vw, Giannakis:2005sa, Reddy:2004my}. 

\subsubsection{Momentum susceptibility}\label{sec:momentum-susceptibility}
The response of the
thermodynamic potential of the 2SC phase to a small momentum of the quark pair can be computed 
absorbing the phase of the condensate into the phase of the quark fields;
the net effect  is a shift of each quark momentum by $\bm q$, see Sec.~\ref{sec:1pwans}. Therefore, the one loop effective action
(in presence of background gauge fields) can be computed
using the same steps that lead to Eq.~\eqref{eq:rambo2}, see also the discussion after Eq.~\eqref{NGspinor2b}. Expanding the
thermodynamic potential around $ q = 0$ one has, at the lowest order,
\begin{equation}
\Omega = \Omega_{\rm 2SC} + \frac{1}{2}\, {\cal K}\, q^2~,
\label{eq:OmegaK}
\end{equation}
where  the momentum susceptibility is given by
\begin{equation}
{\cal K} = \frac{i}{6} \sum_{i=1}^3
\int\frac{d^4 p}{(2\pi)^4}
\text{Tr}\left[\Gamma_i S(p) \Gamma_i S(p)\right]~,
\label{eq:Kdefin}
\end{equation}
where $S(p)$ is the quark propagator in momentum space and $\bm\Gamma = \text{diag}(\bm\gamma, - \bm\gamma)$ is the appropriate vertex factor. Notice that  we are considering an expansion of the thermodynamic potential for small $q$, therefore   the momentum susceptibility does not depend on  $q$, meaning that  $S(p)$ is the 2SC quark propagator, which  can be obtained inverting the expression in Eq.\eqref{eq:Sinvert}. The  momentum susceptibility has an expression similar to the space component of the 2SC polarization tensor, see Eq~\eqref{eq:GBPT},  at vanishing momentum.  In particular  it is possible to show that it is proportional to  the squared Meissner mass of the ${\bar 8}$ mode. The reason is that the only nonvanishing contribution to both the momentum susceptibility and to the Meissner mass of the  ${\bar 8}$ mode are determined from the red-green color sector (blue quarks do not carry  a condensate and do not mix with red and green quarks).  But in this color sector $T_{\bar 8}$ is proportional to the identity and thus the  corresponding vertex factor is proportional to $\bm\Gamma$. A more detailed discussion can be found in the work by \textcite{Giannakis:2004pf}, where it is shown that  
\begin{equation}
{\cal M}_{M,\bar{8}}^2 = \frac{1}{12}\left(g_s^2 + \frac{e^2}{3}\right){\cal K}~,
\label{eq:Krelshp}
\end{equation}
where the mass of the $\bar 8$ mode in the 2SC phase is given in Eq.~\eqref{eq:8til}.

At the transition point between the 2SC phase and the g2SC phase, $\delta\mu = \Delta$,
one finds $M_{M,\bar{8}}^2 < 0$ and  thus, Eq.~\eqref{eq:Krelshp} implies that the system
is unstable towards the formation of pairs with  nonvanishing net momentum; indeed a negative ${\cal K}$ in Eq.~\eqref{eq:OmegaK} implies that  there is a gain in free energy if $q\neq0$. Therefore,
the chromomagnetic instability of the $\bar 8$ mode leads naturally to the FF state. 

Before turning to the computation of the Meissner tensor in the crystalline phase,
we briefly comment  on the absence of total currents induced by the net momentum
of the quark pair. The value of the total momentum is determined by minimizing the free energy, thus $\partial\Omega/\partial q = 0$, and the stationarity condition is equivalent to 
\begin{equation}
\langle \bar\psi\bm\gamma\psi\rangle = 0~,
\label{eq:MattCurVan}
\end{equation}
which implies that no baryon matter current is generated in the ground state. Analogously,
one can show \cite{Giannakis:2004pf} that electric and color currents
vanish as well. In particular, the residual $SU(2)_c\otimes U(1)_{\tilde{Q}}$ symmetry
implies that $\langle\bar\psi\bm\Gamma_A\psi\rangle = 0$ for $A\neq 8$, with
$\bm\Gamma_A$ defined in Eqs.~\eqref{eq:VMcolor} and~\eqref{eq:VMem}; moreover,
for $A=8$ one finds
\begin{equation}
\bm J_8 = \frac{\sqrt{3g_s^2 + e^2}}{6}\langle \bar\psi\bm\gamma\psi\rangle~,
\end{equation}
which vanishes because of the stationary condition, Eq.~\eqref{eq:MattCurVan}. Therefore,
no total current is generated in the FF state.

\subsubsection{Meissner masses in the FF phase}
Since the FF phase has the same gauge symmetry breaking pattern of the 2SC phase, it has five massive gluons. The computation of the Meissner tensor of gluons in the two-flavor FF phase 
has been done by \textcite{Giannakis:2005vw,Giannakis:2005sa},
neglecting neutrality conditions and considering the isospin chemical potential, $\delta\mu =
\mu_e/2$, as a free parameter. \red{Since there exists a privileged direction, the Meissner mass becomes  direction dependent and is in general decomposed into longitudinal and transverse components with respect to $\bm q$.} 
 The Meissner tensors of the gluon fields with adjoint color $a=4, \dots, 7$ are all equal and can be written as 
\be
({\cal M}_{M,4}^2)_{ij} &=& A\left(\delta_{ij} - \frac{q_i q_j}{q^2}\right)
                + B\frac{q_i q_j}{q^2}~.
\ee
 The Meissner tensor of the $\bar 8$ mode can be decomposed in a similar way 
\be
({\cal M}_{M, \bar 8}^{\prime 2})_{ij} = C\left(\delta_{ij} - \frac{q_i q_j}{q^2}\right) + D\frac{q_i q_j}{q^2}\label{eq:CeD}\,. \ee
The coefficients  $A$ and $C$ are called the \textit{transverse} Meissner masses;
similarly, $B$ and $D$ are the \textit{longitudinal} Meissner masses. As discussed in the previous section, the mass of the rotated eighth gluon is related to the variation of the free energy with respect to $q$.  In the FF phase it has been shown by \textcite{Giannakis:2005sa}  that the precise relation is the following
\begin{eqnarray}
C &=& \frac{1}{12}\left(g_s^2 + \frac{e^2}{3}\right)\frac{1}{q}
     \frac{\partial\Omega}{\partial q}~, \\
D &=& \frac{1}{12}\left(g_s^2 + \frac{e^2}{3}\right)\frac{\partial^2\Omega}{\partial q^2}~.
\end{eqnarray}
If the phase with $q\neq 0$ is a minimum of the free energy,
then both the conditions $\partial\Omega/\partial q =0$ and
$\partial^2\Omega/\partial q^2 >0$ must be satisfied. Thus, $C=0$ and $D>0$ at the
minimum. As a consequence,
the Meissner tensor of the $\bar 8$ mode is purely longitudinal
and positively defined.

The coefficients $A$, $B$ and $D$ can be computed analytically. We
refer the interested reader to the original
article \cite{Giannakis:2005vw}. Here, it is enough to consider
the small gap expansion \cite{Giannakis:2005vw,Ciminale:2006sm};
in this approximation scheme one has
\begin{eqnarray}
A &=& \frac{g_s^2\mu^2}{96\pi^2}\frac{\Delta^4}{\delta\mu^4(z_q^{-2}-1)^2}~,\\
B &=&  \frac{g_s^2\mu^2}{8\pi^2}\frac{\Delta^2}{\delta\mu^2(z_q^{-2} -1)}~,\\
D &=& \frac{g_s^2\mu^2}{6\pi^2}\left(1 + \frac{e^2}{3g_s^2}\right)
    \frac{\Delta^2}{\delta\mu^2(z_q^{-2} -1)}~.
\end{eqnarray}
The message of the above equations is that the Meissner tensor is
positively defined for the one plane wave phase, within the small
gap parameter approximation.  
\textcite{Giannakis:2005vw} argue that for a multiple plane wave
structure the situation will be better (within the small gap
expansion). As a matter of fact, at the $\Delta^2/\delta\mu^2$
order, the Meissner tensor is purely longitudinal, and the
longitudinal components are positive at the minimum of the free
energy. Therefore, if the expansion in plane waves contains at
least three linearly independent momenta, the Meissner tensor will
be positive definite, being  additive with respect to different
terms of the plane wave expansion to order $\Delta^2/\delta\mu^2$.
This has been explicitly checked by \textcite{Ciminale:2006sm}.

Besides, in the work by \textcite{Giannakis:2005vw} the numerical computation
of the Meissner masses, beyond the small gap expansion, has been
performed. The results can be summarized as follows. First of all,
the FF phase is found to be more stable than the  2SC
phase in the LOFF window
\begin{equation}
0.706\,\Delta_{\rm 2SC}  \lesssim\delta\mu \lesssim 0.754\,\Delta_{\rm 2SC}~.
\end{equation}
Within this range, the LOFF gap window is 
\begin{equation}
0 < \Delta \lesssim 0.242\,\Delta_{\rm 2SC}~,
\label{eq:LOFFwind}
\end{equation}
and the longitudinal Meissner masses, $B$ and $D$, turn out to be
positive within the range
\begin{equation}
0 < \Delta \lesssim 0.84\,\Delta_{\rm 2SC}~;
\label{eq:interval1}
\end{equation}
moreover, the transverse mass $A$ of the gluons with $a=4,\dots,7$ turns out to be
positive within the  range
\begin{equation}
0 < \Delta \lesssim 0.38\,\Delta_{\rm 2SC}~.
\label{eq:interval2}
\end{equation}
Since the LOFF window for the FF state~\eqref{eq:LOFFwind}
is contained in both the intervals~\eqref{eq:interval1} and~\eqref{eq:interval2},
 the FF state is free from the chromomagnetic instability, as long as it  is energetically favored with respect to the homogeneous phase.

\subsection{Solitonic ground state}\label{sec:solitonic}
The analysis of various crystalline phases has shown that a periodic structure is energetically favored for mismatched Fermi spheres.  In the work by \textcite{Nickel:2008ng} a generalization of the crystalline structure has been proposed for exploring whether the more complicated periodic condensate 
(we do not report the color-flavor structure because it is the same of
Eq.~\eqref{FF-2flavor}):
\begin{equation}
\Delta(z) = \sum_{k\in{\cal Z}}\Delta_{\bm q,k}~e^{2ik q z}~,
\label{eq:BN1}
\end{equation}
may be energetically favored with respect to standard crystalline structures. The wave vector $\bm q$ is taken along the
$z$-axis, thus the condensate corresponds to a band structure along the $z-$direction. Assuming that the condensate is real (which is the assumption considered by \textcite{Nickel:2008ng}), one has $\Delta_{\bm q,k} = \Delta_{\bm
q,-k}^*$ and Eq.~\eqref{eq:BN1} can be rewritten as
\begin{equation}
\Delta(z) = 2\sum_{n=1}^\infty \Delta_n \cos(2nqz)~.
\label{eq:BN2}
\end{equation}
Hereafter we refer to this phase as the solitonic phase. Written in  this form, it is clear that the
ansatz  is a generalization of
the strip structure and amounts to considering higher harmonics contributions to the gap
function, which are not included in the strip.

\begin{figure*}[t]
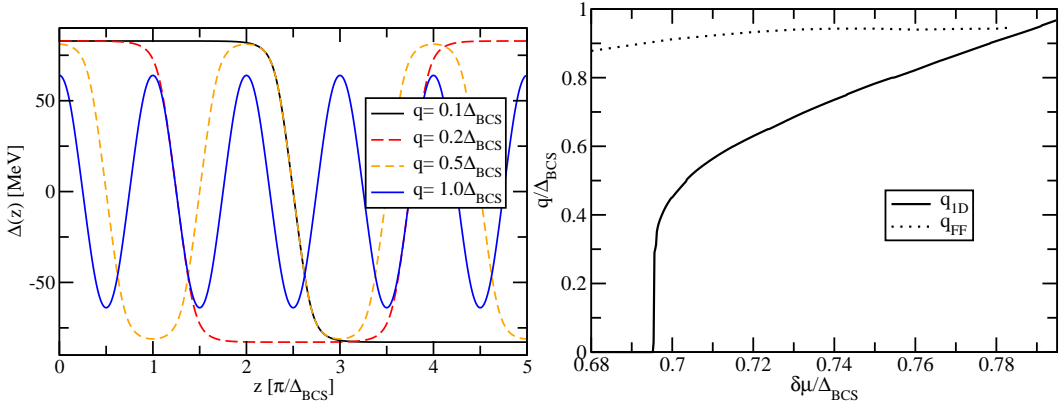

\begin{center}
\subfigure{\includegraphics[width=7cm]{figure9a.eps}}\subfigure{\includegraphics[width=7cm]{figure9b.eps}}
\end{center}
\caption{\label{Fig:soliton}(color online). Left panel: profile for the gap
function, at several values of $q$, obtained from the
self-consistent solution of the gap equation, at $\delta\mu =
0.7\Delta_{\rm BCS}$.  Right panel: physical wave vector magnitude as a
function of $\delta\mu$. In both plots $\Delta_{\rm BCS} \equiv \Delta_{\rm 2SC}$. From \textcite{Nickel:2008ng}.}
\end{figure*}

For any value of $\delta\mu$, the ground state
determined by \textcite{Nickel:2008ng} is obtained using a two-step procedure.
First, the magnitude of $q$ in Eq.~\eqref{eq:BN2} is fixed,
and the profile $\Delta(z)$ is determined by solving the gap
equation. Then, among  several choices for $q$,
the physical value corresponds to the one  minimizing the
free energy. In the left panel of Fig.~\ref{Fig:soliton}, which is
taken from \textcite{Nickel:2008ng}, we plot the profile
$\Delta(z)$ obtained by the numerical solution of the gap equation
at fixed value of $q$. In the plot, $\delta\mu$ is fixed to the
numerical value $0.7\Delta_{\rm 2SC}$ with $\Delta_{\rm 2SC} = 80$ MeV;
but changing $\delta\mu$ does not change the picture
qualitatively. For $q$ of the same order of $\Delta_{\rm 2SC}$, the
shape of the gap function is very close to that of the strip. In
this case, the largest contribution to the gap comes from the
lowest order harmonic $n=1$ in Eq.~\eqref{eq:BN2}. However, for
small values of $q/\Delta_{\rm 2SC}$, the solution of the gap equation
has a solitonic  shape. This is evident in the case
$q=0.1\Delta_{\rm 2SC}$, in which $\Delta(z) \approx \pm\Delta_{\rm 2SC}$
for one half-period, then suddenly changes its sign in a narrow
interval. In this case, the higher order harmonics play a relevant
role in the gap function profile.

In the right panel of Fig.~\ref{Fig:soliton}, we report the plot of the physical
value of $q$ as a function of $\delta\mu$. In the window in which
$q=0$, the ground state is the  homogeneous BCS state. At the
critical value $\delta\mu\equiv\delta\mu_c\approx0.695\Delta_{\rm 2SC}$, the ground state has $q\neq0$. 
This signals the transition to the inhomogeneous
phase. Since $q$ can be arbitrary small in proximity of the
transition point, from the left panel of Fig.~\ref{Fig:soliton} we
read that the ground state will consist of a solitonic structure  for
$\delta\mu\approx\delta\mu_c$. As $\delta\mu$ is increased, the
physical value of $q$ increases as well. Therefore, again from the
left panel of the figure, we read that the solitonic structure will
continuously evolve into the strip structure.

\begin{figure}[b!]
\begin{center}
\includegraphics[width=7cm]{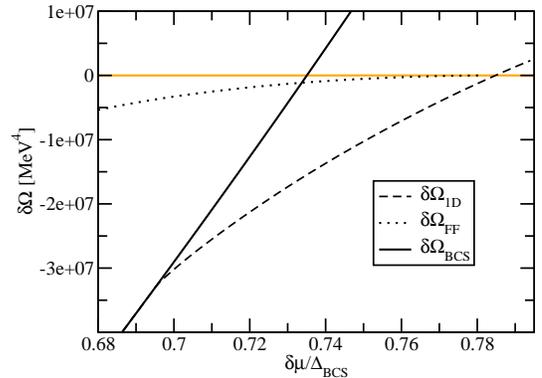}
\end{center}
\caption{\label{Fig:solitonO} Free energy  as a function of
$\delta\mu$, for the BCS phase (solid line), FF phase (dotted
line) and solitonic phase (dashed line). Free energies are
measured with respect to the normal phase. From
\textcite{Nickel:2008ng}.}
\end{figure}

In Fig.~\ref{Fig:solitonO}, taken from
\textcite{Nickel:2008ng}, we report the plot of the BCS free energy 
(solid line) and of the solitonic phase (dashed line) as a
function of $\delta\mu$. The dotted line corresponds to the free
energy of the FF phase. The baseline corresponds to the free energy of the normal phase. With increasing mismatch  a second order phase transition takes place from the
BCS to the solitonic  structure at a value $\delta\mu/\Delta_{\rm BCS} \simeq 0.7$ which is smaller than the value at which
 the transition
from the BCS phase to the FF phase takes place. On the other
hand, the transition from the solitonic phase to the normal
one takes place almost simultaneously with the transition from the
FF phase to the normal one.

In the case of the FF phase, the transition to the BCS state is
first order and to the normal phase is second order. On the other
hand, the transition from the solitonic phase to the BCS phase is second order. This is possible because the gap
function~\eqref{eq:BN2} naturally interpolates between the
homogeneous state, corresponding to a single soliton with infinite
period, and the cosine-shaped solution.

Moreover, the transition to the inhomogeneous state~\eqref{eq:BN2}
to the normal phase is found to be of the first order, in agreement with the results of the smearing procedure for the strip \cite{Casalbuoni:2004wm}, see Table~\ref{tabledeltamu2}. At this transition point, the profile of the gap function is of the cosine type, \textit{i.e.} the strip profile discussed above. 

The results of this analysis are in disagreement with those of the GL expansion  \cite{LO,Bowers:2002xr}, which predicts that the transition from the strip to the normal phase is of the second
order.  Since the GL expansion is expected to be exact in proximity
of a second order phase transition, it is important to understand
the origin of the discrepancy among the GL result and that
by \textcite{Nickel:2008ng} and of the smearing procedure. To this end, \textcite{Nickel:2008ng} compute the free energy
for a gap function $\Delta(z) = \Delta \cos(2qz)$ with a
fixed value of $q$, as it is customary in the GL studies. Their
analysis reveals that in this case, beside the local minimum of
$\Omega$ located at small values of $\Delta$, and which is
captured by the GL expansion, a global minimum appears for larger
values of $\Delta$. Then the authors argue that  to
capture this true minimum, which is responsible for the
first-order transition to the normal phase, terms of at least
eighth order in the GL expansion, which are usually neglected,
should be included. In a subsequent paper \cite{Nickel:2009wj}
Nickel discusses the inhomogeneous phases for lower dimensional modulations in the
NJL model and in the quark-meson model. His results confirm the replacement of the
first order transition of the phase diagram of the homogeneous NJL phase by two
transition lines of second order. These lines are the borders of an inhomogeneous
phase and they intersect at the critical point. An interesting point is also the
relation to the chiral Gross-Neveu model \cite{Gross:1974jv}. A more complete program would require
the inclusion of higher dimensional modulations of the inhomogeneity (as done by \textcite{Abuki:2011pf} for the chiral condensates), and, as pointed out by \textcite{Buballa:2009ct},  simultaneous study of both superconducting and chiral condensates.  

\subsection{Condensed matter  and ultracold fermionic systems}\label{sec:cond-cold}

Fermionic  systems consisting of two different ``flavors" with mismatched Fermi surfaces  and inhomogeneous condensates are quite generic and appear in various contexts.  Particularly interesting are  the population imbalanced superfluid (or superconducting) systems  that can be realized and studied in laboratory.  Examples of these type  are gases of cold atoms, type-II  cuprates and  organic superconductors. The study of these systems is helpful for shedding light on various aspects of superfluidity of asymmetric systems in a framework under  experimental control.

Of considerable importance for their similarity to quark matter  are ultracold systems consisting of fermions of two different species corresponding to two hyperfine states of a fermionic atom, see \textcite{ketterle-review, giorgini-review, Gubbels:2012je, Radzihovsky} for reviews. These fermions have opposite spin and one can change the number of up and down fermions at will. One of the most intriguing aspects of these systems is that the interaction between fermions can be tuned by employing a Feshbach resonance \cite{Feshbach-review}, and therefore the crossover between the BCS and the BEC superfluid phases  can be studied.   This research field has grown in an impressive way in the last two decades and continuous progress in understanding and charactering the properties of these systems 
is under way, see \textit{e.g.} \textcite{ketterle1, ketterle2, ketterle3, Partridge, Muther:2002mc, Bedaque:2003hi, Liu:2002gi, Gubankova:2003uj, Forbes:2004cr, Carlson:2005kg, Castorina:2005kg,
Pao, Son:2005qx, Sheehy:2006qc, Yang:2005,  Bulgac:2006gh, Yang:2006ez,
Gubankova:2006gj, Mannarelli:2006hr, Rizzi, Bulgac:2008tm, Sharma:2008rc}.

A mismatch between the populations of   electrons in a superconductor can also be produced by Zeeman splitting.  The magnetic field that couples with the spins of the electrons can be an external one or an exchange field.  However,  an external magnetic field  couples with the orbital motion of the electrons as well, destroying superconductivity or leading to the creation of  a vortex lattice structure. 

In order to reduce the orbital effect
one employs 2D superconductors, \textit{i.e.} films of superconducting material or systems with a layered structure, and an in-plane magnetic field \cite{Bulaevskii}. Good candidates for LOFF superconductors of this type are  heavy-fermion compounds like CeRu$_2$ \cite{Huxley}. Recently, interesting results have also been obtained with CeCOIN$_5$ \cite{Matsuda:2007}. Quasi-two-dimensional organic superconductors \cite{Uji:2006}  like $\kappa$-(ET) or $\lambda$-(ET) salts, in particular    $\lambda$-(BETS)$_2$F$_e$Cl$_4$, are promising candidates for realizing the LOFF phase, as well. High $T_c$ superconductors are good candidates as well.

As a final remark, we recall that the gapless CSC phases  are the QCD analogue of
a condensed matter phase, known as  Sarma phase \cite{Sarma19631029, Liu:2002gi, Gubankova:2003uj, Forbes:2004cr},  found to be unstable \cite{Pao, Sheehy:2006qc, Mannarelli:2006hr, Gubankova:2006gj,  Gubankova:2008ya, Wu:2003zzh} in the weak coupling limit.

\section{The three-flavor inhomogeneous phases}
\label{sec:Chapter3}
In Sec.~\ref{sec:Chapter2} we have discussed the crystalline inhomogeneous phases that can be realized in two-flavor quark matter. A plethora of possible  structures  have been analyzed by a  Ginzburg-Landau (GL) expansion and by the so-called smearing procedure for determining the favored thermodynamic state. 

Here we extend the analysis to the three-flavor case which is relevant if the effective strange quark mass, $M_s$, is not too heavy.  In compact stars 
$M_s$ will lie somewhere between its current mass of order $100$~MeV and its vacuum
constituent mass of order $500$~MeV, and therefore is of the order of the quark number
chemical potential  $\mu$, which is expected to be in the  range $(400-500)$ MeV.
Furthermore, deconfined quark matter, if
present in compact stars, must be in weak equilibrium and must be
electrically and color neutral\footnote{Actually, as already discussed in Sec.~\ref{sec:intro},  quark matter  must be in a color singlet, however it has been shown by \textcite{Amore:2001uf} that  projecting out color singlet states into color neutral states does not lead to a large change of the free energy.}. All these factors work to separate
the Fermi momenta of the quarks and thus disfavor the cross-species
BCS pairing. 

As a means  to understand how the mismatch among Fermi momenta is linked to $\beta$ equilibrium and neutrality, let us consider   quark matter composed of
 $u$, $d$ and $s$ quarks with no strong interactions. We treat the 
strange quark mass as a parameter considering sufficiently long time scales for which weak equilibrium and electrical neutrality are relevant. 

Since we are assuming that color interactions are absent, the chemical potential of quarks is diagonal in the color indices and the free energy of the system  is given by 
\begin{widetext}
\bea
\Omega_{\rm unpaired}(\mu,\mu_e,M_s) &=&
\frac{3}{\pi^2}\int_0^{P_u^F}(p - \mu_u) p^2 dp +
\frac{3}{\pi^2}\int_0^{P_d^F}(p - \mu_d) p^2 dp \nonumber\\ &+&
\frac{3}{\pi^2}\int_0^{P_s^F}(\sqrt{p^2 + M_s^2} - \mu_s) p^2 dp +
\frac{1}{\pi^2} \int_0^{\mu_e}(p - \mu_e) p^2 dp \,, \eea\end{widetext}
where the Fermi momenta are given by \be\label{fermi-momenta} P_u^F =
\mu_u\,, \,\, P_d^F = \mu_d\,, \,\, P_s^F = \sqrt{\mu_s^2 - M_s^2} \,.
\ee

Weak equilibrium relates the chemical potentials of quarks with different flavors as follows
\be \mu_u = \mu - \frac{2}{3}\mu_e \qquad \mu_d = \mu +
\frac{1}{3}\mu_e \qquad \mu_s = \mu + \frac{1}{3}\mu_e \,,\nonumber\\ \label{eq:chemicals-neutral}\ee and
the electrical chemical potential
is obtained from the neutrality constraint
\begin{equation}
\label{neutrality} \frac{\partial \Omega}{\partial\mu_e} = 0\, .
\end{equation}

\begin{figure}[th]
\begin{center}
\includegraphics[width=8cm,angle=-0]{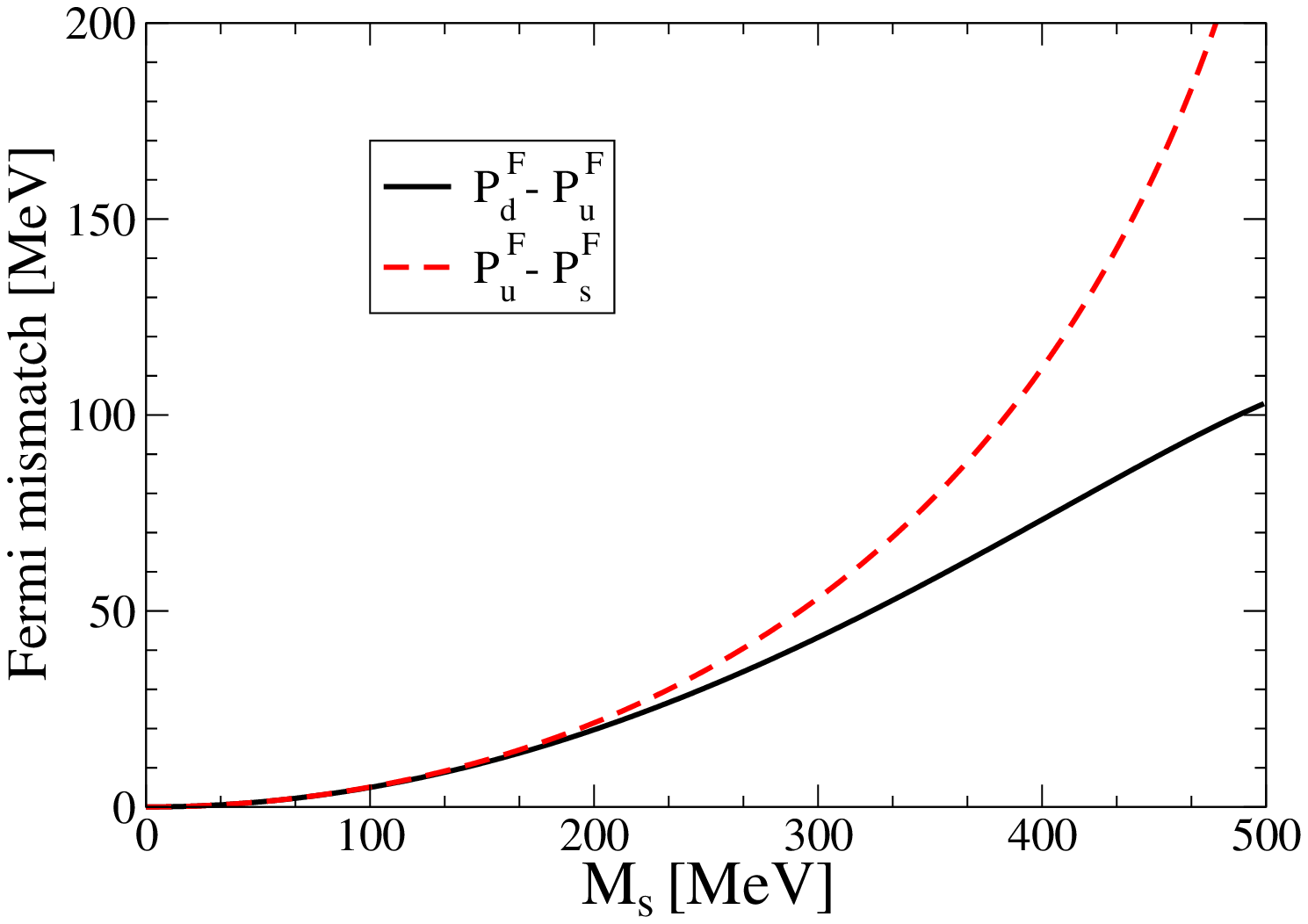}
\end{center}
\caption{(color online) Fermi momenta differences $P_d^F - P_u^F$,
black solid line, and $P_u^F - P_s^F$, red dashed line, as a function
of $M_s$, for  $\mu=500$ MeV. } \label{fig:splitting-neutral}
\end{figure}

Solving  Eq.~\eqref{neutrality}  one determines $\mu_e$ and by
Eqs.~\eqref{fermi-momenta} and \eqref{eq:chemicals-neutral} one obtains  the values of $P_u^F$,
$P_d^F$ and $P_s^F$ for the electrically neutral unpaired phase. The effect of $M_s$ is to reduce the number of strange quarks, which must be compensated by a larger number of down quarks to ensure electrical neutrality. It follows that there is a hierarchy of Fermi momenta
\be P_s^F \le  P_u^F \le   P_d^F \,,
\label{eq:hierarchyPf}
\ee and  in Fig.~\ref{fig:splitting-neutral} we report the  difference of the Fermi momenta $P_d^F - P_u^F$ and $P_u^F - P_s^F$ as a function of $M_s$, for $\mu = 500$ MeV.  

For $M_s \ll \mu$, the effect of a nonzero strange quark mass can be
taken into account by treating the strange quark as massless, but
with a chemical potential that is lowered by $M_s^2/(2\mu)$ from
$\mu+\mu_e/3$. To the order $M_s^2/\mu$ electric neutrality,
Eq.(\ref{neutrality}),  requires that $\mu_e \simeq \frac{M_s^2}{4\mu}$.
In this case  $P_d^F - P_u^F \simeq P_u^F - P_s^F$,
and  we need no longer to be careful about the distinction between
$P_F$ and $\mu$, as we can simply think of the three flavors of
quarks as if they have chemical potentials
\begin{equation}
\mu_d = \mu_u + 2 \,\delta\mu \qquad \mu_u =p_F^u  \qquad \mu_s =
\mu_u - 2 \,\delta\mu\,, \label{pF3}
\end{equation}
with \be \delta\mu \equiv \frac{M_s^2}{8\mu}\label{pF4}\, , \ee and we
can write the chemical potential matrix as
\begin{equation}
\mu_{ij, \alpha\beta} = \delta_{\alpha\beta}\otimes {\rm
diag}\left(\mu_u,\mu_d,\mu_s\right)_{ij}\label{mu matrix}\,.
\end{equation}

\red{The above results are strictly valid for unpaired matter; indeed, if BCS pairing between quarks with different flavors takes place, their Fermi momenta must be  equal and the electron chemical potential vanishes.  As discussed in the two-flavor case, the cross-species homogenous pairing between quarks on split Fermi spheres has a free energy cost proportional to the corresponding chemical potential difference. Since in the CFL phase  quarks of all flavors and of all colors pair according with the condensate in Eq.~\eqref{condensate-CFL}, increasing the value of $\delta\mu$ makes the CFL phase less energetically favored.}
However, small values of the chemical potential difference, \textit{i.e.} of $M_s$,
cannot disrupt the BCS pairing and the CFL phase will be energetically favored. Nevertheless, when $M_s$ is
sufficiently large, the mismatch between the Fermi momenta becomes
large disfavoring  CFL pairing.

As shown in Fig.~\ref{fig:splitting-neutral},   for values of $M_s$ comparable with $\mu$
the strange quarks decouple; in this case  only the 2SC phase with the pairing between up and down quarks seems realizable. However, as pointed out by \textcite{Alford:2002kj}, once  the constraints of electrical and color
neutrality and  $\beta$ equilibrium are imposed, the 2SC phase turns out to be
strongly disfavored or even excluded at least at zero temperature
(for finite temperature evaluations see for example
\textcite{Abuki:2003ut}). \textcite{Alford:2002kj},
performing an expansion in terms of the strange quark mass (at the leading nontrivial order), found that whenever the 2SC phase is more favored than unpaired
matter, then the CFL is even more favored. The point is that the CFL phase is extremely robust, because it allows pairing between all the three flavor  species. 
This analysis has been redone by \textcite{Steiner:2002gx} evaluating the density-dependent
strange quark mass self-consistently. The results by \textcite{Steiner:2002gx} almost confirm previous conclusions, finding the 2SC favored in a very narrow region of density, that is likely to disappear once the hadronic phase boundary is properly taken into account.  Notice that these  results are valid in the weak-coupling approximation. Computations using NJL-like models with a  stronger coupling \cite{Abuki:2005ms, Ruester:2006aj},  see also Sec. \ref{sec:g2SC}, show that a large portion of the phase diagram is occupied by the 2SC phase.

\subsection{The gapless CFL phase}\label{sec:gCFL}
Given that the effect of nonzero strange quark mass and
electrical neutrality constraints is to pull apart the Fermi spheres of
different quark flavors,  there  is  little motivation for assuming the symmetric CFL pairing reported in Eq.~\eqref{condensate-CFL}.
Indeed, one might expect that a mismatch between the Fermi momenta reflects in a reduction of the interaction channel  and thus in a reduction of the corresponding gap parameter. Therefore, 
the gap parameters should be now flavor dependent and  one can consider the generalized pairing ansatz 
\be \langle
0|\psi_{iL}^\alpha\psi_{jL}^\beta|0\rangle=- \langle
0|\psi_{iR}^\alpha\psi_{jR}^\beta|0\rangle \propto \sum_{I=1}^3 \Delta_I\flaeps\coleps\,,
\label{condensate-gCFL2}\nonumber\\ \ee
where $\Delta_{1}$, $\Delta_{2}$ and $\Delta_{3}$, describe $d$-$s$, $u$-$s$ and $u$-$d$ Cooper pairing, respectively. In the work by \textcite{Alford:2003fq,Alford:2004hz} the  superconducting phase characterized
by the ansatz~\eqref{condensate-gCFL2} has been studied by a NJL-like model at zero temperature (see \textcite{Fukushima:2004zq} for a study at nonvanishing temperature), considering the  in-medium strange quark mass as a free parameter, while the light quarks are taken massless. 
The gap equations, coupled to the neutrality conditions, have been solved and the corresponding free energy has been determined. Besides the strange quark mass, the final results depend on the quark chemical potential,
which is fixed to the numerical value $\mu=500$ MeV; moreover, the strength 
of the NJL-like interaction is fixed by the value of the homogenous CFL gap, $\Delta_{\rm CFL}$. The numerical value
$\Delta_{\rm CFL} = 25$ MeV, corresponding to the weak-coupling regime, has been chosen.

\begin{figure*}[t!]
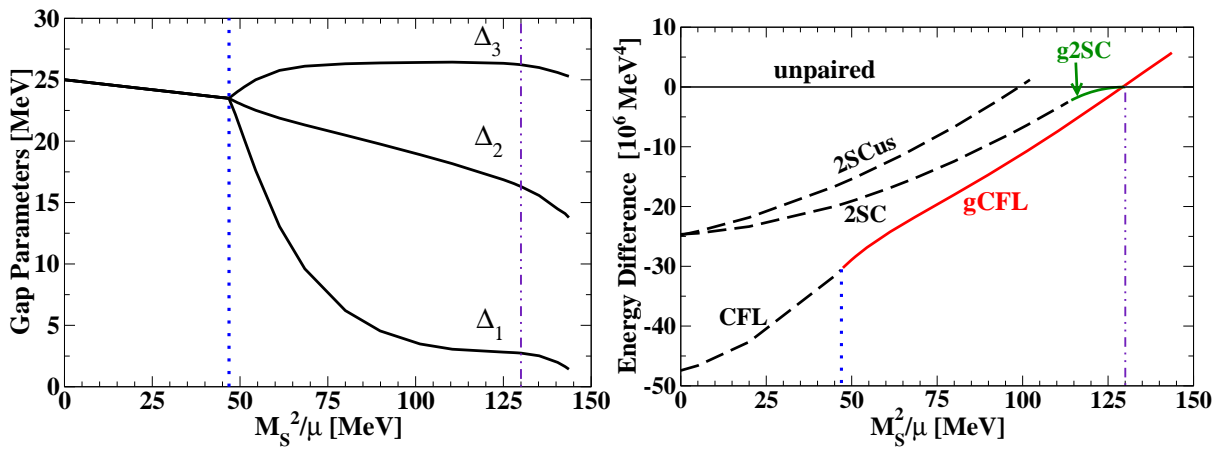
 
\subfigure{\includegraphics[width=8.cm]{figure12a.eps}}\subfigure{\includegraphics[width=8.cm]{figure12b.eps}}
\caption{ (color online) Left panel: gap parameters
$\Delta_3$, $\Delta_2$, and $\Delta_1$ as a function of $M_s^2/\mu$
for $\mu=500$~MeV, in a model where $\Delta_{\rm CFL}=25$~MeV (see text). Right panel: free energy of the
CFL/gCFL phase, relative to that of neutral noninteracting quark
matter and that of the 2SC/g2SC and 2SCus phases. At
$M_s^2/\mu\approx 47.1$ MeV (vertical dotted line) there is a
continuous phase transition between the CFL phase and the gCFL phase.  Above $M_s^2/\mu\approx 130$ MeV (vertical dash-dotted line) unpaired quark
matter has a lower free energy than the gCFL phase. From \textcite{Alford:2004hz}.}\label{fig:gapsCFL}
\end{figure*}

Some of the results by \textcite{Alford:2003fq,Alford:2004hz} are summarized 
in Fig.~\ref{fig:gapsCFL}. On the left panel of the figure, the solutions of the gap equations as a function of $M_s^2/\mu$ are plotted. As it can be inferred from the figure, at
$M_s^2/\mu\approx 47.1$ MeV,  there is a
continuous phase transition between the CFL phase and a phase characterized by
$\Delta_{3}>\Delta_{2}>\Delta_{1}>0$. Note that the hierarchy  of the gap parameters is in agreement with the hierarchy of the Fermi momenta  splitting reported in Fig. \ref{fig:splitting-neutral}, which substantiates the reasoning presented above about the relation between mismatched Fermi momenta and gap parameters.
Actually, the mismatches of  chemical potential  are not only due to the finite value of the strange quark mass, but as well as to the nonzero values of the color chemical potentials $\mu_3$ and $\mu_8$, which however do not overturn the hierarchy in Fig. \ref{fig:splitting-neutral}.

The   phase appearing at $M_s^2/\mu\approx 47.1$ MeV is dubbed gapless CFL, or gCFL, phase, and is the three-flavor analog of the g2SC phase discussed in Sec.~\ref{sec:g2SC}.
The symmetry breaking  pattern is the same of the CFL phase, see Eq.~\eqref{eq:breakingCFL},  but the  spectra of some fermionic excitations are gapless \cite{Alford:2003fq,Alford:2004hz, Kryjevski:2004jw}.  The possibility of a gapless CFL superconductor was first argued by \textcite{Alford:1999xc}, where  a toy model was used to infer the effect of a heavy $M_s$ on the quasiparticle spectrum and it was found that if the
condensates of light and strange quarks are below a certain
critical  value  there is not  a minimum excitation energy in the quasiparticle
spectrum. As discussed in Sec.~\ref{sec:mismatched},
the mechanism at the origin of gapless modes in a superconductor is quite general, and is due to a 
 mismatch $\delta\mu$  between condensing fermions  of the order of $\Delta$.
  
In  three-flavor  quark matter the situation is complicated 
by the presence of several interaction channels and  of three  chemical potential differences proportional to  $\mu_e$, $\mu_3$ and $\mu_8$. Thus,  in the  gCFL phase the dispersion laws are more complicated than those reported in Eq.~\eqref{eq:H2}.  However, the qualitative result is rather similar, as gapless excitations appear in the spectrum for sufficiently large mismatches between the Fermi momenta of the various species. 

Comparing the free energy of the gCFL with other candidate phases,
\textit{e.g.} the 2SC phase, or the 2SC$+s$, which is a three-flavor  model in
which it is taken into account the decoupled strange quark, or the so-called 2SCus
in which  only the $u$-$s$ pairing survives (\textit{i.e.} $\Delta_{2}>0,
\Delta_{1}=\Delta_{3}=0$) or the g2SC  discussed in
the Sec.~\ref{sec:g2SC},  the gCFL phase turns out to be 
energetically favored  in a quite large window of the control parameter \cite{Alford:2003fq,Alford:2004hz}. The comparison of the gCFL free energy with the free energy of some of theses phases is reported in  Fig.~\ref{fig:gapsCFL}. Besides the considered phases, other pairing patterns have been proposed \cite{Iida:2003cc}, corresponding to the  uSC phase (with $\Delta_1=0$ and $\Delta_2, \Delta_3 \neq 0$) and the dSC phase (with $\Delta_2=0$ and $\Delta_1, \Delta_3 \neq 0$), see \cite{Fukushima:2004zq, Ruester:2006aj} for a comparison of the thermodynamic potentials of the various phases. 

However, as discussed in the two-flavor case, the fact  that a phase is energetically favored with respect to other phases  is not a sufficient condition to ensuring its stability. Indeed,  the analysis of fluctuations around the mean field solutions may reveal that the phase does not correspond to a minimum of the free energy. 
As in the case of the g2SC phase, the gCFL phase is indeed
chromomagnetically unstable, because four of the eight  Meissner masses of gluons become imaginary when  gapless modes appear \cite{Casalbuoni:2004tb,Fukushima:2005cm}.  Further increasing the mismatch among the Fermi surfaces, the masses of the remaining four gluons become imaginary, as well.  The effects of temperature on the Meissner masses in the   gCFL phase has been studied by \textcite{Fukushima:2005cm}, in which it is found that for sufficiently high temperatures (of about $10$ MeV for the parameter choice by \textcite{Fukushima:2005cm}), the gCFL phase becomes  stable. However, such a  temperature is much larger than the typical temperature of  compact stars and therefore cannot be used as an argument in favor of the gCFL phase. 

\red{Since the gCFL is chromomagnetically unstable but it is energetically favored with respect to both unpaired quark matter and  the two-flavor CSC  phases,  there must exist a phase with a lower free energy.  Interestingly, at vanishing temperature the gCFL phase is favored with respect to the other phases even at weak coupling, whereas the g2SC phase is energetically favored with respect to unpaired quark matter and the 2SC phase only in the intermediate coupling regime, see Sec. \ref{sec:g2SC}.}
 
\red{As discussed in the two-flavor case, the chromomagnetic instability   can be associated with the tendency of the system to develop supercurrents, that is to realize a FF-like state.  
Then, as done in the two-flavor case, one can generalize the FF state to more complicated structures trying to single out the favored ground state among various crystalline phases. We shall investigate various crystalline patterns in the following sections.}

\subsection{Three-flavor  crystalline phase: Two plane waves} \label{sec:3flavor_1PW}
In this section we present
the evaluation of the simplest nontrivial three-flavor CCSC phase
using a modified Nambu-Gorkov formalism. The approach presented is
based on the high-density effective theory (HDET) by
\textcite{Casalbuoni:2001gt,Nardulli:2002ma} 
and the evaluation is performed with and without making a GL
expansion.

\subsubsection{Nambu-Gorkov and HDET formalisms for the three-flavor crystalline phase} \label{sec:NG-HDET}
\red{As already discussed in Sec.~\ref{sec:1pwans}  one can properly choose the Nambu-Gorkov basis for simplifying the calculation in  the  FF  phase. Here we extend that reasoning to  the three-flavor case.} In the three-flavor  case the
simplest form of inhomogeneous pairing corresponds to the condensate 
\begin{equation}
\Delta_{ij}^{\alpha \beta}= \sum_{I=1}^{3}\,\Delta_I e^{2i{\bm
q_I}\cdot{\red{\bm x}}}\,\flaeps \coleps ~,\label{Delta-3pw}
\end{equation}
meaning that   for each  pairing channel we assume a
FF ansatz, with $2{\bm q_I}$ representing the momentum of the
 pair.

In this case it is not possible  to diagonalize the
propagator in ${\bm p}$-space, because three different fields are locked together in pairs and thus one can only eliminate in the off-diagonal terms of the propagator two independent momenta.
However, as shown in Eqs.~\eqref{pF3} and \eqref{pF4},  the separation between the $s$ and $d$ Fermi spheres
 is twice the separation between the $d$ and
$s$  Fermi spheres and the   $u$  and
$s$  Fermi spheres, thus it is reasonable to expect $\Delta_1 \ll
\Delta_2,\Delta_3$. As a first approximation one can consider  $\Delta_1=0$
and in this case it is possible to diagonalize the propagator in momentum space.
 
  The form of the two-flavor Nambu-Gorkov spinor
(\ref{NGspinor2b})
 immediately suggests that we analyze the three-flavor crystalline phase
with condensate (\ref{Delta-3pw}) with $\Delta_1$ set to zero by
introducing the Nambu-Gorkov spinor \be
\chi({\bm p}) = \left(\begin{array}{l} \psi_u({\bm p})\\ \psi_d({\bm p}-2{\bm q}_3) \\
\psi_s({\bm p}-2{\bm q}_2)\\
\psi_u^C(-{\bm p})\\ \psi_d^C(-{\bm p}+2{\bm q}_3) \\
\psi_s^C(-{\bm p}+2{\bm q}_2)
\end{array}\right)\ .
\label{NGspinor} \ee It is the clear that it would  not be possible to
use this method of calculation if $\Delta_1$ were kept nonzero,
except for the special case in which ${\bm q}_1={\bm q}_2 - {\bm
q}_3$. (That is, except in this special case which is far from
sufficiently generic, it will not be possible to choose a
Nambu-Gorkov basis such that one obtains a propagator that is
diagonal in some momentum variable ${\bm p}$.) Moreover, it seems
unlikely that this method can be employed to analyze more
complicated crystal structures. Indeed, it seems that the pairing with
  $\Delta_1=0$ and $\Delta_2$ and
$\Delta_3$ each multiplying a single plane wave, is the most complex
example that is currently known how to analyze without using the
GL expansion.

We now implement the calculation in the basis \eqref{NGspinor} using
the HDET formalism.  We shall generalize the discussion presented in Sec.~\ref{sec:2flavor-smearing}  including the shifts of the quasiparticle momenta. To this end we Fourier decompose
the fermionic fields as follows:
\begin{equation} \psi_i^{\alpha}(x)= e^{-i {\bm k_i \cdot \red{\bm x}}}\int\frac{d \Omega}{4\pi}e^{-i\mu{\bm v}\cdot{\bm x}}\,\left(\psi^\alpha_{{i},\bm
v}(x)+\psi^{\alpha-}_{{i},\bm v}(x)\right)\label{decomp} \, ,
\end{equation}
which differs from Eq.~\eqref{decomp-2flavor} for the presence of 
the three  vectors ${\bm k_i}$, one for each
flavor, that we shall specify below.  In the standard HDET
approximation   ${\bm k}_i \equiv 0$ and the field $\psi^\alpha_{{i},\bm v}(x)$ is used to describe a quark in a patch in momentum space in the vicinity of  momentum
${\bm P}=\mu{\bm v}$. The introduction of the $\bm k_i$ vectors  means
that now $\psi^\alpha_{{i},\bm v}(x)$ describes a quark with  momentum
in a patch in the vicinity of momentum $\mu{\bm v} +{\bm k}_i$ and the chemical potential differences with respect to the average value $\mu$ are then given by
\be
\delta\mu_i(\bm v)= \mu_i - \mu -\bm k_i \cdot \bm v\,.
\ee

 At the leading order in $1/\mu$
(\textit{i.e}. neglecting the contribution of antiparticles)
the \red{free} Lagrangian can be written as in Eq.~\eqref{L110}, but with a velocity dependent chemical potential difference:
\begin{equation} 
{\cal L}= \int\frac{d{\Omega}}{4\pi}\, \left[
\psi^{\alpha \dagger}_{{i},\bm
v}\left(i V \cdot \de
+ \delta\mu_{i}({\bm v})\right)
\psi^{\alpha}_{{i},\bm
v} \right]~. \label{L11} 
\end{equation}

To take into account diquark condensation, we add to Eq.~\eqref{L11} a term similar to Eq.~\eqref{eq:LD-2FL}, which however takes into account the $\bm k_i$ shifts in the momenta, that is
\begin{equation} 
{\cal L}_{\Delta}= -\frac{1}{2}\int\frac{d\Omega}{4\pi}\,\Delta_{ij}^{\alpha \beta}
\psi^{\alpha T}_{{i},
-\bm v} \,C\,\gamma_5\, \psi^\beta_{j, {\bm v}}\, e^{-i  {\bm
({\bm k}_i+{\bm k}_j) \cdot {\bm x} }}~+ {\rm h.c.}~,
\end{equation}
where the condensate $\Delta_{ij}^{\alpha \beta}$ is given in
\eqref{Delta-3pw} and we have omitted the term proportional to $\Delta^2$. In the three-flavor case it is convenient to work in a new basis for the
spinor fields defined by
\be\label{FA-basis}
\psi^\alpha_{i}= \sum_{A=1}^9 (F_A)^\alpha_i \psi_A\,,
\ee
where the   matrices $F_A$ are given by
\begin{align*}
F_1 &= \frac{1}{3} I +T_3 + \frac{1}{\sqrt{3}} T_8\,, \qquad  F_2 = \frac{1}{3} I -T_3 + \frac{1}{\sqrt{3}} T_8\,,  \\
F_3 &= \frac{1}{3} I  - \frac{2}{\sqrt{3}} T_8\,,  \qquad 
F_{4,5} = T_1 \pm i T_2\,,  \\ F_{6,7} &= T_4 \pm i T_5\,, \qquad F_{8,9} = T_6 \pm i T_7\,,
\label{eq:FA}\end{align*}
with $T_a$ the $SU(3)$ generators defined in Eq.~\eqref{eq:TA} and $I$ the $3 \times 3$ identity matrix. Introducing the velocity dependent  Nambu-Gorkov fields
\begin{equation}
\chi^A_{\bm v}=\frac{1}{\sqrt{2}}\left(\begin{array}{c}
  \psi_{\bm v} \\
  C\,\psi^*_{- \bm v}
\end{array}\right)_A \,,\label{nambu-gorkov}
\end{equation}
and replacing  any matrix $M$ in color-flavor space with $M_{AB} = \text{Tr} [F_A^T M F_B^{\phantom{T}}]$,  the NJL Lagrangian density can be written in the compact form
\begin{equation}
{\cal L}= \frac{1}{2}\int\frac{d\Omega}{4\pi} \chi^{A\dagger}_{\bm v}~S^{-1}_{AB}({\bm v})~\chi_{\bm v}^B
\, ,\label{total}
\end{equation}
where the inverse propagator is given by
\begin{equation}
S^{-1}_{AB}=\left(
\begin{array}{cc}
  \left( V\cdot \ell\, + \delta\mu_{A}({\bm  v}) \right) \delta_{AB} & -\Delta_{AB} \\
  -\Delta_{AB} &  (\bar V\cdot \ell\, - \delta\mu_{A}({-\bm v}))\delta_{AB}
\end{array}
\right),\label{inverse}
\end{equation}
where $\ell^\nu = (p_0,\xi {\bm v})$. As discussed in Sec.~\ref{sec:2flavor-smearing}
for  the standard HDET, the
integration over momentum space is separated into an angular
integration over ${\bm v}$ and a radial integration over $-\delta \le
\xi \le \delta$. The cutoff $\delta$ must be taken  smaller
than $\mu$ but  much larger than the homogeneous gap parameter (that is $\Delta_{\rm CFL}$ in the three-flavor case) and $\delta\mu$, see Fig.\ref{hierarchy}.

If $\Delta_1=0$,  the space
dependence in the anomalous terms of the propagator  can be eliminated by choosing 
 \be
\bm{k_u}+{\bm k_d} = 2 {\bm q}_3\,,   \qquad
\bm{k_u}+{\bm k_s} = 2{\bm q}_2\,, \label{choiceofk2} \ee
and the calculation is technically simplified. It has been numerically checked  by \textcite{Mannarelli:2006fy} that different
choices of ${\bm k}_u$, ${\bm k}_d$ and ${\bm k}_s$ satisfying
(\ref{choiceofk2}) yield the same results for the gap parameter and
free energy. The different choices yield quite different
intermediate stages to the calculation; the fact that  the final results are the same
 is a nontrivial check of the
numerics.

From the Lagrangian  \eqref{total}, following a derivation
analogous to that by \textcite{Alford:2004hz}, the free energy  can be evaluated to be
 \be
 \Omega =
- \frac{\mu^2}{4 \pi^2}\sum_{a=1}^{18}\int_{-\delta}^{+\delta} d
\xi \int \frac{d\Omega}{4\pi} ~|E_a({\bm v},\xi)|
+\frac{2\Delta^2}{G}- \frac{\mu_{e}^4}{12 \pi^2},\nonumber\\ \label{Omega}
\ee 
 where we have set $\Delta_2=\Delta_3=\Delta$ and where $G$ is
the NJL coupling constant. 
The dependence on  $G$  can be eliminated by using  the CFL gap equation, 
\begin{equation}
\Delta_{\rm CFL} = 2^{2/3}\delta\,\exp\left\{-\frac{\pi^2}{2 G\mu^2}\right\}~,
\label{eq:CFL-gap}
\end{equation}
where $\Delta_{\rm CFL}$  is the CFL gap parameter for $M_{s}=0$ and $\mu_{e}=0$.  \red{As discussed for example by \textcite{Nardulli:2002ma},  although the value of the free energy and of the gap parameter depend on $\delta$, the energetically favored phase is independent of it. Alternatively, one can use a coupling constant which explicitly depends on the momentum cutoff as in the work by \textcite{Ippolito:2007uz}.}

In Eq.~\eqref{Omega}, the $E_a$  are the energies of the fermionic 
quasiparticles, which  are given by the $18$ roots of
$\det S^{-1}=0$,  given in Eq. \eqref{inverse}. The doubling of degrees of freedom in the Nambu-Gorkov
formalism means that the $18$ roots come in pairs whose energies are
related by $E_a({\bm v},\xi) = E_b(-{\bm
v},\xi)$. One set of nine roots describes $(\psi_{d,{\bm
v}},\psi_{u,-{\bm v}})$ and $(\psi_{s,{\bm v}},\psi_{u,-{\bm v}})$
pairing, while the other set describes $(\psi_{u.{\bm
v}},\psi_{d,-{\bm v}})$ and $(\psi_{u,{\bm v}},\psi_{s,-{\bm v}})$
pairing (color indices have been omitted for simplicity). Since ${\bm v}$ is integrated over, the free energy can be
evaluated by doing the sum in (\ref{Omega}) over either set of nine
roots, instead of over all 18, and multiplying the sum by two.

The lowest free energy state is determined minimizing the free energy  given in Eq.~(\ref{Omega}) with respect to the gap parameter $\Delta$ and with respect to ${\bm q}_2$ and ${\bm q}_3$.   One could
also determine self-consistently the values of $\mu_e$, $\mu_3$ and
$\mu_8$ which ensure electrical and color neutrality.  
Moreover, one could allow $\Delta_2\neq \Delta_3$ and minimize
with respect to the two gap parameters separately. However, in the
results that we shall present in the next section we shall fix
$\Delta_2=\Delta_3=\Delta$, $\mu_e=M_s^2/(4\mu)$ and
$\mu_3=\mu_8=0$, as is correct for small $\Delta$. 

\subsubsection{Ginzburg-Landau analysis}\label{sec:3flavor-GL}
As discussed in the two-flavor case,   the GL 
expansion  allows us to determine the gap parameter and the free energy for a generic crystalline 
structure. We shall discuss the GL results for various crystalline phases in Section \ref{sec:3flavor-crystals}; in the present section we restrict to the case of the condensate reported in Eq.~\eqref{Delta-3pw}.   
In the three-flavor case  the GL expansion is controlled by the ratio $\Delta/\Delta_{\rm CFL}\approx \Delta/\delta\mu$ and is reliable  in the vicinity of a second order transition;  this restriction implies that pairing does not
significantly change any number density, and thus 
one can assume $\mu_3 = \mu_8 = 0$ and $\mu_e \approx M_s^2/4\mu$ as in the normal phase. We shall briefly comment  in Sec.~\ref{sec:1overmu} on the latter approximation.

For the condensate in  Eq.~\eqref{Delta-3pw}, the GL expansion of  the free energy is given by \cite{Casalbuoni:2005zp, Mannarelli:2006fy}
\begin{equation}
\Omega =\Omega_n+ \sum_{I=1}^3\left( \alpha_I \,\Delta_I^2
+ \frac{\beta_I}{2}\,\Delta_I^4 + \sum_{J\neq
I}\frac{\beta_{IJ}}{2}\,\Delta_I^2\Delta_J^2 \right) + O(\Delta^6) \,,
\label{eq:OmegaDelta11}
\end{equation}
\red{which can be seen as an extension of the ${\cal O}(\Delta^4)$   two-flavor GL free energy, \textit{cf.} Eq.~\eqref{eq:OGL}, to the three-flavor case. We neglect the sextic (and higher order) terms because for the simple  inhomogeneous condensates considered here the quartic terms are positive. Note that in contrast to the two-flavor case,  the presence of three $\Delta_I$'s requires the introduction of various coefficients with indices referring to the particular condensate $\Delta_I$. In particular, the term proportional to $\beta_{IJ}$, with $I \neq J$, corresponds to the interaction term between different condensates.  }

Using the HDET formalism, it has been shown by  \textcite{Casalbuoni:2005zp} that  $\alpha_I$ and $\beta_I$ have the same 
formal expression derived  in the GL analysis of the two-flavor FF
structure. This   means that
\be
 \alpha_I \equiv &\, \alpha_{\rm FF}(q_I,\delta\mu_I)
\label{alfa} \qquad
\beta_I\equiv &\,
\beta_{\rm FF}(q_I,\delta\mu_{I})\,, \ee
with $\alpha_{\rm FF}$ defined in Eq.~\eqref{alfaFF} (with $\Delta_{\rm 2SC} \to 2^{1/3} \Delta_{\rm CFL}$, for replacing the 2SC gap parameter with the CFL gap parameter, compare Eq.~\eqref{eq:2SC-gap} with Eq.~\eqref{eq:CFL-gap}) and $\beta_{\rm FF}$   defined in Eq.~\eqref{betaFF}. 
 \red{We have indicated with $\delta\mu_I$ the chemical potential difference between  quark whose  flavor is not $I$,  \textit{e.g.} $2 \delta\mu_1 = \mu_d-\mu_s$.} In the weak-coupling limit  the chemical potential differences are the same of the unpaired phase, thus from  Eq.~\eqref{pF3} one obtains \be \delta\mu_{1} = -\mu_e\,, \qquad  \delta\mu_{2} = -\delta\mu_{3} = - \frac{\mu_e}{2}\,.
\label{eq:deltamu123}
\ee

For the interaction term between different condensates,  it turns out that 
\begin{equation}
\beta_{12}\!=\!-\!\frac{\mu^2}{4\pi^2} \int\frac{d{\bm v
}}{4\pi}\,\frac{1}{(i\epsilon-2{\bm q_1}\cdot{\bm v
}-2\delta\mu_{1})\,(i\epsilon-2{\bm q_2}\cdot{\bm v
}-2\delta\mu_{2})}\label{beta12},\end{equation}
and $\beta_{13}$ is
obtained from $\beta_{12}$  by changing ${\bm
q_2\to \bm q_3}$ and $\mu_s\leftrightarrow\mu_d$; in a similar way $\beta_{23}$ is obtained  from
$\beta_{12}$ by changing ${\bm q_1\to \bm q_3}$ and
$\mu_s\leftrightarrow\mu_u$. These are the
only terms in which a dependence of the free energy from the
relative orientation of the ${\bm q_I}$ can arise at this order.

One should fix the norms
 $q_I$ and the relative orientation of the  three vectors ${\bm q_I}$ by a minimization procedure.
This is a complex task  requiring  the
simultaneous  minimization of the free energy with respect to many parameters.  What is usually done to
circumvent this complication is to propose different definite
structures for the ${\bm q_I}$, selecting among
them the one with the lowest free energy. In the present simple case we can do something better, using the angle 
$\phi$  between $\bm q_2$ and $\bm q_3$ as a variational parameter. Indeed, it has been found by \textcite{Casalbuoni:2005zp} that the
energetically favored solution corresponds to
$\Delta_1=0$ and $\Delta_2 \approx \Delta_3$ and clearly in this case  the free energy is  independent of $\bm q_1$. 
As to the norms $q_2$ and $q_3$, since we work in the GL
approximation, it is possible to neglect the ${\cal O}(\Delta^2)$ terms
in the minimization of $\Omega$. Thus,  one has that
$\partial\alpha_I/\partial q_I =0 $ for $I=2,3$, which is identical
to the condition for two flavors  giving the result
  $ q_I=z_q^{-1}|\delta\mu_I|$,
 with $z_q$ defined in Eq.~\eqref{eq:zetaq}. Considering this condition and  Eq.~\eqref{eq:deltamu123} one has that      \be |\delta\mu_2| = |\delta\mu_3| \equiv \delta\mu\,,\qquad  \vert{\bm q_2}\vert=\vert{\bm q_3}\vert \equiv q  \label{eq:ualq2q3}\,, \ee
therefore implies that the $u$-$s$
and $d$-$u$ quark pairs momenta have the same modulus. \red{A pictorial representation of the Fermi surfaces and of the pairing regions for this simple configuration is reported in Fig. \ref{ribbonsfigure} for two different values of $\phi$.}

Given  Eq.~\eqref{eq:ualq2q3} and taking $\Delta_1=0$,  the free energy including up to ${\cal O}(\Delta^4)$ terms can be written as
\begin{equation}
\Omega=\Omega_n+2\,\alpha_{FF}\Delta^2+(\beta_{FF}+\beta_{23})\Delta^4
\,,
\label{eq:OGL2PWs}
\end{equation}
where $\alpha_{FF} \equiv  \alpha_{FF} (q,\delta\mu)$, $\beta_{FF} \equiv  \beta_{FF} (q,\delta\mu)$ and
   \be \Delta\equiv \Delta_2=\Delta_3\,,
\label{eq:equalD2D3}
\ee
is the only independent gap parameter. Minimizing the free energy expression with respect to $\Delta$ we obtain  for values of $\delta\mu$ where $\alpha_{FF}$ is negative, the solution  
\be
\Delta^2=\frac{|\alpha_{FF}|}{\beta_{FF}+\beta_{23}}\,, \label{DeltaGL} \ee and the free energy at the minimum is given by
\be \Omega=\Omega_n - \frac{\alpha_{FF}^2}{\beta_{FF} + \beta_{23}}\, ,
\label{OmegaGL} \ee 
which still depends on the angle $\phi$  by $\beta_{23}$.  

\subsubsection{Testing the Ginzburg-Landau approximation}\label{sec:testing}
\begin{figure*}[ht]
\subfigure{\includegraphics[width=5.5cm,angle=-90]{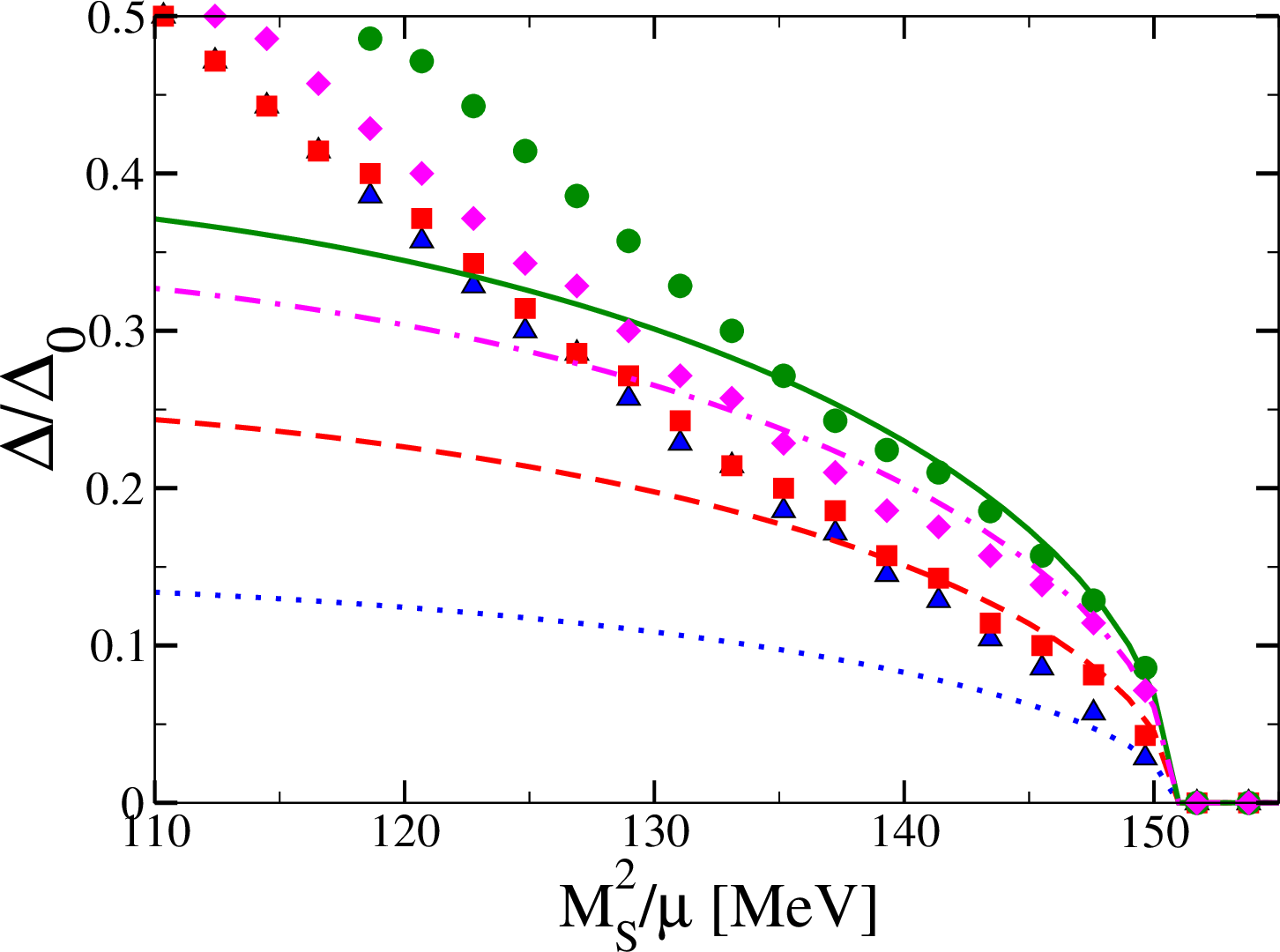}}\hspace{1.cm}
\subfigure{\includegraphics[width=5.5cm,angle=-90]{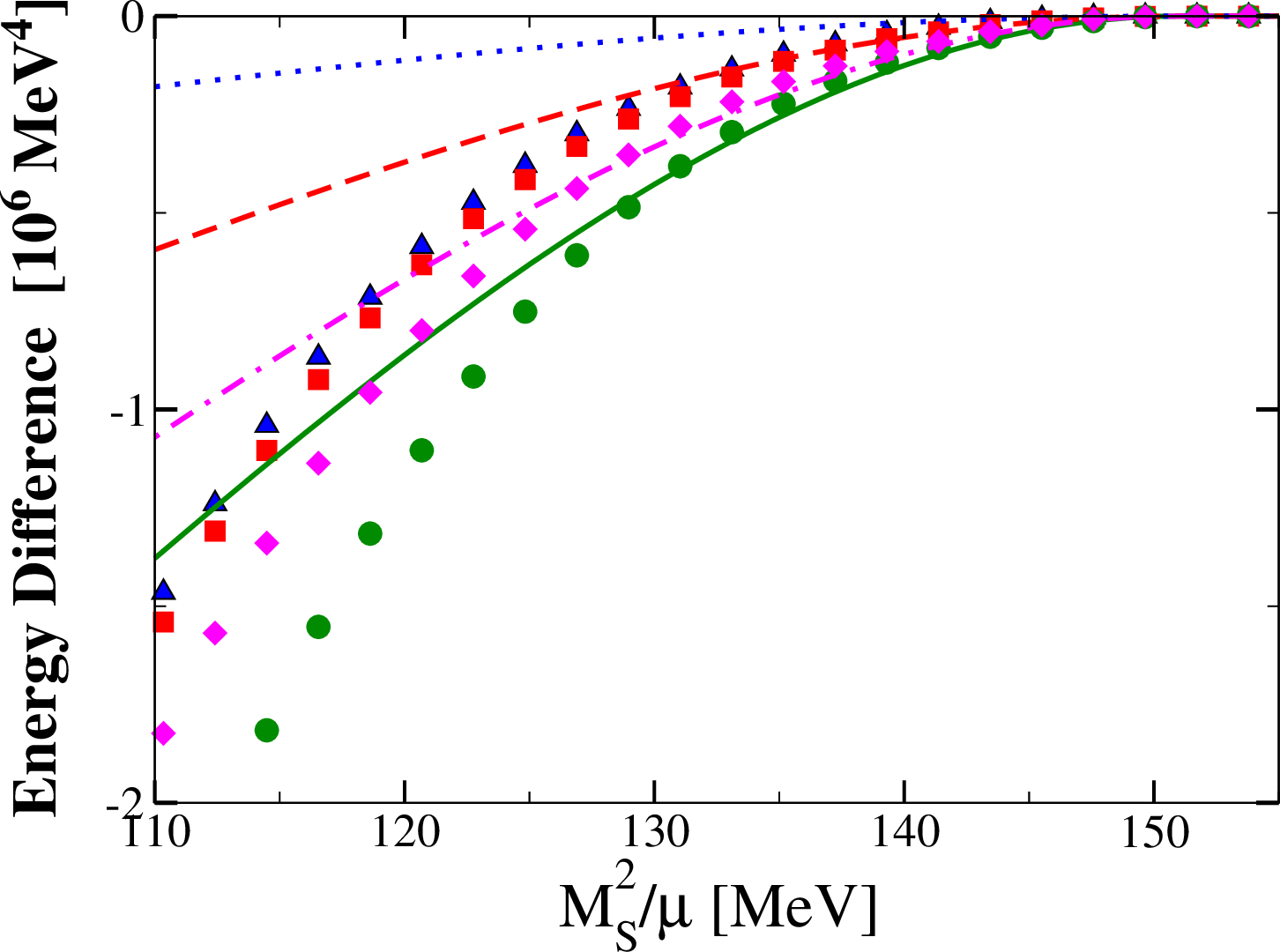}}
\caption{(color online) Plot of $\Delta/\Delta_0$ (left panel), with $\Delta_0=\Delta_{\rm CFL}$, and of  the free
energy relative to neutral non interacting quark matter (right
panel) as a function of $M_s^2/\mu$ for four  values of the angle
$\phi$ between $\bm q_2$ and $\bm q_3$. The various lines correspond
to the calculations done in the GL approximation
  whereas dots correspond to the NJL
calculation, done without making a GL approximation.
The full lines (green online)  and circles correspond to $\phi=0$,
the dashed-dotted lines (magenta online) and diamonds correspond to
$\phi=2\pi /3$, the dashed lines (red online) and squares correspond
to $\phi=7\pi/8$,  the dotted lines (blue online) and triangles
correspond to $\phi=31\pi/32$. From \textcite{Mannarelli:2006fy}.} \label{figgap}
\end{figure*}

We are now able to evaluate the gap parameter and the
free energy for  the ``crystalline" color superconducting phase
\begin{equation}
\Delta_{ij}^{\alpha \beta}= \Delta ( e^{2i{
q \bm n_2}\cdot{\bm {\red x}}}  \varepsilon^{\alpha\beta
2}\,\epsilon_{ij2} +  e^{2i{
q \bm n_3}\cdot{\bm {\red x}}} \varepsilon^{\alpha\beta
3}\,\epsilon_{ij3} ) \,,\label{eq:Delta-2pw}
\end{equation}
 with and without the GL approximation
\cite{Mannarelli:2006fy}; here we have written $\bm q_2 = q\, \bm n_2$ and $\bm q_3 = q\, \bm n_3$. \red{In Fig.~\ref{figgap}} we report the results  obtained for
$\mu=500$~MeV and  $\Delta_{\rm CFL}= 25$~MeV. The calculations
are made varying  $M_s$, but we plot quantities versus $M_s^2/\mu$
because the most important effect of a nonzero $M_s$ is the splitting
between the $d$, $u$ and $s$ Fermi momenta given in Eq.~\eqref{pF3}. We report the results for four values of the angle
between $\bm n_2$ and $\bm n_3$: $\phi= 0$, $2\pi/3$, $7\pi/8$ and $
31\pi/32$. The lines correspond to the GL analysis, with
 $\Delta$  and $\Omega-\Omega_n$ determined from Eqs.~\eqref{DeltaGL} and
\eqref{OmegaGL}, respectively,  using Eq.~\eqref{alfa} to relate $\alpha_{FF}$ to
$\delta\mu_I$ and hence to $M_s^2/\mu$. The points correspond to the
NJL calculation without the GL expansion, hereafter full NJL,  obtained by minimizing the free energy of
Eq.~\eqref{Omega} with respect to $\Delta$.

For all values of the angle $\phi$, both the full NJL calculation and the GL expansion have a second
order transition to the normal phase at $M_s^2/\mu \simeq 151$~MeV, clearly visible in the left panel of  Fig.~\ref{figgap}, corresponding to $\delta\mu \simeq 0.75\Delta_{\rm CFL}$.
In the GL calculation the independence of the critical value of  $\delta\mu$ on  $\phi$ occurs
because  the location of the phase transition depends only on $\alpha_I$,
which is independent of $\phi$. Near the phase
transition, where $\Delta/\Delta_{\rm CFL}$ and hence $\Delta/\delta\mu$ are
small, there is good agreement between the full NJL calculation and the
GL approximation, as expected. 
When
the GL approximation breaks down, it does so
conservatively, under-predicting both $\Delta$ and
$|\Omega-\Omega_n|$. Furthermore, even considering a situation in which
 the GL approximation has broken down
quantitatively, it correctly predicts the qualitative feature that
at all values of $M_s^2/\mu$ the favored crystal structure is
that with $\phi=0$. Therefore, the GL approximation is
useful as a qualitative guide even in a case in which it has broken down
quantitatively.

It is evident from Fig.~\ref{figgap} that the extent of the regime
in which the GL approximation is quantitatively
reliable is strongly $\phi$-dependent.  In the best case, which it
turns out is $\phi=0$, the results of the GL
calculation are in good agreement with those of the full NJL
calculation as long as $\Delta/\Delta_{\rm CFL} \lesssim 0.25$,
corresponding to $\Delta/\delta\mu \lesssim 0.35$. For larger
$\phi$, the GL approximation yields quantitatively
reliable results only for much smaller $\Delta$. For example, with
$\phi=31\pi/32$  the GL calculation
gives results in quantitative agreement with the full NJL
calculation only for $\Delta/\Delta_{\rm CFL} \lesssim 0.04$,
corresponding to $\Delta/\delta\mu \lesssim 0.05$ \cite{Mannarelli:2006fy}.

\red{A remarkable result is that the free energy of the favored  phase, corresponding to $\bm n_2 = \bm n_3$,  has a lower free energy of the CFL and gCFL phases 
in the window \cite{Casalbuoni:2005zp, Mannarelli:2006fy} 
\be128~{\rm MeV}\lesssim \frac{M_s^2}{\mu}\lesssim 151~{\rm
MeV}\,,
\label{eq:window2PW}
\ee 
in which the GL expansion of the free energy for the favored  structure is in  excellent 
agreement with the result of  the full NJL calculation, see the right panel of Fig.~\ref{figgap}.  
This result strongly motivates the study of more complicated crystalline structures employing the GL approximation.}

\red{Note that the range in Eq.~\eqref{eq:window2PW} has been obtained  considering the fixed value of the gap parameter $\Delta_{\rm CFL}=25$ MeV. However, the CFL gap parameter and the constituent strange quark mass should depend on $\mu$.  We shall discuss this point in Sec.~\ref{sec:LOFFwindow} considering the dependence of both  $\Delta_{\rm CFL}$ and $M_s$ on $\mu$ within a NJL model.}

\begin{figure}[t]
\begin{center}
\includegraphics[width=8cm,angle=0]{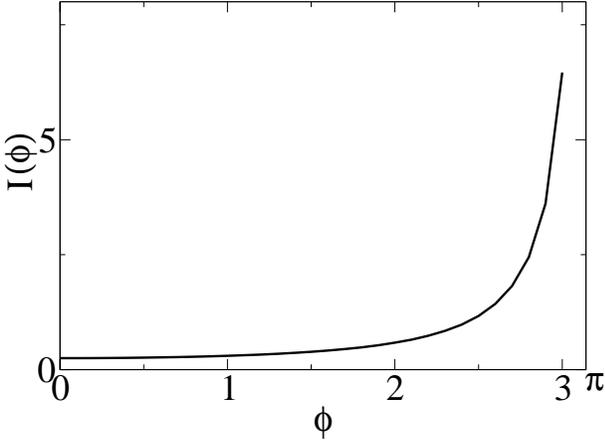}
\end{center}
\caption{ $I(\phi)$, defined in Eq.~(\ref{Iresult}), versus the
angle $\phi$ between the wave vectors ${\bm q}_2$ and ${\bm q}_3$. This function is proportional to the free energy  cost for having overlap between ribbons associated with different wave vectors. For   ${\bm q}_2$ parallel to  ${\bm q}_3$, corresponding to $\phi=0$, there is no overlap between ribbons and no free energy cost has to be paid. The free energy cost diverges for perfectly overlapping ribbons, corresponding to $\phi=\pi$. }
\label{funcbeta23}
\end{figure}

We can now get a deeper insight about the favored  
orientation of the ${\bm
q_I}$ from the analysis of $\beta_{23}$. As explained above, this is the only term in the GL expansion that depends on the relative orientation between $\bm n_2$ and $\bm n_3$ and can be rewritten as

\be \beta_{23}=\frac{
\mu^2}{\pi^2\delta\mu^2}I( \phi)\label{beta23} \,,\ee
where
\be I(\phi) = \Re e
\int\frac{d\Omega}{4\pi}\frac{-1}{(i\epsilon-z_q^{-1}
\bm v \cdot \bm n_2-1)  (i\epsilon-z_q^{-1} \bm v \cdot \bm n_3+1)}\,,\nonumber\\ \label{Iresult} \ee
can be evaluated numerically  and the result is plotted in
Fig.~\ref{funcbeta23} showing that the minimum value  occurs
at $\phi=0$, that is for $\bm n_1 = \bm n_2$, in agreement with the fact that  the configuration that minimizes $I(\phi)$ does as well minimize the free energy, see Eq. \eqref{OmegaGL}.
 We note that although $I(\phi)$ is an increasing function of $\phi$ it depends very
slightly on the angle except close to the value $\phi=\pi$, \textit{i.e.}
when the two vectors are antiparallel and $I(\phi)$
diverges.
\begin{figure*}[t!]
\subfigure{\includegraphics[width=7.cm,angle=0]{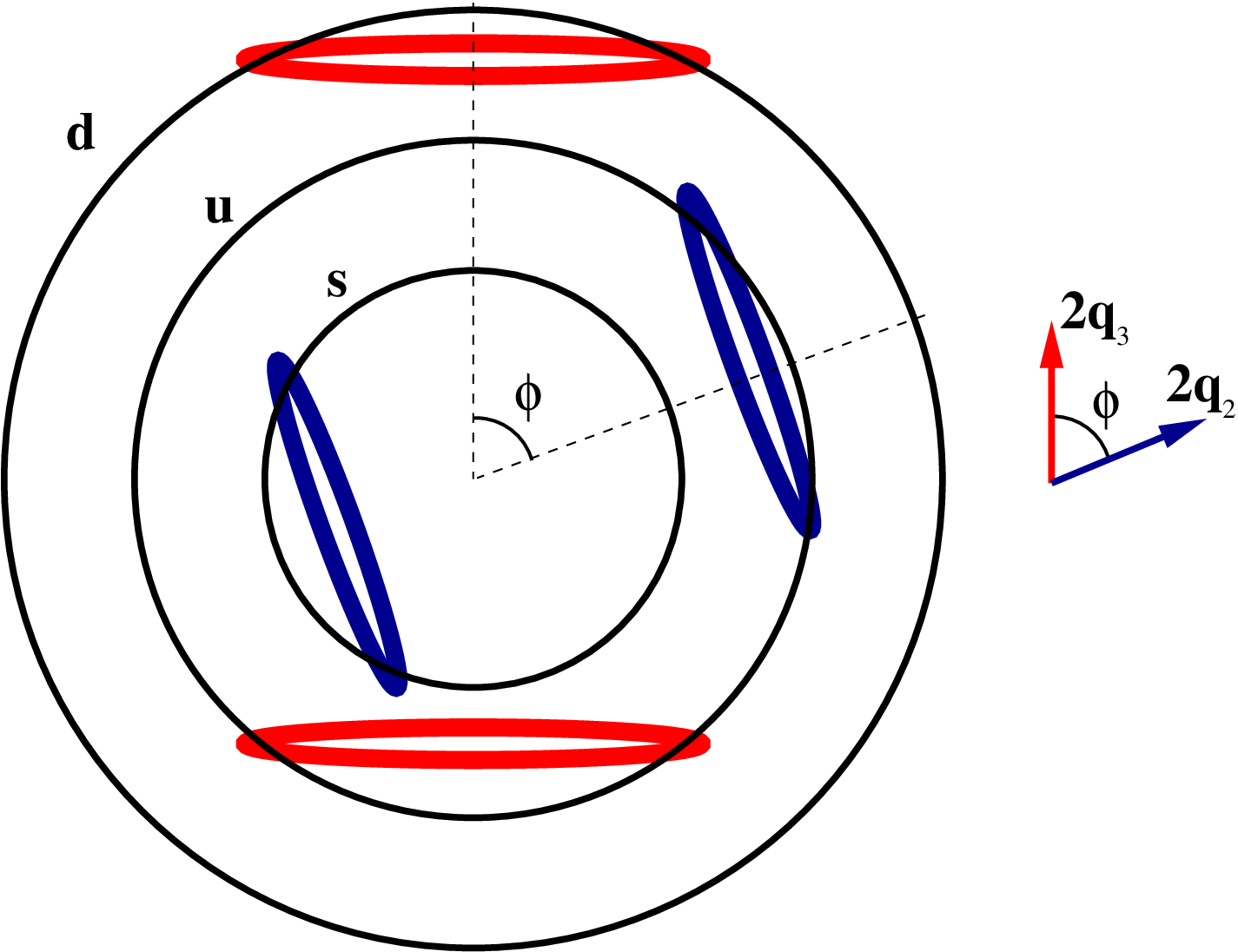}}\hspace{1cm}
\subfigure{\includegraphics[width=7.cm,angle=0]{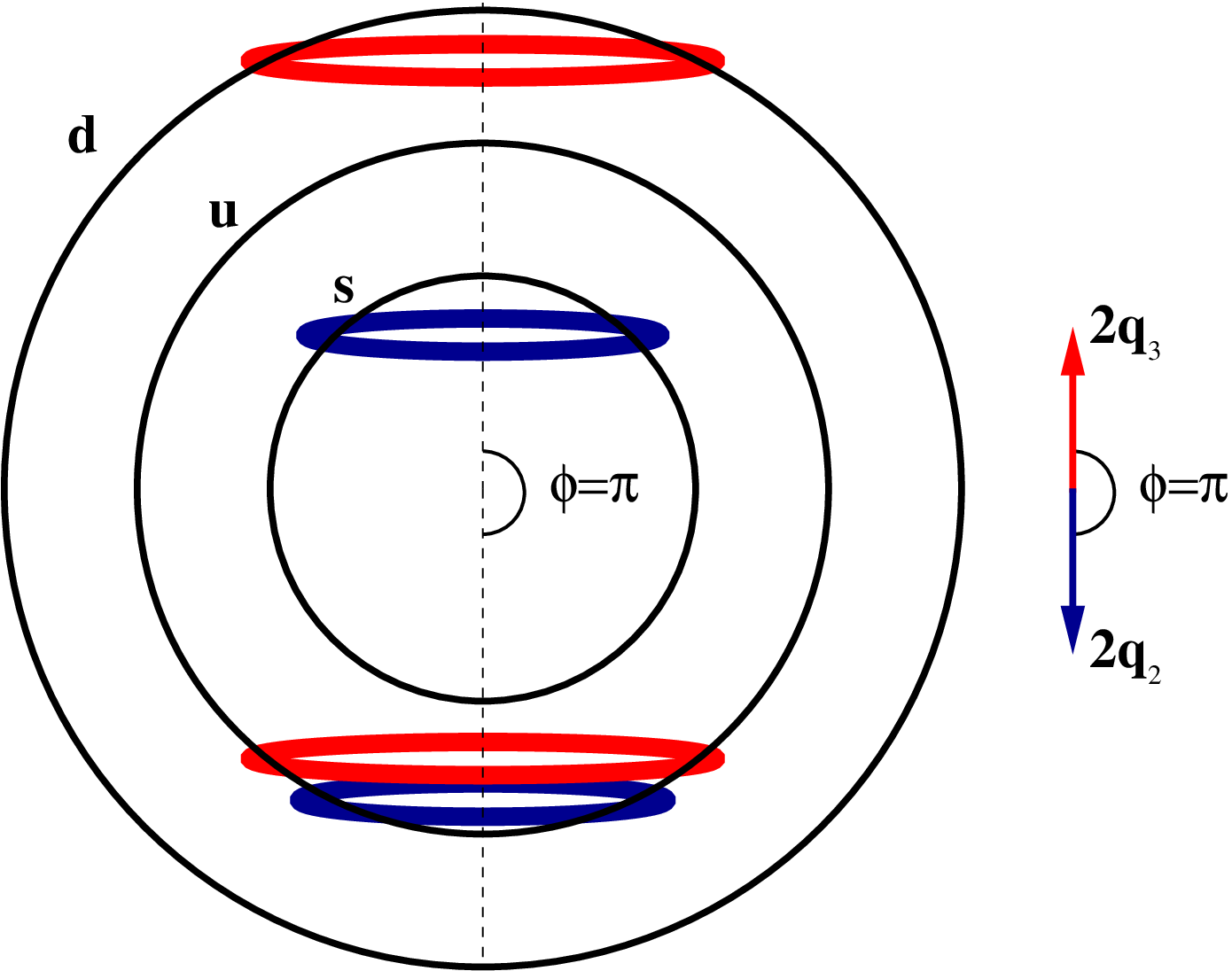}}
\caption{(color online) Sketch showing where on the Fermi surfaces pairing occurs
for condensates in which $\bm q_2$ and $\bm q_3$ are at an angle $\phi$.  The light gray (red online) 
ribbons on the $d$ and $u$ Fermi surfaces indicate those quarks that
contribute the most to the $\langle ud \rangle$ condensate with gap
parameter $\Delta_3$ and wave vector $\bm q_3$, which points upward
in both panels. The dark gray (blue online) ribbons on the $u$ and
$s$ Fermi surfaces indicate those quarks that contribute most to the
$\langle us \rangle$ condensate with gap parameter $\Delta_2$ and
wave vector $\bm q_2$. For descriptive reasons, in the left panel it is  shown the case $\phi = 70^{\circ}$, and in the right panel the antipodal case, with the two vectors pointing in opposite directions.  We have
greatly exaggerated the splitting between the Fermi surfaces, relative to the values used in the
calculations reported in Sec. \ref{sec:3flavor_1PW}.} \label{ribbonsfigure}
\end{figure*}

Although the divergence of $\beta_{23}$ at $\phi=\pi$ is not
physically relevant, since it is not the value that minimizes the free energy, it is 
worth considering this case to gain a qualitative understanding of the behavior of the GL approximation.  We  see in Fig.~\ref{ribbonsfigure} that there are two
pairing rings on the up quark Fermi surface, because some up quarks
pair with down quarks forming Cooper pairs with wave vector $2
\bm{q}_3$ and other up quarks pair with strange quarks forming
Cooper pairs with wave vector $2\bm{q}_2$. However, as shown in the
right panel of Fig.~\ref{ribbonsfigure}, if  $\phi=\pi$ the two
pairing rings on the up quark Fermi surface are close to be coincident.
In the weak-coupling limit in which $\delta\mu/\mu\rightarrow 0$
(and $\Delta_{\rm CFL}\rightarrow 0$ with $\delta\mu/\Delta_{\rm CFL}$ fixed) these
two rings become precisely coincident.  We attribute the divergence
in $\beta_{23}$ to the fact that  antiparallel wave vectors pay
an infinite free energy price and hence are forbidden, because of the
coincidence of these two pairing rings.  In contrast, if $\phi=0$, the two pairing rings on the up
Fermi surface are as far apart as they can be, and $\beta_{23}$ and
the free energy of the state are minimized. This qualitative
understanding also highlights that it is  only in the strict
GL and weak-coupling limits that the cost of choosing
antiparallel wave vectors diverges.  If $\Delta/\delta\mu$ is small
but nonzero, the pairing regions are ribbons on the Fermi surfaces
instead of lines. And, if $\delta\mu/\mu$ is small but not taken to
zero (as of course is the case in Fig.~\ref{ribbonsfigure}) then the
two ribbons on the up Fermi surface will have slightly different
diameter, as the figure indicates. This means that we expect that if
we do a calculation at small but nonzero $\Delta_{\rm CFL}\sim\delta\mu$,
and do not make a GL expansion, we should find some
free energy penalty for choosing $\phi=\pi$, but not a divergent
one. This is indeed the result that one obtains without using the GL expansion, see Fig.~\ref{figgap}.
These results also explain why the breakdown of the GL expansion in Fig.~\ref{figgap} happens for very small values of $\Delta$ for $\phi \approx \pi$, while only for larger values of $\Delta$ for $\phi = 0$. Indeed, increasing $\phi$  from $0$ to $\pi$, the  $\beta_{23}$ term in the GL expansion increases,  and therefore the radius of convergence of the GL expansion decreases.

A more quantitative study of the radius of convergence of
the GL approximation would require evaluating (at
least) the $\Delta^6$ terms, whose coefficients we shall generically
call $\gamma$. Because we are working in the vicinity of a point
where $\alpha_I=0$, the first estimator of the radius of convergence
that we can construct comes by requiring $\gamma \Delta^6 \lesssim
(\beta_I+\beta_{23})\Delta^4$. Thus, the results of the comparison in
Fig.~\ref{figgap} are not conclusive on this point, but they
indicate that the radius of convergence in $\Delta$
decreases with increasing $\phi$, and tends towards zero for
$\phi\rightarrow \pi$.  

\subsubsection{Chromomagnetic stability of the three-flavor crystalline phase}\label{sec:3stability}

\begin{figure}[t!]
\includegraphics[width=8cm]{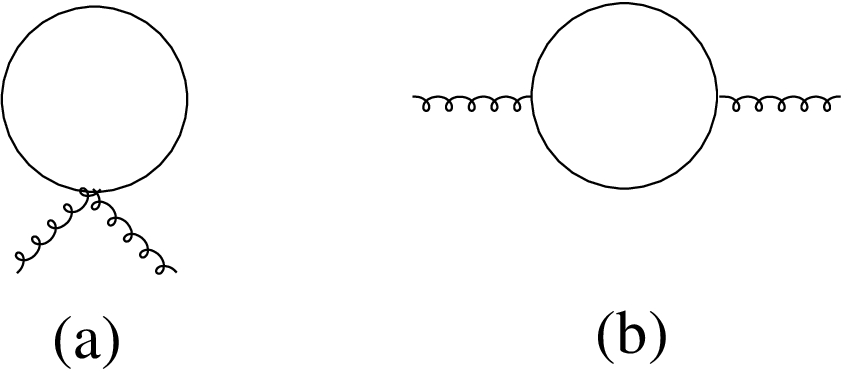}
\caption{Tadpole-like diagram (a) and charm-like diagram (b)  contributing to the gluon masses in the HDET. The curly lines correspond to the gluon fields; the solid lines correspond to the positive energy quark  fields, see Eq.~\eqref{decomp}. The interaction vertices can be obtained from the expansion of the effective Lagrangian in  Eq.~\eqref{L11bis}, including terms of order $1/\mu$.
 \label{fig:2point}}\end{figure}

Given that the three-flavor  CCSC phase in Eq.~\eqref{eq:Delta-2pw} is thermodynamically favored with respect to the CFL and normal phases, it remains to 
be proven that it is chromomagnetically stable.
This issue has been discussed within the GL expansion by \textcite{Ciminale:2006sm}. It has been shown that gauging the NJL Lagrangian  the Meissner masses of gluons are real and positive, meaning that this phase is chromomagnetically stable.
To \red{properly} take into account gluons,   the HDET Lagrangian has to be extended including  the nonlocal terms arising from the integration over the negative energy fields \cite{Casalbuoni:2003sa,Casalbuoni:2002my}, obtaining in momentum space
\begin{equation} {\cal L}= \psi_{i,{\bm v}}^{\alpha\dagger}({\ell})\left(V \cdot
\ell_{ij}^{\alpha\beta}  + \mu_{ij}^{\alpha\beta}
 + P_{\mu\nu} \left[
\frac{\ell_{\mu} \ell_{\nu}}{  \tilde V \cdot \ell + 2 \mu
}\right]_{ij}^{\alpha\beta} \right)\psi_{j,{\bm v}}^\beta({\ell})
\label{L11bis}\, ,
\end{equation} where \be(\ell^\mu)_{ij}^{\alpha\beta}=\ell^\mu\delta_{ij}\delta^{\alpha\beta}-g_s A_aT_a^{\alpha\beta}
\delta_{ij}\,,\ee and
\be P^{\mu\nu}= g^{\mu \nu} -
\frac{\left(V^{\mu}\tilde V^{\nu}+ \tilde
  V^{\mu} V^{\nu} \right)}{2}\,.\ee 
\red{In the HDET there is an infinite series of  effective vertices suppressed by increasing powers of $1/\mu$, describing the
coupling between gluons and quarks.  These interaction vertices  arise from the expansion of the nonlocal term in the square bracket of \eqref{L11bis}.
For the leading-order evaluation of the gluon self-energy  two vertices are relevant: the one-gluon coupling to two quarks (three-body vertex, coupling $\sim g_s$); the two-gluon coupling to two
quarks (four-body vertex, coupling $\sim g_s^2$). 
These two vertices originate from the terms in \red{Eq.~\eqref{L11bis}} with one and two
momenta $\ell$, respectively.  One may naively think that the contribution to the gluon self-energy  due to  the  four-body vertex would be negligible with respect to the contribution arising from the  three-body vertex, because of the $1/\mu$ suppression. Instead, the two vertices  lead to  the two one-loop diagrams shown in Fig.~\ref{fig:2point}, whose contribution  to the Meissner mass are clearly of the same order $g_s^2\mu^2$. Remarkably, these are the  only order  $g_s^2$ contributions to the Meissner mass not suppressed in the $\mu \rightarrow \infty$ limit, see  \textcite{Casalbuoni:2001ha} for more details.}

More in detail, the four-body coupling gives rise to the tadpole-like Feynman diagram, in Fig.~\ref{fig:2point}(a), contributing  $g_s^2\mu^2/(2\pi^2)$ to the Meissner mass. This  contribution is  momentum independent,  is the same for all the eight gluons and is the same one has in the CFL phase.  The three-body coupling gives rise to the charm-like Feynman diagram, in Fig.~\ref{fig:2point}(b), contributing  \be i\Pi_{ab}^{\mu\nu}(x,y)=\,-\,{\rm Tr}[\,i\, S(x,y)\,i\,
H_{a}^{\mu}\,i\, S(y,x)\,i\, H_{b}^{\nu}]\nonumber\\\label{eq:Pol}\ee 
to the polarization tensor. Here the trace is over all the internal indexes; $S(x,y)$ is the quark
propagator, and 
 \be H_{a}^{\mu}=\,i\,\frac{g_s}2\,\left(\begin{array}{cc}i
V^{\mu}\red{T_{a}}& 0
\\ 0 & -i\tilde{V}^{\mu}\red{T^{\ast}_{a}}\end{array}\right)\,,\ee
is the vertex matrix in the HDET formalism. 

\begin{figure*}[t!] \begin{center}
\subfigure{\includegraphics[width=8.1cm]{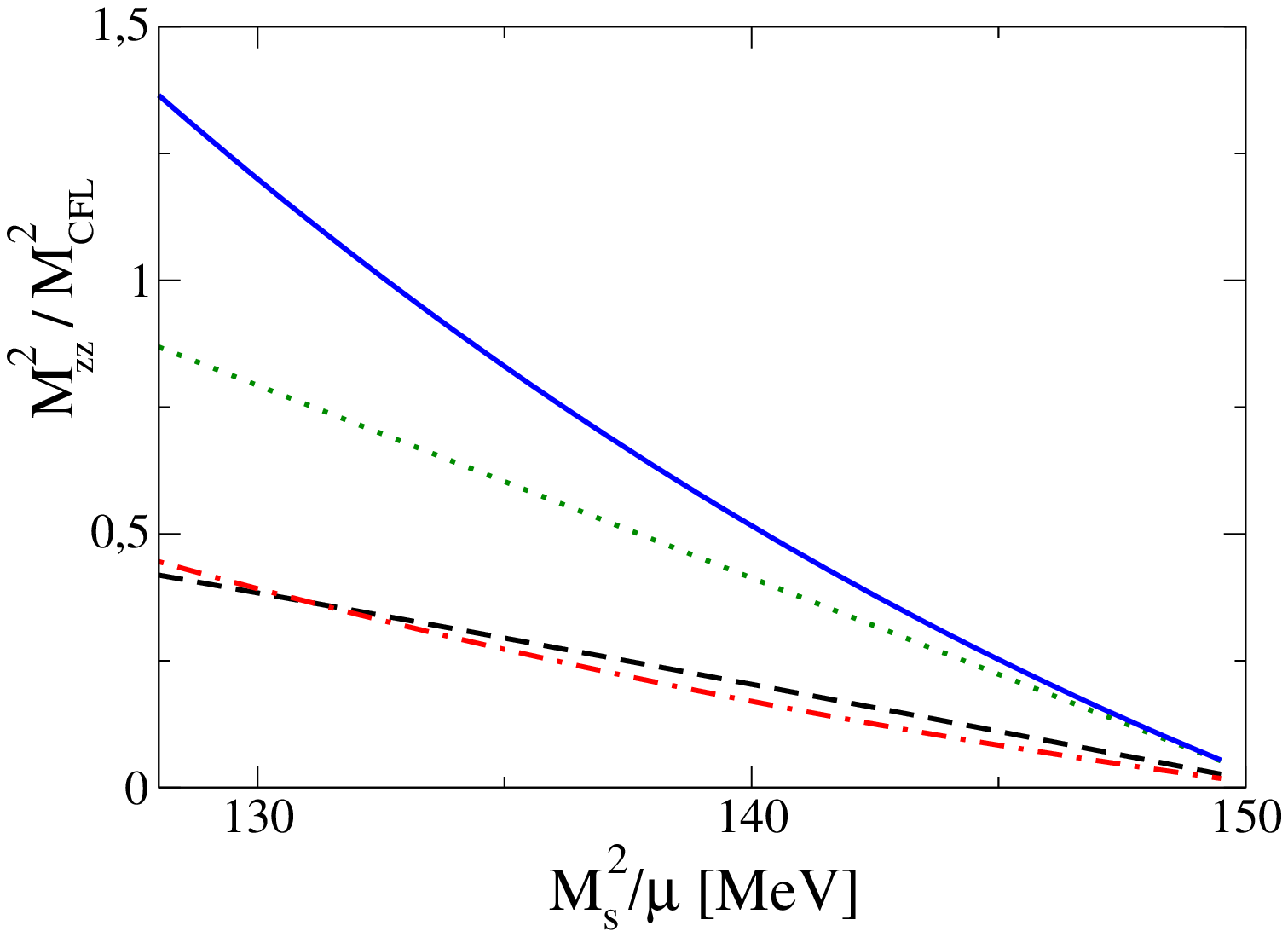}}\subfigure{\includegraphics[width=8.1cm]{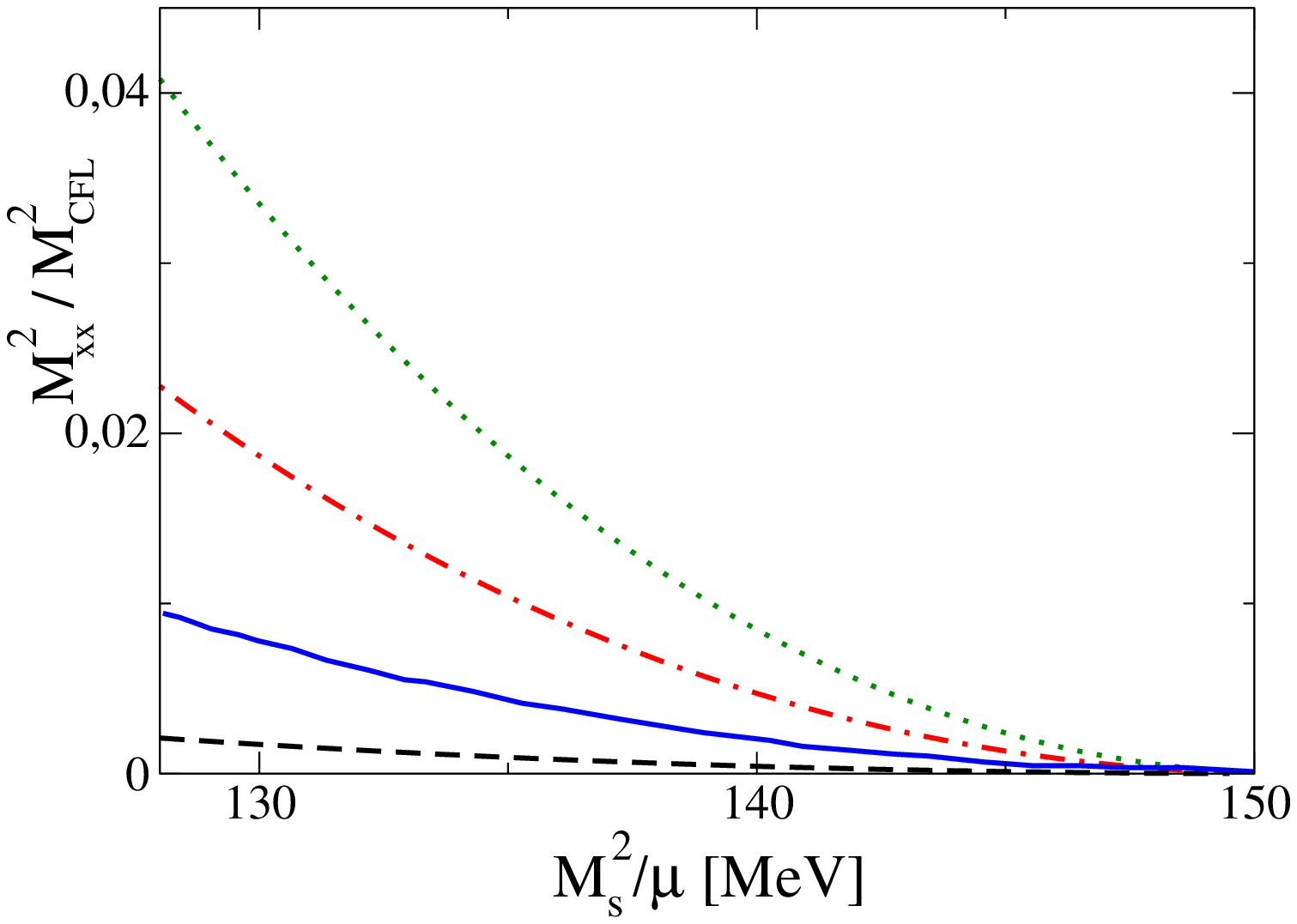}}
\end{center}
\caption{\label{masselong}{(color online) Squared Meissner masses of the gluons  $\tilde A_3$ (solid blue line), $A_6$ (dotted green line), $\tilde A_8$ (dashed black line), and $A_1$ (dot-dashed red line), in units of the CFL squared Meissner mass, vs $M_s^2 /\mu$. Left panel: longitudinal masses. Right panel: transverse masses. Adapted from \textcite{Ciminale:2006sm}.}}
\end{figure*}
The Meissner masses are defined in terms of the gluon self-energy 
in momentum space, see Eq. \eqref{eq:meissner-generic}, and in this case 
the mass matrix of gluons in the adjoint sector  $a=3,8$ is not diagonal. Therefore we introduce the two linear
combinations
\begin{equation}
\tilde A_{i3} = \cos\theta_i A_{i3} + \sin\theta_i
A_{i8}~,~~~~~\tilde A_{i8} = -\sin\theta_i A_{i3} + \cos\theta_i
A_{i8} \,, \label{eq:mixing}
\end{equation}
which are eigenstates of the mass matrix. In Eq.~\eqref{eq:mixing} the
subscript $i$ denotes the spatial component of the gluon field; it
is easily shown that the mixing angle satisfies the equation
\begin{equation}
\tan2\theta_i = \frac{2{\cal M}_{ii,38}^2}{{\cal M}_{ii,33}^2-{\cal
M}_{ii,88}^2}~;
\end{equation} the corresponding Meissner masses are the eigenvalues
of the matrix\be\left(\dd
                  \begin{array}{ccc}
                    {\cal M}^2_{ij,33} &&{\cal M}^2_{ij,38} \\ &&\\
                    {\cal M}^2_{ij,38} && {\cal M}^2_{ij,88} \\
                  \end{array}
                \right)~,
\ee
which turn out to be positive and are reported in
Fig.~\ref{masselong}.

On the left panel of Fig.~\ref{masselong} we report the longitudinal
(\textit{i.e.} $zz$) components of the squared Meissner masses against
$M_s^2/\mu$, in units of the CFL squared
Meissner mass \cite{Son:2000tu,Rischke:2000ra},
at the ${\cal O}(\Delta^4)$;
on the right
panel the results for the transverse (\textit{i.e.} $xx$) squared Meissner
masses are given. 

These results are the analogs of those obtained in the two-flavor
case by \textcite{Giannakis:2005vw}. Note that 
the transverse mass of $\tilde A_8$, although positive, is
almost zero, being three orders of magnitude smaller than the
other ones.
The conclusion is that the  three-flavor  condensate in Eq.~\eqref{eq:Delta-2pw} has no chromomagnetic instability, at least within the GL approximation. The results by \textcite{Ciminale:2006sm} have not been extended to more complicated structures, however,  for 
a generic CCSC structure   the Meissner tensor should be positive definite for small values of
$\Delta$, since it is additive with respect to different terms
of order $\Delta^2$ in the GL expansion \cite{Giannakis:2005vw}. These
considerations suggest that  the CCSC phase can be a good candidate
in removing the chromomagnetic instability of the homogeneous gapless
CSC phases of QCD. 

\subsubsection{Influence of ${\cal O}(1/\mu)$ corrections}\label{sec:1overmu}
\begin{figure*}[t]
\subfigure{\includegraphics[width=7.25cm]{figure18a.eps}}
\subfigure{\includegraphics[width=8cm]{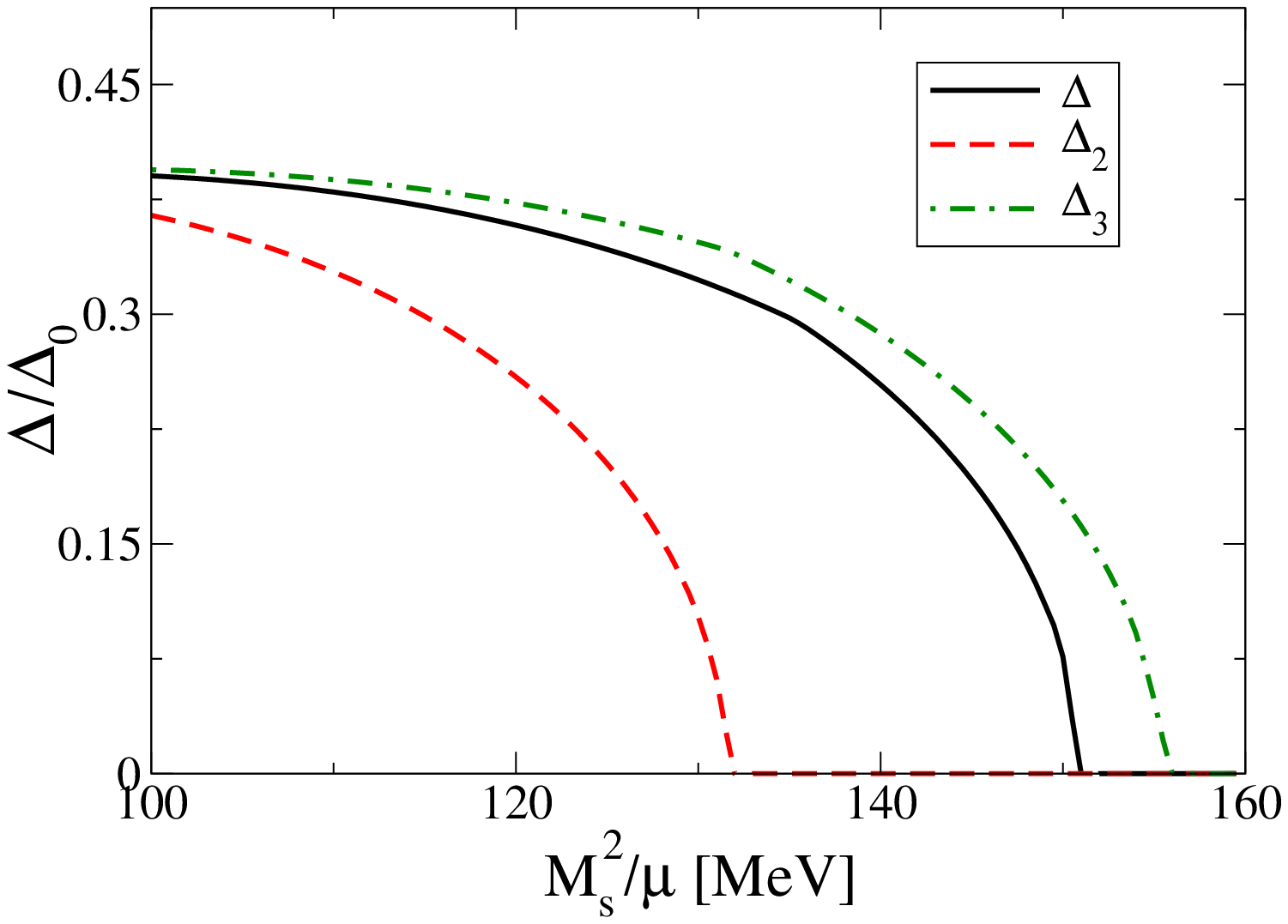}}
\caption{(color online) Left panel: Quark chemical potential difference as a function of $M_{s}^2/\mu$.
The dashed \red{red} line corresponds to   $|\delta\mu_{2}|$
and  the dash-dotted \red{green} line corresponds to $|\delta\mu_{3}|$;
their common value $M_{s}^2/(8\mu)$,
obtained neglecting ${\cal O}(1/\mu)$ corrections, corresponds to the solid black line. Right panel: gap parameters as a function of $M_{s}^2/\mu$. The solid
black line represents the  solution $\Delta=\Delta_{2}=\Delta_{3}$ obtained neglecting ${\cal O}(1/\mu)$ corrections; the
dashed \red{red} line and the dash-dotted \red{green} line represent 
respectively $\Delta_{2}$ and $\Delta_{3}$ and are obtained including  ${\cal O}(1/\mu)$ corrections.
All the gap parameters are normalized to the CFL gap
$\Delta_0=\Delta_{\rm CFL}=25$ MeV. Adapted from \textcite{Casalbuoni:2006zs}.
 \label{splitting:fig1}}\end{figure*}

Since the Meissner masses have been determined including the ${\cal O}(1/\mu)$ corrections in the HDET Lagrangian, for a consistent calculation it is necessary considering the effect of ${\cal O}(1/\mu)$ corrections on the gap parameters and the free energy \cite{Casalbuoni:2006zs}.  These corrections amount to a shift of the  strange quark Fermi momentum to lower values with respect to the corresponding chemical potential 
\begin{equation}
{p}^F_{s}\simeq
\mu_{s}-\frac{M_{s}^2}{2\mu_{s}}-\frac{1}{2\mu}
\left(\frac{M_{s}^2}{2\mu}\right)^2~, \label{eq:MsMsMs}
\end{equation} 
and thus increasing the difference between the $u$ and $s$ strange chemical potential, without affecting the 
$u$-$d$ chemical potential difference. Therefore (neglecting  corrections proportional to $\mu_3$ and $\mu_8$ which are of order $\Delta^6$)  one has that
\be\delta\mu_{2}=
\frac1 2\left[\mu_{e}-\frac{M_s^2}{2\mu}-\frac{1}{3\mu}\left
(\frac{M_s^2}{2\mu}\right)^2 \right]\,,\,\, 
\delta\mu_{3}=\frac{\mu_{e}}{2}\,,\nonumber\\ \ee
and  $|\delta\mu_2|>|\delta\mu_3|$ (neglecting ${\cal O}(1/\mu)$ corrections one would get 
$|\delta\mu_2|=|\delta\mu_3|$ as in Eq.~\eqref{eq:deltamu123}, indeed in this limit $\displaystyle
\mu_e=M_s^2/4\mu$). The results for the splitting of chemical
potentials are reported in the left panel of Fig.~\ref{splitting:fig1}, together with the result obtained
in the large $\mu$ limit.

As a consequence of the ${\cal O}(1/\mu)$ corrections,  the
two gap parameters are not equal and $\Delta_2<\Delta_3$, as shown in the right panel of Fig.
\ref{splitting:fig1}. This effect is akin to the one considered in the gCFL phase, see Fig.~\ref{fig:gapsCFL}, 
but in that case the chromomagnetic instability prevented the splitting  of the gap parameters.
The effect of the corrections considered here is to enlarge the LOFF window,  as it can be seen from the right panel of Fig. \ref{splitting:fig1}. Neglecting ${\cal O}(1/\mu)$ corrections  the crystalline phase
 was characterized by one gap, and by condensation in two
channels: $u$-$s$ and $u$-$d$. Including  ${\cal O}(1/\mu)$ corrections there is pairing in both channels
 only for small values of $M_s^2/\mu$, but the crystalline phase extends to larger values of $M_s^2/\mu$ where  only $\Delta_3\neq 0$ and therefore there is no $u$-$s$ pairing.
This phase is the two-flavor FF phase, with a chemical potential difference determined  by the ratio of the strange quark mass to  the average quark chemical potential.

\subsection{Ginzburg-Landau analysis of crystalline structures}\label{sec:3flavor-crystals}
As in the two-flavor case, more complicated crystalline structure can be considered and 
analyzed by the GL expansion \cite{Rajagopal:2006ig}. 
The most general   two-flavor CCSC condensate was given in Eq.~\eqref{1cs}
and in the three-flavor case it can be extended to 
\begin{equation}
\begin{split}
\langle
0|\psi_{iL}^\alpha\psi_{jL}^\beta|0\rangle &= - \langle
0|\psi_{iR}^\alpha\psi_{jR}^\beta|0\rangle\\ \propto
 &\sum_{I=1}^3\
   \coleps\flaeps \!\!
\sum_{\q{I}{m}\in\setq{I}{}} \Delta_{I,\q{I}{m}} e^{2i\q{I}{m}\cdot\bm x} \,.
 \label{condensate-crystal-2}\end{split}
\end{equation}
This condensate is antisymmetric in color $(\alpha,\beta)$, spin, flavor $(i,j)$ indices,
and can be viewed as well as a  generalization of the CFL ansatz. As in two-flavor quark matter, we shall  consider the simplified case 
with $\Delta_{I,\q{I}{m}}  \equiv \Delta_I$  independent of $\q{I}{m}$. The resulting condensate is the one reported in Eq.~\eqref{condensate-crystal}, where  each gap parameter has  a  periodic modulation in space corresponding to a crystalline structure, \textit{e.g.} $\Delta_2 $ is associated with a crystal due to  $u$-$s$ pairing which is described by the vectors ${\bm q}_2^m$,  where $m$ is the index which identifies the elements of the  set  $\setq{2}{}$, meaning that
the ${\bm q}_2^m$'s are the reciprocal vectors which define the crystal structure of the $u$-$s$ condensate. 

Even within this simplified case, self-consistent computations
based on Eq.~\eqref{condensate-crystal} are very complicated.
As a consequence, several additional assumptions and simplifications have been used by \textcite{Rajagopal:2006ig}. 
As discussed in the previous section,  the pairing between $d$ and $s$ quarks can been neglected; this is a very reasonable approximation, since the difference of chemical potentials between $d$ and $s$ quarks is approximately twice the imbalance of chemical potentials between $u$-$s$ and $u$-$d$. Therefore, $\Delta_1 = 0$, and  one can take $\Delta_2=\Delta_3$, neglecting  ${\cal O}(1/\mu)$ corrections.
As a further  simplification,   only  crystalline structures with wave vectors $\{\bm q_2^m\}$ and $\{\bm q_3^m\}$ with equal modulus are considered. This means that any  phase is determined by one gap parameter,  by the modulus of the total pair momentum, $q$, and by the sets of unit vectors which determine the two crystalline structures. The corresponding pairing ansatz simplifies to 
\begin{equation}\begin{split}
\langle
0|\psi_{iL}^\alpha\psi_{jL}^\beta|0\rangle& = - \langle
0|\psi_{iR}^\alpha\psi_{jR}^\beta|0\rangle\\ \propto&
 \Delta \sum_{I=2}^3\
  \coleps\flaeps \
\sum_{\bm n_{I}^{m}\in \{\bm n_I \}} e^{2i q \bm n_I^m\cdot\bm x} \,.
 \label{condensate-crystal-3}
\end{split}
\end{equation}

Finally, \textcite{Rajagopal:2006ig}  consider only structures 
which are exchange symmetric, which means that $\{\bm n_2\}$ and $\{\bm n_3\}$ can be exchanged
by some combination of rotations and reflections applied simultaneously to all the wave vectors.\\

The thermodynamic potential within the GL expansion up to the 
sextic order in the gap parameters is given by a generalization of Eq.~\eqref{eq:OGL}, see also Eqs.~\eqref{eq:OmegaDelta11} and \eqref{eq:OGL2PWs}, that is
\begin{widetext}
\begin{eqnarray}
\Omega&=& \Omega_n + \left[P\alpha(|\Delta_2|^2 + |\Delta_3|^2) 
 + \frac{\beta}{2}(|\Delta_2|^4 + |\Delta_3|^4)
 + \frac{\gamma}{3}(|\Delta_2|^6 + |\Delta_3|^6)\right] \nonumber \\
  &&+ 
  \left[\frac{\beta_{23}}{2}|\Delta_2|^2 |\Delta_3|^2 
     +  \frac{\gamma_{233}}{3}|\Delta_2|^2 |\Delta_3|^4 + \frac{\gamma_{322}}{3}|\Delta_2|^4 |\Delta_3|^2\right] + {\cal O}(\Delta^8)~,
     \label{eq:GL3fzzz}
\end{eqnarray}
\end{widetext}
where  we have explicitly diversified $\Delta_2$ and $\Delta_3$ as a means  to
show the interaction between the condensates in the $u$-$d$ and the $u$-$s$
channels. The coefficients $\alpha$,
$\beta$ and $\gamma$ in the first square bracket are the same coefficients computed for the two-flavor case by \textcite{Bowers:2002xr}.  The terms in the second square bracket are peculiar of the three-flavor case.
In the numerical computations the free energy
is minimized with respect to 
 $\Delta$ and for any crystalline structure considered it is found that the thermodynamic potential is always bounded from below. Therefore,  it is possible  to compute the numerical value of $\Delta$ and the value of the free energy density at the minimum.

Among the many crystalline structures considered by \textcite{Rajagopal:2006ig}, two of them 
have been found to have the lowest free energy: they are called the CubeX and the
2Cube45z structures. \red{In the CubeX crystal  each set contains four vectors, that is $\{\bm n_2\} = \{\bm n_2^1,\bm n_2^2, \bm n_2^3,\bm n_2^4\}$ and $\{\bm n_3\} = \{\bm n_3^1,\bm n_3^2, \bm n_3^3,\bm n_3^4\} $ with 
\bea
&\bm n_2^1=\sqrt{\frac{1}3}(1,1,1) =  -\bm n_2^2\,, \,& \bm n_2^3=\sqrt{\frac{1}3}(-1,-1,1) =  -\bm n_2^4 \,,\nonumber \\ 
&\bm n_3^1=\sqrt{\frac{1}3}(-1,1,1) =  -\bm n_3^2\,, \, & \bm n_3^3=\sqrt{\frac{1}3}(1,-1,1) =  -\bm n_3^4 \,.\nonumber
\eea
Thus, the vectors of each set point to the vertices of a rectangle}; the eight vectors together point toward
the vertices of a cube. In the 2Cube45z crystal, $\{\bm n_2\}$ and $\{\bm n_3\}$ each contains
eight wave vectors, pointing to the corners of a cube; the two cubes are rotated 
by $45$ degrees about the $z-$axis. The computation by \textcite{Rajagopal:2006ig} shows that the CubeX crystal is favored over all the considered structures in the range
\begin{equation}
2.9\Delta_{\rm CFL} < \frac{M_s^2}{\mu} < 6.4\Delta_{\rm CFL}~; \label{eq:CubeXhhh}
\end{equation}
similarly for the 2Cube45z crystal it is found that it is energetically favored for
\begin{equation}
6.4\Delta_{\rm CFL} < \frac{M_s^2}{\mu} < 10.4\Delta_{\rm CFL}~. \label{eq:2Cube45Zhhh}
\end{equation}
Putting together the above results, the CCSC phase is
favored with respect to the homogeneous CFL, gCFL and unpaired phases in the range
\begin{equation}
2.9\Delta_{\rm CFL} < \frac{M_s^2}{\mu} < 10.4\Delta_{\rm CFL}~. \label{eq:CUBErange}
\end{equation}
The above condition can be translated to a condition on $\mu$ 
only if the constituent strange quark mass as a function of $\mu$ is known.
This topic will be discussed in the next section.

We remark that from the quantitative point of view, the result in Eq.~\eqref{eq:CUBErange} 
should be taken with care. The GL expansion is reliable
only if the gap parameter $\Delta$ is small, compared to $\delta\mu$. On the
other hand, the numerical results by \textcite{Rajagopal:2006ig} show that
within this expansion, the ratio $\Delta/\delta\mu$ turns out to be of order one. 
This clearly signals that the results lie beyond the validity of the expansion itself.
Moreover, the coefficients in the expansion in Eq.~\eqref{eq:GL3fzzz} depend on the 
microscopic model. Therefore, Eq.~\eqref{eq:CUBErange} is certainly model dependent.
Nevertheless, the qualitative picture which arises in the work by \textcite{Rajagopal:2006ig}
seems to be quite robust: the crystalline superconductor has lower free energy
with respect to the single-plane-wave state. This conclusion seems very reasonable and
model independent: in fact, the phase space
for pairing in the case of CubeX and 2Cube45z is larger than the one corresponding to
the single-plane-wave structure; this suggests that the free energy gain for multiple plane waves
is larger than the one obtained for the single-plane-wave state.

\subsubsection{LOFF window in the QCD phase diagram}\label{sec:LOFFwindow}
\red{It is important to identify, within a self-consistent computation, the
range of chemical potential in which the CCSC phase is expected to be thermodynamically favored. This would correspond to a chemical potential LOFF window in Fig. \ref{fig-phase1}.  Then, 
using an appropriate  equation of state this chemical potential window can be translated into a density window.}
So far   we have reported the LOFF window
in terms of   $M_s^2/\mu$ for the two-plane-waves structure and for the favored crystalline phases, respectively. For translating this window
to a window in $\mu$, it is necessary to consider the chemical potential dependence of both the CFL gap parameter and 
of the constituent quark masses. \red{This investigation has been started by \textcite{Ippolito:2007uz} considering vanishing temperature. In that article, the in-medium quark masses are computed
self-consistently within a NJL model and the CFL gap has been evaluated in the chiral limit  by means of Eq.~\eqref{eq:CFL-gap}, where the considered hard momentum cutoff is taken as  $\delta=\Lambda- \mu$.   \textcite{Ippolito:2007uz}   considered $\Lambda  \simeq 643 $ MeV with a corresponding  NJL coupling constant  given by $G(\Lambda) \simeq 13.2$ GeV$^{-2}$.} 

In the simple case of two  plane waves, Eq.~\eqref{eq:Delta-2pw},   
the state with $\bm n_2 = \bm n_3$ is energetically favored in the range given in Eq.~\eqref{eq:window2PW}. This range has been obtained for a fixed value of the CFL gap parameter. Considering the dependence of the CFL gap on the chemical potential given in Eq.~\eqref{eq:CFL-gap} and the  results for the constituent strange quark mass $M_s(\mu)$ computed by \textcite{Ippolito:2007uz},  that range can be  transformed  in
\begin{equation}
4.8\Delta_{\rm CFL}(\mu) \lesssim \frac{M_s(\mu)^2}{\mu} \lesssim 7.6\Delta_{\rm CFL}(\mu)\,,\label{eq:window1}
\end{equation}
and then to the quark chemical potential window 
\be 467~\text{MeV} \lesssim \mu\lesssim 488~\text{MeV} ~.\label{una}\ee Therefore, there
exists a small but finite window in $\mu$ in which the structure \eqref{eq:Delta-2pw}  has a lower free energy
than both gCFL quark matter and  normal quark matter. 

One might expect that such a small window is considerably enlarged in more complicated CCSC phases. As discussed in the previous section, the analysis by \textcite{Rajagopal:2006ig} shows
that more complicated crystalline structures are favored in the
interval reported in Eq.~\eqref{eq:CUBErange},
which using the self-consistent treatment by  \textcite{Ippolito:2007uz} transforms  in 
\be
442\,~\text{MeV}\lesssim \mu\lesssim 515~\text{MeV} ~.\label{cubo}\ee  
\red{This result is certainly model dependent, however it shows that the actual extension of the LOFF window in Fig. \ref{fig-phase1} might not be very large. Therefore, it is likely that the CCSC phase  occupies only a fraction of the quark core of hybrid compact stars. We discuss this topic in Sec.~\ref{sec:mass-radius}.}

\subsection{Shear modulus and Nambu-Goldstone modes}\label{sec:shear}
The three-flavor  CCSC  phase has several low-energy excitations;
besides gapless quasifermions, there are the bosonic modes related to  the spontaneous
breaking of translational symmetry, the three phonons, of the chiral symmetry,
the eight  pseudoscalar NGBs, and to the breaking of  $U(1)_B$ symmetry, the so-called $H$ phonon.  The phonon Lagrangian for  the two-flavor CCSC phase has been discussed in \ref{phonons}; here we extend the results to the  three-flavor case \cite{Mannarelli:2007bs} deriving the low-energy coefficients of the GL expansion from a NJL-like model.  Regarding the eight  pseudoscalar NGBs, we  briefly discuss the results by \textcite{Anglani:2007aa} obtained for the three-flavor CCSC phase. 

\subsubsection{Phonon effective action and shear modulus}\label{sec:shear1}

According to  the basic theory of elastic media \cite{Landau:Elastic},  the elastic moduli  are related to the potential energy cost of small deformations of the crystal. Therefore, the  evaluation of  the shear modulus requires the knowledge of the low-energy Lagrangian for the displacement fields
\cite{Mannarelli:2007bs}.   Since the three condensates that characterize the crystalline phase can oscillate independently,  three are three sets of displacement fields $\bm u_I(x)$. Thus, we can extend the definition of the phonon fields in Eq.~\eqref{phon2sc} by
\begin{equation}
\Delta_I(\rr)\rightarrow
\Delta_I^u(x)=\Delta_I(\rr-\vu_I(x))~,\label{cond phon1}
\end{equation}
and the  Lagrangian  that includes fluctuation on the top of the mean field solution can be written
as
\begin{equation}
{\cal L} = \ha \bar{\chi} \left( \begin{array}{cc}
i\cross{\partial}+\cross{\mu} & \Delta^u(x)    \\
\bar{\Delta}^u(x) &   (i\cross{\partial}-\cross{\mu})^T
\end{array} \right) \chi \,+ \frac{1}{16G}
{\rm tr}\bigl((\bar{\Delta}^u)^T\Delta^u\bigr)\,,
\label{fullLagrangian}
\end{equation}
where ${\rm tr}$ represents the trace over color, flavor and Dirac indices.

To find the low-energy effective action describing the phonons
and the gapless fermionic excitations, one should integrate out the
high-energy fermion fields. The procedure is detailed in the work by  \textcite{Mannarelli:2007bs}, where the effective action is derived starting from a NJL-like microscopic model. The final form of the effective action of the phonon fields is given by
\begin{equation}\begin{split}
i{\cal{S}}[\vu]=\log(Z[\vu])= & i\intspace{x} \Bigl[
 \frac{1}{16G}{\rm tr}
\bigl((\bar{\Delta}^u)^T\Delta^u\bigr)\Bigr] \\ &+ \ha {\rm Tr}_{{\rm
ng}}\log\left(S^{-1}\right)\label{Z2}\;,\end{split}
\end{equation}
where ${\rm Tr}_{{\rm ng}}$ stands for the trace over the
Nambu-Gorkov, color, flavor and Dirac indices and a further trace over a set of functions on space-time containing all
energy modes. The full inverse propagator,  is given by
\begin{equation}
S^{-1}=\left( \begin{array}{cc}
i\cross{\partial}+\cross{\mu} & \Delta^u(x)    \\
\bar{\Delta}^u(x) &   (i\cross{\partial}-\cross{\mu})^T
\end{array} \right)\label{inv prop}\;,
\end{equation}
and it includes interactions of the phonon fields.

For the crystal structures CubeX and 2Cube45z, the full inverse
propagator cannot be inverted, so a GL expansion has
been performed in order to obtain the effective action for the
phonon field, first separating the full inverse propagator into the
free part, $S^{-1}_0$, and a part containing the condensate,
$\Sigma$, as follows:
${S^{-1}}=S^{-1}_0+\Sigma$,
where
\begin{equation}
{S^{-1}_0}=\left( \begin{array}{cc}
i\cross{\partial}+\cross{\mu} & 0    \\
0 &   (i\cross{\partial}-\cross{\mu})^T
\end{array} \right)\,,
\end{equation}
and
\begin{equation}
\Sigma=\left( \begin{array}{cc}
0 & \Delta^u(\rr)    \\
\bar{\Delta}^u(\rr) &   0
\end{array} \right)\, .
\label{split propagator}
\end{equation}
Then, one can expand the term $\log(S^{-1})$ that appears on the
right-hand side of Eq.~(\ref{Z2}) as

\begin{equation}\begin{split}
{\rm Tr}_{{\rm ng}}\bigl(\log({S^{-1}_0}+\Sigma\bigr))=&{\rm
Tr}_{{\rm ng}}\bigl(\log{S^{-1}_0}\bigr) +{\rm Tr}_{{\rm
ng}}\bigl({S_0}\Sigma\bigr)\\ &-\ha {\rm Tr}_{{\rm
ng}}\bigl({S_0}\Sigma\bigr)^2 +...\label{log expansion}\,.\end{split}
\end{equation}
The ${\rm Tr}_{{\rm ng}}\bigl(\log{S^{-1}_0}\bigr)$ term  is related
to the free energy of unpaired (normal) quark matter. Furthermore,
it is easily found that only even powers of $\bigl(S_0\Sigma\bigr)$
contribute to the trace over Nambu-Gorkov indices. 
Expanding the effective action in powers of $\phi_I^m = 2 \bm q^m \cdot \bm u_I$ and including the first nontrivial quadratic term,
   which is calculated to order $\Delta^2$, one obtains 
\begin{equation}
{\cal{S}}^{\phi^2\Delta^2}=\sum_I
\sum_{\q{I}{m}}\fourier{k}\fiak{I}{m}{k}\fiak{I}{m}{-k}
\Delta_I^*\Delta_I\piak{I}{m}{k}\label{Seff1}\;,
\end{equation}
where $k=k_2-k_1$ is the four momentum of the phonon and
\begin{equation}
\begin{split}
\piak{I}{m}{k} = i\sum_{{j\neq k}\atop{\neq I}}&\fourier{p}\\ \times&{\rm
Tr}\Biggl[
\frac{1}{(\cross{p}+\cross{q}_I^m+\cross{k}_1+\cross{\mu}_j)
(\cross{p}-\cross{q}_I^m+\cross{k}_2-\cross{\mu}_k)} \\
-&\frac{1}{(\cross{p}+\cross{q}_I^m+\cross{\mu}_j)
(\cross{p}-\cross{q}_I^m-\cross{\mu}_k)}\Biggr]\label{eq:piak}\;,
\end{split}
\end{equation}
where the trace is over Dirac indices. 
Integrating the expression above one obtains 
\begin{widetext}
\begin{equation}
{\cal{S}}^{\Delta^2}[{\bm u}]=\ha\intspace{x}
 \sum_I \kappa_I \sum_{\bm n_I^m}\left[
 \partial_0(\bm n_I^m \cdot \vu_I)\partial_0(\bm n_I^m \cdot\vu_I) -
 (\bm n_I^m\cdot{\bm \partial})(\bm n_I^m\cdot\vu_I) (\bm n_I^m\cdot{\bm\partial})(\bm n_I^m\cdot\vu_I)
 \right]\label{eq:action_ph}\;,
\end{equation}
\end{widetext}
where 
\begin{equation}
\kappa_I\equiv
\frac{2\mu^2|\Delta_I|^2}{\pi^2(1-z_q^2)}\;,
\label{lambdapw}
\end{equation}
with $z_q $ defined in Eq.~\eqref{eq:zetaq}. This action generalizes Eq.~\eqref{eq:lagrangian_GL_ph} to the three-flavor case and is  the LO low-energy effective action valid for phonons in any CCSC phase, indeed it has a general expression depending on the $\bm n_I^m$ vectors. Only terms of the second order in derivatives and in
the phonon fields are reported, but higher order terms can be obtained in a similar way. 

In Eq.~\eqref{eq:action_ph} there are no terms
that ``mix'' the different $\vu_I(k)$, meaning that at this order the displacement  of the various crystals can be considered separately. This follows
from the fact that the Lagrangian conserves
particle number for every flavor of quarks, which corresponds to
symmetry under independent global phase rotations of quark fields of
the three flavors, meaning independent phase
rotations of the three  $\Delta_I$'s. The effective action should be
invariant under these rotations and hence $\Delta_I$ can only occur
in the combination $\Delta_I^*\Delta_I$. Since the effective action has been obtained up to the second order in the gap parameters $\Delta_I$'s   the mixing terms do not appear; mixing between different fluctuations can only appear in higher order terms, \textit{e.g.} by terms like
$\mu^2|\Delta_I\Delta_J|^2 \partial\vu_I\partial\vu_J/\dm{}^2$.

The coefficients $\kappa_I$ determine the potential energy cost of a fluctuation and therefore  are related to the shear modulus.  In particular, \textcite{Mannarelli:2007bs} found  that for the the two favored  structures, 2Cube45z and CubeX, the shear modulus is a $3 \times 3$ non diagonal matrix in coordinate space with entries proportional to
\begin{equation}
\nu_{\rm CQM} = 2.47\, {\rm MeV}/{\rm fm}^3
\left(\frac{\Delta}{10~{\rm MeV}}\right)^2 \left(\frac{\mu}{400~\rm{MeV}}\right)^2\,.
\label{nunumerical}
\end{equation}
Considering typical values of the quark chemical potential in the range
\begin{equation}
350~{\rm MeV}<\mu<500~{\rm MeV}\,,
\label{murange}
\end{equation}
and 
\begin{equation}
5~{\rm MeV} < \Delta < 25~{\rm MeV}\ ,
\label{Deltarange}
\end{equation}
one has that 
\begin{equation}
0.47~{\rm MeV}/{\rm fm}^3 < \nu_{\rm CQM} < 24~{\rm MeV}/{\rm fm}^3\ .
\label{nuCQMrange}
\end{equation}

The standard neutron star crust, which is a conventional
crystal of positively charged ions immersed in a fluid of
electrons (and, at sufficient depth, a fluid of neutrons) has a 
shear modulus approximately given by \cite{Strohmayer, Mannarelli:2007bs}
\begin{equation}
0.092~{\rm keV}/{\rm fm}^3 < \nu_{\rm NM}<23~{\rm keV}/{\rm fm}^3 \,,
\label{nuNMrange}
\end{equation}
thus, the crystalline quark matter
is more rigid than the conventional neutron star crust by at least a factor of 20-1000.
Note that the three-flavor CCSC phase is also a superfluid,  by
picking a phase its order parameter does indeed break the quark-number $U(1)_B$
symmetry spontaneously. These results demonstrate that this
superfluid phase of matter is at the same time a rigid solid and a superfluid.

\subsubsection{Goldstone modes}\label{sec:counting}

The superfluid property of the CCSC phase derives from the existence of a massless NGB associated with the breaking of $U(1)_B$. Actually, the crystalline condensate  breaks the same global (and local) symmetries of the 
 CFL phase \cite{Alford:1998mk},  leaving unbroken a global symmetry group: $SU(3)_{c+L+R} \times Z_2$, see 
 Eq.~\eqref{eq:breakingCFL}. Therefore, there are nine NGBs  due to the spontaneous breaking of  $U(1)_B$ and of the chiral symmetry, but only 
the $U(1)_B$ boson ($H-$phonon) is massless. Indeed, the  pseudo-scalars associated with chiral symmetry have
 mass, because this group is explicitly broken by quark mass terms and by chemical potential differences.  

Alike the phonons described in the previous section, the NGBs discussed here describe  the fluctuations of the condensate, see for example \textcite{Eguchi:1976iz}. The pseudo-NGBs are related to fluctuations in flavor space, while the  $H-$phonon, $\varphi$, describes fluctuations in baryonic number. The form of the low-energy Lagrangian describing both phonons and the $H-$phonon can be determined from symmetry arguments alone following the  discussion by \textcite{Leutwyler:1996er,Son:2005ak}, see also \textcite{Cirigliano:2011tj}. Here we shall focus on the microscopic  derivation of the     $H-$phonon Lagrangian and introduce it by  means of the transformation
$\psi\rightarrow U^\dagger \psi$ with $U =
\exp\left\{i\varphi/f_\varphi\right\}$, where $f_\varphi$ is its decay constant.  Taking into account the unitary rotations, the HDET Lagrangian takes the form \cite{Anglani:2007aa}
\begin{widetext}
\begin{equation}
{\cal L}=\frac{1}{2}\int\!\frac{d \Omega}{4\pi}~\chi^\dagger_A
\left(\begin{array}{cc}
        (V\cdot\ell +\delta\mu_A(\bm v))~\delta_{AB} & -\Xi_{BA}^\star \\
        -\Xi_{AB} & (V\cdot\ell -\delta\mu_A(-\bm v))~\delta_{AB}
      \end{array}
\right)\chi_B ~,\label{eq:Lagr1bis}
\end{equation}
\end{widetext}
where all the quantities have been
defined in Sec. \ref{sec:NG-HDET}, see Eq.\eqref{inverse},  except $\Xi_{AB}$, which is given
by
\begin{equation}
\Xi_{AB}=\Delta_I^ \star({\bm x}) \Tr[\epsilon_I (F_A U^\dagger)^T
\varepsilon^I F_B U^\dagger]~,
\end{equation}
with $\epsilon_I \equiv \flaeps $ and $\varepsilon^I \equiv \coleps.$

\red{Expanding the Lagrangian in the scalar field one derives  three-body and  four-body interaction vertices. 
At the leading order in $\Delta$ these couplings provide the dominant contribution to the $H-$phonon self-energy.
We have encountered similar interaction terms in  the evaluation of the Meissner masses  of gluons discussed in Sec.~\ref{sec:3stability}. As in that case
the two leading contributions are given by the  tadpole-like  and charm-like Feynman diagrams reported respectively in Fig.~\ref{fig:2point}(a)  and Fig.~\ref{fig:2point}(b), but with the gluon lines replaced by $H-$phonon lines.  Technically the  calculation is  very similar to the one sketched in  Sec.~\ref{sec:3stability} }and the result is that the charm-like diagram contributes to the self-energy by \begin{equation}
{\cal S}_{\rm c.l.} = -i
\frac{2}{f_\varphi^2}\sum_{I=2}^3\Delta_I^2~\sum_{\bm q_I^m} \int\!\frac{d^4
k}{(2\pi)^4}~\varphi(-k) \varphi(k) ~{\cal P}_{I}^m(k_0,{\bm
k})~,\label{eq:se1}
\end{equation}
with $k = (k_0,{\bm k})$ and ${\cal P}_{I}^m(k_0,{\bm k})$ corresponds to the HDET version of
Eq.~\eqref{eq:piak},
\begin{widetext}
\begin{eqnarray}
{\cal P}_{I}^m(k_0,{\bm k}) &=& -2 \mu^2\int\!\frac{d\Omega}{4\pi}
\int\!\frac{d p_0 d\xi}{(2\pi)^4} \left[\frac{1}{(\tilde
V\cdot\ell+\delta\mu-{\bm q}_I^m\cdot{\bm v})[V\cdot(\ell + k) + \delta\mu-{\bm q}_I^m\cdot{\bm v}] } \right. \nonumber \\
&& ~~~~~~~~~~~~~~~~~~ + \left.\frac{1}{( V\cdot\ell - \delta\mu-{\bm
q}_I^m\cdot{\bm v})[\tilde V\cdot(\ell + k)  -
\delta\mu-{\bm q}_I^m\cdot{\bm v}] } \right] + \delta\mu \rightarrow -\delta\mu\,. \label{eq:PaI}
\end{eqnarray}
\end{widetext}
The tadpole-like contribution is given by
\begin{equation}
{\cal S}_{\rm t.l.} = i
\frac{2}{f_\varphi^2}\sum_{I=2}^3\Delta_I^2~\sum_{\bm q_I^m} \int\!\frac{d^4
k}{(2\pi)^4}\varphi(-k) \varphi(k) {\cal P}_{I}^m(k_0=0,{\bm
k}=0).\label{eq:tad1}
\end{equation}

 We do not provide further details of the calculation which can be found in the work by \textcite{Anglani:2007aa}, and we only quote the final result valid at small  momenta:
\begin{equation}
{\cal L}_\varphi(k) =  \frac{1}{2}\varphi(-k)\left[k_0^2 {\cal I}_0 - k_i k_j
V_{ij}\right]\varphi(k)~,\label{eq:Trallalla}
\end{equation}
where
\begin{eqnarray}
{\cal I}_0& =& -\frac{\mu^2}{\pi^2 f_\varphi^2}  \sum_{I=2}^3 \Delta_I^2\sum_{\bm q_I^m}~\Re e\nonumber\\ && \times
\int\!\frac{d \Omega}{4\pi}\frac{1}{(\delta\mu - {\bm
q}_I^m\cdot {\bm v} + i0^+)^2}+ (\delta\mu \rightarrow -
\delta\mu)~,\ \nonumber \\
\\
V_{ij} &=& -\frac{\mu^2}{\pi^2 f_\varphi^2}\sum_{I=2}^3 \Delta_I^2  \sum_{\bm q_I^m}~\Re
e\nonumber \\ && \times \int\!\frac{d \Omega}{4\pi}\frac{\bm v_i \bm v_j}{(\delta\mu - {\bm
q}_I^m\cdot {\bm v} + i0^+)^2}+ (\delta\mu \rightarrow
-\delta\mu)~.\nonumber \\ \label{eq:cici}
\end{eqnarray}

Specializing these  results to the  case  $\Delta_2 = \Delta_3 = \Delta$, $|{\bm
q}_2^a| = |{\bm q}_3^a| = q$, and requiring canonical normalization of the
Lagrangian in Eq.~\eqref{eq:Trallalla}, leads to
\begin{equation}
f_\varphi^2 = \frac{4 P
\mu^2}{\pi^2}\frac{z_q^2 \Delta^2}{\delta\mu^2(1-z_q^2)}~,\label{eq:fSqPHI}
\end{equation}
where $P$ is the number of plane waves comprising each crystal (we assume that the number of plane waves in the two crystals are equal). Upon substituting this result in Eq.~\eqref{eq:cici}, and specializing it to   $P=1$, we find $V_{ij}= [\text{diag}(0,0,1)]_{ij}$, where we have chosen ${\bm q}_2$ and ${\bm q}_3$  along the positive $z$-axis. For the two cubic
structures corresponding to the values $P=4$ (CubeX) and $P=8$ (2Cube45z) we find
$V_{ij} = \delta_{ij}/3$, \textit{i.e.} the velocity is isotropic and has the value $1/\sqrt 3$.

The extension of the above procedure to the octet of pseudo-NGBs is
straightforward, see \textcite{Anglani:2007aa}.
Here we just recall  that one of the important result is that  the squared masses of the
pseudo-NGBs are always positive, thus  kaon condensation
for the considered CCSC phases is excluded, at least at the order
$\Delta^2$.

\red{As we shall see in the next sections, it is of astrophysical interest the study of rotating quark matter.  In particular, in Sec.~\ref{sec:glitches} we shall see how superfluid vortices may influence the spinning evolution of compact stars. The response of the crystalline phase to rotation should indeed result in the formation of vortices. The naive expectation is the formation of $U(1)_B$ vortices \cite{Iida:2002ev, Forbes:2001gj}, but nonabelian vortices \cite{Auzzi:2003fs, Hanany:2003hp} might be energetically favored \cite{Balachandran:2005ev, Eto:2009tr}. }

\section{Astrophysics}
\label{sec:Astrophysics}
The artificial creation of the low-temperature and high-density conditions appropriate for testing the properties of color superconductors is one of the challenging aims of high-energy experiments \cite{Klahn:2012uq}. \red{To date, however, the unique \textit{laboratory} in which  these extreme conditions can be realized is the core of a compact stellar object (CSO).}\footnote{
Note that even if  the relevant densities and  low temperatures conditions were reached in a terrestrial  laboratory, the conditions realized in a CSO are different from those produced in an accelerator, because in the former case  quark matter is long-lived, charge neutral and in $\beta$ equilibrium.}
In the inner core of a CSO the extremely high densities and low temperatures may favor the transition from  nuclear to quark matter \cite{Ivanenko1965, Ivanenko1969, Collins, Baym} and in turn, to the color superconducting (CSC) phase. Indeed, if matter is compressed at densities about a factor $5$ larger than the density of an ordinary nucleus, a simple geometrical reasoning suggests that   baryons are likely to lose their identity and dissolve into deconfined quarks \cite{Weber-book}. In this case  compact (hybrid) stars featuring quark cores  would exist providing a window on the properties of QCD at high baryon densities. Assuming that deconfined matter is present, it should be in a CSC phase because its critical temperature (in weak coupling) is given  by $T_{c} \simeq 0.57 \Delta$, and, as we have seen in the previous sections   $\Delta\sim 5 -100$ MeV, although lower values of the gap parameter, of the order of the keV up to few MeV, can be realized in the spin-1 single flavor pairing phase \cite{Alford:2002rz, Alford:1997zt, Buballa:2002wy, Schafer:2000tw, Schmitt:2004et, Schmitt:2002sc}. In any case,  for the greatest part of the compact star lifetime, the temperature is much lower than this critical temperature and  the CSC state is thermodynamically favored.

\red{Clearly, the basic question is whether gravity   in the interior of CSOs  is able to compress matter to such extreme densities. This question is still open, but some progress has been done in the last years.} In particular, recent astronomical observation of very massive CSOs \cite{Demorest:2010bx} seem to disfavor this possibility \cite{Logoteta:2012ms}, but the results depend on the poorly known equation of state (EoS) of matter at high density, and the possibility that hybrid stars of about  $2 M_\odot$ have a quark matter core \cite{Alford:2004pf, Bonanno:2011ch} or a  crystalline color superconducting (CCSC) core  \cite{Ippolito:2007hn} cannot be excluded.  

Besides the stellar mass, the presence of deconfined quark matter  in the interior of a CSO can be probed by a number of astrophysical observables linked to the microscopic properties of CSC matter.  In particular, the 
 mass-radius relation,  various transport properties, the $r$-mode evolution, glitches and very strong magnetic fields are under scrutiny for ruling in or out CSC matter. Unfortunately, the available observational data do not allow us to infer in a unique way  the internal structures of CSOs, but  the investigation of the astrophysical signatures can help to pave  the path connecting theoretical models to astronomical observations.
In the following we shall present a brief state-of-the-art of some astrophysical implications related to the presence of  CCSC matter in CSOs.

\subsection{Gravitational waves}\label{sec:gwaves}
As reported in Section \ref{sec:shear}  the CCSC phase  is extremely rigid, with  a shear modulus larger than the  one of standard neutron star crust  by at least a factor 20. The existence of such a rigid  core within compact stars may have a variety of observable consequences. A  large deformation of the core,   initially produced for example by magnetic fields not aligned with the rotation axis, can be sustained  by the rigid structure provided by the crystalline condensate. If the deformation of the  core is in a shape that has a nonzero quadrupole moment and if this axis-asymmetric mass distribution is not compensated by the overlying nuclear envelope,  the spinning CSO  would efficiently emit  gravitational waves  \cite{Lin:2007rz, Haskell:2007zz}. The size of the distortion of the mass distribution can be measured by the equatorial ellipticity
\be
\epsilon =\frac{I_{xx} - I_{yy}}{I_{zz}} \,,
\ee
which is defined in terms of the star principal moments of inertia. Standard neutron stars are expected to have maximum ellipticity of the order of $\epsilon_{\rm max} \approx 10^{-6} $, meaning that this is the largest deformation sustainable by the crust before breaking.  For the three-flavor crystalline phases considered in Section \ref{sec:shear}, deformed to the maximum extent allowed by the shear modulus,  the maximum equatorial ellipticity sustainable  could be as large as $\epsilon_{\rm max}  \approx10^{-2}$ \cite{Lin:2007rz, Haskell:2007zz}. Different models of exotic stars composed of quark-clusters \cite{Xu:2003xe} have a magnitude of the maximum ellipticity which is smaller than the CCSC phase by two orders of magnitude \cite{Owen:2005fn}.

\begin{figure*}[t]
\begin{center}
\subfigure{
\includegraphics[width=8.cm]{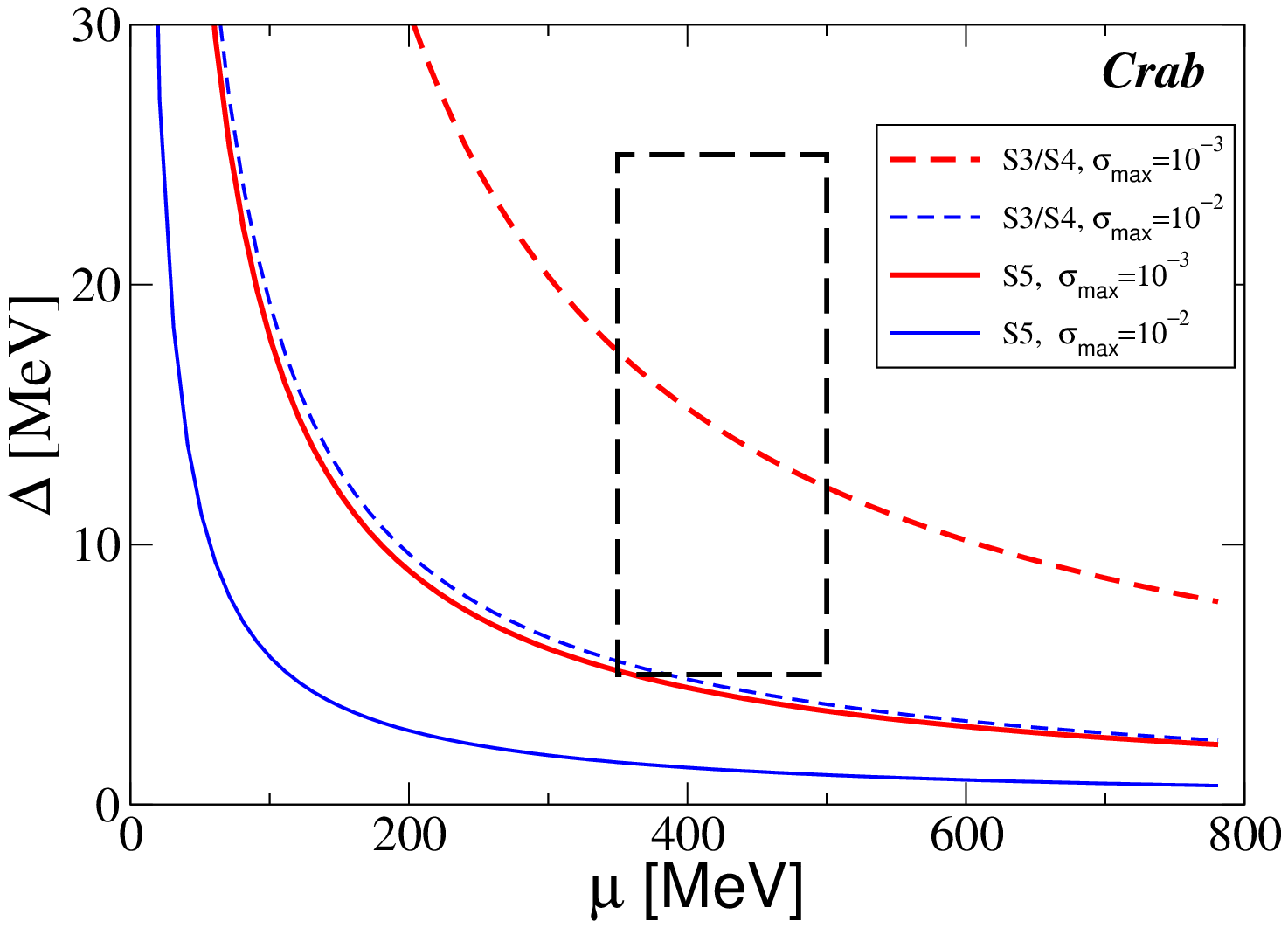}} \subfigure{\includegraphics[width=8.cm]{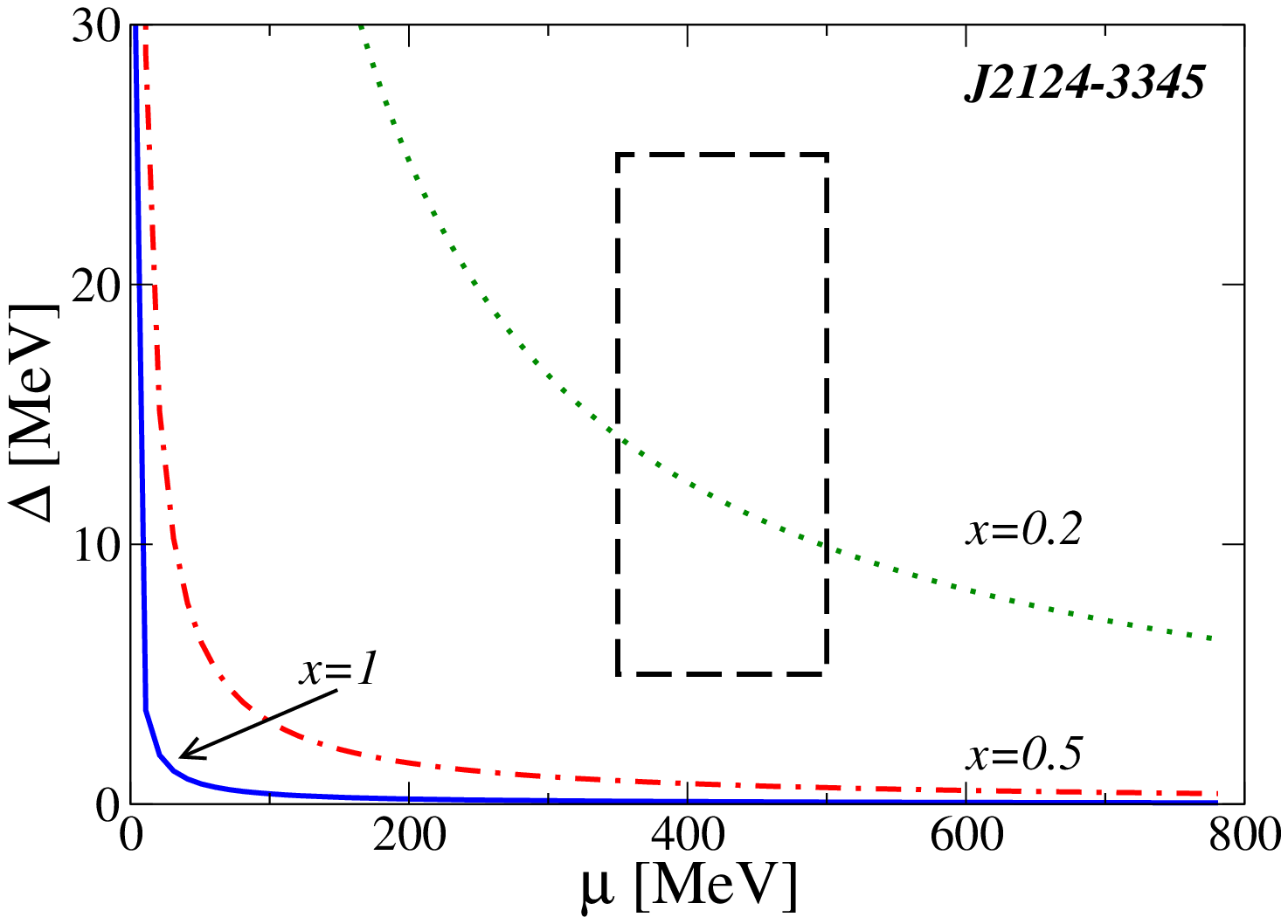}}
\end{center}
\caption{(color online). Exclusion plots in the $\Delta$-$\mu$ plane for the Crab (left panel) and the pulsar J2124-3358 (right panel) obtained assuming that these stars have mass $M=1.4$ M$_\odot$, radius $R=10$ km, uniform mass distribution and are maximally strained.  The area above each line is excluded by the corresponding model. The rectangular  box is the theoretically allowed region of $\mu$ and $\Delta$ for the CCSC phase, see \textcite{Mannarelli:2007bs}. \textit{Left panel}: The  dashed lines correspond to the results by \textcite{Lin:2007rz} which considers  the  S3/S4 LIGO data of the Crab pulsar, for the case $\sigma_{\rm max} = 10^{-3}$  (upper heavy red dashed line) and $\sigma_{\rm max} = 10^{-2}$ (lower light blue dashed line)  and assuming that the whole star consists of maximally strained crystalline color superconducting matter. The solid lines have  been determined using Eq. (7) by  \textcite{Lin:2007rz}, but with the  S5 LIGO data \cite{Collaboration:2009rfa} and considering the case of $\sigma_{\rm max} = 10^{-3}$  (upper heavy red solid line) and $\sigma_{\rm max} = 10^{-2}$ (lower light blue solid line).  \textit{Right panel}: 
Dependence of the exclusion region on the size of the crystalline core for the pulsar J2124-3358 obtained using the S5 LIGO data \cite{Collaboration:2009rfa} and considering the case of $\sigma_{\rm max} = 10^{-3}$. The three lines correspond to different values of $x=R_c/R$, where $R_c$ is the radius of the crystalline color superconducting core. }\label{fig:exclusion-Delta-mu}
\end{figure*}

The gravitational waves emitted by an optimally oriented compact star spinning at a frequency $\nu$, would produce a strain on Earth-based interferometric detectors (such as 
LIGO, VIRGO, GEO600 and TAMA300)  given by 
\be
h_0 = \frac{16 \pi^2 G}{c^2} \frac{\epsilon I_{zz} \nu^2}{r}\,,
\ee 
where   $G$ is the gravitational constant and $r$ is the distance.   If the standard value  $I_{zz} = 10^{38} $ kg m$^2$ is assumed, the above equation can be used to relate the ellipticity to the spinning  frequency and distance.  The LIGO nondetection of  gravity waves from nearby neutron stars already limits the parameter space of CCSC matter. A first analysis of CSOs which include the crystalline phase was made by \textcite{Lin:2007rz, Haskell:2007zz} using the results of the S3/S4   LIGO scientific run \cite{Abbott:2007ce} and by \textcite{Knippel:2009st} using the results   of the S5   LIGO scientific run \cite{Abbott:2008fx} on gravitational wave emission from the Crab neutron star. This is a young neutron star, first observed the year 1054,  which is a pulsating emitter of electromagnetic radiation (generically known as pulsars). In Fig. \ref{fig:exclusion-Delta-mu} we present the exclusion plots for the Crab (left panel) and for the pulsar J2124-3358 (right panel), assuming maximum strain and that these stars are constituted by incompressible and uniformly distributed matter. The dashed lines on the left panel are those obtained by  \textcite{Lin:2007rz} for the Crab pulsar using the S3/S4 LIGO data assuming that the whole star is in the crystalline color superconducting phase. We have determined the solid lines using the S5 LIGO data by \textcite{Collaboration:2009rfa}. The parameter space appears to be extremely reduced and even more severe constraints can be obtained from the pulsar J2124-3358, which leads to the exclusion plots reported on the right panel of Fig. \ref{fig:exclusion-Delta-mu}. 

Unfortunately, the maximum deformation of a star depends as well on the breaking strain, $\sigma_{\rm max}$, which measures the largest shear stress deformation sustainable by a rigid body before breaking. For standard  neutron stars it is assumed that    $10^{-5} \le \sigma_{\rm max} \le 10^{-2}$. The results shown on the left panel of Fig. \ref{fig:exclusion-Delta-mu} have been obtained with $\sigma_{\rm max} = 10^{-3}$  (upper red dashed line) and $\sigma_{\rm max} = 10^{-2}$ (lower blue dashed line). The plots on the right panel are instead obtained assuming   $\sigma_{\rm max} = 10^{-3}$ and considering three different values of $x=R_c/R$, where $R_c$ is the CCSC core radius. For lower values of $\sigma_{\rm max}$ the various curves would move  up. \red{It is important to remark that there seems to be little  motivation for assuming that the breaking strain  of the CCSC phase be of the order discussed above,  because the breaking mechanism (if it exists) of the crystalline pattern of CCSC matter is still unknown.}

From Fig. \ref{fig:exclusion-Delta-mu}  we infer  that  the S5 LIGO data give some hints about the properties of  quark matter (under the assumption discussed above), because if it is present  it should have a breaking strain $\sigma_{\rm max} \le10^{-3}$ and that if   $\sigma_{\rm max} \sim 10^{-3}$,  it is unlikely that the pulsar J2124-3358 is a crystalline quark star; although, it could still contain a small crystalline quark core. 
With Advanced LIGO and Virgo detectors the spin-down limit on gravitational wave (GW) emission from known CSOs will reach $\epsilon \approx 10^{-5}$, and objects with ellipticity of the order of $10^{-6}$ would be detectable up to the Galactic center \cite{Palomba:2012wn}. These observation will put even more severe limitations on the parameters of the crystalline phases and on the EoS of highly compressed matter.

\red{Different constraints on the CCSC phase can be obtained from the study of $r-$mode oscillations, which, if not damped, would lead to a rapid slowdown of a spinning compact star by emission of gravitational waves \cite{Andersson:1997xt, Friedman:1997uh}. The analysis by \textcite{Rupak:2012wk} shows that a quark star with a CCSC crust can provide  sufficiently large damping for preventing the growth of these oscillations even for  rapidly rotating CSOs.}

\subsection{Glitches}\label{sec:glitches}
The rigidity of the crystalline phase may also be put in relation with the anomalies in the frequency of rotation of CSOs observed as pulsars \cite{Alford:2000ze, Mannarelli:2007bs}.  Pulsars  steadily spin down,  because they lose rotational energy by emitting electromagnetic radiation, but occasionally the angular velocity at the crust of the star  suddenly spins up in a dramatic event called a glitch.  Pulsar glitches are rare events first observed in the Vela pulsar \cite{Radhakrishnan1969, Reichley1969} and in the Crab  pulsar \cite{Boynton1969, Richards1969} which manifest in a variety of sizes \cite{Espinoza:2011pq}, with an  activity changing with time and  peaking for pulsars with a characteristic age of about  $10$ kyr.  Although a number of models have been proposed including two-fluid models \cite{Baym1969n, Anderson:1975,Alpar:1977} and ``crust-quake" models \cite{Ruderman1969, Baym1971} (for a review of early models, see \textcite{Ruderman1972}),   these phenomena still remain to be well understood, and  their underlying mechanism  is not yet completely clear.  

The standard explanation of pulsar glitches \cite{Anderson:1975,Alpar:1977}  requires the presence of two basic ingredients: a superfluid in some region of the star and a  rigid structure that can pin the vortex lines  without deforming when the vortices are under tension. The spinning superfluid is indeed threaded by superfluid vortex lines with quantized circulation and as the crust of the spinning pulsar slowly loses angular momentum by radiation emission, the superfluid vortices tend  to move toward the surface of the star to compensate the rotational energy bias. The reason  is that   the angular momentum of a superfluid is proportional to the density of vortices and therefore the only way in which a superfluid can reduce the rotational energy is by \textit{vortex creep} towards the boundary of the superfluid region. But vortex lines are topological objects and if they are  pinned to a rigid structure they cannot move outward and therefore the superfluid cannot spin down. Thus, as  time passes, the frequency lag between the superfluid component of the star and the rest of the star increases. This state persists until the stress exerted by the pinned vortices on the rigid structure reaches a critical value, equal to the pinning force.  Then, there is a sudden avalanche in which many vortices unpin from their original sites, move  outwards, in part reaching the superfluid boundary  and in part repinning.  Both processes reduce the angular momentum of the superfluid component,  meaning that the rest of the star, including in particular the surface whose angular velocity is observed, speeds up. 

The two above-mentioned basic ingredients  are both present in the CCSC matter, indeed the crystalline phases are rigid as well as superfluid. Since the core of superfluid vortices consists of unpaired matter,  it is reasonable to expect that the superfluid vortex lines  will have lower free energy if they are centered along the intersections of the nodal planes of the underlying crystal structure, \textit{i.e.} along lines along which the condensate vanishes even in the absence of a vortex. An  estimate of the pinning force on vortices within  CCSC  quark matter \cite{Mannarelli:2007bs} indicates that it is  comparable to that on neutron superfluid vortices within a conventional neutron star crust \cite{Alparcreep1, Alparcreep2}. Although the basic requirements for explaining glitches are both present, several  issues  remain to be addressed and a great deal  of theoretical work remains before the hypothesis that pulsar glitches originate within a CCSC core is developed fully enough to allow it to confront  the observational  data. \red{The roadmap for achieving this goal includes: the explicit construction of a vortex line on the top of the crystalline pattern;  the understanding of the pinning mechanism and the calculation of the corresponding  pinning force;  the investigation of  the mechanism which allows angular momentum transfer between  the CCSC core and the crust, presumably by means of  the common electron fluid or by coupling through the magnetic field.} 

The same mechanism outlined above might work for a star with a CFL inner core and an outer core in the CCSC phase. In this case superfluid vortices lying in the CFL phase are pinned to the periodic structure of the CCSC phase.

\subsection{Cooling and Urca processes}\label{sec:cooling}
Neutron stars cool down by neutrino emission and by photon emission, the latter dominating at late ages ($t \gtrsim10^6$ yr). The cooling rate of a CSO may give information on its  interior constitution, because different phases of hadronic matter cool down in a rather different way, see for example \textcite{Pethick-review-Urca} for a brief review.

After a very short epoch, when the temperature of a compact star is of the
order of $\sim 10^{11}$K and neutrinos are trapped in the stellar
core \cite{Shapiro-Teukolsky, Prakash:2000jr, Steiner:2002gx, Ruester:2005ib}, the temperature drops and the mean free path of neutrinos becomes larger than the star radius. Then, the neutrinos emitted from any part of the  neutron star are free to escape and the cooling is governed by the following differential equation: \be
\frac{dT}{dt} = - \frac{ L_\nu+L_\gamma}{V_{\rm n}c_V^{\rm n} +
V_{\rm q}c_V^{\rm q}} = - \frac{V_{\rm n}\varepsilon_\nu^{\rm  n} +
V_{\rm q}\varepsilon_\nu^{\rm q} + L_\gamma} {V_{\rm n}c_V^{\rm n} +
V_{\rm q}c_V^{\rm q}}, \label{ANGLANI1} \ee where $T$ is the inner
temperature at time $t$ and  $L_\nu$ and $L_\gamma$ are respectively neutrino
and photon  luminosities, \textit{i.e.} heat losses per unit time; with
 $c_V^{n}$ and $\varepsilon_\nu^{\rm n}$  (resp.    $c_V^{q}$ and $\varepsilon_\nu^{\rm q}$) we denote the specific heat and  the emissivity  of nuclear matter  (resp. of quark matter).  \red{Here it is assumed that  the specific heats and the emissivities  do not depend on the local value of the density and thus the pertinent volumes of nuclear matter, $V_n$, and quark matter, $V_q$, factorize. In principle, for obtaining the heat loss one should  integrate the density dependent quantities over the corresponding volume.  In our simplified treatment we neglect this dependence,  considering toy models of CSOs with a constant density. Although this approximation is rather rough, especially for nuclear matter, it serves to show the qualitative effect due to  the presence of the crystalline phase. We  assume, as well, a common inner temperature $T$, which is appropriate for sufficiently old compact stars \cite{Lattimer1994}, see for example \textcite{Ho:2011aa} for a  recent discussion. }

Since the mean free path of photons is very short, the luminosity by  photon emission is  a surface effect and  can be estimated by the  black-body expression
\begin{equation} L_\gamma \simeq 4\pi R^2 \sigma T^4_s\ ,
\end{equation}
where $R$ is the radius of the star, $\sigma$ is the
Stefan-Boltzmann constant and  the
surface temperature  is given by    \be T_s \simeq 0.87\times 10^6
\left(\frac{g_s}{10^{14} {\rm cm}/{\rm s}^2}\right)^{1/4}
\left(\frac{T_b}{10^8 {\rm K}}\right)^{0.55}\!\!\text{K}\,,\nonumber \\ \label{surf}\ee see \textcite{Gudmundsson:1982, Page:2004fy}, where $T_b \simeq T$ is the temperature at the basis of the stellar envelope and  $g_s= G_N
M/R^2$ is the surface gravity.  \red{The above relation between $T_s$ and $T_b$ holds for CSOs with a standard crust. In the following we shall always assume that the matter at the basis of the envelope consists of nuclear matter with deconfined matter eventually present within the core of the star.}

For what concerns  the neutrino luminosity  one has to consider the 
relevant  weak processes. When kinematically  allowed, direct Urca processes
 are the most efficient  cooling mechanism for a CSO
in the early stage of its lifetime \cite{Shapiro-Teukolsky}. However,  the  neutrino emission via the (nuclear) direct  Urca processes, $n \to p + e + \bar\nu_e$ and $e^- +p \to n + \nu_e$,  is
only allowed for certain EoS \cite{Lattimer:1991ib} having a sufficiently large  proton abundance  to guarantee  energy-momentum  conservation \cite{Chiu1964, Bachall1965}. Therefore, considering nuclear matter, only modified Urca processes are in general considered, where a bystander particle
allows energy-momentum conservation. The resulting cooling is less
rapid and the emissivity turns out to be 
\be
\varepsilon_\nu^{\rm n}=\left(1.2\times 10^4\, {\rm erg~cm}^{-3}{\rm
s}^{-1} \right) \left(\frac{n}{n_0}\right)^{2/3}
\left(\frac{T}{10^7~{\rm K}}\right)^8 \,,\nonumber \\\ee
much smaller than the emission rate $\varepsilon_{\nu}^{\rm n} \sim T^6$ due to direct Urca processes \cite{Shapiro-Teukolsky}. Here  $n$ is the number density and $n_0 = 0.16$ fm$^{-3}$ is the nuclear equilibrium density.

 These considerations apply to stars containing standard nuclear
matter;  faster cooling can be determined by the presence of a pion condensate \cite{Bachall1965, Maxwell:1977zz, Muto:1988vw} or a kaon  condensate \cite{Brown:1988ik}. If  the central region of the star consists
of deconfined quark matter direct Urca processes involving quarks, \textit{i.e.} the processes $d \to
u + e^- +\bar\nu_e$ and $u + e^- \to d +\nu_e$, may take place and
largely contribute to the cooling rate. It has been shown
by Iwamoto \cite{Iwamoto1, Iwamoto2, Iwamoto1982An} that quark direct Urca processes are
kinematically allowed and the corresponding emission rate for
massless quarks is of the order $\alpha_s T^6$, where $\alpha_s$
is the strong coupling constant. This result is valid if quark
matter is a normal Fermi liquid, but  in  the CSC phase the expression above is not correct  because quarks form  Cooper pairs  and  fermionic excitations are
gapped. If the color superconductor is cold (\textit{i.e.} $T \ll T_c$) and homogeneous (\textit{i.e.} the
quasiparticle gap, $\Delta$, does not depend on the spatial coordinate), the corresponding neutrino emissivity and  specific heat
are suppressed by a factor $e^{-\Delta/T}$. This suppression is particularly strong in the  CFL phase, because $\Delta$ is large and all quarks are paired.

However,  the ground state of quark matter in realistic  conditions might not be the CFL phase, but a phase with a less symmetric pairing pattern. In this case  not all quarks form Cooper pairs, with unpaired quarks giving the leading contribution to the neutrino emissivity. It is also possible that the temperature of the CSO is close to $T_c$, in that case the emissivity of quark matter is not exponentially suppressed \cite{Jaikumar:2005hy}. In the following we shall evaluate the contribution of the CCSC  phase to the neutrino emissivity and specific heat  showing how the cooling curve is modified.

\subsubsection{Neutrino emissivity \label{Emissivity}}The
transition rate for  the $\beta$ decay of a down quark  $d_\alpha$, of
color $\al=r,g,b$, into an up quark $u_\al$ \be d_\al(p_1)~\to~
\bar\nu_e(p_2)~+~u_\al(p_3)~+~ e^-(p_4)\,,\label{process}\ee
 is \be
W_{\rm fi}=V(2\pi)^4\delta^4(p_1-p_2-p_3-p_4)|\mathcal{M}|^2\prod_{i=1}^4
\frac{1}{2E_iV}~,\nonumber \\\ee where $V$ is the available volume and
$\mathcal{M}$ is the invariant amplitude. Neglecting quark masses
the squared invariant amplitude averaged over the initial spins and
summed over spins in the final state
is\begin{equation}\label{eq:feyn_ampl}
  |\mathcal{M}|^2=64 G^2_F \cos^2\theta_c (p_1 \cdot p_2)(p_3 \cdot
  p_4)\,,
\end{equation}
where $G_F$ is the Fermi constant and $\theta_c$ the Cabibbo angle;
we  neglect the strange-quark $\beta$ decay whose contribution
is smaller by a factor of $\tan^2\theta_c$.  Since for  relatively aging stars there is no
neutrino trapping, the neutrino
momentum and energy are both of the order $T$. The magnitude of the other momenta is
of the order of the corresponding Fermi momenta $p_1 \sim p_{F}^1 \sim \mu $, $p_3 \sim p_{F}^3 \sim \mu $
and $p_4 \sim p_{F}^4 \sim \mu_e$, which is smaller, but still sizable (see the discussion in Sec.~\ref{sec:Chapter3}). It follows that
the momentum conservation can be implemented neglecting $\bm p_2$
and one can depict the 3-momentum conservation for the decay
\eqref{process} as a triangle \cite{Iwamoto1, Iwamoto2, Iwamoto1982An} having for sides ${\bm
p_1}$, ${\bm p_3}$ and ${\bm p_4}$.  It follows that we can
approximate \be (p_1 \cdot p_2)(p_3 \cdot
  p_4)\simeq E_1E_2E_3E_4(1-\cos\theta_{12})
  (1-\cos\theta_{34})\,,\ee where $E_j$ are the energies and  $\theta_{12}$ (resp. $\theta_{34}$)
  is the  angle between momenta of the  down quark  and the neutrino
  (resp. between the up quark and the electron). 
  
  In the  CSC phase
 one has to take into account that  the neutrino emissivity 
\begin{eqnarray}
\varepsilon_{\nu}^{\rm q}&=&\sum_{\al=r,g,b}\varepsilon_{\nu}^\alpha=\sum_{\al=r,g,b}
\frac{2}{V}\left[\prod\limits_{i=1}^4
\int\frac{d^3p_i}{(2\pi)^3}\right] E_2\,W_{\rm fi}n({\bm p_{1}})\nonumber \\ && \times [1-n({\bm p_{3}})]\,[1-n({\bm
p_4})]\,B_{d_\alpha }^2({\bm p_{1}})B_{ u_\alpha}^2({\bm p_{3}})\,,
\label{eq:emissDef}
\end{eqnarray}  depends on the Bogolyubov coefficients $B_{ u_\alpha}$ and $B_{ d_\alpha}$
which are functions of the quasiparticle  dispersion laws \cite{Alford:2004zr}. In Eq.~\eqref{eq:emissDef} the quark thermal equilibrium Fermi
distributions   
\begin{equation}
n({\bm p_{j}})=\left(1+ e^{\frac{E_j({\bm p_j})-\mu_j}T}\right)^{-1}
\,,\label{thermal-Fermi}\end{equation}
appears  because strong
and electromagnetic processes establish thermal equilibrium 
much faster than weak interactions. The overall factor of $2$ in Eq.~\eqref{eq:emissDef} keeps
into account the electron capture process.

The cooling of the CCSC matter with condensate \eqref{eq:Delta-2pw} was studied by \textcite{Anglani:2006br}, and in this case the largest contribution to the emissivity stems from the phase space region around the quark gapless modes,  while the relevant momentum for the electron is its Fermi momentum, thus we have that
 \begin{equation}\int\!\! d^3p_1\!\int\!\! d^3p_3\!\!\int \!\!d^3p_4\approx\!\!\int\mu_e^2\,dp_4 d\Omega_4\,P^2_{1}\,dp_1\,d\Omega_1P^2_{3}\,dp_3\,d\Omega_3\,,\end{equation} with $d\Omega_j=\sin\vartheta_j\,
  d\vartheta_j~d\phi_j$ and $P_1$ (resp. $P_3$) is the
quark down (resp. quark up) momentum where the corresponding
quasiparticle energy vanishes. The gapless momenta $P_1$ and $P_3$ depend on
  the angle $\vartheta_j$  that quark momenta form with the pair momentum $\bm q$.
   The expression of the integral in Eq.~\eqref{eq:emissDef} can be simplified  expanding  around the gapless modes,     $E_j(p)\simeq \mu_j+v_j\,(p-P_j)$ (for $ j=1$ and $3$), with
the quasiparticle velocity given by \be\label{velocity}
  v_j=\left.\frac{\partial E_j}{\partial
  p}\right|_{p=P_j}~.
\ee In the three-flavor case  the dispersion law of
each quasiparticle has from one to three gapless modes, thus
one has to expand the corresponding dispersion laws around each
gapless momentum. 

Using the above approximations the neutrino emissivity for each
pair of gapless momenta $P_1,P_3$, can be written as
\begin{widetext}
\begin{equation}
\varepsilon_\nu^\alpha \simeq\frac{G_F^2\,\cos^2\theta_c\,\mu_e^2\,T^6}{32\pi^8}
\,\mathcal{I}\, \prod_{j=1}^4\int
d\Omega_j\frac{P_1^2P_3^2\,B_{d_\alpha }^2({P_{1}})B_{ u_\alpha}^2
({P_{3}})}{|v_1|\,|v_3|}\delta^{(3)}({\bm p_1}-{\bm p_3}-{\bm p_4}-{\bm q}) (1-\cos\theta_{12})
  (1-\cos\theta_{34})\,,
\label{emissivity}
\end{equation}
\end{widetext}
where $ \mathcal{I}=~\frac{457\pi^6}{5040}$.  Some of the
angular integrations appearing in \eqref{emissivity}
 can be performed analytically, see \textcite{Anglani:2006br}  for more details, and the numerical computation can be reduced observing that even if each quasiparticle dispersion law is characterized by
various gapless momenta, not all of them satisfy the conservation of momentum
${\bm p_1}-{\bm p_3}-{\bm p_4}-{\bm q}=0$.

From Eq.~\eqref{emissivity} one can deduce that the largest contribution to the emissivity is due to blue quarks, that is to the process in Eq.~\eqref{process} with $\alpha=b$. The reason is that  according to the   
Ginzburg-Landau analysis by  \textcite{Casalbuoni:2005zp},  the gap parameter $\Delta_1$ vanishes and therefore the down blue quark is ungapped (see the discussion after Eq.~\eqref{condensate-gCFL2}).  \red{Quarks of different flavor or color are instead gapped;  neglecting ${\cal O}(1/\mu)$ corrections (see Sec.~\ref{sec:1overmu}) one has $\Delta_2=\Delta_3=\Delta$.} 
The dispersion of down blue quarks  is $E_1(p)=p$, with
 gapless momentum $P_1\simeq \mu$ independent of $\vartheta_1$, mixing coefficient $B_{d_b}(P_1)=1$
 and $v_1=1$. The up blue quark is instead paired and
has two  gapless momenta depending on $\vartheta_3$, see \textcite{Anglani:2006br} for an explicit  expression. Of these two gapless momenta one satisfies the momentum conservation and thus the decay of a down blue quark in an up blue quark is not suppressed with respect to the analogous decay in unpaired quark matter, giving the leading contribution to the emissivity.  The contributions of  quarks with red and green colors can be treated in a similar way, but for these colors neither the down nor the up quarks are unpaired and the corresponding processes are thus suppressed. \red{The sum of the contributions of all quarks turns out to be
\be
\varepsilon_{\nu} \approx 4.3 \times 10^{13} \left(\frac{T}{10^7 \text{K}}\right)^6 \text{erg cm}^{-3}\text{s}^{-1}\,,
\ee
and is comparable with the emissivity of unpaired quark matter, see for example \textcite{Iwamoto1982An}.}

\subsubsection{Specific heats \label{specificheat}}

As already discussed in Sec. \ref{sec:disp2flavor} for the two-flavor case, at low temperature the largest contribution to  the specific heat is determined  by the  fermionic
quasiparticles.   The specific heat of the three-flavor crystalline phase  is given by the same formal expression given in Eq.~\eqref{eq:CvDef}, but with  the quasiparticle dispersion laws of the three-flavor CCSC phase. The computation can be simplified following the same reasoning used above:  the contributions of gapped modes are exponentially suppressed and each gapless mode  contributes by a
factor $\propto T$. This result follows from the evaluation of the
integral in Eq.~\eqref{eq:CvDef} employing  the saddle point method
and assuming that the quasiparticle dispersion laws are linear close
the gapless momenta. Then, also the angular integral can be simplified because the  dispersion laws are gapless only in a restricted angular region. The numerical analysis  confirms the above results, see  \textcite{Anglani:2006br} for a discussion and for an expression of the specific heat.

For unpaired nuclear matter  and unpaired quark matter the contribution of each fermionic species can  be approximated
by the fermionic ideal gas result \be c_V=\frac{T}{3}
P^F\sqrt{m^2 c^2 +(P^F)^2} \ , \ee where $m$ and $P^F$ are the appropriate  fermionic mass
and   Fermi momentum, respectively. For unpaired nuclear matter,
the three species are neutrons, protons and electrons with Fermi
momenta evaluated as in neutral matter in weak equilibrium
\cite{Shapiro-Teukolsky}: \be P^F_n \simeq \left(340\text{ MeV}\right)\left(\frac{n}{n_0}\right)^{1/3}\!\!\!,
P^F_p=P^f_e \simeq \left(60~{\rm
MeV}\right)\left(\frac{n}{n_0}\right)^{2/3}\!.\nonumber\\  \ee For 
unpaired quarks, considering electric neutrality and weak equilibrium, the  nine quark
species have  Fermi momenta  independent of color  given by $
P^F_d=  \mu+\displaystyle\frac{M_s^2}{12\mu}$, $P^F_u =
 \mu-\displaystyle\frac{M_s^2}{6\mu}$ and  $P^F_s =  \mu-\displaystyle\frac{5 M_s^2}{12\mu}$, see the discussion in Sec. \ref{sec:Chapter3}.

\subsubsection{Cooling  by neutrino emission \label{cooling}}

For evaluating the effect on the compact star cooling of the CCSC phase  we consider three different  toy models comprising  neutral and $\beta$ equilibrated matter \cite{Anglani:2006br}.
 Model I is a star consisting of unpaired
``nuclear" matter (neutrons, protons and electrons) with mass
$M=1.4M_\odot$ (where $M_\odot$ is the solar mass), radius $R=12$ km and uniform number density $n=1.5\,n_0$. Model II is a star containing a core of radius $R_c=5$ km of  unpaired quark matter with $\mu=500$ MeV, with an outer part of unpaired
nuclear matter with uniform density $n$. Assuming a star mass
$M=1.4\,M_\odot$ from the solution of the Tolman-Oppenheimer-Volkov (TOV)
equations \cite{Tolman, Oppenheimer-Volkoff} one gets a star radius $R=10 $ km. Model III
is a compact star containing a core of electric and color neutral three-flavor quark matter in 
the  CCSC phase with  gap parameter given in Eq.~\eqref{eq:Delta-2pw},
with $ \Delta\simeq 6$ MeV and $\mu=500$ MeV, $M_s^2/\mu=140$ MeV. The outer part of the star is made of unpaired nuclear matter. 
Since the value of the gap parameter in the model III is small, the radius of the CSO and
of the quark core do not differ appreciably from those of a star
with a core of unpaired quark matter, \textit{i.e.} $R_c=5$ km and $R=10$ km
(also in these cases $M=1.4 M_\odot$).

\begin{figure}[t]
\begin{center}
\includegraphics[width=8cm]{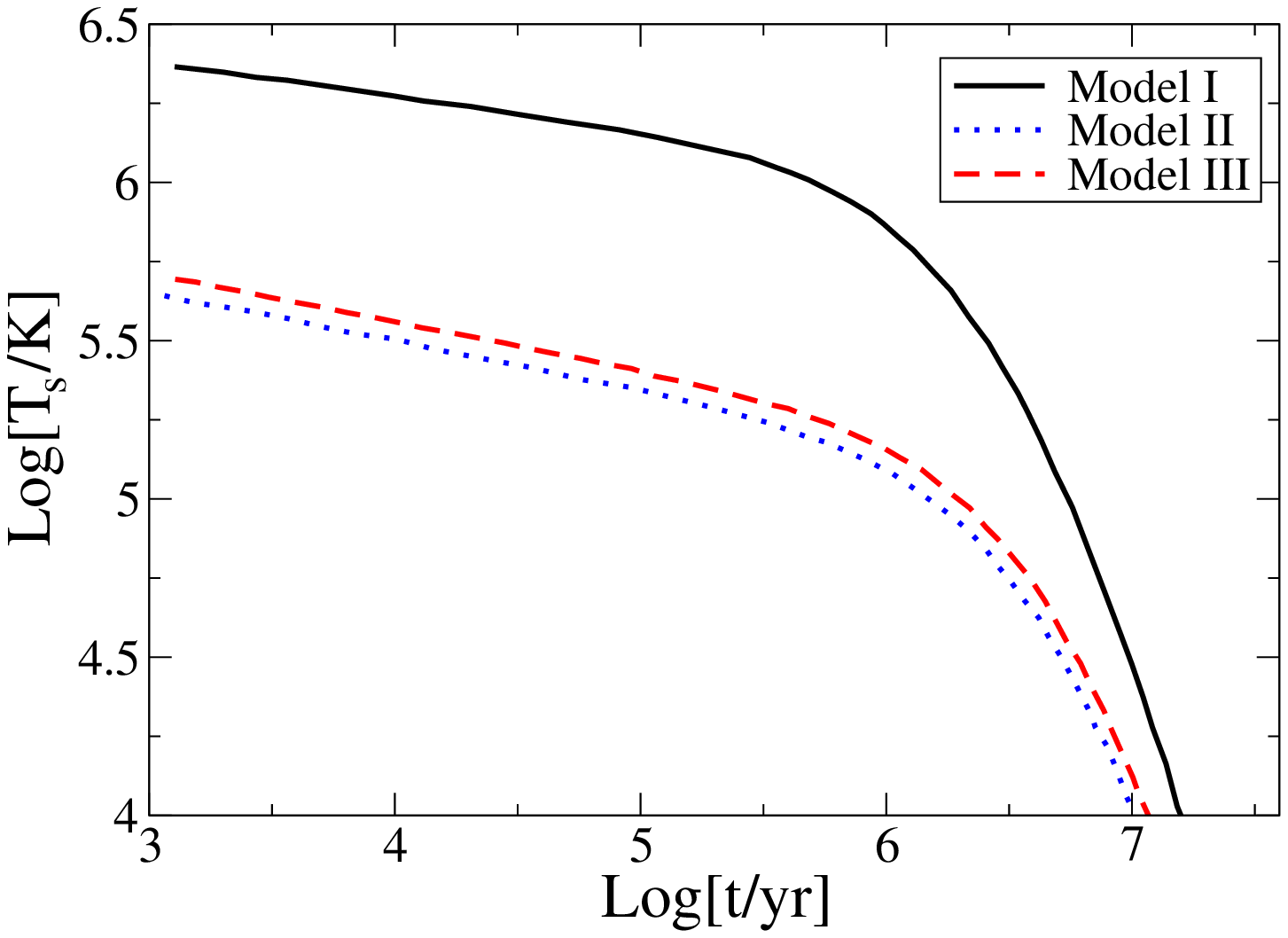}
\end{center}
\caption{(color online). Surface temperature, in Kelvin, as a function of time, in years, for various  models of compact stars \cite{Anglani:2006br}. 
Model I is a neutron star of  nuclear matter with uniform density $n=0.24$ fm$^{-3}$ and radius $R=12$ Km; model II corresponds to a star with $R=10$ km, having an outer part of nuclear matter and a core of radius $R_c=5$ Km of unpaired quark matter; model III is like model II, but in the core there is quark matter in a crystalline phase, see the text for more details. For all stars $M=1.4\,M_\odot$ and the temperature at $t=1\,$yr was set to $T=10^9$K. Parameters for the core are $\mu=500$ MeV and $M^2_s/\mu$=140 MeV.  Adapted from \textcite{Anglani:2006br}.}\label{fig:coolings}
\end{figure}

In Fig. \ref{fig:coolings}  the
cooling curves of the surface temperature are shown as a function of time  for the various models of star.  
For unpaired quark matter the coupling $\alpha_s\simeq 1$ is used, corresponding to $\mu=500$ MeV and $\Lambda_{\rm QCD}=250$ MeV. The use of perturbative QCD at such small momentum scales is however questionable. Therefore the results for the model II should be considered with some caution and the curve is plotted only to allow a comparison with the other models. The  similarity between the \red{dashed  curve and the dotted curve} follows from the fact that in the CCSC phase the quasiparticle dispersion laws are linear and gapless, so that the scaling laws $c_V\sim T$ and $\varepsilon_\nu\sim T^6$ are analogous to those of the unpaired quark matter. \red{Thus, the two models cannot easily be distinguished.}  From this figure we can see  that stars with a CCSC core (or an unpaired quark core)
cool down faster than ordinary neutron stars. \red{In particular, for $10^3$ yr $< t < 10^6$ yr the cooling is dominated by the neutrino emission and one has that for model I,   $T \sim t^{-1/6}$, while  $T \sim t^{-1/4}$ for  model II and model III. At later times photon emission dominates and the three cooling curves move closer.}

\begin{figure*}[t]
\begin{center}
\includegraphics[width=10cm]{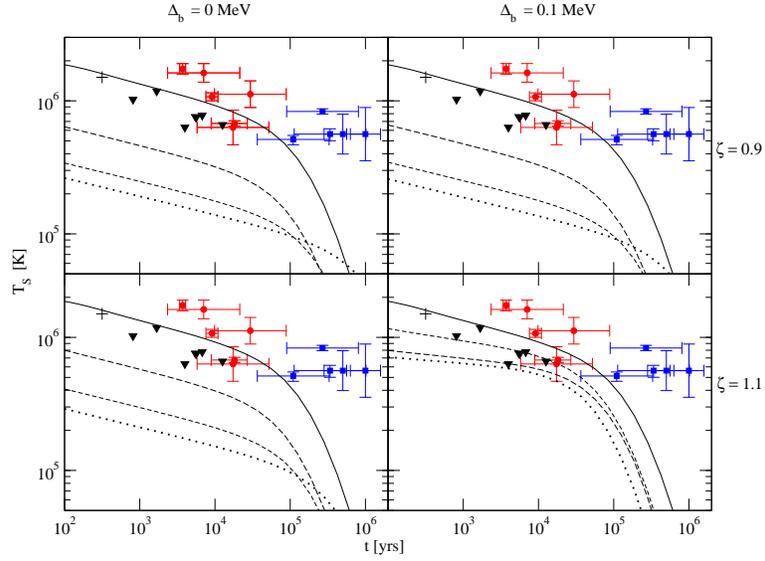}
\end{center}
\caption{(color online). Surface temperature, in Kelvin, as a function of time, in years, for various models of compact stars \cite{Hess:2011qw} comprising a two-flavor LOFF phase between $u$ and $d$ red and green quarks, with gap parameter $\Delta$, and  pairing between $u$ and $d$ blue quarks, with gap parameter $\Delta_b$. Various values of the  central density, $\rho_c$, of $\zeta=\Delta/\delta\mu$ and of $\Delta_b$, are considered. The solid  curve  refers to a model with no quark matter, central density $5.1 \rho_{14}$, where $\rho_{14} = 10 ^{14}$ g cm$^{-3}$, corresponding to a star with $M\simeq0.54 M_\odot$.  The long dashed  curve corresponds to a model with $\rho_c = 10.8 \rho_{14}$, $M\simeq1.91 M_\odot$ and a quark core with radius $R_c\simeq 0.68$\,km. The short dashed curve corresponds to a model with $\rho_c = 11.8 \rho_{14}$, $M\simeq1.93 M_\odot$ and $R_c\simeq 3.41$\,km. The dotted curve corresponds to a model with $\rho_c = 21.0 \rho_{14}$, $M\simeq2.05 M_\odot$ and $R_c\simeq 6.77$\,km.  The symbols correspond to the observational data points, see  \textcite{Hess:2011qw} for an explanation; here we only remark that the $+$ sing on the top left corner corresponds to Cassiopeia A \cite{Heinke:2010cr, Shternin:2010qi}.  From \textcite{Hess:2011qw}.}\label{fig:coolings-HS}
\end{figure*}

Similar results, with a more refined  analysis, have been obtained by \textcite{Hess:2011qw}, using a more realistic EoS   to model compact stars with different masses. Their results are reported in Fig.~\ref{fig:coolings-HS}. The considered models  have an envelope of standard nuclear matter, cooling by the modified Urca process and by neutral current bremsstrahlung processes, and a core of quark matter comprising a two-flavor inhomogeneous phase, cooling by the direct Urca process.  The solid curve corresponds to standard nuclear matter, the other curves correspond to hybrid stars with a quark core of different radii. In order to reproduce, with an hybrid model, the measured surface temperatures of compact stars, the \red{BCS-like pairing}  between $u$ and $d$ blue quarks has to be included, with a pairing gap, $\Delta_b$, of the order of $0.1$ MeV. In this case models with a  quark core radius $R_c \lesssim 1$\,km seem  to be in good agreement with part of the observational data.

All these results have
interesting phenomenological consequences because  observational measurements on the cooling of compact stars are being accumulated at an increasing rate. 
Some data indicate that stars with an age in the range $10^3-10^4$
years have a temperature significantly smaller than expected
on the basis of the modified Urca processes, see Fig.  \ref{fig:coolings-HS}. It is difficult however
to infer, from these data, predictions on the star composition, as
theses stars may have different masses. But, quite recently, the  thermal evolution of  the neutron star in Cassiopeia A \cite{Heinke:2010cr, Shternin:2010qi}, corresponding to the $+$ sign in  Fig. \ref{fig:coolings-HS}, has been observed  (note that we do not report in  Fig. \ref{fig:coolings-HS} the detailed observational data by \textcite{Heinke:2010cr, Shternin:2010qi}) and  explained as an effect of neutron superfluidity  \cite{Page:2004fy, Page:2010aw, Shternin:2010qi}. From Fig.~\ref{fig:coolings-HS} we see that models with a LOFF-like two-flavor pairing  seem to fail to reproduce the initial observed values of age-temperature of this compact star. \red{However, it has been recently shown that the rapid cooling of Cassiopeia A might be explained as a phase transition from a gapped to a gapless (possibly crystalline) phase in a two-flavor CSC \cite{Sedrakian:2013pva}. The caveat is that the considered CSO should have a mass $\simeq 2 M_\odot$}.    It would be interesting to  check whether the observed behavior can be reproduced with the thermodynamically favored CCSC  phase, for which, however,  the  identification of the quasiparticle dispersion laws is still lacking. 

The rapid cooling of neutron superfluids  determined by \textcite{Page:2004fy, Page:2010aw, Shternin:2010qi} is due to difermion pair-breaking effect, which enhances the emissivity of nuclear matter.   The presence of  CCSC matter has a very similar effect. Indeed, the fast cooling of relatively young compact stars with a CCSC  core should be a consequence of the scaling laws for neutrino emissivity and
specific heat, which in turn  strongly depend on the existence of gapless points  \cite{Anglani:2006br, Hess:2011qw}, present at the edge of the pairing regions in momentum space. Since this property is typical of any CCSC phase, independently of
detailed form of the condensate, when a CCSC core nucleates inside a compact star it should be followed by a rapid cooling, pretty much as the formation of a neutron superfluid determines the rapid cooling  of a young neutron star.  %

Although the  above-reported analysis is not conclusive, because a detailed treatment of the thermodynamically favored CCSC phase is still missing,  some qualitative assessments can be made from the obtained results.  Slow cooling is typical of stars containing only nonsuperfluid standard nuclear matter or
of stars with a uniform CSC phase (like CFL);  the observed cooling rate of  the neutron star in Cassiopeia A \cite{Heinke:2010cr,Shternin:2010qi}, seems to disfavor both possibilities, leaving superfluid nuclear matter and CCSC matter as candidate phases.  The latter possibility is compatible with the result that for intermediate densities the quark normal state and the CFL phase are less favored than the CCSC phase  in a certain range of values of the quark chemical potential.

Note  that the results reported above do not  properly take into account the heat transport  inside the compact star.  It would be interesting to include in the analysis the various transport mechanisms to have  detailed simulations of the CSO cooling, see for example \textcite{Ho:2011aa}.

\subsection{Mass-radius relation}\label{sec:mass-radius}

Since any phase transition leads to a softening of the EoS \cite{Lattimer:2000nx, Haensel2003}, \red{at one time it was thought} that hybrid stars --- having a quark matter core and an envelope of baryonic matter --- should have mass $M \lesssim 1.7 M_\odot$, see for example \textcite{Maieron:2004af, Alford:2002rj, Buballa:2003et}.   For CSOs with larger masses, the deconfinement phase transition from baryonic to quark matter would reduce the central pressure to the point of instability towards black hole formation, unless some repulsive interaction between quarks prevents the collapse. 
 
Some evidence for massive neutron stars with $M\sim 2M_{\odot}$ has
been inferred from various astronomical observations:  A  compact
star may exist in the LMXB (Low Mass X-ray Binary) 4U 1636-536 with
$2.0\pm 0.1 M_{\odot}$ \cite{barret2005}. A measurement on the
pulsar PSR B1516+02B in the Globular Cluster M5 gave $M  =2.08\pm
0.19 M_{\odot}$ \cite{Freire2008}. The  millisecond pulsar J1614-2230  
has a mass $(1.97 \pm 0.04) M_\odot$ accurately measured by Shapiro delay \cite{Demorest:2010bx}.
Although these observations  seem to disfavor the presence of quark  matter \cite{Logoteta:2012ms},  the details of a stability analysis depend on the theoretical model employed for the description of the hadronic phase, of deconfinement, and of the CSC matter. We shall show below that in the presence of the crystalline phase large masses can indeed be reached \cite{Ippolito:2007hn}. The drawback, as we shall see, is that if $M \simeq 2 M_\odot$ CSOs have a CCSC core, then ordinary $M \simeq 1.4 M_\odot$ CSOs are unlikely to  have a CCSC core. 

Since a first principle calculation of the high-density EoS is not feasible,  the description of quark matter relies on several different  models including the MIT bag model, the NJL model and the chromodielectric model. The results obtained within these three models may differ in a sizable way, and do as well depend on the detailed form of the model considered.  
Recent phenomenological studies of hybrid stars based on the MIT bag model where carried out using a generic parameterization of the quark matter EoS by \textcite{Alford:2004pf} and it was shown that hybrid stars may actually ``masquerade" as neutron stars.   In the work by \textcite{Alford:2004pf} nonperturbative QCD corrections to the EoS of the Fermi gas were parameterized  in a rather general way, with reasonable estimates \cite{Fraga:2001id}. Taking into account these corrections, stable hybrid stars containing CFL quark matter may exist with maximum mass of about $2 M_{\odot}$.

Various studies of  very massive hybrid  stars within the three-flavor NJL model
displayed a general instability  towards  collapse
into a black hole \cite{Buballa:2003et}. Stable stars featuring
the 2SC phase were obtained with typical maximum masses $M\sim 1.7 M_\odot$ assuming reasonable values of the constituent quark masses \cite{Shovkovy:2003ps} or by replacing the hard NJL
cut-offs by soft form factors with parameters fitted to a certain set
of data \cite{Grigorian:2003vi,Blaschke:2007ri}. Heavier objects can
be obtained if a repulsive vector interaction is introduced in the NJL
Lagrangian \cite{Klahn:2006iw,Bonanno:2011ch, Orsaria:2012je}, which makes the EoS stiffer,
but at the same time, reducing the amount of deconfined quark matter in the core of the star, or with no pure quark matter at all \cite{Orsaria:2012je}, \textit{i.e.}   with a core comprising a mixed phase of quarks and hadrons.

Another source of  uncertainty comes from the nuclear EoS at high densities, which  can be constructed starting from a number of different principles
\cite{Weber-book,Sedrakian:2006mq} and a large number of EoS
have been proposed for hybrid star configurations. 
In the analysis presented below, of the considered EoS,  only the stiffest were found to be  admissible for phase equilibrium between nuclear and CCSC matter.

The inclusion of  hyperonic matter  in the EoS of compact stars although reasonable is certainly troublesome,  because  the nucleon-hyperon and hyperon-hyperon interactions are not well known, even though some progress has been done mainly by lattice simulations \cite{Aoki:2012tk}.  Hyperons would certainly soften the EoS \cite{Glendenning:1998ag}, making the comparison with the observed $2 M_\odot$ of some CSOs  problematic \cite{Lattimer:2006xb}.

\subsubsection{Matching the equation of state} \label{sec:matching}

The self-consistent computation of the strange quark mass given by
\textcite{Ippolito:2007uz} allows the evaluation of  pressure as a function
of the quark chemical potential $\mu$. Thus,  varying the quark chemical potential, the
 phase equilibrium  between the confined and the CSC phase can be constructed. 

A  possible normalization of the quark pressure in the NJL
model is obtained by requiring that  the pressure  vanishes at zero density and
temperature \cite{Buballa:1998pr,Sandin:2007zr}. In the terminology
of the MIT bag model, this is equivalent to a subtraction of the bag
constant from the thermodynamic potential \cite{Alford:2004pf}.
Since the value of the bag constant is related to confinement, which
is absent in the NJL model, it appears reasonable 
changing  its value, and hence the normalization of the
pressure. The simplest option is to consider the  case of a constant shift in
the asymptotic value of the pressure; an alternative is the use
of form factors for the bag constant \cite{Grigorian:2003vi}.

Regarding the matching between the nuclear and quark matter EoS, it can be performed in the \textit{strong
coupling} limit, corresponding to $G_D/G_S= 1$ \cite{Sandin:2007zr, Ruester:2005jc} (see Eq.~\eqref{eq:Lint} and the discussion after Eq.~\eqref{eq:NC}). The transition from the confined phase to quark matter happens at the baryonic chemical potential at which the pressures of the two phases are equal, meaning that the chemical potential curves $P(\mu)$ for these phases cross. At intermediate coupling and weak coupling, corresponding respectively to  $G_D/G_S \sim 0.75$ and   $G_D/G_S \lesssim 0.7$,  the $P(\mu)$ curves for  nuclear and quark matter do not cross; the models are thus incompatible, meaning that they cannot describe the desired transition between nuclear and quark matter.

In the following we shall discuss two simple models of CSOs~\cite{Ippolito:2007hn}. In the first model, hereafter named Nuclear, we consider standard nuclear matter described by the Dirac-Brueckner-Hartree-Fock theory. The  selected EoS is the hardest one in the collection by \textcite{Weber-book, Weber:2007ch}. The second model, hereafter Model A, has the same low-density EoS of Nuclear matched at an interface via the Maxwell construction to the  high-density  EoS of CCSC matter. In order to achieve this matching  the zero density pressure is shifted by an amount $\delta p = 10$ MeV/fm$^3$. This is equivalent
to a variation of the bag constant. Another possibility is to set $\delta p = 0$, but varying
the value of the constituent masses of the light quarks in the fit
of the parameters of the NJL model; for small values of the light
quark masses the matching between  quark and nuclear equations of
state is facilitated \cite{Buballa:2003et}.

\begin{figure*}
\begin{center}
\subfigure{\includegraphics[width=8.cm,angle=0]{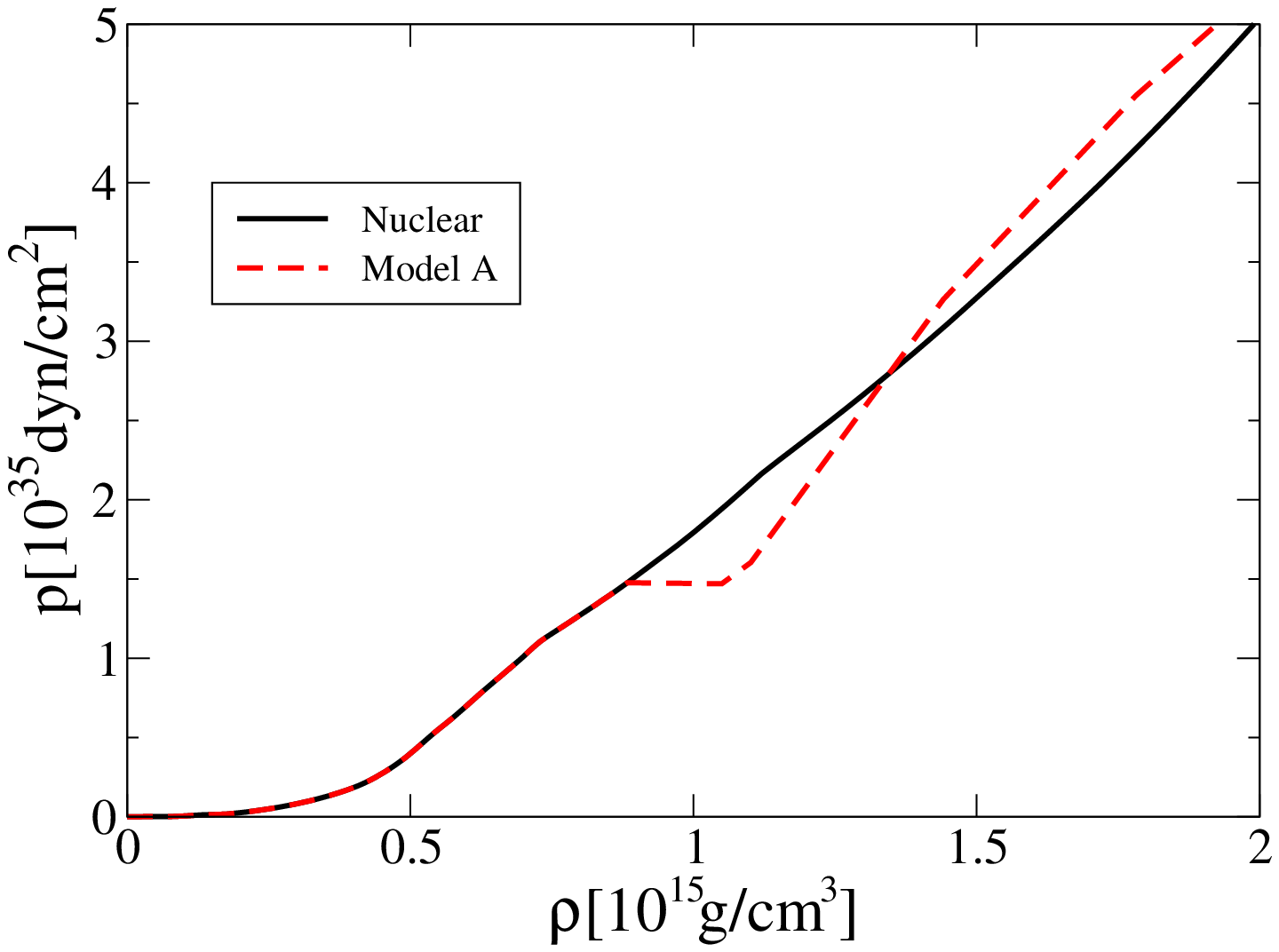}}
\subfigure{\includegraphics[width=8.cm,angle=0]{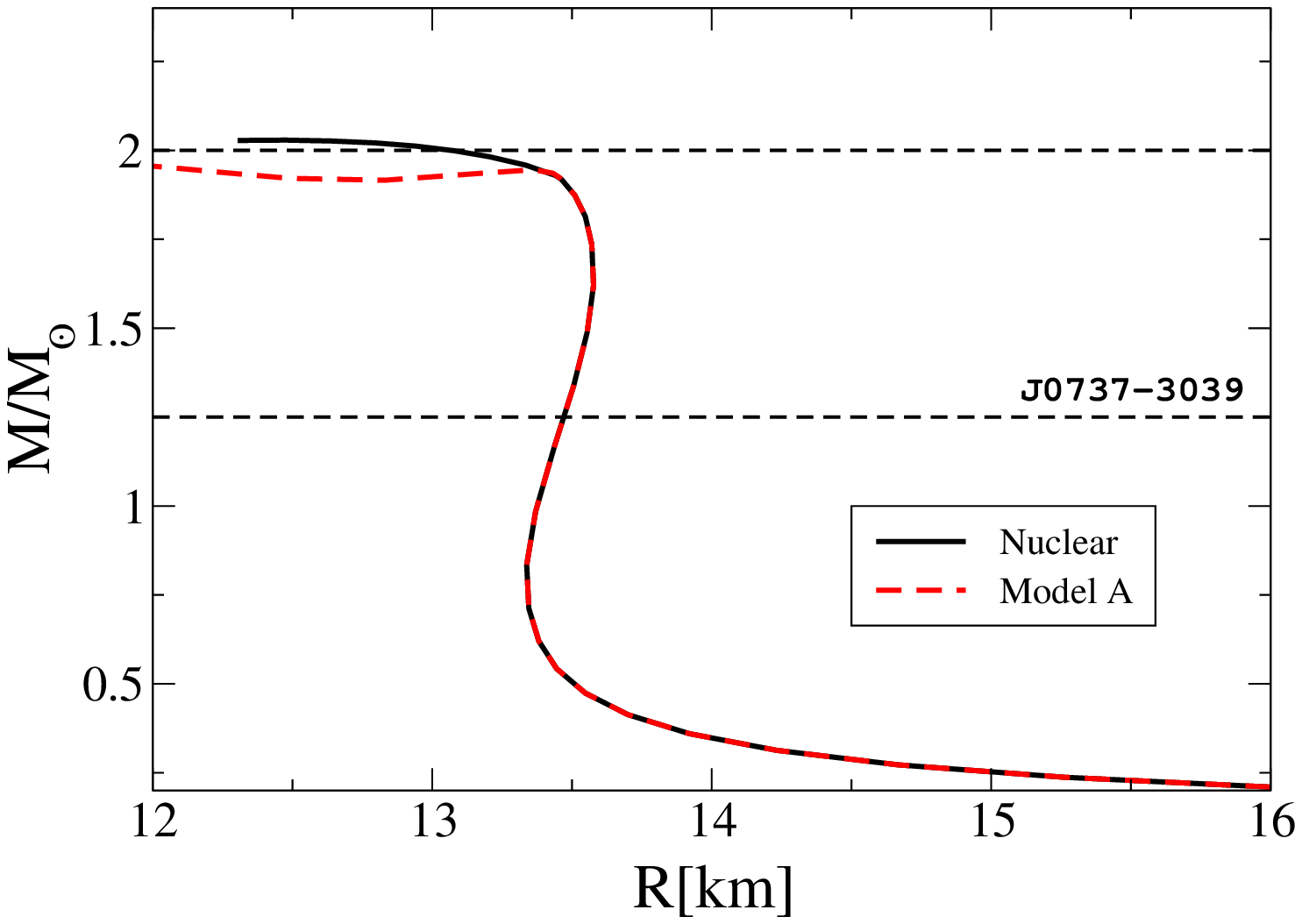}}
\end{center}
\caption{(color online). Left panel: Pressure versus matter density for the two  considered models.
The solid black line labeled as Nuclear refers to the EoS based on the Dirac-Brueckner-Hartree-Fock approach \cite{Weber-book, Weber:2007ch}.  The dashed red line corresponds to model  A.  The EoS of Model A  has the low-density EoS corresponding to Nuclear and the high-density EoS obtained by a NJL model describing CCSC matter.  At the deconfinement phase transition there is a jump in the density at constant pressure. Right panel: Mass-radius diagram for nonrotating configurations. The dashed lines correspond to 
the lower pulsar mass bound from J0737-3039 and the  $2 M_\odot$ upper bound from various astronomical observations. }\label{fig:p_vs_rho}
\end{figure*}

Note that   because of   the
Maxwell construction of the deconfinement phase transition, there is
a jump in the density at constant pressure as illustrated in the left panel of
Fig.~\ref{fig:p_vs_rho}.

\subsubsection{Results for nonrotating configurations} \label{sec:results}

Given the EoS, the spherically symmetric solutions of Einstein's equations for
nonrotating self-gravitating fluids are obtained by the TOV equations.  For simplicity we only  discuss the two nonrotating configurations corresponding to the Model A and Nuclear; see~\cite{Ippolito:2007hn} for different models including rotation. A generic feature of the TOV solutions is the existence of a maximum mass for any
EoS; as the central density is increased beyond the
value corresponding to the maximum mass, the star becomes unstable
towards collapse to a black hole. One criterion for the stability of
a sequence of configurations is the requirement that 
$dM/d\rho_c >0$, meaning that the mass of the star should be an
increasing function of the central density. 
At the point of instability the fundamental (pulsation) modes become unstable. 
If stability is regained at higher central densities, the modes by
which the stars become unstable towards the eventual collapse belong
to higher-order harmonics.

The right panel of Fig. \ref{fig:p_vs_rho} shows that for configurations constructed from the Nuclear EoS, the stable sequence extends up to a maximum mass of the order $2 M_{\odot}$; the value of the
maximum mass is large, since the chosen EoS is
stiff. The hybrid configurations branch off from the nuclear
configurations when the central density reaches that of the
deconfinement phase transition.  
We also report  the astronomical bounds on the
masses  of CSOs (horizontal dashed lines). The upper bound corresponds to 
$M = 2 M_{\odot}$, while the  lower bound to a mass of about $(1.249\pm 0.001) M_{\odot}$ 
 inferred from the millisecond binary
J0737-3039 \cite{Lyne:2004cj}.  Note that both the hybrid stars and their
nuclear counterparts have masses and radii within these bounds. However, the hybrid configurations are more compact than their nuclear counterparts, \textit{i.e.} they have smaller radii. Contrary to the case of self-bound quark
stars \cite{Alcock:1986hz}, whose radii could be much smaller than the radii of purely
nuclear stars, the differences between the radii of hybrid  and
nuclear stars are tiny and cannot be used to distinguish
these two classes by means of current astronomical observations.
A remarkable result is that although the hybrid Model A is roughly consistent with the bound of $M\sim 2 M_\odot$, it should be noted that canonical  $1.4 M_{\odot}$ CSOs will be purely nuclear if they are described  by this  model. According to the presented analysis when the CCSC matter  appears the star first gets unstable and then it becomes stable only at higher values of the central density (corresponding to the so-called third family compact stars~\cite{Gerlach:1968zz, Glendenning:1998ag, Schertler:2000xq}). Moreover, only sufficiently massive CSOs, with mass    $M \gtrsim 1.8 M_\odot$, can contain a fraction of quark matter. Different hybrid EoS constructed in a similar way lead to analogous results ~\cite{Ippolito:2007hn}.

%
\section{Conclusion}
\label{sec:Conclusion}

The investigation of the properties of matter in extreme conditions is one of the most fascinating and challenging frontiers in high energy physics.  The aim is to understanding the fundamental properties of matter when the  relevant degrees of freedom are deconfined quarks and gluons. According to QCD, cold and dense matter at asymptotic densities should be in the color superconducting (CSC) phase, with quarks forming Cooper pairs. 
  
Ongoing research in cold and dense quark matter is now confronting astrophysical  data, allowing us  to exclude some CSC models and/or to restrict the parameter space of some phases. The most important consideration for a comparison with the astrophysical data is certainly that the conditions realized in compact stellar objects may determine a mismatch between the Fermi momenta of quarks, disfavoring the homogeneous CSC phases, and favoring less symmetric diquark pairing.

In this Review, we focused on  the inhomogeneous CSC  phases, chiefly on the crystalline color superconducting (CCSC) phase. This phase  might be the CSC phase realized  in the core of compact stars, if deconfined quark matter is present,  because it allows pairing between quarks on mismatched Fermi surfaces. 

First proposed in  the two-flavor case by \textcite{Alford:2000ze}, the CCSC phase has been studied by a large number of authors both for two- and three-flavor quark matter. We have reported on the various results obtained showing that nowadays it is a well-developed  subject with  several open problems and fascinating applications to compact star astrophysics.
In this respect,  this phase of matter has very appealing features, residing in its extraordinary properties. The CCSC is characterized by a periodic modulation of the condensate which is extremely resistant to deformations. Thus, compact stars featuring CCSC matter may sustain large ``mountains",  
meaning that spinning compact stars might be strong  sources of gravitational waves. Moreover, since the three-flavor CCSC phase is also superfluid, when rotating it will develop quantized vortices which might be  pinned to the periodic condensate, thus it satisfies the basic requirements for explaining stellar glitches. Finally, we have also discussed how the presence of CCSC matter might  influence the mass-radius relation and  the cooling curve of a compact star, although in this case the presence of a crystalline modulation appears to be  less  relevant.

Whether the CCSC phase will stand the test of the increasing observational data on  compact stellar objects it is still unclear. It seems to us that among the astrophysical data, probably those on stellar glitches and those provided by  the next-generation gravitational wave detectors might give the most stringent constraints on the CCSC matter. 
However, to properly confront the astrophysical data, we still need to understand several fundamental properties of the CCSC phase.  Among the open issues,  we remark that the individuation of the favored crystalline structure does so far rely on a Ginzburg-Landau (GL) expansion. Although  the GL analysis does certainly give very useful qualitative informations on the  favored periodic modulations, the resulting gap parameter and free energy  are under  poor quantitative control.  Moreover, in order to study stellar glitches, a thoughtful understanding  of  superfluid vortices in CCSC matter has to be developed. So far only  an order of magnitude estimate of the pinning force  has been obtained, which is not enough for a full description of the vortex dynamics and of the associated stellar spinning evolution.

\section*{List of symbols and abbreviations}
For clarity we report below a list of the most used acronyms and symbols.\\

\begin{description}
\item[bcc] Body-centered cube
\item[BCS] Bardeen-Cooper-Schrieffer 
\item[BEC] Bose-Einstein condensate 
\item[CCSC] Crystalline color superconducting 
\item[CFL ] Color-flavor locked 
\item[CC] Chandrasekhar-Clogston
\item[CSC] Color superconducting  
\item[CSO] Compact stellar object 
\item[EoS] Equation of state 
\item[fcc] Face-centered cube
\item[FF] Fulde-Ferrell
\item[GL] Ginzburg-Landau 
\item[GW] Gravitational waves 
\item[HDET] High-density effective theory 
\item[LO] Larkin-Ochinnikov 
\item[LOFF] Larkin-Ochinnikov-Fulde-Ferrell  
\item[NGB] Nambu-Goldstone boson
\item[NJL] Nambu-Jona Lasinio
\item[QCD] Quantum chromodynamics 
\item[QGP] Quark-gluon plasma 
\item[TOV] Tolman-Opennheimer-Volkoff
\item[2SC] Two-flavor color superconducting  
\item[gCFL] Gapless color-flavor locked
\item[g2SC] Gapless two-flavor color superconducting 
\end{description}
\begin{description}
\item[$a,b$] Adjoint color indices
\item[$i,j,k$] Flavor indices
\item[$s,t$] Spin indices
\item[$G_D$] Quark-quark coupling constant
\item[$G_S$] Quark-antiquark coupling constant
\item[$P^F$] Fermi momentum
\item[$M_s$] Constituent strange quark mass
\item[$M_\odot$] Solar mass
\item[$\alpha,\beta,\gamma$] Fundamental color indices
\item[$\delta\mu$] Chemical potential difference
\item[$\epsilon_{ijk}$] Levi-Civita symbol in flavor space
\item[$\varepsilon^{\alpha\beta\gamma}$] Levi-Civita symbol in color space
\item[$\mu$] Quark chemical potential
\item[$\mu_e$] Electron chemical potential
\item[$\Delta$] Pairing gap 
\item[$\Omega$] Free energy 
\end{description}

\begin{acknowledgments}
The  review has benefited of many fruitful discussions with several colleagues of various institutions. We would like to thank M.G.~Alford, K.~Fukushima, A.~Sedrakian and R.~Sharma for useful suggestions. We thank A.~Sedrakian for providing us with the data points  in Fig.~\ref{fig:p_vs_rho}. 
\end{acknowledgments}

%


\begin{thebibliography}{276}%
\makeatletter
\providecommand \@ifxundefined [1]{%
 \@ifx{#1\undefined}
}%
\providecommand \@ifnum [1]{%
 \ifnum #1\expandafter \@firstoftwo
 \else \expandafter \@secondoftwo
 \fi
}%
\providecommand \@ifx [1]{%
 \ifx #1\expandafter \@firstoftwo
 \else \expandafter \@secondoftwo
 \fi
}%
\providecommand \natexlab [1]{#1}%
\providecommand \enquote  [1]{``#1''}%
\providecommand \bibnamefont  [1]{#1}%
\providecommand \bibfnamefont [1]{#1}%
\providecommand \citenamefont [1]{#1}%
\providecommand \href@noop [0]{\@secondoftwo}%
\providecommand \href [0]{\begingroup \@sanitize@url \@href}%
\providecommand \@href[1]{\@@startlink{#1}\@@href}%
\providecommand \@@href[1]{\endgroup#1\@@endlink}%
\providecommand \@sanitize@url [0]{\catcode `\\12\catcode `\$12\catcode
  `\&12\catcode `\#12\catcode `\^12\catcode `\_12\catcode `\%12\relax}%
\providecommand \@@startlink[1]{}%
\providecommand \@@endlink[0]{}%
\providecommand \url  [0]{\begingroup\@sanitize@url \@url }%
\providecommand \@url [1]{\endgroup\@href {#1}{\urlprefix }}%
\providecommand \urlprefix  [0]{URL }%
\providecommand \Eprint [0]{\href }%
\providecommand \doibase [0]{http://dx.doi.org/}%
\providecommand \selectlanguage [0]{\@gobble}%
\providecommand \bibinfo  [0]{\@secondoftwo}%
\providecommand \bibfield  [0]{\@secondoftwo}%
\providecommand \translation [1]{[#1]}%
\providecommand \BibitemOpen [0]{}%
\providecommand \bibitemStop [0]{}%
\providecommand \bibitemNoStop [0]{.\EOS\space}%
\providecommand \EOS [0]{\spacefactor3000\relax}%
\providecommand \BibitemShut  [1]{\csname bibitem#1\endcsname}%
\let\auto@bib@innerbib\@empty
\bibitem [{\citenamefont {Abbott}\ \emph {et~al.}(2007)\citenamefont {Abbott}
  \emph {et~al.}}]{Abbott:2007ce}%
  \BibitemOpen
  \bibfield  {author} {\bibinfo {author} {\bibnamefont {Abbott}, \bibfnamefont
  {B.}},  \emph {et~al.} (\bibinfo {collaboration} {LIGO Scientific
  Collaboration})} (\bibinfo {year} {2007}),\ \href {\doibase
  10.1103/PhysRevD.76.042001} {\bibfield  {journal} {\bibinfo  {journal}
  {Phys.Rev.}\ }\textbf {\bibinfo {volume} {D76}},\ \bibinfo {pages}
  {042001}},\ \Eprint {http://arxiv.org/abs/gr-qc/0702039} {arXiv:gr-qc/0702039
  [gr-qc]} \BibitemShut {NoStop}%
\bibitem [{\citenamefont {Abbott}\ \emph {et~al.}(2008)\citenamefont {Abbott}
  \emph {et~al.}}]{Abbott:2008fx}%
  \BibitemOpen
  \bibfield  {author} {\bibinfo {author} {\bibnamefont {Abbott}, \bibfnamefont
  {B.}},  \emph {et~al.} (\bibinfo {collaboration} {LIGO Scientific
  Collaboration})} (\bibinfo {year} {2008}),\ \href {\doibase
  10.1088/0004-637X/706/1/L203, 10.1086/591526} {\bibfield  {journal} {\bibinfo
   {journal} {Astrophys.J.}\ }\textbf {\bibinfo {volume} {683}},\ \bibinfo
  {pages} {L45}},\ \Eprint {http://arxiv.org/abs/0805.4758} {arXiv:0805.4758
  [astro-ph]} \BibitemShut {NoStop}%
\bibitem [{\citenamefont {Abbott}\ \emph {et~al.}(2010)\citenamefont {Abbott}
  \emph {et~al.}}]{Collaboration:2009rfa}%
  \BibitemOpen
  \bibfield  {author} {\bibinfo {author} {\bibnamefont {Abbott}, \bibfnamefont
  {B.}},  \emph {et~al.} (\bibinfo {collaboration} {Virgo Collaboration})}
  (\bibinfo {year} {2010}),\ \href {\doibase 10.1088/0004-637X/713/1/671}
  {\bibfield  {journal} {\bibinfo  {journal} {Astrophys.J.}\ }\textbf {\bibinfo
  {volume} {713}},\ \bibinfo {pages} {671}},\ \Eprint
  {http://arxiv.org/abs/0909.3583} {arXiv:0909.3583 [astro-ph.HE]} \BibitemShut
  {NoStop}%
\bibitem [{\citenamefont {Abuki}(2003)}]{Abuki:2003ut}%
  \BibitemOpen
  \bibfield  {author} {\bibinfo {author} {\bibnamefont {Abuki}, \bibfnamefont
  {H.}}} (\bibinfo {year} {2003}),\ \href {\doibase 10.1143/PTP.110.937}
  {\bibfield  {journal} {\bibinfo  {journal} {Prog.Theor.Phys.}\ }\textbf
  {\bibinfo {volume} {110}},\ \bibinfo {pages} {937}},\ \Eprint
  {http://arxiv.org/abs/hep-ph/0306074} {arXiv:hep-ph/0306074 [hep-ph]}
  \BibitemShut {NoStop}%
\bibitem [{\citenamefont {Abuki}\ \emph {et~al.}(2012)\citenamefont {Abuki},
  \citenamefont {Ishibashi},\ and\ \citenamefont {Suzuki}}]{Abuki:2011pf}%
  \BibitemOpen
  \bibfield  {author} {\bibinfo {author} {\bibnamefont {Abuki}, \bibfnamefont
  {H.}}, \bibinfo {author} {\bibfnamefont {D.}~\bibnamefont {Ishibashi}}, \
  and\ \bibinfo {author} {\bibfnamefont {K.}~\bibnamefont {Suzuki}}} (\bibinfo
  {year} {2012}),\ \href {\doibase 10.1103/PhysRevD.85.074002} {\bibfield
  {journal} {\bibinfo  {journal} {Phys.Rev.}\ }\textbf {\bibinfo {volume}
  {D85}},\ \bibinfo {pages} {074002}},\ \Eprint
  {http://arxiv.org/abs/1109.1615} {arXiv:1109.1615 [hep-ph]} \BibitemShut
  {NoStop}%
\bibitem [{\citenamefont {Abuki}\ and\ \citenamefont
  {Kunihiro}(2006)}]{Abuki:2005ms}%
  \BibitemOpen
  \bibfield  {author} {\bibinfo {author} {\bibnamefont {Abuki}, \bibfnamefont
  {H.}}, \ and\ \bibinfo {author} {\bibfnamefont {T.}~\bibnamefont {Kunihiro}}}
  (\bibinfo {year} {2006}),\ \href {\doibase 10.1016/j.nuclphysa.2005.12.019}
  {\bibfield  {journal} {\bibinfo  {journal} {Nucl.Phys.}\ }\textbf {\bibinfo
  {volume} {A768}},\ \bibinfo {pages} {118}},\ \Eprint
  {http://arxiv.org/abs/hep-ph/0509172} {arXiv:hep-ph/0509172 [hep-ph]}
  \BibitemShut {NoStop}%
\bibitem [{\citenamefont {Alcock}\ \emph {et~al.}(1986)\citenamefont {Alcock},
  \citenamefont {Farhi},\ and\ \citenamefont {Olinto}}]{Alcock:1986hz}%
  \BibitemOpen
  \bibfield  {author} {\bibinfo {author} {\bibnamefont {Alcock}, \bibfnamefont
  {C.}}, \bibinfo {author} {\bibfnamefont {E.}~\bibnamefont {Farhi}}, \ and\
  \bibinfo {author} {\bibfnamefont {A.}~\bibnamefont {Olinto}}} (\bibinfo
  {year} {1986}),\ \href {\doibase 10.1086/164679} {\bibfield  {journal}
  {\bibinfo  {journal} {Astrophys.J.}\ }\textbf {\bibinfo {volume} {310}},\
  \bibinfo {pages} {261}}\BibitemShut {NoStop}%
\bibitem [{\citenamefont {Alford}(2001)}]{Alford:2001dt}%
  \BibitemOpen
  \bibfield  {author} {\bibinfo {author} {\bibnamefont {Alford}, \bibfnamefont
  {M.~G.}}} (\bibinfo {year} {2001}),\ \href {\doibase
  10.1146/annurev.nucl.51.101701.132449} {\bibfield  {journal} {\bibinfo
  {journal} {Ann.Rev.Nucl.Part.Sci.}\ }\textbf {\bibinfo {volume} {51}},\
  \bibinfo {pages} {131}},\ \Eprint {http://arxiv.org/abs/hep-ph/0102047}
  {arXiv:hep-ph/0102047 [hep-ph]} \BibitemShut {NoStop}%
\bibitem [{\citenamefont {Alford}\ \emph
  {et~al.}(1999{\natexlab{a}})\citenamefont {Alford}, \citenamefont {Berges},\
  and\ \citenamefont {Rajagopal}}]{Alford:1999pa}%
  \BibitemOpen
  \bibfield  {author} {\bibinfo {author} {\bibnamefont {Alford}, \bibfnamefont
  {M.~G.}}, \bibinfo {author} {\bibfnamefont {J.}~\bibnamefont {Berges}}, \
  and\ \bibinfo {author} {\bibfnamefont {K.}~\bibnamefont {Rajagopal}}}
  (\bibinfo {year} {1999}{\natexlab{a}}),\ \href {\doibase
  10.1016/S0550-3213(99)00410-1} {\bibfield  {journal} {\bibinfo  {journal}
  {Nucl.Phys.}\ }\textbf {\bibinfo {volume} {B558}},\ \bibinfo {pages} {219}},\
  \Eprint {http://arxiv.org/abs/hep-ph/9903502} {arXiv:hep-ph/9903502 [hep-ph]}
  \BibitemShut {NoStop}%
\bibitem [{\citenamefont {Alford}\ \emph
  {et~al.}(2000{\natexlab{a}})\citenamefont {Alford}, \citenamefont {Berges},\
  and\ \citenamefont {Rajagopal}}]{Alford:1999xc}%
  \BibitemOpen
  \bibfield  {author} {\bibinfo {author} {\bibnamefont {Alford}, \bibfnamefont
  {M.~G.}}, \bibinfo {author} {\bibfnamefont {J.}~\bibnamefont {Berges}}, \
  and\ \bibinfo {author} {\bibfnamefont {K.}~\bibnamefont {Rajagopal}}}
  (\bibinfo {year} {2000}{\natexlab{a}}),\ \href {\doibase
  10.1103/PhysRevLett.84.598} {\bibfield  {journal} {\bibinfo  {journal}
  {Phys.Rev.Lett.}\ }\textbf {\bibinfo {volume} {84}},\ \bibinfo {pages}
  {598}},\ \Eprint {http://arxiv.org/abs/hep-ph/9908235} {arXiv:hep-ph/9908235
  [hep-ph]} \BibitemShut {NoStop}%
\bibitem [{\citenamefont {Alford}\ \emph
  {et~al.}(2000{\natexlab{b}})\citenamefont {Alford}, \citenamefont {Berges},\
  and\ \citenamefont {Rajagopal}}]{Alford:1999pb}%
  \BibitemOpen
  \bibfield  {author} {\bibinfo {author} {\bibnamefont {Alford}, \bibfnamefont
  {M.~G.}}, \bibinfo {author} {\bibfnamefont {J.}~\bibnamefont {Berges}}, \
  and\ \bibinfo {author} {\bibfnamefont {K.}~\bibnamefont {Rajagopal}}}
  (\bibinfo {year} {2000}{\natexlab{b}}),\ \href {\doibase
  10.1016/S0550-3213(99)00830-5} {\bibfield  {journal} {\bibinfo  {journal}
  {Nucl.Phys.}\ }\textbf {\bibinfo {volume} {B571}},\ \bibinfo {pages} {269}},\
  \Eprint {http://arxiv.org/abs/hep-ph/9910254} {arXiv:hep-ph/9910254 [hep-ph]}
  \BibitemShut {NoStop}%
\bibitem [{\citenamefont {Alford}\ \emph {et~al.}(2003)\citenamefont {Alford},
  \citenamefont {Bowers}, \citenamefont {Cheyne},\ and\ \citenamefont
  {Cowan}}]{Alford:2002rz}%
  \BibitemOpen
  \bibfield  {author} {\bibinfo {author} {\bibnamefont {Alford}, \bibfnamefont
  {M.~G.}}, \bibinfo {author} {\bibfnamefont {J.~A.}\ \bibnamefont {Bowers}},
  \bibinfo {author} {\bibfnamefont {J.~M.}\ \bibnamefont {Cheyne}}, \ and\
  \bibinfo {author} {\bibfnamefont {G.~A.}\ \bibnamefont {Cowan}}} (\bibinfo
  {year} {2003}),\ \href {\doibase 10.1103/PhysRevD.67.054018} {\bibfield
  {journal} {\bibinfo  {journal} {Phys.Rev.}\ }\textbf {\bibinfo {volume}
  {D67}},\ \bibinfo {pages} {054018}},\ \Eprint
  {http://arxiv.org/abs/hep-ph/0210106} {arXiv:hep-ph/0210106 [hep-ph]}
  \BibitemShut {NoStop}%
\bibitem [{\citenamefont {Alford}\ \emph {et~al.}(2001)\citenamefont {Alford},
  \citenamefont {Bowers},\ and\ \citenamefont {Rajagopal}}]{Alford:2000ze}%
  \BibitemOpen
  \bibfield  {author} {\bibinfo {author} {\bibnamefont {Alford}, \bibfnamefont
  {M.~G.}}, \bibinfo {author} {\bibfnamefont {J.~A.}\ \bibnamefont {Bowers}}, \
  and\ \bibinfo {author} {\bibfnamefont {K.}~\bibnamefont {Rajagopal}}}
  (\bibinfo {year} {2001}),\ \href {\doibase 10.1103/PhysRevD.63.074016}
  {\bibfield  {journal} {\bibinfo  {journal} {Phys.Rev.}\ }\textbf {\bibinfo
  {volume} {D63}},\ \bibinfo {pages} {074016}},\ \Eprint
  {http://arxiv.org/abs/hep-ph/0008208} {arXiv:hep-ph/0008208 [hep-ph]}
  \BibitemShut {NoStop}%
\bibitem [{\citenamefont {Alford}\ \emph
  {et~al.}(2005{\natexlab{a}})\citenamefont {Alford}, \citenamefont {Braby},
  \citenamefont {Paris},\ and\ \citenamefont {Reddy}}]{Alford:2004pf}%
  \BibitemOpen
  \bibfield  {author} {\bibinfo {author} {\bibnamefont {Alford}, \bibfnamefont
  {M.~G.}}, \bibinfo {author} {\bibfnamefont {M.}~\bibnamefont {Braby}},
  \bibinfo {author} {\bibfnamefont {M.~W.}\ \bibnamefont {Paris}}, \ and\
  \bibinfo {author} {\bibfnamefont {S.}~\bibnamefont {Reddy}}} (\bibinfo {year}
  {2005}{\natexlab{a}}),\ \href {\doibase 10.1086/430902} {\bibfield  {journal}
  {\bibinfo  {journal} {Astrophys.J.}\ }\textbf {\bibinfo {volume} {629}},\
  \bibinfo {pages} {969}},\ \Eprint {http://arxiv.org/abs/nucl-th/0411016}
  {arXiv:nucl-th/0411016 [nucl-th]} \BibitemShut {NoStop}%
\bibitem [{\citenamefont {Alford}\ \emph
  {et~al.}(2005{\natexlab{b}})\citenamefont {Alford}, \citenamefont {Jotwani},
  \citenamefont {Kouvaris}, \citenamefont {Kundu},\ and\ \citenamefont
  {Rajagopal}}]{Alford:2004zr}%
  \BibitemOpen
  \bibfield  {author} {\bibinfo {author} {\bibnamefont {Alford}, \bibfnamefont
  {M.~G.}}, \bibinfo {author} {\bibfnamefont {P.}~\bibnamefont {Jotwani}},
  \bibinfo {author} {\bibfnamefont {C.}~\bibnamefont {Kouvaris}}, \bibinfo
  {author} {\bibfnamefont {J.}~\bibnamefont {Kundu}}, \ and\ \bibinfo {author}
  {\bibfnamefont {K.}~\bibnamefont {Rajagopal}}} (\bibinfo {year}
  {2005}{\natexlab{b}}),\ \href {\doibase 10.1103/PhysRevD.71.114011}
  {\bibfield  {journal} {\bibinfo  {journal} {Phys.Rev.}\ }\textbf {\bibinfo
  {volume} {D71}},\ \bibinfo {pages} {114011}},\ \Eprint
  {http://arxiv.org/abs/astro-ph/0411560} {arXiv:astro-ph/0411560 [astro-ph]}
  \BibitemShut {NoStop}%
\bibitem [{\citenamefont {Alford}\ \emph
  {et~al.}(1999{\natexlab{b}})\citenamefont {Alford}, \citenamefont
  {Kapustin},\ and\ \citenamefont {Wilczek}}]{Alford:1998sd}%
  \BibitemOpen
  \bibfield  {author} {\bibinfo {author} {\bibnamefont {Alford}, \bibfnamefont
  {M.~G.}}, \bibinfo {author} {\bibfnamefont {A.}~\bibnamefont {Kapustin}}, \
  and\ \bibinfo {author} {\bibfnamefont {F.}~\bibnamefont {Wilczek}}} (\bibinfo
  {year} {1999}{\natexlab{b}}),\ \href {\doibase 10.1103/PhysRevD.59.054502}
  {\bibfield  {journal} {\bibinfo  {journal} {Phys.Rev.}\ }\textbf {\bibinfo
  {volume} {D59}},\ \bibinfo {pages} {054502}},\ \Eprint
  {http://arxiv.org/abs/hep-lat/9807039} {arXiv:hep-lat/9807039 [hep-lat]}
  \BibitemShut {NoStop}%
\bibitem [{\citenamefont {Alford}\ \emph {et~al.}(2004)\citenamefont {Alford},
  \citenamefont {Kouvaris},\ and\ \citenamefont {Rajagopal}}]{Alford:2003fq}%
  \BibitemOpen
  \bibfield  {author} {\bibinfo {author} {\bibnamefont {Alford}, \bibfnamefont
  {M.~G.}}, \bibinfo {author} {\bibfnamefont {C.}~\bibnamefont {Kouvaris}}, \
  and\ \bibinfo {author} {\bibfnamefont {K.}~\bibnamefont {Rajagopal}}}
  (\bibinfo {year} {2004}),\ \href {\doibase 10.1103/PhysRevLett.92.222001}
  {\bibfield  {journal} {\bibinfo  {journal} {Phys.Rev.Lett.}\ }\textbf
  {\bibinfo {volume} {92}},\ \bibinfo {pages} {222001}},\ \Eprint
  {http://arxiv.org/abs/hep-ph/0311286} {arXiv:hep-ph/0311286 [hep-ph]}
  \BibitemShut {NoStop}%
\bibitem [{\citenamefont {Alford}\ \emph
  {et~al.}(2005{\natexlab{c}})\citenamefont {Alford}, \citenamefont
  {Kouvaris},\ and\ \citenamefont {Rajagopal}}]{Alford:2004hz}%
  \BibitemOpen
  \bibfield  {author} {\bibinfo {author} {\bibnamefont {Alford}, \bibfnamefont
  {M.~G.}}, \bibinfo {author} {\bibfnamefont {C.}~\bibnamefont {Kouvaris}}, \
  and\ \bibinfo {author} {\bibfnamefont {K.}~\bibnamefont {Rajagopal}}}
  (\bibinfo {year} {2005}{\natexlab{c}}),\ \href {\doibase
  10.1103/PhysRevD.71.054009} {\bibfield  {journal} {\bibinfo  {journal}
  {Phys.Rev.}\ }\textbf {\bibinfo {volume} {D71}},\ \bibinfo {pages}
  {054009}},\ \Eprint {http://arxiv.org/abs/hep-ph/0406137}
  {arXiv:hep-ph/0406137 [hep-ph]} \BibitemShut {NoStop}%
\bibitem [{\citenamefont {Alford}\ and\ \citenamefont
  {Rajagopal}(2002)}]{Alford:2002kj}%
  \BibitemOpen
  \bibfield  {author} {\bibinfo {author} {\bibnamefont {Alford}, \bibfnamefont
  {M.~G.}}, \ and\ \bibinfo {author} {\bibfnamefont {K.}~\bibnamefont
  {Rajagopal}}} (\bibinfo {year} {2002}),\ \href@noop {} {\bibfield  {journal}
  {\bibinfo  {journal} {JHEP}\ }\textbf {\bibinfo {volume} {0206}},\ \bibinfo
  {pages} {031}},\ \Eprint {http://arxiv.org/abs/hep-ph/0204001}
  {arXiv:hep-ph/0204001 [hep-ph]} \BibitemShut {NoStop}%
\bibitem [{\citenamefont {Alford}\ \emph {et~al.}(1998)\citenamefont {Alford},
  \citenamefont {Rajagopal},\ and\ \citenamefont {Wilczek}}]{Alford:1997zt}%
  \BibitemOpen
  \bibfield  {author} {\bibinfo {author} {\bibnamefont {Alford}, \bibfnamefont
  {M.~G.}}, \bibinfo {author} {\bibfnamefont {K.}~\bibnamefont {Rajagopal}}, \
  and\ \bibinfo {author} {\bibfnamefont {F.}~\bibnamefont {Wilczek}}} (\bibinfo
  {year} {1998}),\ \href {\doibase 10.1016/S0370-2693(98)00051-3} {\bibfield
  {journal} {\bibinfo  {journal} {Phys.Lett.}\ }\textbf {\bibinfo {volume}
  {B422}},\ \bibinfo {pages} {247}},\ \Eprint
  {http://arxiv.org/abs/hep-ph/9711395} {arXiv:hep-ph/9711395 [hep-ph]}
  \BibitemShut {NoStop}%
\bibitem [{\citenamefont {Alford}\ \emph
  {et~al.}(1999{\natexlab{c}})\citenamefont {Alford}, \citenamefont
  {Rajagopal},\ and\ \citenamefont {Wilczek}}]{Alford:1998mk}%
  \BibitemOpen
  \bibfield  {author} {\bibinfo {author} {\bibnamefont {Alford}, \bibfnamefont
  {M.~G.}}, \bibinfo {author} {\bibfnamefont {K.}~\bibnamefont {Rajagopal}}, \
  and\ \bibinfo {author} {\bibfnamefont {F.}~\bibnamefont {Wilczek}}} (\bibinfo
  {year} {1999}{\natexlab{c}}),\ \href {\doibase 10.1016/S0550-3213(98)00668-3}
  {\bibfield  {journal} {\bibinfo  {journal} {Nucl.Phys.}\ }\textbf {\bibinfo
  {volume} {B537}},\ \bibinfo {pages} {443}},\ \Eprint
  {http://arxiv.org/abs/hep-ph/9804403} {arXiv:hep-ph/9804403 [hep-ph]}
  \BibitemShut {NoStop}%
\bibitem [{\citenamefont {Alford}\ and\ \citenamefont
  {Reddy}(2003)}]{Alford:2002rj}%
  \BibitemOpen
  \bibfield  {author} {\bibinfo {author} {\bibnamefont {Alford}, \bibfnamefont
  {M.~G.}}, \ and\ \bibinfo {author} {\bibfnamefont {S.}~\bibnamefont {Reddy}}}
  (\bibinfo {year} {2003}),\ \href {\doibase 10.1103/PhysRevD.67.074024}
  {\bibfield  {journal} {\bibinfo  {journal} {Phys.Rev.}\ }\textbf {\bibinfo
  {volume} {D67}},\ \bibinfo {pages} {074024}},\ \Eprint
  {http://arxiv.org/abs/nucl-th/0211046} {arXiv:nucl-th/0211046 [nucl-th]}
  \BibitemShut {NoStop}%
\bibitem [{\citenamefont {Alford}\ \emph {et~al.}(2008)\citenamefont {Alford},
  \citenamefont {Schmitt}, \citenamefont {Rajagopal},\ and\ \citenamefont
  {Schafer}}]{Alford:2007xm}%
  \BibitemOpen
  \bibfield  {author} {\bibinfo {author} {\bibnamefont {Alford}, \bibfnamefont
  {M.~G.}}, \bibinfo {author} {\bibfnamefont {A.}~\bibnamefont {Schmitt}},
  \bibinfo {author} {\bibfnamefont {K.}~\bibnamefont {Rajagopal}}, \ and\
  \bibinfo {author} {\bibfnamefont {T.}~\bibnamefont {Schafer}}} (\bibinfo
  {year} {2008}),\ \href {\doibase 10.1103/RevModPhys.80.1455} {\bibfield
  {journal} {\bibinfo  {journal} {Rev.Mod.Phys.}\ }\textbf {\bibinfo {volume}
  {80}},\ \bibinfo {pages} {1455}},\ \Eprint {http://arxiv.org/abs/0709.4635}
  {arXiv:0709.4635 [hep-ph]} \BibitemShut {NoStop}%
\bibitem [{\citenamefont {Alford}\ and\ \citenamefont
  {Wang}(2005)}]{Alford:2005qw}%
  \BibitemOpen
  \bibfield  {author} {\bibinfo {author} {\bibnamefont {Alford}, \bibfnamefont
  {M.~G.}}, \ and\ \bibinfo {author} {\bibfnamefont {Q.-h.}\ \bibnamefont
  {Wang}}} (\bibinfo {year} {2005}),\ \href {\doibase
  10.1088/0954-3899/31/7/017} {\bibfield  {journal} {\bibinfo  {journal}
  {J.Phys.}\ }\textbf {\bibinfo {volume} {G31}},\ \bibinfo {pages} {719}},\
  \Eprint {http://arxiv.org/abs/hep-ph/0501078} {arXiv:hep-ph/0501078 [hep-ph]}
  \BibitemShut {NoStop}%
\bibitem [{\citenamefont {Alford}\ and\ \citenamefont
  {Wang}(2006)}]{Alford:2005kj}%
  \BibitemOpen
  \bibfield  {author} {\bibinfo {author} {\bibnamefont {Alford}, \bibfnamefont
  {M.~G.}}, \ and\ \bibinfo {author} {\bibfnamefont {Q.-h.}\ \bibnamefont
  {Wang}}} (\bibinfo {year} {2006}),\ \href {\doibase
  10.1088/0954-3899/32/2/001} {\bibfield  {journal} {\bibinfo  {journal}
  {J.Phys.}\ }\textbf {\bibinfo {volume} {G32}},\ \bibinfo {pages} {63}},\
  \Eprint {http://arxiv.org/abs/hep-ph/0507269} {arXiv:hep-ph/0507269 [hep-ph]}
  \BibitemShut {NoStop}%
\bibitem [{\citenamefont {Allton}\ \emph {et~al.}(2003)\citenamefont {Allton},
  \citenamefont {Ejiri}, \citenamefont {Hands}, \citenamefont {Kaczmarek},
  \citenamefont {Karsch} \emph {et~al.}}]{Allton:2003vx}%
  \BibitemOpen
  \bibfield  {author} {\bibinfo {author} {\bibnamefont {Allton}, \bibfnamefont
  {C.~R.}}, \bibinfo {author} {\bibfnamefont {S.}~\bibnamefont {Ejiri}},
  \bibinfo {author} {\bibfnamefont {S.~J.}\ \bibnamefont {Hands}}, \bibinfo
  {author} {\bibfnamefont {O.}~\bibnamefont {Kaczmarek}}, \bibinfo {author}
  {\bibfnamefont {F.}~\bibnamefont {Karsch}},  \emph {et~al.}} (\bibinfo {year}
  {2003}),\ \href {\doibase 10.1103/PhysRevD.68.014507} {\bibfield  {journal}
  {\bibinfo  {journal} {Phys.Rev.}\ }\textbf {\bibinfo {volume} {D68}},\
  \bibinfo {pages} {014507}},\ \Eprint {http://arxiv.org/abs/hep-lat/0305007}
  {arXiv:hep-lat/0305007 [hep-lat]} \BibitemShut {NoStop}%
\bibitem [{\citenamefont {{Alpar}}(1977)}]{Alpar:1977}%
  \BibitemOpen
  \bibfield  {author} {\bibinfo {author} {\bibnamefont {{Alpar}}, \bibfnamefont
  {M.~A.}}} (\bibinfo {year} {1977}),\ \href {\doibase 10.1086/155183}
  {\bibfield  {journal} {\bibinfo  {journal} {\apj}\ }\textbf {\bibinfo
  {volume} {213}},\ \bibinfo {pages} {527}}\BibitemShut {NoStop}%
\bibitem [{\citenamefont {{Alpar}}\ \emph
  {et~al.}(1984{\natexlab{a}})\citenamefont {{Alpar}}, \citenamefont
  {{Anderson}}, \citenamefont {{Pines}},\ and\ \citenamefont
  {{Shaham}}}]{Alparcreep2}%
  \BibitemOpen
  \bibfield  {author} {\bibinfo {author} {\bibnamefont {{Alpar}}, \bibfnamefont
  {M.~A.}}, \bibinfo {author} {\bibfnamefont {P.~W.}\ \bibnamefont
  {{Anderson}}}, \bibinfo {author} {\bibfnamefont {D.}~\bibnamefont {{Pines}}},
  \ and\ \bibinfo {author} {\bibfnamefont {J.}~\bibnamefont {{Shaham}}}}
  (\bibinfo {year} {1984}{\natexlab{a}}),\ \href {\doibase 10.1086/161849}
  {\bibfield  {journal} {\bibinfo  {journal} {\apj}\ }\textbf {\bibinfo
  {volume} {278}},\ \bibinfo {pages} {791}}\BibitemShut {NoStop}%
\bibitem [{\citenamefont {{Alpar}}\ \emph
  {et~al.}(1984{\natexlab{b}})\citenamefont {{Alpar}}, \citenamefont {{Pines}},
  \citenamefont {{Anderson}},\ and\ \citenamefont {{Shaham}}}]{Alparcreep1}%
  \BibitemOpen
  \bibfield  {author} {\bibinfo {author} {\bibnamefont {{Alpar}}, \bibfnamefont
  {M.~A.}}, \bibinfo {author} {\bibfnamefont {D.}~\bibnamefont {{Pines}}},
  \bibinfo {author} {\bibfnamefont {P.~W.}\ \bibnamefont {{Anderson}}}, \ and\
  \bibinfo {author} {\bibfnamefont {J.}~\bibnamefont {{Shaham}}}} (\bibinfo
  {year} {1984}{\natexlab{b}}),\ \href {\doibase 10.1086/161616} {\bibfield
  {journal} {\bibinfo  {journal} {\apj}\ }\textbf {\bibinfo {volume} {276}},\
  \bibinfo {pages} {325}}\BibitemShut {NoStop}%
\bibitem [{\citenamefont {Amore}\ \emph {et~al.}(2002)\citenamefont {Amore},
  \citenamefont {Birse}, \citenamefont {McGovern},\ and\ \citenamefont
  {Walet}}]{Amore:2001uf}%
  \BibitemOpen
  \bibfield  {author} {\bibinfo {author} {\bibnamefont {Amore}, \bibfnamefont
  {P.}}, \bibinfo {author} {\bibfnamefont {M.~C.}\ \bibnamefont {Birse}},
  \bibinfo {author} {\bibfnamefont {J.~A.}\ \bibnamefont {McGovern}}, \ and\
  \bibinfo {author} {\bibfnamefont {N.~R.}\ \bibnamefont {Walet}}} (\bibinfo
  {year} {2002}),\ \href {\doibase 10.1103/PhysRevD.65.074005} {\bibfield
  {journal} {\bibinfo  {journal} {Phys.Rev.}\ }\textbf {\bibinfo {volume}
  {D65}},\ \bibinfo {pages} {074005}},\ \Eprint
  {http://arxiv.org/abs/hep-ph/0110267} {arXiv:hep-ph/0110267 [hep-ph]}
  \BibitemShut {NoStop}%
\bibitem [{\citenamefont {{Anderson}}\ and\ \citenamefont
  {{Itoh}}(1975)}]{Anderson:1975}%
  \BibitemOpen
  \bibfield  {author} {\bibinfo {author} {\bibnamefont {{Anderson}},
  \bibfnamefont {P.~W.}}, \ and\ \bibinfo {author} {\bibfnamefont
  {N.}~\bibnamefont {{Itoh}}}} (\bibinfo {year} {1975}),\ \href {\doibase
  10.1038/256025a0} {\bibfield  {journal} {\bibinfo  {journal} {\nat}\ }\textbf
  {\bibinfo {volume} {256}},\ \bibinfo {pages} {25}}\BibitemShut {NoStop}%
\bibitem [{\citenamefont {Andersson}(1998)}]{Andersson:1997xt}%
  \BibitemOpen
  \bibfield  {author} {\bibinfo {author} {\bibnamefont {Andersson},
  \bibfnamefont {N.}}} (\bibinfo {year} {1998}),\ \href {\doibase
  10.1086/305919} {\bibfield  {journal} {\bibinfo  {journal} {Astrophys.J.}\
  }\textbf {\bibinfo {volume} {502}},\ \bibinfo {pages} {708}},\ \Eprint
  {http://arxiv.org/abs/gr-qc/9706075} {arXiv:gr-qc/9706075 [gr-qc]}
  \BibitemShut {NoStop}%
\bibitem [{\citenamefont {Anglani}\ \emph {et~al.}(2007)\citenamefont
  {Anglani}, \citenamefont {Gatto}, \citenamefont {Ippolito}, \citenamefont
  {Nardulli},\ and\ \citenamefont {Ruggieri}}]{Anglani:2007aa}%
  \BibitemOpen
  \bibfield  {author} {\bibinfo {author} {\bibnamefont {Anglani}, \bibfnamefont
  {R.}}, \bibinfo {author} {\bibfnamefont {R.}~\bibnamefont {Gatto}}, \bibinfo
  {author} {\bibfnamefont {N.~D.}\ \bibnamefont {Ippolito}}, \bibinfo {author}
  {\bibfnamefont {G.}~\bibnamefont {Nardulli}}, \ and\ \bibinfo {author}
  {\bibfnamefont {M.}~\bibnamefont {Ruggieri}}} (\bibinfo {year} {2007}),\
  \href {\doibase 10.1103/PhysRevD.76.054007} {\bibfield  {journal} {\bibinfo
  {journal} {Phys.Rev.}\ }\textbf {\bibinfo {volume} {D76}},\ \bibinfo {pages}
  {054007}},\ \Eprint {http://arxiv.org/abs/0706.1781} {arXiv:0706.1781
  [hep-ph]} \BibitemShut {NoStop}%
\bibitem [{\citenamefont {Anglani}\ \emph {et~al.}(2011)\citenamefont
  {Anglani}, \citenamefont {Mannarelli},\ and\ \citenamefont
  {Ruggieri}}]{Anglani:2011cw}%
  \BibitemOpen
  \bibfield  {author} {\bibinfo {author} {\bibnamefont {Anglani}, \bibfnamefont
  {R.}}, \bibinfo {author} {\bibfnamefont {M.}~\bibnamefont {Mannarelli}}, \
  and\ \bibinfo {author} {\bibfnamefont {M.}~\bibnamefont {Ruggieri}}}
  (\bibinfo {year} {2011}),\ \href {\doibase 10.1088/1367-2630/13/5/055002}
  {\bibfield  {journal} {\bibinfo  {journal} {New J.Phys.}\ }\textbf {\bibinfo
  {volume} {13}},\ \bibinfo {pages} {055002}},\ \Eprint
  {http://arxiv.org/abs/1101.4277} {arXiv:1101.4277 [hep-ph]} \BibitemShut
  {NoStop}%
\bibitem [{\citenamefont {Anglani}\ \emph {et~al.}(2006)\citenamefont
  {Anglani}, \citenamefont {Nardulli}, \citenamefont {Ruggieri},\ and\
  \citenamefont {Mannarelli}}]{Anglani:2006br}%
  \BibitemOpen
  \bibfield  {author} {\bibinfo {author} {\bibnamefont {Anglani}, \bibfnamefont
  {R.}}, \bibinfo {author} {\bibfnamefont {G.}~\bibnamefont {Nardulli}},
  \bibinfo {author} {\bibfnamefont {M.}~\bibnamefont {Ruggieri}}, \ and\
  \bibinfo {author} {\bibfnamefont {M.}~\bibnamefont {Mannarelli}}} (\bibinfo
  {year} {2006}),\ \href {\doibase 10.1103/PhysRevD.74.074005} {\bibfield
  {journal} {\bibinfo  {journal} {Phys.Rev.}\ }\textbf {\bibinfo {volume}
  {D74}},\ \bibinfo {pages} {074005}},\ \Eprint
  {http://arxiv.org/abs/hep-ph/0607341} {arXiv:hep-ph/0607341 [hep-ph]}
  \BibitemShut {NoStop}%
\bibitem [{\citenamefont {Aoki}\ \emph {et~al.}(2012)\citenamefont {Aoki} \emph
  {et~al.}}]{Aoki:2012tk}%
  \BibitemOpen
  \bibfield  {author} {\bibinfo {author} {\bibnamefont {Aoki}, \bibfnamefont
  {S.}},  \emph {et~al.} (\bibinfo {collaboration} {HAL QCD Collaboration})}
  (\bibinfo {year} {2012}),\ \href@noop {} {\ }\Eprint
  {http://arxiv.org/abs/1206.5088} {arXiv:1206.5088 [hep-lat]} \BibitemShut
  {NoStop}%
\bibitem [{\citenamefont {Auzzi}\ \emph {et~al.}(2003)\citenamefont {Auzzi},
  \citenamefont {Bolognesi}, \citenamefont {Evslin}, \citenamefont {Konishi},\
  and\ \citenamefont {Yung}}]{Auzzi:2003fs}%
  \BibitemOpen
  \bibfield  {author} {\bibinfo {author} {\bibnamefont {Auzzi}, \bibfnamefont
  {R.}}, \bibinfo {author} {\bibfnamefont {S.}~\bibnamefont {Bolognesi}},
  \bibinfo {author} {\bibfnamefont {J.}~\bibnamefont {Evslin}}, \bibinfo
  {author} {\bibfnamefont {K.}~\bibnamefont {Konishi}}, \ and\ \bibinfo
  {author} {\bibfnamefont {A.}~\bibnamefont {Yung}}} (\bibinfo {year} {2003}),\
  \href {\doibase 10.1016/j.nuclphysb.2003.09.029} {\bibfield  {journal}
  {\bibinfo  {journal} {Nucl.Phys.}\ }\textbf {\bibinfo {volume} {B673}},\
  \bibinfo {pages} {187}},\ \Eprint {http://arxiv.org/abs/hep-th/0307287}
  {arXiv:hep-th/0307287 [hep-th]} \BibitemShut {NoStop}%
\bibitem [{\citenamefont {Bahcall}\ and\ \citenamefont
  {Wolf}(1965)}]{Bachall1965}%
  \BibitemOpen
  \bibfield  {author} {\bibinfo {author} {\bibnamefont {Bahcall}, \bibfnamefont
  {J.~N.}}, \ and\ \bibinfo {author} {\bibfnamefont {R.~A.}\ \bibnamefont
  {Wolf}}} (\bibinfo {year} {1965}),\ \href {\doibase
  10.1103/PhysRevLett.14.343} {\bibfield  {journal} {\bibinfo  {journal} {Phys.
  Rev. Lett.}\ }\textbf {\bibinfo {volume} {14}},\ \bibinfo {pages}
  {343}}\BibitemShut {NoStop}%
\bibitem [{\citenamefont {Bailin}\ and\ \citenamefont
  {Love}(1979)}]{Bailin:1979nh}%
  \BibitemOpen
  \bibfield  {author} {\bibinfo {author} {\bibnamefont {Bailin}, \bibfnamefont
  {D.}}, \ and\ \bibinfo {author} {\bibfnamefont {A.}~\bibnamefont {Love}}}
  (\bibinfo {year} {1979}),\ \href {\doibase 10.1088/0305-4470/12/10/009}
  {\bibfield  {journal} {\bibinfo  {journal} {J.Phys.}\ }\textbf {\bibinfo
  {volume} {A12}},\ \bibinfo {pages} {L283}}\BibitemShut {NoStop}%
\bibitem [{\citenamefont {Bailin}\ and\ \citenamefont
  {Love}(1984)}]{Bailin:1983bm}%
  \BibitemOpen
  \bibfield  {author} {\bibinfo {author} {\bibnamefont {Bailin}, \bibfnamefont
  {D.}}, \ and\ \bibinfo {author} {\bibfnamefont {A.}~\bibnamefont {Love}}}
  (\bibinfo {year} {1984}),\ \href {\doibase 10.1016/0370-1573(84)90145-5}
  {\bibfield  {journal} {\bibinfo  {journal} {Phys.Rept.}\ }\textbf {\bibinfo
  {volume} {107}},\ \bibinfo {pages} {325}}\BibitemShut {NoStop}%
\bibitem [{\citenamefont {Balachandran}\ \emph {et~al.}(2006)\citenamefont
  {Balachandran}, \citenamefont {Digal},\ and\ \citenamefont
  {Matsuura}}]{Balachandran:2005ev}%
  \BibitemOpen
  \bibfield  {author} {\bibinfo {author} {\bibnamefont {Balachandran},
  \bibfnamefont {A.}}, \bibinfo {author} {\bibfnamefont {S.}~\bibnamefont
  {Digal}}, \ and\ \bibinfo {author} {\bibfnamefont {T.}~\bibnamefont
  {Matsuura}}} (\bibinfo {year} {2006}),\ \href {\doibase
  10.1103/PhysRevD.73.074009} {\bibfield  {journal} {\bibinfo  {journal}
  {Phys.Rev.}\ }\textbf {\bibinfo {volume} {D73}},\ \bibinfo {pages}
  {074009}},\ \Eprint {http://arxiv.org/abs/hep-ph/0509276}
  {arXiv:hep-ph/0509276 [hep-ph]} \BibitemShut {NoStop}%
\bibitem [{\citenamefont {{Bardeen}}\ \emph
  {et~al.}(1957{\natexlab{a}})\citenamefont {{Bardeen}}, \citenamefont
  {{Cooper}},\ and\ \citenamefont {{Schrieffer}}}]{Bardeen:1957-2}%
  \BibitemOpen
  \bibfield  {author} {\bibinfo {author} {\bibnamefont {{Bardeen}},
  \bibfnamefont {J.}}, \bibinfo {author} {\bibfnamefont {L.~N.}\ \bibnamefont
  {{Cooper}}}, \ and\ \bibinfo {author} {\bibfnamefont {J.~R.}\ \bibnamefont
  {{Schrieffer}}}} (\bibinfo {year} {1957}{\natexlab{a}}),\ \href {\doibase
  10.1103/PhysRev.106.162} {\bibfield  {journal} {\bibinfo  {journal} {Phys.
  Rev.}\ }\textbf {\bibinfo {volume} {106}},\ \bibinfo {pages}
  {162}}\BibitemShut {NoStop}%
\bibitem [{\citenamefont {{Bardeen}}\ \emph
  {et~al.}(1957{\natexlab{b}})\citenamefont {{Bardeen}}, \citenamefont
  {{Cooper}},\ and\ \citenamefont {{Schrieffer}}}]{Bardeen:1957-1}%
  \BibitemOpen
  \bibfield  {author} {\bibinfo {author} {\bibnamefont {{Bardeen}},
  \bibfnamefont {J.}}, \bibinfo {author} {\bibfnamefont {L.~N.}\ \bibnamefont
  {{Cooper}}}, \ and\ \bibinfo {author} {\bibfnamefont {J.~R.}\ \bibnamefont
  {{Schrieffer}}}} (\bibinfo {year} {1957}{\natexlab{b}}),\ \href {\doibase
  10.1103/PhysRev.108.1175} {\bibfield  {journal} {\bibinfo  {journal} {Phys.
  Rev.}\ }\textbf {\bibinfo {volume} {108}},\ \bibinfo {pages}
  {1175}}\BibitemShut {NoStop}%
\bibitem [{\citenamefont {{Barret}}\ \emph {et~al.}(2005)\citenamefont
  {{Barret}}, \citenamefont {{Olive}},\ and\ \citenamefont
  {{Miller}}}]{barret2005}%
  \BibitemOpen
  \bibfield  {author} {\bibinfo {author} {\bibnamefont {{Barret}},
  \bibfnamefont {D.}}, \bibinfo {author} {\bibfnamefont {J.-F.}\ \bibnamefont
  {{Olive}}}, \ and\ \bibinfo {author} {\bibfnamefont {M.~C.}\ \bibnamefont
  {{Miller}}}} (\bibinfo {year} {2005}),\ \href {\doibase
  10.1111/j.1365-2966.2005.09214.x} {\bibfield  {journal} {\bibinfo  {journal}
  {\mnras}\ }\textbf {\bibinfo {volume} {361}},\ \bibinfo {pages} {855}},\
  \Eprint {http://arxiv.org/abs/arXiv:astro-ph/0505402}
  {arXiv:astro-ph/0505402} \BibitemShut {NoStop}%
\bibitem [{\citenamefont {Barrois}(1977)}]{Barrois:1977xd}%
  \BibitemOpen
  \bibfield  {author} {\bibinfo {author} {\bibnamefont {Barrois}, \bibfnamefont
  {B.~C.}}} (\bibinfo {year} {1977}),\ \href {\doibase
  10.1016/0550-3213(77)90123-7} {\bibfield  {journal} {\bibinfo  {journal}
  {Nucl.Phys.}\ }\textbf {\bibinfo {volume} {B129}},\ \bibinfo {pages}
  {390}}\BibitemShut {NoStop}%
\bibitem [{\citenamefont {{Baym}}\ and\ \citenamefont {{Chin}}(1976)}]{Baym}%
  \BibitemOpen
  \bibfield  {author} {\bibinfo {author} {\bibnamefont {{Baym}}, \bibfnamefont
  {G.}}, \ and\ \bibinfo {author} {\bibfnamefont {S.~A.}\ \bibnamefont
  {{Chin}}}} (\bibinfo {year} {1976}),\ \href {\doibase
  10.1016/0370-2693(76)90517-7} {\bibinfo  {journal} {Phys.Lett.}\ ,\ \bibinfo
  {pages} {241}}\BibitemShut {NoStop}%
\bibitem [{\citenamefont {Baym}\ \emph {et~al.}(1969)\citenamefont {Baym},
  \citenamefont {Pethick}, \citenamefont {Pines},\ and\ \citenamefont
  {Ruderman}}]{Baym1969n}%
  \BibitemOpen
\bibfield  {journal} {  }\bibfield  {author} {\bibinfo {author} {\bibnamefont
  {Baym}, \bibfnamefont {G.}}, \bibinfo {author} {\bibfnamefont
  {C.}~\bibnamefont {Pethick}}, \bibinfo {author} {\bibfnamefont
  {D.}~\bibnamefont {Pines}}, \ and\ \bibinfo {author} {\bibfnamefont
  {M.}~\bibnamefont {Ruderman}}} (\bibinfo {year} {1969}),\ \href {\doibase
  10.1038/224872a0} {\bibfield  {journal} {\bibinfo  {journal} {Nature}\
  }\textbf {\bibinfo {volume} {224}},\ \bibinfo {pages} {872}}\BibitemShut
  {NoStop}%
\bibitem [{\citenamefont {{Baym}}\ and\ \citenamefont
  {{Pines}}(1971)}]{Baym1971}%
  \BibitemOpen
  \bibfield  {author} {\bibinfo {author} {\bibnamefont {{Baym}}, \bibfnamefont
  {G.}}, \ and\ \bibinfo {author} {\bibfnamefont {D.}~\bibnamefont {{Pines}}}}
  (\bibinfo {year} {1971}),\ \href {\doibase 10.1016/0003-4916(71)90084-4}
  {\bibfield  {journal} {\bibinfo  {journal} {Annals Phys.}\ }\textbf {\bibinfo
  {volume} {66}},\ \bibinfo {pages} {816}}\BibitemShut {NoStop}%
\bibitem [{\citenamefont {Beane}\ \emph {et~al.}(2000)\citenamefont {Beane},
  \citenamefont {Bedaque},\ and\ \citenamefont {Savage}}]{Beane:2000ms}%
  \BibitemOpen
  \bibfield  {author} {\bibinfo {author} {\bibnamefont {Beane}, \bibfnamefont
  {S.~R.}}, \bibinfo {author} {\bibfnamefont {P.~F.}\ \bibnamefont {Bedaque}},
  \ and\ \bibinfo {author} {\bibfnamefont {M.~J.}\ \bibnamefont {Savage}}}
  (\bibinfo {year} {2000}),\ \href {\doibase 10.1016/S0370-2693(00)00606-7}
  {\bibfield  {journal} {\bibinfo  {journal} {Phys.Lett.}\ }\textbf {\bibinfo
  {volume} {B483}},\ \bibinfo {pages} {131}},\ \Eprint
  {http://arxiv.org/abs/hep-ph/0002209} {arXiv:hep-ph/0002209 [hep-ph]}
  \BibitemShut {NoStop}%
\bibitem [{\citenamefont {Bedaque}\ \emph {et~al.}(2003)\citenamefont
  {Bedaque}, \citenamefont {Caldas},\ and\ \citenamefont
  {Rupak}}]{Bedaque:2003hi}%
  \BibitemOpen
  \bibfield  {author} {\bibinfo {author} {\bibnamefont {Bedaque}, \bibfnamefont
  {P.~F.}}, \bibinfo {author} {\bibfnamefont {H.}~\bibnamefont {Caldas}}, \
  and\ \bibinfo {author} {\bibfnamefont {G.}~\bibnamefont {Rupak}}} (\bibinfo
  {year} {2003}),\ \href {\doibase 10.1103/PhysRevLett.91.247002} {\bibfield
  {journal} {\bibinfo  {journal} {Phys.Rev.Lett.}\ }\textbf {\bibinfo {volume}
  {91}},\ \bibinfo {pages} {247002}},\ \Eprint
  {http://arxiv.org/abs/cond-mat/0306694} {arXiv:cond-mat/0306694 [cond-mat]}
  \BibitemShut {NoStop}%
\bibitem [{\citenamefont {Bedaque}\ and\ \citenamefont
  {Schafer}(2002)}]{Bedaque:2001je}%
  \BibitemOpen
  \bibfield  {author} {\bibinfo {author} {\bibnamefont {Bedaque}, \bibfnamefont
  {P.~F.}}, \ and\ \bibinfo {author} {\bibfnamefont {T.}~\bibnamefont
  {Schafer}}} (\bibinfo {year} {2002}),\ \href {\doibase
  10.1016/S0375-9474(01)01272-6} {\bibfield  {journal} {\bibinfo  {journal}
  {Nucl.Phys.}\ }\textbf {\bibinfo {volume} {A697}},\ \bibinfo {pages} {802}},\
  \Eprint {http://arxiv.org/abs/hep-ph/0105150} {arXiv:hep-ph/0105150 [hep-ph]}
  \BibitemShut {NoStop}%
\bibitem [{\citenamefont {Berges}\ and\ \citenamefont
  {Rajagopal}(1999)}]{Berges:1998rc}%
  \BibitemOpen
  \bibfield  {author} {\bibinfo {author} {\bibnamefont {Berges}, \bibfnamefont
  {J.}}, \ and\ \bibinfo {author} {\bibfnamefont {K.}~\bibnamefont
  {Rajagopal}}} (\bibinfo {year} {1999}),\ \href {\doibase
  10.1016/S0550-3213(98)00620-8} {\bibfield  {journal} {\bibinfo  {journal}
  {Nucl.Phys.}\ }\textbf {\bibinfo {volume} {B538}},\ \bibinfo {pages} {215}},\
  \Eprint {http://arxiv.org/abs/hep-ph/9804233} {arXiv:hep-ph/9804233 [hep-ph]}
  \BibitemShut {NoStop}%
\bibitem [{\citenamefont {Blaschke}\ \emph {et~al.}(2007)\citenamefont
  {Blaschke}, \citenamefont {Gomez~Dumm}, \citenamefont {Grunfeld},
  \citenamefont {Klahn},\ and\ \citenamefont {Scoccola}}]{Blaschke:2007ri}%
  \BibitemOpen
  \bibfield  {author} {\bibinfo {author} {\bibnamefont {Blaschke},
  \bibfnamefont {D.}}, \bibinfo {author} {\bibfnamefont {D.}~\bibnamefont
  {Gomez~Dumm}}, \bibinfo {author} {\bibfnamefont {A.~G.}\ \bibnamefont
  {Grunfeld}}, \bibinfo {author} {\bibfnamefont {T.}~\bibnamefont {Klahn}}, \
  and\ \bibinfo {author} {\bibfnamefont {N.~N.}\ \bibnamefont {Scoccola}}}
  (\bibinfo {year} {2007}),\ \href {\doibase 10.1103/PhysRevC.75.065804}
  {\bibfield  {journal} {\bibinfo  {journal} {Phys.Rev.}\ }\textbf {\bibinfo
  {volume} {C75}},\ \bibinfo {pages} {065804}},\ \Eprint
  {http://arxiv.org/abs/nucl-th/0703088} {arXiv:nucl-th/0703088 [nucl-th]}
  \BibitemShut {NoStop}%
\bibitem [{\citenamefont {Bonanno}\ and\ \citenamefont
  {Sedrakian}(2012)}]{Bonanno:2011ch}%
  \BibitemOpen
  \bibfield  {author} {\bibinfo {author} {\bibnamefont {Bonanno}, \bibfnamefont
  {L.}}, \ and\ \bibinfo {author} {\bibfnamefont {A.}~\bibnamefont
  {Sedrakian}}} (\bibinfo {year} {2012}),\ \href@noop {} {\bibfield  {journal}
  {\bibinfo  {journal} {Astron.Astrophys.}\ }\textbf {\bibinfo {volume}
  {539}},\ \bibinfo {pages} {A16}},\ \Eprint {http://arxiv.org/abs/1108.0559}
  {arXiv:1108.0559 [astro-ph.SR]} \BibitemShut {NoStop}%
\bibitem [{\citenamefont {Bowers}\ \emph {et~al.}(2001)\citenamefont {Bowers},
  \citenamefont {Kundu}, \citenamefont {Rajagopal},\ and\ \citenamefont
  {Shuster}}]{Bowers:2001ip}%
  \BibitemOpen
  \bibfield  {author} {\bibinfo {author} {\bibnamefont {Bowers}, \bibfnamefont
  {J.~A.}}, \bibinfo {author} {\bibfnamefont {J.}~\bibnamefont {Kundu}},
  \bibinfo {author} {\bibfnamefont {K.}~\bibnamefont {Rajagopal}}, \ and\
  \bibinfo {author} {\bibfnamefont {E.}~\bibnamefont {Shuster}}} (\bibinfo
  {year} {2001}),\ \href {\doibase 10.1103/PhysRevD.64.014024} {\bibfield
  {journal} {\bibinfo  {journal} {Phys.Rev.}\ }\textbf {\bibinfo {volume}
  {D64}},\ \bibinfo {pages} {014024}},\ \Eprint
  {http://arxiv.org/abs/hep-ph/0101067} {arXiv:hep-ph/0101067 [hep-ph]}
  \BibitemShut {NoStop}%
\bibitem [{\citenamefont {Bowers}\ and\ \citenamefont
  {Rajagopal}(2002)}]{Bowers:2002xr}%
  \BibitemOpen
  \bibfield  {author} {\bibinfo {author} {\bibnamefont {Bowers}, \bibfnamefont
  {J.~A.}}, \ and\ \bibinfo {author} {\bibfnamefont {K.}~\bibnamefont
  {Rajagopal}}} (\bibinfo {year} {2002}),\ \href {\doibase
  10.1103/PhysRevD.66.065002} {\bibfield  {journal} {\bibinfo  {journal}
  {Phys.Rev.}\ }\textbf {\bibinfo {volume} {D66}},\ \bibinfo {pages}
  {065002}},\ \Eprint {http://arxiv.org/abs/hep-ph/0204079}
  {arXiv:hep-ph/0204079 [hep-ph]} \BibitemShut {NoStop}%
\bibitem [{\citenamefont {{Boynton}}\ \emph {et~al.}(1969)\citenamefont
  {{Boynton}}, \citenamefont {{Groth}}, \citenamefont {{Partridge}},\ and\
  \citenamefont {{Wilkinson}}}]{Boynton1969}%
  \BibitemOpen
  \bibfield  {author} {\bibinfo {author} {\bibnamefont {{Boynton}},
  \bibfnamefont {P.~E.}}, \bibinfo {author} {\bibfnamefont {E.~J.}\
  \bibnamefont {{Groth}}, \bibfnamefont {III}}, \bibinfo {author}
  {\bibfnamefont {R.~B.}\ \bibnamefont {{Partridge}}}, \ and\ \bibinfo {author}
  {\bibfnamefont {D.~T.}\ \bibnamefont {{Wilkinson}}}} (\bibinfo {year}
  {1969}),\ \href@noop {} {\bibfield  {journal} {\bibinfo  {journal}
  {\iaucirc}\ }\textbf {\bibinfo {volume} {2179}},\ \bibinfo {pages}
  {1}}\BibitemShut {NoStop}%
\bibitem [{\citenamefont {Brown}\ \emph {et~al.}(1988)\citenamefont {Brown},
  \citenamefont {Kubodera}, \citenamefont {Page},\ and\ \citenamefont
  {Pizzochero}}]{Brown:1988ik}%
  \BibitemOpen
  \bibfield  {author} {\bibinfo {author} {\bibnamefont {Brown}, \bibfnamefont
  {G.~E.}}, \bibinfo {author} {\bibfnamefont {K.}~\bibnamefont {Kubodera}},
  \bibinfo {author} {\bibfnamefont {D.}~\bibnamefont {Page}}, \ and\ \bibinfo
  {author} {\bibfnamefont {P.}~\bibnamefont {Pizzochero}}} (\bibinfo {year}
  {1988}),\ \href {\doibase 10.1103/PhysRevD.37.2042} {\bibfield  {journal}
  {\bibinfo  {journal} {Phys.Rev.}\ }\textbf {\bibinfo {volume} {D37}},\
  \bibinfo {pages} {2042}}\BibitemShut {NoStop}%
\bibitem [{\citenamefont {Brown}\ \emph {et~al.}(2000)\citenamefont {Brown},
  \citenamefont {Liu},\ and\ \citenamefont {Ren}}]{Brown:1999aq}%
  \BibitemOpen
  \bibfield  {author} {\bibinfo {author} {\bibnamefont {Brown}, \bibfnamefont
  {W.~E.}}, \bibinfo {author} {\bibfnamefont {J.~T.}\ \bibnamefont {Liu}}, \
  and\ \bibinfo {author} {\bibfnamefont {H.-C.}\ \bibnamefont {Ren}}} (\bibinfo
  {year} {2000}),\ \href {\doibase 10.1103/PhysRevD.61.114012} {\bibfield
  {journal} {\bibinfo  {journal} {Phys.Rev.}\ }\textbf {\bibinfo {volume}
  {D61}},\ \bibinfo {pages} {114012}},\ \Eprint
  {http://arxiv.org/abs/hep-ph/9908248} {arXiv:hep-ph/9908248 [hep-ph]}
  \BibitemShut {NoStop}%
\bibitem [{\citenamefont {Buballa}(2005)}]{Buballa:2003qv}%
  \BibitemOpen
  \bibfield  {author} {\bibinfo {author} {\bibnamefont {Buballa}, \bibfnamefont
  {M.}}} (\bibinfo {year} {2005}),\ \href {\doibase
  10.1016/j.physrep.2004.11.004} {\bibfield  {journal} {\bibinfo  {journal}
  {Phys.Rept.}\ }\textbf {\bibinfo {volume} {407}},\ \bibinfo {pages} {205}},\
  \Eprint {http://arxiv.org/abs/hep-ph/0402234} {arXiv:hep-ph/0402234 [hep-ph]}
  \BibitemShut {NoStop}%
\bibitem [{\citenamefont {Buballa}\ \emph {et~al.}(2003)\citenamefont
  {Buballa}, \citenamefont {Hosek},\ and\ \citenamefont
  {Oertel}}]{Buballa:2002wy}%
  \BibitemOpen
  \bibfield  {author} {\bibinfo {author} {\bibnamefont {Buballa}, \bibfnamefont
  {M.}}, \bibinfo {author} {\bibfnamefont {J.}~\bibnamefont {Hosek}}, \ and\
  \bibinfo {author} {\bibfnamefont {M.}~\bibnamefont {Oertel}}} (\bibinfo
  {year} {2003}),\ \href {\doibase 10.1103/PhysRevLett.90.182002} {\bibfield
  {journal} {\bibinfo  {journal} {Phys.Rev.Lett.}\ }\textbf {\bibinfo {volume}
  {90}},\ \bibinfo {pages} {182002}},\ \Eprint
  {http://arxiv.org/abs/hep-ph/0204275} {arXiv:hep-ph/0204275 [hep-ph]}
  \BibitemShut {NoStop}%
\bibitem [{\citenamefont {Buballa}\ \emph {et~al.}(2004)\citenamefont
  {Buballa}, \citenamefont {Neumann}, \citenamefont {Oertel},\ and\
  \citenamefont {Shovkovy}}]{Buballa:2003et}%
  \BibitemOpen
  \bibfield  {author} {\bibinfo {author} {\bibnamefont {Buballa}, \bibfnamefont
  {M.}}, \bibinfo {author} {\bibfnamefont {F.}~\bibnamefont {Neumann}},
  \bibinfo {author} {\bibfnamefont {M.}~\bibnamefont {Oertel}}, \ and\ \bibinfo
  {author} {\bibfnamefont {I.}~\bibnamefont {Shovkovy}}} (\bibinfo {year}
  {2004}),\ \href {\doibase 10.1016/j.physletb.2004.05.064} {\bibfield
  {journal} {\bibinfo  {journal} {Phys.Lett.}\ }\textbf {\bibinfo {volume}
  {B595}},\ \bibinfo {pages} {36}},\ \Eprint
  {http://arxiv.org/abs/nucl-th/0312078} {arXiv:nucl-th/0312078 [nucl-th]}
  \BibitemShut {NoStop}%
\bibitem [{\citenamefont {Buballa}\ and\ \citenamefont
  {Nickel}(2010)}]{Buballa:2009ct}%
  \BibitemOpen
  \bibfield  {author} {\bibinfo {author} {\bibnamefont {Buballa}, \bibfnamefont
  {M.}}, \ and\ \bibinfo {author} {\bibfnamefont {D.}~\bibnamefont {Nickel}}}
  (\bibinfo {year} {2010}),\ \href@noop {} {\bibfield  {journal} {\bibinfo
  {journal} {Acta Phys.Polon.Supp.}\ }\textbf {\bibinfo {volume} {3}},\
  \bibinfo {pages} {523}},\ \Eprint {http://arxiv.org/abs/0911.2333}
  {arXiv:0911.2333 [hep-ph]} \BibitemShut {NoStop}%
\bibitem [{\citenamefont {Buballa}\ and\ \citenamefont
  {Oertel}(1999)}]{Buballa:1998pr}%
  \BibitemOpen
  \bibfield  {author} {\bibinfo {author} {\bibnamefont {Buballa}, \bibfnamefont
  {M.}}, \ and\ \bibinfo {author} {\bibfnamefont {M.}~\bibnamefont {Oertel}}}
  (\bibinfo {year} {1999}),\ \href {\doibase 10.1016/S0370-2693(99)00533-X}
  {\bibfield  {journal} {\bibinfo  {journal} {Phys.Lett.}\ }\textbf {\bibinfo
  {volume} {B457}},\ \bibinfo {pages} {261}},\ \Eprint
  {http://arxiv.org/abs/hep-ph/9810529} {arXiv:hep-ph/9810529 [hep-ph]}
  \BibitemShut {NoStop}%
\bibitem [{\citenamefont {{Bulaevski{\v \i}}}(1973)}]{Bulaevskii}%
  \BibitemOpen
  \bibfield  {author} {\bibinfo {author} {\bibnamefont {{Bulaevski{\v \i}}},
  \bibfnamefont {L.~N.}}} (\bibinfo {year} {1973}),\ \href@noop {} {\bibfield
  {journal} {\bibinfo  {journal} {Soviet Journal of Experimental and
  Theoretical Physics}\ }\textbf {\bibinfo {volume} {37}},\ \bibinfo {pages}
  {1133}}\BibitemShut {NoStop}%
\bibitem [{\citenamefont {Bulgac}\ and\ \citenamefont
  {Forbes}(2008)}]{Bulgac:2008tm}%
  \BibitemOpen
  \bibfield  {author} {\bibinfo {author} {\bibnamefont {Bulgac}, \bibfnamefont
  {A.}}, \ and\ \bibinfo {author} {\bibfnamefont {M.~M.}\ \bibnamefont
  {Forbes}}} (\bibinfo {year} {2008}),\ \href {\doibase
  10.1103/PhysRevLett.101.215301} {\bibfield  {journal} {\bibinfo  {journal}
  {Phys.Rev.Lett.}\ }\textbf {\bibinfo {volume} {101}},\ \bibinfo {pages}
  {215301}},\ \Eprint {http://arxiv.org/abs/0804.3364} {arXiv:0804.3364
  [cond-mat.supr-con]} \BibitemShut {NoStop}%
\bibitem [{\citenamefont {Bulgac}\ \emph {et~al.}(2006)\citenamefont {Bulgac},
  \citenamefont {Forbes~McNeil},\ and\ \citenamefont
  {Schwenk}}]{Bulgac:2006gh}%
  \BibitemOpen
  \bibfield  {author} {\bibinfo {author} {\bibnamefont {Bulgac}, \bibfnamefont
  {A.}}, \bibinfo {author} {\bibfnamefont {M.}~\bibnamefont {Forbes~McNeil}}, \
  and\ \bibinfo {author} {\bibfnamefont {A.}~\bibnamefont {Schwenk}}} (\bibinfo
  {year} {2006}),\ \href {\doibase 10.1103/PhysRevLett.97.020402} {\bibfield
  {journal} {\bibinfo  {journal} {Phys.Rev.Lett.}\ }\textbf {\bibinfo {volume}
  {97}},\ \bibinfo {pages} {020402}},\ \Eprint
  {http://arxiv.org/abs/cond-mat/0602274} {arXiv:cond-mat/0602274 [cond-mat]}
  \BibitemShut {NoStop}%
\bibitem [{\citenamefont {Carlson}\ and\ \citenamefont
  {Reddy}(2005)}]{Carlson:2005kg}%
  \BibitemOpen
  \bibfield  {author} {\bibinfo {author} {\bibnamefont {Carlson}, \bibfnamefont
  {J.}}, \ and\ \bibinfo {author} {\bibfnamefont {S.}~\bibnamefont {Reddy}}}
  (\bibinfo {year} {2005}),\ \href {\doibase 10.1103/PhysRevLett.95.060401}
  {\bibfield  {journal} {\bibinfo  {journal} {Phys.Rev.Lett.}\ }\textbf
  {\bibinfo {volume} {95}},\ \bibinfo {pages} {060401}},\ \Eprint
  {http://arxiv.org/abs/cond-mat/0503256} {arXiv:cond-mat/0503256 [cond-mat]}
  \BibitemShut {NoStop}%
\bibitem [{\citenamefont {Carter}\ and\ \citenamefont
  {Diakonov}(1999)}]{Carter:1998ji}%
  \BibitemOpen
  \bibfield  {author} {\bibinfo {author} {\bibnamefont {Carter}, \bibfnamefont
  {G.~W.}}, \ and\ \bibinfo {author} {\bibfnamefont {D.}~\bibnamefont
  {Diakonov}}} (\bibinfo {year} {1999}),\ \href {\doibase
  10.1103/PhysRevD.60.016004} {\bibfield  {journal} {\bibinfo  {journal}
  {Phys.Rev.}\ }\textbf {\bibinfo {volume} {D60}},\ \bibinfo {pages}
  {016004}},\ \Eprint {http://arxiv.org/abs/hep-ph/9812445}
  {arXiv:hep-ph/9812445 [hep-ph]} \BibitemShut {NoStop}%
\bibitem [{\citenamefont {Casalbuoni}\ \emph {et~al.}(2006)\citenamefont
  {Casalbuoni}, \citenamefont {Ciminale}, \citenamefont {Gatto}, \citenamefont
  {Nardulli},\ and\ \citenamefont {Ruggieri}}]{Casalbuoni:2006zs}%
  \BibitemOpen
  \bibfield  {author} {\bibinfo {author} {\bibnamefont {Casalbuoni},
  \bibfnamefont {R.}}, \bibinfo {author} {\bibfnamefont {M.}~\bibnamefont
  {Ciminale}}, \bibinfo {author} {\bibfnamefont {R.}~\bibnamefont {Gatto}},
  \bibinfo {author} {\bibfnamefont {G.}~\bibnamefont {Nardulli}}, \ and\
  \bibinfo {author} {\bibfnamefont {M.}~\bibnamefont {Ruggieri}}} (\bibinfo
  {year} {2006}),\ \href {\doibase 10.1016/j.physletb.2006.09.051} {\bibfield
  {journal} {\bibinfo  {journal} {Phys.Lett.}\ }\textbf {\bibinfo {volume}
  {B642}},\ \bibinfo {pages} {350}},\ \Eprint
  {http://arxiv.org/abs/hep-ph/0606242} {arXiv:hep-ph/0606242 [hep-ph]}
  \BibitemShut {NoStop}%
\bibitem [{\citenamefont {Casalbuoni}\ \emph {et~al.}(2004)\citenamefont
  {Casalbuoni}, \citenamefont {Ciminale}, \citenamefont {Mannarelli},
  \citenamefont {Nardulli}, \citenamefont {Ruggieri} \emph
  {et~al.}}]{Casalbuoni:2004wm}%
  \BibitemOpen
  \bibfield  {author} {\bibinfo {author} {\bibnamefont {Casalbuoni},
  \bibfnamefont {R.}}, \bibinfo {author} {\bibfnamefont {M.}~\bibnamefont
  {Ciminale}}, \bibinfo {author} {\bibfnamefont {M.}~\bibnamefont
  {Mannarelli}}, \bibinfo {author} {\bibfnamefont {G.}~\bibnamefont
  {Nardulli}}, \bibinfo {author} {\bibfnamefont {M.}~\bibnamefont {Ruggieri}},
  \emph {et~al.}} (\bibinfo {year} {2004}),\ \href {\doibase
  10.1103/PhysRevD.70.054004} {\bibfield  {journal} {\bibinfo  {journal}
  {Phys.Rev.}\ }\textbf {\bibinfo {volume} {D70}},\ \bibinfo {pages}
  {054004}},\ \Eprint {http://arxiv.org/abs/hep-ph/0404090}
  {arXiv:hep-ph/0404090 [hep-ph]} \BibitemShut {NoStop}%
\bibitem [{\citenamefont {Casalbuoni}\ \emph
  {et~al.}(2002{\natexlab{a}})\citenamefont {Casalbuoni}, \citenamefont
  {De~Fazio}, \citenamefont {Gatto}, \citenamefont {Nardulli},\ and\
  \citenamefont {Ruggieri}}]{Casalbuoni:2002st}%
  \BibitemOpen
  \bibfield  {author} {\bibinfo {author} {\bibnamefont {Casalbuoni},
  \bibfnamefont {R.}}, \bibinfo {author} {\bibfnamefont {F.}~\bibnamefont
  {De~Fazio}}, \bibinfo {author} {\bibfnamefont {R.}~\bibnamefont {Gatto}},
  \bibinfo {author} {\bibfnamefont {G.}~\bibnamefont {Nardulli}}, \ and\
  \bibinfo {author} {\bibfnamefont {M.}~\bibnamefont {Ruggieri}}} (\bibinfo
  {year} {2002}{\natexlab{a}}),\ \href {\doibase 10.1016/S0370-2693(02)02757-0}
  {\bibfield  {journal} {\bibinfo  {journal} {Phys.Lett.}\ }\textbf {\bibinfo
  {volume} {B547}},\ \bibinfo {pages} {229}},\ \Eprint
  {http://arxiv.org/abs/hep-ph/0209105} {arXiv:hep-ph/0209105 [hep-ph]}
  \BibitemShut {NoStop}%
\bibitem [{\citenamefont {Casalbuoni}\ \emph {et~al.}(2000)\citenamefont
  {Casalbuoni}, \citenamefont {Duan},\ and\ \citenamefont
  {Sannino}}]{Casalbuoni:2000cn}%
  \BibitemOpen
  \bibfield  {author} {\bibinfo {author} {\bibnamefont {Casalbuoni},
  \bibfnamefont {R.}}, \bibinfo {author} {\bibfnamefont {Z.-y.}\ \bibnamefont
  {Duan}}, \ and\ \bibinfo {author} {\bibfnamefont {F.}~\bibnamefont
  {Sannino}}} (\bibinfo {year} {2000}),\ \href {\doibase
  10.1103/PhysRevD.62.094004} {\bibfield  {journal} {\bibinfo  {journal}
  {Phys.Rev.}\ }\textbf {\bibinfo {volume} {D62}},\ \bibinfo {pages}
  {094004}},\ \Eprint {http://arxiv.org/abs/hep-ph/0004207}
  {arXiv:hep-ph/0004207 [hep-ph]} \BibitemShut {NoStop}%
\bibitem [{\citenamefont {Casalbuoni}\ \emph
  {et~al.}(2002{\natexlab{b}})\citenamefont {Casalbuoni}, \citenamefont
  {Fabiano}, \citenamefont {Gatto}, \citenamefont {Mannarelli},\ and\
  \citenamefont {Nardulli}}]{Casalbuoni:2002my}%
  \BibitemOpen
  \bibfield  {author} {\bibinfo {author} {\bibnamefont {Casalbuoni},
  \bibfnamefont {R.}}, \bibinfo {author} {\bibfnamefont {E.}~\bibnamefont
  {Fabiano}}, \bibinfo {author} {\bibfnamefont {R.}~\bibnamefont {Gatto}},
  \bibinfo {author} {\bibfnamefont {M.}~\bibnamefont {Mannarelli}}, \ and\
  \bibinfo {author} {\bibfnamefont {G.}~\bibnamefont {Nardulli}}} (\bibinfo
  {year} {2002}{\natexlab{b}}),\ \href {\doibase 10.1103/PhysRevD.66.094006}
  {\bibfield  {journal} {\bibinfo  {journal} {Phys.Rev.}\ }\textbf {\bibinfo
  {volume} {D66}},\ \bibinfo {pages} {094006}},\ \Eprint
  {http://arxiv.org/abs/hep-ph/0208121} {arXiv:hep-ph/0208121 [hep-ph]}
  \BibitemShut {NoStop}%
\bibitem [{\citenamefont {Casalbuoni}\ and\ \citenamefont
  {Gatto}(1999)}]{Casalbuoni:1999wu}%
  \BibitemOpen
  \bibfield  {author} {\bibinfo {author} {\bibnamefont {Casalbuoni},
  \bibfnamefont {R.}}, \ and\ \bibinfo {author} {\bibfnamefont
  {R.}~\bibnamefont {Gatto}}} (\bibinfo {year} {1999}),\ \href {\doibase
  10.1016/S0370-2693(99)01032-1} {\bibfield  {journal} {\bibinfo  {journal}
  {Phys.Lett.}\ }\textbf {\bibinfo {volume} {B464}},\ \bibinfo {pages} {111}},\
  \Eprint {http://arxiv.org/abs/hep-ph/9908227} {arXiv:hep-ph/9908227 [hep-ph]}
  \BibitemShut {NoStop}%
\bibitem [{\citenamefont {Casalbuoni}\ \emph
  {et~al.}(2005{\natexlab{a}})\citenamefont {Casalbuoni}, \citenamefont
  {Gatto}, \citenamefont {Ippolito}, \citenamefont {Nardulli},\ and\
  \citenamefont {Ruggieri}}]{Casalbuoni:2005zp}%
  \BibitemOpen
  \bibfield  {author} {\bibinfo {author} {\bibnamefont {Casalbuoni},
  \bibfnamefont {R.}}, \bibinfo {author} {\bibfnamefont {R.}~\bibnamefont
  {Gatto}}, \bibinfo {author} {\bibfnamefont {N.}~\bibnamefont {Ippolito}},
  \bibinfo {author} {\bibfnamefont {G.}~\bibnamefont {Nardulli}}, \ and\
  \bibinfo {author} {\bibfnamefont {M.}~\bibnamefont {Ruggieri}}} (\bibinfo
  {year} {2005}{\natexlab{a}}),\ \href {\doibase
  10.1016/j.physletb.2005.08.123, 10.1016/j.physletb.2006.01.057} {\bibfield
  {journal} {\bibinfo  {journal} {Phys.Lett.}\ }\textbf {\bibinfo {volume}
  {B627}},\ \bibinfo {pages} {89}},\ \Eprint
  {http://arxiv.org/abs/hep-ph/0507247} {arXiv:hep-ph/0507247 [hep-ph]}
  \BibitemShut {NoStop}%
\bibitem [{\citenamefont {Casalbuoni}\ \emph
  {et~al.}(2001{\natexlab{a}})\citenamefont {Casalbuoni}, \citenamefont
  {Gatto}, \citenamefont {Mannarelli},\ and\ \citenamefont
  {Nardulli}}]{Casalbuoni:2001gt}%
  \BibitemOpen
  \bibfield  {author} {\bibinfo {author} {\bibnamefont {Casalbuoni},
  \bibfnamefont {R.}}, \bibinfo {author} {\bibfnamefont {R.}~\bibnamefont
  {Gatto}}, \bibinfo {author} {\bibfnamefont {M.}~\bibnamefont {Mannarelli}}, \
  and\ \bibinfo {author} {\bibfnamefont {G.}~\bibnamefont {Nardulli}}}
  (\bibinfo {year} {2001}{\natexlab{a}}),\ \href {\doibase
  10.1016/S0370-2693(01)00645-1} {\bibfield  {journal} {\bibinfo  {journal}
  {Phys.Lett.}\ }\textbf {\bibinfo {volume} {B511}},\ \bibinfo {pages} {218}},\
  \Eprint {http://arxiv.org/abs/hep-ph/0101326} {arXiv:hep-ph/0101326 [hep-ph]}
  \BibitemShut {NoStop}%
\bibitem [{\citenamefont {Casalbuoni}\ \emph
  {et~al.}(2002{\natexlab{c}})\citenamefont {Casalbuoni}, \citenamefont
  {Gatto}, \citenamefont {Mannarelli},\ and\ \citenamefont
  {Nardulli}}]{Casalbuoni:2002pa}%
  \BibitemOpen
  \bibfield  {author} {\bibinfo {author} {\bibnamefont {Casalbuoni},
  \bibfnamefont {R.}}, \bibinfo {author} {\bibfnamefont {R.}~\bibnamefont
  {Gatto}}, \bibinfo {author} {\bibfnamefont {M.}~\bibnamefont {Mannarelli}}, \
  and\ \bibinfo {author} {\bibfnamefont {G.}~\bibnamefont {Nardulli}}}
  (\bibinfo {year} {2002}{\natexlab{c}}),\ \href {\doibase
  10.1103/PhysRevD.66.014006} {\bibfield  {journal} {\bibinfo  {journal}
  {Phys.Rev.}\ }\textbf {\bibinfo {volume} {D66}},\ \bibinfo {pages}
  {014006}},\ \Eprint {http://arxiv.org/abs/hep-ph/0201059}
  {arXiv:hep-ph/0201059 [hep-ph]} \BibitemShut {NoStop}%
\bibitem [{\citenamefont {Casalbuoni}\ \emph
  {et~al.}(2002{\natexlab{d}})\citenamefont {Casalbuoni}, \citenamefont
  {Gatto}, \citenamefont {Mannarelli},\ and\ \citenamefont
  {Nardulli}}]{Casalbuoni:2001ha}%
  \BibitemOpen
  \bibfield  {author} {\bibinfo {author} {\bibnamefont {Casalbuoni},
  \bibfnamefont {R.}}, \bibinfo {author} {\bibfnamefont {R.}~\bibnamefont
  {Gatto}}, \bibinfo {author} {\bibfnamefont {M.}~\bibnamefont {Mannarelli}}, \
  and\ \bibinfo {author} {\bibfnamefont {G.}~\bibnamefont {Nardulli}}}
  (\bibinfo {year} {2002}{\natexlab{d}}),\ \href {\doibase
  10.1016/S0370-2693(01)01370-3} {\bibfield  {journal} {\bibinfo  {journal}
  {Phys.Lett.}\ }\textbf {\bibinfo {volume} {B524}},\ \bibinfo {pages} {144}},\
  \Eprint {http://arxiv.org/abs/hep-ph/0107024} {arXiv:hep-ph/0107024 [hep-ph]}
  \BibitemShut {NoStop}%
\bibitem [{\citenamefont {Casalbuoni}\ \emph
  {et~al.}(2005{\natexlab{b}})\citenamefont {Casalbuoni}, \citenamefont
  {Gatto}, \citenamefont {Mannarelli}, \citenamefont {Nardulli},\ and\
  \citenamefont {Ruggieri}}]{Casalbuoni:2004tb}%
  \BibitemOpen
  \bibfield  {author} {\bibinfo {author} {\bibnamefont {Casalbuoni},
  \bibfnamefont {R.}}, \bibinfo {author} {\bibfnamefont {R.}~\bibnamefont
  {Gatto}}, \bibinfo {author} {\bibfnamefont {M.}~\bibnamefont {Mannarelli}},
  \bibinfo {author} {\bibfnamefont {G.}~\bibnamefont {Nardulli}}, \ and\
  \bibinfo {author} {\bibfnamefont {M.}~\bibnamefont {Ruggieri}}} (\bibinfo
  {year} {2005}{\natexlab{b}}),\ \href {\doibase
  10.1016/j.physletb.2004.11.045, 10.1016/j.physletb.2005.04.025} {\bibfield
  {journal} {\bibinfo  {journal} {Phys.Lett.}\ }\textbf {\bibinfo {volume}
  {B605}},\ \bibinfo {pages} {362}},\ \Eprint
  {http://arxiv.org/abs/hep-ph/0410401} {arXiv:hep-ph/0410401 [hep-ph]}
  \BibitemShut {NoStop}%
\bibitem [{\citenamefont {Casalbuoni}\ \emph {et~al.}(2003)\citenamefont
  {Casalbuoni}, \citenamefont {Gatto}, \citenamefont {Mannarelli},
  \citenamefont {Nardulli}, \citenamefont {Ruggieri} \emph
  {et~al.}}]{Casalbuoni:2003sa}%
  \BibitemOpen
  \bibfield  {author} {\bibinfo {author} {\bibnamefont {Casalbuoni},
  \bibfnamefont {R.}}, \bibinfo {author} {\bibfnamefont {R.}~\bibnamefont
  {Gatto}}, \bibinfo {author} {\bibfnamefont {M.}~\bibnamefont {Mannarelli}},
  \bibinfo {author} {\bibfnamefont {G.}~\bibnamefont {Nardulli}}, \bibinfo
  {author} {\bibfnamefont {M.}~\bibnamefont {Ruggieri}},  \emph {et~al.}}
  (\bibinfo {year} {2003}),\ \href {\doibase 10.1016/j.physletb.2003.09.071}
  {\bibfield  {journal} {\bibinfo  {journal} {Phys.Lett.}\ }\textbf {\bibinfo
  {volume} {B575}},\ \bibinfo {pages} {181}},\ \Eprint
  {http://arxiv.org/abs/hep-ph/0307335} {arXiv:hep-ph/0307335 [hep-ph]}
  \BibitemShut {NoStop}%
\bibitem [{\citenamefont {Casalbuoni}\ \emph
  {et~al.}(2001{\natexlab{b}})\citenamefont {Casalbuoni}, \citenamefont
  {Gatto},\ and\ \citenamefont {Nardulli}}]{Casalbuoni:2000na}%
  \BibitemOpen
  \bibfield  {author} {\bibinfo {author} {\bibnamefont {Casalbuoni},
  \bibfnamefont {R.}}, \bibinfo {author} {\bibfnamefont {R.}~\bibnamefont
  {Gatto}}, \ and\ \bibinfo {author} {\bibfnamefont {G.}~\bibnamefont
  {Nardulli}}} (\bibinfo {year} {2001}{\natexlab{b}}),\ \href {\doibase
  10.1016/S0370-2693(00)01390-3} {\bibfield  {journal} {\bibinfo  {journal}
  {Phys.Lett.}\ }\textbf {\bibinfo {volume} {B498}},\ \bibinfo {pages} {179}},\
  \Eprint {http://arxiv.org/abs/hep-ph/0010321} {arXiv:hep-ph/0010321 [hep-ph]}
  \BibitemShut {NoStop}%
\bibitem [{\citenamefont {Casalbuoni}\ and\ \citenamefont
  {Nardulli}(2004)}]{Casalbuoni:2003wh}%
  \BibitemOpen
  \bibfield  {author} {\bibinfo {author} {\bibnamefont {Casalbuoni},
  \bibfnamefont {R.}}, \ and\ \bibinfo {author} {\bibfnamefont
  {G.}~\bibnamefont {Nardulli}}} (\bibinfo {year} {2004}),\ \href {\doibase
  10.1103/RevModPhys.76.263} {\bibfield  {journal} {\bibinfo  {journal}
  {Rev.Mod.Phys.}\ }\textbf {\bibinfo {volume} {76}},\ \bibinfo {pages}
  {263}},\ \Eprint {http://arxiv.org/abs/hep-ph/0305069} {arXiv:hep-ph/0305069
  [hep-ph]} \BibitemShut {NoStop}%
\bibitem [{\citenamefont {Castorina}\ \emph {et~al.}(2005)\citenamefont
  {Castorina}, \citenamefont {Grasso}, \citenamefont {Oertel}, \citenamefont
  {Urban},\ and\ \citenamefont {Zappala}}]{Castorina:2005kg}%
  \BibitemOpen
  \bibfield  {author} {\bibinfo {author} {\bibnamefont {Castorina},
  \bibfnamefont {P.}}, \bibinfo {author} {\bibfnamefont {M.}~\bibnamefont
  {Grasso}}, \bibinfo {author} {\bibfnamefont {M.}~\bibnamefont {Oertel}},
  \bibinfo {author} {\bibfnamefont {M.}~\bibnamefont {Urban}}, \ and\ \bibinfo
  {author} {\bibfnamefont {D.}~\bibnamefont {Zappala}}} (\bibinfo {year}
  {2005}),\ \href {\doibase 10.1103/PhysRevA.72.025601} {\bibfield  {journal}
  {\bibinfo  {journal} {Phys.Rev.}\ }\textbf {\bibinfo {volume} {A72}},\
  \bibinfo {pages} {025601}},\ \Eprint {http://arxiv.org/abs/cond-mat/0504391}
  {arXiv:cond-mat/0504391 [cond-mat]} \BibitemShut {NoStop}%
\bibitem [{\citenamefont {Chandrasekhar}(1962)}]{Chandrasekhar}%
  \BibitemOpen
  \bibfield  {author} {\bibinfo {author} {\bibnamefont {Chandrasekhar},
  \bibfnamefont {B.~S.}}} (\bibinfo {year} {1962}),\ \href {\doibase
  10.1063/1.1777362} {\bibfield  {journal} {\bibinfo  {journal} {Appl. Phys.
  Lett.}\ }\textbf {\bibinfo {volume} {1}},\ \bibinfo {pages} {7}}\BibitemShut
  {NoStop}%
\bibitem [{\citenamefont {Chin}\ \emph {et~al.}(2010)\citenamefont {Chin},
  \citenamefont {Grimm}, \citenamefont {Julienne},\ and\ \citenamefont
  {Tiesinga}}]{Feshbach-review}%
  \BibitemOpen
  \bibfield  {author} {\bibinfo {author} {\bibnamefont {Chin}, \bibfnamefont
  {C.}}, \bibinfo {author} {\bibfnamefont {R.}~\bibnamefont {Grimm}}, \bibinfo
  {author} {\bibfnamefont {P.}~\bibnamefont {Julienne}}, \ and\ \bibinfo
  {author} {\bibfnamefont {E.}~\bibnamefont {Tiesinga}}} (\bibinfo {year}
  {2010}),\ \href {\doibase 10.1103/RevModPhys.82.1225} {\bibfield  {journal}
  {\bibinfo  {journal} {Rev. Mod. Phys.}\ }\textbf {\bibinfo {volume} {82}},\
  \bibinfo {pages} {1225}}\BibitemShut {NoStop}%
\bibitem [{\citenamefont {Chiu}\ and\ \citenamefont
  {Salpeter}(1964)}]{Chiu1964}%
  \BibitemOpen
  \bibfield  {author} {\bibinfo {author} {\bibnamefont {Chiu}, \bibfnamefont
  {H.-Y.}}, \ and\ \bibinfo {author} {\bibfnamefont {E.~E.}\ \bibnamefont
  {Salpeter}}} (\bibinfo {year} {1964}),\ \href {\doibase
  10.1103/PhysRevLett.12.413} {\bibfield  {journal} {\bibinfo  {journal} {Phys.
  Rev. Lett.}\ }\textbf {\bibinfo {volume} {12}},\ \bibinfo {pages}
  {413}}\BibitemShut {NoStop}%
\bibitem [{\citenamefont {Ciminale}\ \emph {et~al.}(2006)\citenamefont
  {Ciminale}, \citenamefont {Nardulli}, \citenamefont {Ruggieri},\ and\
  \citenamefont {Gatto}}]{Ciminale:2006sm}%
  \BibitemOpen
  \bibfield  {author} {\bibinfo {author} {\bibnamefont {Ciminale},
  \bibfnamefont {M.}}, \bibinfo {author} {\bibfnamefont {G.}~\bibnamefont
  {Nardulli}}, \bibinfo {author} {\bibfnamefont {M.}~\bibnamefont {Ruggieri}},
  \ and\ \bibinfo {author} {\bibfnamefont {R.}~\bibnamefont {Gatto}}} (\bibinfo
  {year} {2006}),\ \href {\doibase 10.1016/j.physletb.2006.03.075} {\bibfield
  {journal} {\bibinfo  {journal} {Phys.Lett.}\ }\textbf {\bibinfo {volume}
  {B636}},\ \bibinfo {pages} {317}},\ \Eprint
  {http://arxiv.org/abs/hep-ph/0602180} {arXiv:hep-ph/0602180 [hep-ph]}
  \BibitemShut {NoStop}%
\bibitem [{\citenamefont {Cirigliano}\ \emph {et~al.}(2011)\citenamefont
  {Cirigliano}, \citenamefont {Reddy},\ and\ \citenamefont
  {Sharma}}]{Cirigliano:2011tj}%
  \BibitemOpen
  \bibfield  {author} {\bibinfo {author} {\bibnamefont {Cirigliano},
  \bibfnamefont {V.}}, \bibinfo {author} {\bibfnamefont {S.}~\bibnamefont
  {Reddy}}, \ and\ \bibinfo {author} {\bibfnamefont {R.}~\bibnamefont
  {Sharma}}} (\bibinfo {year} {2011}),\ \href {\doibase
  10.1103/PhysRevC.84.045809} {\bibfield  {journal} {\bibinfo  {journal}
  {Phys.Rev.}\ }\textbf {\bibinfo {volume} {C84}},\ \bibinfo {pages}
  {045809}},\ \Eprint {http://arxiv.org/abs/1102.5379} {arXiv:1102.5379
  [nucl-th]} \BibitemShut {NoStop}%
\bibitem [{\citenamefont {Clogston}(1962)}]{Clogston}%
  \BibitemOpen
  \bibfield  {author} {\bibinfo {author} {\bibnamefont {Clogston},
  \bibfnamefont {A.~M.}}} (\bibinfo {year} {1962}),\ \href {\doibase
  10.1103/PhysRevLett.9.266} {\bibfield  {journal} {\bibinfo  {journal} {Phys.
  Rev. Lett.}\ }\textbf {\bibinfo {volume} {9}},\ \bibinfo {pages}
  {266}}\BibitemShut {NoStop}%
\bibitem [{\citenamefont {Collins}\ and\ \citenamefont
  {Perry}(1975)}]{Collins}%
  \BibitemOpen
  \bibfield  {author} {\bibinfo {author} {\bibnamefont {Collins}, \bibfnamefont
  {J.~C.}}, \ and\ \bibinfo {author} {\bibfnamefont {M.~J.}\ \bibnamefont
  {Perry}}} (\bibinfo {year} {1975}),\ \href {\doibase
  10.1103/PhysRevLett.34.1353} {\bibfield  {journal} {\bibinfo  {journal}
  {Phys. Rev. Lett.}\ }\textbf {\bibinfo {volume} {34}},\ \bibinfo {pages}
  {1353}}\BibitemShut {NoStop}%
\bibitem [{\citenamefont {{Cooper}}(1956)}]{Cooper:1956}%
  \BibitemOpen
  \bibfield  {author} {\bibinfo {author} {\bibnamefont {{Cooper}},
  \bibfnamefont {L.~N.}}} (\bibinfo {year} {1956}),\ \href {\doibase
  10.1103/PhysRev.104.1189} {\bibfield  {journal} {\bibinfo  {journal} {Phys.
  Rev.}\ }\textbf {\bibinfo {volume} {104}},\ \bibinfo {pages}
  {1189}}\BibitemShut {NoStop}%
\bibitem [{\citenamefont {Demorest}\ \emph {et~al.}(2010)\citenamefont
  {Demorest}, \citenamefont {Pennucci}, \citenamefont {Ransom}, \citenamefont
  {Roberts},\ and\ \citenamefont {Hessels}}]{Demorest:2010bx}%
  \BibitemOpen
  \bibfield  {author} {\bibinfo {author} {\bibnamefont {Demorest},
  \bibfnamefont {P.}}, \bibinfo {author} {\bibfnamefont {T.}~\bibnamefont
  {Pennucci}}, \bibinfo {author} {\bibfnamefont {S.}~\bibnamefont {Ransom}},
  \bibinfo {author} {\bibfnamefont {M.}~\bibnamefont {Roberts}}, \ and\
  \bibinfo {author} {\bibfnamefont {J.}~\bibnamefont {Hessels}}} (\bibinfo
  {year} {2010}),\ \href {\doibase 10.1038/nature09466} {\bibfield  {journal}
  {\bibinfo  {journal} {Nature}\ }\textbf {\bibinfo {volume} {467}},\ \bibinfo
  {pages} {1081}},\ \Eprint {http://arxiv.org/abs/1010.5788} {arXiv:1010.5788
  [astro-ph.HE]} \BibitemShut {NoStop}%
\bibitem [{\citenamefont {Eguchi}(1976)}]{Eguchi:1976iz}%
  \BibitemOpen
  \bibfield  {author} {\bibinfo {author} {\bibnamefont {Eguchi}, \bibfnamefont
  {T.}}} (\bibinfo {year} {1976}),\ \href {\doibase 10.1103/PhysRevD.14.2755}
  {\bibfield  {journal} {\bibinfo  {journal} {Phys.Rev.}\ }\textbf {\bibinfo
  {volume} {D14}},\ \bibinfo {pages} {2755}}\BibitemShut {NoStop}%
\bibitem [{\citenamefont {Espinoza}\ \emph {et~al.}(2011)\citenamefont
  {Espinoza}, \citenamefont {Lyne}, \citenamefont {Stappers},\ and\
  \citenamefont {Kramer}}]{Espinoza:2011pq}%
  \BibitemOpen
  \bibfield  {author} {\bibinfo {author} {\bibnamefont {Espinoza},
  \bibfnamefont {C.~M.}}, \bibinfo {author} {\bibfnamefont {A.~G.}\
  \bibnamefont {Lyne}}, \bibinfo {author} {\bibfnamefont {B.~W.}\ \bibnamefont
  {Stappers}}, \ and\ \bibinfo {author} {\bibfnamefont {M.}~\bibnamefont
  {Kramer}}} (\bibinfo {year} {2011}),\ \href {\doibase
  10.1111/j.1365-2966.2011.18503.x} {\bibfield  {journal} {\bibinfo  {journal}
  {Mon.Not.Roy.Astron.Soc.}\ }\textbf {\bibinfo {volume} {414}},\ \bibinfo
  {pages} {1679}},\ \Eprint {http://arxiv.org/abs/1102.1743} {arXiv:1102.1743
  [astro-ph.HE]} \BibitemShut {NoStop}%
\bibitem [{\citenamefont {Eto}\ \emph {et~al.}(2010)\citenamefont {Eto},
  \citenamefont {Nitta},\ and\ \citenamefont {Yamamoto}}]{Eto:2009tr}%
  \BibitemOpen
  \bibfield  {author} {\bibinfo {author} {\bibnamefont {Eto}, \bibfnamefont
  {M.}}, \bibinfo {author} {\bibfnamefont {M.}~\bibnamefont {Nitta}}, \ and\
  \bibinfo {author} {\bibfnamefont {N.}~\bibnamefont {Yamamoto}}} (\bibinfo
  {year} {2010}),\ \href {\doibase 10.1103/PhysRevLett.104.161601} {\bibfield
  {journal} {\bibinfo  {journal} {Phys.Rev.Lett.}\ }\textbf {\bibinfo {volume}
  {104}},\ \bibinfo {pages} {161601}},\ \Eprint
  {http://arxiv.org/abs/0912.1352} {arXiv:0912.1352 [hep-ph]} \BibitemShut
  {NoStop}%
\bibitem [{\citenamefont {Evans}\ \emph {et~al.}(2000)\citenamefont {Evans},
  \citenamefont {Hormuzdiar}, \citenamefont {Hsu},\ and\ \citenamefont
  {Schwetz}}]{Evans:1999at}%
  \BibitemOpen
  \bibfield  {author} {\bibinfo {author} {\bibnamefont {Evans}, \bibfnamefont
  {N.~J.}}, \bibinfo {author} {\bibfnamefont {J.}~\bibnamefont {Hormuzdiar}},
  \bibinfo {author} {\bibfnamefont {S.~D.~H.}\ \bibnamefont {Hsu}}, \ and\
  \bibinfo {author} {\bibfnamefont {M.}~\bibnamefont {Schwetz}}} (\bibinfo
  {year} {2000}),\ \href {\doibase 10.1016/S0550-3213(00)00253-4} {\bibfield
  {journal} {\bibinfo  {journal} {Nucl.Phys.}\ }\textbf {\bibinfo {volume}
  {B581}},\ \bibinfo {pages} {391}},\ \Eprint
  {http://arxiv.org/abs/hep-ph/9910313} {arXiv:hep-ph/9910313 [hep-ph]}
  \BibitemShut {NoStop}%
\bibitem [{\citenamefont {Evans}\ \emph {et~al.}(1999)\citenamefont {Evans},
  \citenamefont {Hsu},\ and\ \citenamefont {Schwetz}}]{Evans:1998nf}%
  \BibitemOpen
  \bibfield  {author} {\bibinfo {author} {\bibnamefont {Evans}, \bibfnamefont
  {N.~J.}}, \bibinfo {author} {\bibfnamefont {S.~D.~H.}\ \bibnamefont {Hsu}}, \
  and\ \bibinfo {author} {\bibfnamefont {M.}~\bibnamefont {Schwetz}}} (\bibinfo
  {year} {1999}),\ \href {\doibase 10.1016/S0370-2693(99)00093-3} {\bibfield
  {journal} {\bibinfo  {journal} {Phys.Lett.}\ }\textbf {\bibinfo {volume}
  {B449}},\ \bibinfo {pages} {281}},\ \Eprint
  {http://arxiv.org/abs/hep-ph/9810514} {arXiv:hep-ph/9810514 [hep-ph]}
  \BibitemShut {NoStop}%
\bibitem [{\citenamefont {Ferrer}\ and\ \citenamefont {de~la
  Incera}(2007)}]{Ferrer:2007uw}%
  \BibitemOpen
  \bibfield  {author} {\bibinfo {author} {\bibnamefont {Ferrer}, \bibfnamefont
  {E.~J.}}, \ and\ \bibinfo {author} {\bibfnamefont {V.}~\bibnamefont {de~la
  Incera}}} (\bibinfo {year} {2007}),\ \href {\doibase
  10.1103/PhysRevD.76.114012} {\bibfield  {journal} {\bibinfo  {journal}
  {Phys.Rev.}\ }\textbf {\bibinfo {volume} {D76}},\ \bibinfo {pages}
  {114012}},\ \Eprint {http://arxiv.org/abs/0705.2403} {arXiv:0705.2403
  [hep-ph]} \BibitemShut {NoStop}%
\bibitem [{\citenamefont {Fodor}\ and\ \citenamefont
  {Katz}(2002)}]{Fodor:2001au}%
  \BibitemOpen
  \bibfield  {author} {\bibinfo {author} {\bibnamefont {Fodor}, \bibfnamefont
  {Z.}}, \ and\ \bibinfo {author} {\bibfnamefont {S.~D.}\ \bibnamefont {Katz}}}
  (\bibinfo {year} {2002}),\ \href {\doibase 10.1016/S0370-2693(02)01583-6}
  {\bibfield  {journal} {\bibinfo  {journal} {Phys.Lett.}\ }\textbf {\bibinfo
  {volume} {B534}},\ \bibinfo {pages} {87}},\ \Eprint
  {http://arxiv.org/abs/hep-lat/0104001} {arXiv:hep-lat/0104001 [hep-lat]}
  \BibitemShut {NoStop}%
\bibitem [{\citenamefont {Forbes}\ and\ \citenamefont
  {Zhitnitsky}(2002)}]{Forbes:2001gj}%
  \BibitemOpen
  \bibfield  {author} {\bibinfo {author} {\bibnamefont {Forbes}, \bibfnamefont
  {M.~M.}}, \ and\ \bibinfo {author} {\bibfnamefont {A.~R.}\ \bibnamefont
  {Zhitnitsky}}} (\bibinfo {year} {2002}),\ \href {\doibase
  10.1103/PhysRevD.65.085009} {\bibfield  {journal} {\bibinfo  {journal}
  {Phys.Rev.}\ }\textbf {\bibinfo {volume} {D65}},\ \bibinfo {pages}
  {085009}},\ \Eprint {http://arxiv.org/abs/hep-ph/0109173}
  {arXiv:hep-ph/0109173 [hep-ph]} \BibitemShut {NoStop}%
\bibitem [{\citenamefont {Forbes~McNeil}\ \emph {et~al.}(2005)\citenamefont
  {Forbes~McNeil}, \citenamefont {Gubankova}, \citenamefont {Liu~Vincent},\
  and\ \citenamefont {Wilczek}}]{Forbes:2004cr}%
  \BibitemOpen
  \bibfield  {author} {\bibinfo {author} {\bibnamefont {Forbes~McNeil},
  \bibfnamefont {M.}}, \bibinfo {author} {\bibfnamefont {E.}~\bibnamefont
  {Gubankova}}, \bibinfo {author} {\bibfnamefont {W.}~\bibnamefont
  {Liu~Vincent}}, \ and\ \bibinfo {author} {\bibfnamefont {F.}~\bibnamefont
  {Wilczek}}} (\bibinfo {year} {2005}),\ \href {\doibase
  10.1103/PhysRevLett.94.017001} {\bibfield  {journal} {\bibinfo  {journal}
  {Phys.Rev.Lett.}\ }\textbf {\bibinfo {volume} {94}},\ \bibinfo {pages}
  {017001}},\ \Eprint {http://arxiv.org/abs/hep-ph/0405059}
  {arXiv:hep-ph/0405059 [hep-ph]} \BibitemShut {NoStop}%
\bibitem [{\citenamefont {Fraga}\ \emph {et~al.}(2001)\citenamefont {Fraga},
  \citenamefont {Pisarski},\ and\ \citenamefont
  {Schaffner-Bielich}}]{Fraga:2001id}%
  \BibitemOpen
  \bibfield  {author} {\bibinfo {author} {\bibnamefont {Fraga}, \bibfnamefont
  {E.~S.}}, \bibinfo {author} {\bibfnamefont {R.~D.}\ \bibnamefont {Pisarski}},
  \ and\ \bibinfo {author} {\bibfnamefont {J.}~\bibnamefont
  {Schaffner-Bielich}}} (\bibinfo {year} {2001}),\ \href {\doibase
  10.1103/PhysRevD.63.121702} {\bibfield  {journal} {\bibinfo  {journal}
  {Phys.Rev.}\ }\textbf {\bibinfo {volume} {D63}},\ \bibinfo {pages}
  {121702}},\ \Eprint {http://arxiv.org/abs/hep-ph/0101143}
  {arXiv:hep-ph/0101143 [hep-ph]} \BibitemShut {NoStop}%
\bibitem [{\citenamefont {Frautschi}(1978)}]{Frautschi:1978rz}%
  \BibitemOpen
  \bibfield  {author} {\bibinfo {author} {\bibnamefont {Frautschi},
  \bibfnamefont {S.~C.}}} (\bibinfo {year} {1978}),{\emph
  {\bibinfo {title} { Asymptotic Freedom and Color Superconductivity in Dense Quark Matter, in Hadronic Matter at Extreme Energy Density,~Ettore Majorana International Science Series, Springer US}}}\href@noop{}{\ }\BibitemShut {NoStop}%
\bibitem [{\citenamefont {{Freire}}\ \emph {et~al.}(2008)\citenamefont
  {{Freire}}, \citenamefont {{Wolszczan}}, \citenamefont {{van den Berg}},\
  and\ \citenamefont {{Hessels}}}]{Freire2008}%
  \BibitemOpen
  \bibfield  {author} {\bibinfo {author} {\bibnamefont {{Freire}},
  \bibfnamefont {P.~C.~C.}}, \bibinfo {author} {\bibfnamefont {A.}~\bibnamefont
  {{Wolszczan}}}, \bibinfo {author} {\bibfnamefont {M.}~\bibnamefont {{van den
  Berg}}}, \ and\ \bibinfo {author} {\bibfnamefont {J.~W.~T.}\ \bibnamefont
  {{Hessels}}}} (\bibinfo {year} {2008}),\ \href {\doibase 10.1086/587832}
  {\bibfield  {journal} {\bibinfo  {journal} {\apj}\ }\textbf {\bibinfo
  {volume} {679}},\ \bibinfo {pages} {1433}},\ \Eprint
  {http://arxiv.org/abs/0712.3826} {arXiv:0712.3826} \BibitemShut {NoStop}%
\bibitem [{\citenamefont {Friedman}\ and\ \citenamefont
  {Morsink}(1998)}]{Friedman:1997uh}%
  \BibitemOpen
  \bibfield  {author} {\bibinfo {author} {\bibnamefont {Friedman},
  \bibfnamefont {J.~L.}}, \ and\ \bibinfo {author} {\bibfnamefont {S.~M.}\
  \bibnamefont {Morsink}}} (\bibinfo {year} {1998}),\ \href {\doibase
  10.1086/305920} {\bibfield  {journal} {\bibinfo  {journal} {Astrophys.J.}\
  }\textbf {\bibinfo {volume} {502}},\ \bibinfo {pages} {714}},\ \Eprint
  {http://arxiv.org/abs/gr-qc/9706073} {arXiv:gr-qc/9706073 [gr-qc]}
  \BibitemShut {NoStop}%
\bibitem [{\citenamefont {Fromm}\ \emph {et~al.}(2012)\citenamefont {Fromm},
  \citenamefont {Langelage}, \citenamefont {Lottini},\ and\ \citenamefont
  {Philipsen}}]{Fromm:2011qi}%
  \BibitemOpen
  \bibfield  {author} {\bibinfo {author} {\bibnamefont {Fromm}, \bibfnamefont
  {M.}}, \bibinfo {author} {\bibfnamefont {J.}~\bibnamefont {Langelage}},
  \bibinfo {author} {\bibfnamefont {S.}~\bibnamefont {Lottini}}, \ and\
  \bibinfo {author} {\bibfnamefont {O.}~\bibnamefont {Philipsen}}} (\bibinfo
  {year} {2012}),\ \href {\doibase 10.1007/JHEP01(2012)042} {\bibfield
  {journal} {\bibinfo  {journal} {JHEP}\ }\textbf {\bibinfo {volume} {1201}},\
  \bibinfo {pages} {042}},\ \Eprint {http://arxiv.org/abs/1111.4953}
  {arXiv:1111.4953 [hep-lat]} \BibitemShut {NoStop}%
\bibitem [{\citenamefont {Fukushima}(2005)}]{Fukushima:2005cm}%
  \BibitemOpen
  \bibfield  {author} {\bibinfo {author} {\bibnamefont {Fukushima},
  \bibfnamefont {K.}}} (\bibinfo {year} {2005}),\ \href {\doibase
  10.1103/PhysRevD.72.074002} {\bibfield  {journal} {\bibinfo  {journal}
  {Phys.Rev.}\ }\textbf {\bibinfo {volume} {D72}},\ \bibinfo {pages}
  {074002}},\ \Eprint {http://arxiv.org/abs/hep-ph/0506080}
  {arXiv:hep-ph/0506080 [hep-ph]} \BibitemShut {NoStop}%
\bibitem [{\citenamefont {Fukushima}(2006)}]{Fukushima:2006su}%
  \BibitemOpen
  \bibfield  {author} {\bibinfo {author} {\bibnamefont {Fukushima},
  \bibfnamefont {K.}}} (\bibinfo {year} {2006}),\ \href {\doibase
  10.1103/PhysRevD.73.094016} {\bibfield  {journal} {\bibinfo  {journal}
  {Phys.Rev.}\ }\textbf {\bibinfo {volume} {D73}},\ \bibinfo {pages}
  {094016}},\ \Eprint {http://arxiv.org/abs/hep-ph/0603216}
  {arXiv:hep-ph/0603216 [hep-ph]} \BibitemShut {NoStop}%
\bibitem [{\citenamefont {Fukushima}\ \emph {et~al.}(2005)\citenamefont
  {Fukushima}, \citenamefont {Kouvaris},\ and\ \citenamefont
  {Rajagopal}}]{Fukushima:2004zq}%
  \BibitemOpen
  \bibfield  {author} {\bibinfo {author} {\bibnamefont {Fukushima},
  \bibfnamefont {K.}}, \bibinfo {author} {\bibfnamefont {C.}~\bibnamefont
  {Kouvaris}}, \ and\ \bibinfo {author} {\bibfnamefont {K.}~\bibnamefont
  {Rajagopal}}} (\bibinfo {year} {2005}),\ \href {\doibase
  10.1103/PhysRevD.71.034002} {\bibfield  {journal} {\bibinfo  {journal}
  {Phys.Rev.}\ }\textbf {\bibinfo {volume} {D71}},\ \bibinfo {pages}
  {034002}},\ \Eprint {http://arxiv.org/abs/hep-ph/0408322}
  {arXiv:hep-ph/0408322 [hep-ph]} \BibitemShut {NoStop}%
\bibitem [{\citenamefont {Fulde}\ and\ \citenamefont {Ferrell}(1964)}]{FF}%
  \BibitemOpen
  \bibfield  {author} {\bibinfo {author} {\bibnamefont {Fulde}, \bibfnamefont
  {P.}}, \ and\ \bibinfo {author} {\bibfnamefont {R.~A.}\ \bibnamefont
  {Ferrell}}} (\bibinfo {year} {1964}),\ \href {\doibase
  10.1103/PhysRev.135.A550} {\bibfield  {journal} {\bibinfo  {journal} {Phys.
  Rev.}\ }\textbf {\bibinfo {volume} {135}},\ \bibinfo {pages}
  {A550}}\BibitemShut {NoStop}%
\bibitem [{\citenamefont {Gatto}\ and\ \citenamefont
  {Ruggieri}(2007)}]{Gatto:2007ja}%
  \BibitemOpen
  \bibfield  {author} {\bibinfo {author} {\bibnamefont {Gatto}, \bibfnamefont
  {R.}}, \ and\ \bibinfo {author} {\bibfnamefont {M.}~\bibnamefont {Ruggieri}}}
  (\bibinfo {year} {2007}),\ \href {\doibase 10.1103/PhysRevD.75.114004}
  {\bibfield  {journal} {\bibinfo  {journal} {Phys.Rev.}\ }\textbf {\bibinfo
  {volume} {D75}},\ \bibinfo {pages} {114004}},\ \Eprint
  {http://arxiv.org/abs/hep-ph/0703276} {arXiv:hep-ph/0703276 [hep-ph]}
  \BibitemShut {NoStop}%
\bibitem [{\citenamefont {de~Gennes}(1966)}]{gen66}%
  \BibitemOpen
  \bibfield  {author} {\bibinfo {author} {\bibnamefont {de~Gennes},
  \bibfnamefont {P.~G.}}} (\bibinfo {year} {1966}),\ \href@noop {} {\emph
  {\bibinfo {title} {Superconductivity of Metals and Alloys}}}\ (\bibinfo
  {publisher} {Benjamin},\ \bibinfo {address} {New York})\BibitemShut {NoStop}%
\bibitem [{\citenamefont {Gerlach}(1968)}]{Gerlach:1968zz}%
  \BibitemOpen
  \bibfield  {author} {\bibinfo {author} {\bibnamefont {Gerlach}, \bibfnamefont
  {U.~H.}}} (\bibinfo {year} {1968}),\ \href {\doibase
  10.1103/PhysRev.172.1325} {\bibfield  {journal} {\bibinfo  {journal}
  {Phys.Rev.}\ }\textbf {\bibinfo {volume} {172}},\ \bibinfo {pages}
  {1325}}\BibitemShut {NoStop}%
\bibitem [{\citenamefont {Giannakis}\ \emph {et~al.}(2005)\citenamefont
  {Giannakis}, \citenamefont {Hou},\ and\ \citenamefont
  {Ren}}]{Giannakis:2005sa}%
  \BibitemOpen
  \bibfield  {author} {\bibinfo {author} {\bibnamefont {Giannakis},
  \bibfnamefont {I.}}, \bibinfo {author} {\bibfnamefont {D.-f.}\ \bibnamefont
  {Hou}}, \ and\ \bibinfo {author} {\bibfnamefont {H.-C.}\ \bibnamefont {Ren}}}
  (\bibinfo {year} {2005}),\ \href {\doibase 10.1016/j.physletb.2005.10.001}
  {\bibfield  {journal} {\bibinfo  {journal} {Phys.Lett.}\ }\textbf {\bibinfo
  {volume} {B631}},\ \bibinfo {pages} {16}},\ \Eprint
  {http://arxiv.org/abs/hep-ph/0507306} {arXiv:hep-ph/0507306 [hep-ph]}
  \BibitemShut {NoStop}%
\bibitem [{\citenamefont {Giannakis}\ and\ \citenamefont
  {Ren}(2005{\natexlab{a}})}]{Giannakis:2004pf}%
  \BibitemOpen
  \bibfield  {author} {\bibinfo {author} {\bibnamefont {Giannakis},
  \bibfnamefont {I.}}, \ and\ \bibinfo {author} {\bibfnamefont {H.-C.}\
  \bibnamefont {Ren}}} (\bibinfo {year} {2005}{\natexlab{a}}),\ \href {\doibase
  10.1016/j.physletb.2005.02.020} {\bibfield  {journal} {\bibinfo  {journal}
  {Phys.Lett.}\ }\textbf {\bibinfo {volume} {B611}},\ \bibinfo {pages} {137}},\
  \Eprint {http://arxiv.org/abs/hep-ph/0412015} {arXiv:hep-ph/0412015 [hep-ph]}
  \BibitemShut {NoStop}%
\bibitem [{\citenamefont {Giannakis}\ and\ \citenamefont
  {Ren}(2005{\natexlab{b}})}]{Giannakis:2005vw}%
  \BibitemOpen
  \bibfield  {author} {\bibinfo {author} {\bibnamefont {Giannakis},
  \bibfnamefont {I.}}, \ and\ \bibinfo {author} {\bibfnamefont {H.-C.}\
  \bibnamefont {Ren}}} (\bibinfo {year} {2005}{\natexlab{b}}),\ \href {\doibase
  10.1016/j.nuclphysb.2005.06.008} {\bibfield  {journal} {\bibinfo  {journal}
  {Nucl.Phys.}\ }\textbf {\bibinfo {volume} {B723}},\ \bibinfo {pages} {255}},\
  \Eprint {http://arxiv.org/abs/hep-th/0504053} {arXiv:hep-th/0504053 [hep-th]}
  \BibitemShut {NoStop}%
\bibitem [{\citenamefont {{Giorgini}}\ \emph {et~al.}(2008)\citenamefont
  {{Giorgini}}, \citenamefont {{Pitaevskii}},\ and\ \citenamefont
  {{Stringari}}}]{giorgini-review}%
  \BibitemOpen
  \bibfield  {author} {\bibinfo {author} {\bibnamefont {{Giorgini}},
  \bibfnamefont {S.}}, \bibinfo {author} {\bibfnamefont {L.~P.}\ \bibnamefont
  {{Pitaevskii}}}, \ and\ \bibinfo {author} {\bibfnamefont {S.}~\bibnamefont
  {{Stringari}}}} (\bibinfo {year} {2008}),\ \href {\doibase
  10.1103/RevModPhys.80.1215} {\bibfield  {journal} {\bibinfo  {journal}
  {Reviews of Modern Physics}\ }\textbf {\bibinfo {volume} {80}},\ \bibinfo
  {pages} {1215}},\ \Eprint {http://arxiv.org/abs/0706.3360} {arXiv:0706.3360
  [cond-mat.other]} \BibitemShut {NoStop}%
\bibitem [{\citenamefont {Glendenning}\ and\ \citenamefont
  {Kettner}(2000)}]{Glendenning:1998ag}%
  \BibitemOpen
  \bibfield  {author} {\bibinfo {author} {\bibnamefont {Glendenning},
  \bibfnamefont {N.~K.}}, \ and\ \bibinfo {author} {\bibfnamefont
  {C.}~\bibnamefont {Kettner}}} (\bibinfo {year} {2000}),\ \href@noop {}
  {\bibfield  {journal} {\bibinfo  {journal} {Astron.Astrophys.}\ }\textbf
  {\bibinfo {volume} {353}},\ \bibinfo {pages} {L9}},\ \Eprint
  {http://arxiv.org/abs/astro-ph/9807155} {arXiv:astro-ph/9807155 [astro-ph]}
  \BibitemShut {NoStop}%
\bibitem [{\citenamefont {Gorbar}(2000)}]{Gorbar:2000ms}%
  \BibitemOpen
  \bibfield  {author} {\bibinfo {author} {\bibnamefont {Gorbar}, \bibfnamefont
  {E.}}} (\bibinfo {year} {2000}),\ \href {\doibase 10.1103/PhysRevD.62.014007}
  {\bibfield  {journal} {\bibinfo  {journal} {Phys.Rev.}\ }\textbf {\bibinfo
  {volume} {D62}},\ \bibinfo {pages} {014007}},\ \Eprint
  {http://arxiv.org/abs/hep-ph/0001211} {arXiv:hep-ph/0001211 [hep-ph]}
  \BibitemShut {NoStop}%
\bibitem [{\citenamefont {Gorbar}\ \emph {et~al.}(2007)\citenamefont {Gorbar},
  \citenamefont {Hashimoto},\ and\ \citenamefont {Miransky}}]{Gorbar:2007vx}%
  \BibitemOpen
  \bibfield  {author} {\bibinfo {author} {\bibnamefont {Gorbar}, \bibfnamefont
  {E.}}, \bibinfo {author} {\bibfnamefont {M.}~\bibnamefont {Hashimoto}}, \
  and\ \bibinfo {author} {\bibfnamefont {V.}~\bibnamefont {Miransky}}}
  (\bibinfo {year} {2007}),\ \href {\doibase 10.1103/PhysRevD.75.085012}
  {\bibfield  {journal} {\bibinfo  {journal} {Phys.Rev.}\ }\textbf {\bibinfo
  {volume} {D75}},\ \bibinfo {pages} {085012}},\ \Eprint
  {http://arxiv.org/abs/hep-ph/0701211} {arXiv:hep-ph/0701211 [hep-ph]}
  \BibitemShut {NoStop}%
\bibitem [{\citenamefont {Gorbar}\ \emph
  {et~al.}(2006{\natexlab{a}})\citenamefont {Gorbar}, \citenamefont
  {Hashimoto}, \citenamefont {Miransky},\ and\ \citenamefont
  {Shovkovy}}]{Gorbar:2006up}%
  \BibitemOpen
  \bibfield  {author} {\bibinfo {author} {\bibnamefont {Gorbar}, \bibfnamefont
  {E.}}, \bibinfo {author} {\bibfnamefont {M.}~\bibnamefont {Hashimoto}},
  \bibinfo {author} {\bibfnamefont {V.}~\bibnamefont {Miransky}}, \ and\
  \bibinfo {author} {\bibfnamefont {I.}~\bibnamefont {Shovkovy}}} (\bibinfo
  {year} {2006}{\natexlab{a}}),\ \href {\doibase 10.1103/PhysRevD.73.111502}
  {\bibfield  {journal} {\bibinfo  {journal} {Phys.Rev.}\ }\textbf {\bibinfo
  {volume} {D73}},\ \bibinfo {pages} {111502}},\ \Eprint
  {http://arxiv.org/abs/hep-ph/0602251} {arXiv:hep-ph/0602251 [hep-ph]}
  \BibitemShut {NoStop}%
\bibitem [{\citenamefont {Gorbar}\ \emph
  {et~al.}(2006{\natexlab{b}})\citenamefont {Gorbar}, \citenamefont
  {Hashimoto},\ and\ \citenamefont {Miransky}}]{Gorbar:2005rx}%
  \BibitemOpen
  \bibfield  {author} {\bibinfo {author} {\bibnamefont {Gorbar}, \bibfnamefont
  {E.~V.}}, \bibinfo {author} {\bibfnamefont {M.}~\bibnamefont {Hashimoto}}, \
  and\ \bibinfo {author} {\bibfnamefont {V.~A.}\ \bibnamefont {Miransky}}}
  (\bibinfo {year} {2006}{\natexlab{b}}),\ \href {\doibase
  10.1016/j.physletb.2005.10.063} {\bibfield  {journal} {\bibinfo  {journal}
  {Phys.Lett.}\ }\textbf {\bibinfo {volume} {B632}},\ \bibinfo {pages} {305}},\
  \Eprint {http://arxiv.org/abs/hep-ph/0507303} {arXiv:hep-ph/0507303 [hep-ph]}
  \BibitemShut {NoStop}%
\bibitem [{\citenamefont {Gorbar}\ \emph
  {et~al.}(2006{\natexlab{c}})\citenamefont {Gorbar}, \citenamefont
  {Hashimoto},\ and\ \citenamefont {Miransky}}]{Gorbar:2005tx}%
  \BibitemOpen
  \bibfield  {author} {\bibinfo {author} {\bibnamefont {Gorbar}, \bibfnamefont
  {E.~V.}}, \bibinfo {author} {\bibfnamefont {M.}~\bibnamefont {Hashimoto}}, \
  and\ \bibinfo {author} {\bibfnamefont {V.~A.}\ \bibnamefont {Miransky}}}
  (\bibinfo {year} {2006}{\natexlab{c}}),\ \href {\doibase
  10.1103/PhysRevLett.96.022005} {\bibfield  {journal} {\bibinfo  {journal}
  {Phys.Rev.Lett.}\ }\textbf {\bibinfo {volume} {96}},\ \bibinfo {pages}
  {022005}},\ \Eprint {http://arxiv.org/abs/hep-ph/0509334}
  {arXiv:hep-ph/0509334 [hep-ph]} \BibitemShut {NoStop}%
\bibitem [{\citenamefont {{Gradshteyn}}\ and\ \citenamefont
  {{Ryzhik}}(1980)}]{Gradshteyn}%
  \BibitemOpen
  \bibfield  {author} {\bibinfo {author} {\bibnamefont {{Gradshteyn}},
  \bibfnamefont {I.~S.}}, \ and\ \bibinfo {author} {\bibfnamefont {I.~M.}\
  \bibnamefont {{Ryzhik}}}} (\bibinfo {year} {1980}),\ \href@noop {} {\emph
  {\bibinfo {title} {Table of integrals, series and products, New York: Academic Press, 1980, 5th corr.~and
  enl.~ed.}}}\BibitemShut {Stop}%
\bibitem [{\citenamefont {Grigorian}\ \emph {et~al.}(2004)\citenamefont
  {Grigorian}, \citenamefont {Blaschke},\ and\ \citenamefont
  {Aguilera}}]{Grigorian:2003vi}%
  \BibitemOpen
  \bibfield  {author} {\bibinfo {author} {\bibnamefont {Grigorian},
  \bibfnamefont {H.}}, \bibinfo {author} {\bibfnamefont {D.}~\bibnamefont
  {Blaschke}}, \ and\ \bibinfo {author} {\bibfnamefont {D.~N.}\ \bibnamefont
  {Aguilera}}} (\bibinfo {year} {2004}),\ \href {\doibase
  10.1103/PhysRevC.69.065802} {\bibfield  {journal} {\bibinfo  {journal}
  {Phys.Rev.}\ }\textbf {\bibinfo {volume} {C69}},\ \bibinfo {pages}
  {065802}},\ \Eprint {http://arxiv.org/abs/astro-ph/0303518}
  {arXiv:astro-ph/0303518 [astro-ph]} \BibitemShut {NoStop}%
\bibitem [{\citenamefont {Gross}\ and\ \citenamefont
  {Neveu}(1974)}]{Gross:1974jv}%
  \BibitemOpen
  \bibfield  {author} {\bibinfo {author} {\bibnamefont {Gross}, \bibfnamefont
  {D.~J.}}, \ and\ \bibinfo {author} {\bibfnamefont {A.}~\bibnamefont {Neveu}}}
  (\bibinfo {year} {1974}),\ \href {\doibase 10.1103/PhysRevD.10.3235}
  {\bibfield  {journal} {\bibinfo  {journal} {Phys.Rev.}\ }\textbf {\bibinfo
  {volume} {D10}},\ \bibinfo {pages} {3235}}\BibitemShut {NoStop}%
\bibitem [{\citenamefont {Gubankova}\ \emph {et~al.}(2010)\citenamefont
  {Gubankova}, \citenamefont {Mannarelli},\ and\ \citenamefont
  {Sharma}}]{Gubankova:2008ya}%
  \BibitemOpen
  \bibfield  {author} {\bibinfo {author} {\bibnamefont {Gubankova},
  \bibfnamefont {E.}}, \bibinfo {author} {\bibfnamefont {M.}~\bibnamefont
  {Mannarelli}}, \ and\ \bibinfo {author} {\bibfnamefont {R.}~\bibnamefont
  {Sharma}}} (\bibinfo {year} {2010}),\ \href {\doibase
  10.1016/j.aop.2010.05.001} {\bibfield  {journal} {\bibinfo  {journal} {Annals
  Phys.}\ }\textbf {\bibinfo {volume} {325}},\ \bibinfo {pages} {1987}},\
  \Eprint {http://arxiv.org/abs/0804.0782} {arXiv:0804.0782
  [cond-mat.supr-con]} \BibitemShut {NoStop}%
\bibitem [{\citenamefont {Gubankova}\ \emph {et~al.}(2006)\citenamefont
  {Gubankova}, \citenamefont {Schmitt},\ and\ \citenamefont
  {Wilczek}}]{Gubankova:2006gj}%
  \BibitemOpen
  \bibfield  {author} {\bibinfo {author} {\bibnamefont {Gubankova},
  \bibfnamefont {E.}}, \bibinfo {author} {\bibfnamefont {A.}~\bibnamefont
  {Schmitt}}, \ and\ \bibinfo {author} {\bibfnamefont {F.}~\bibnamefont
  {Wilczek}}} (\bibinfo {year} {2006}),\ \href {\doibase
  10.1103/PhysRevB.74.064505} {\bibfield  {journal} {\bibinfo  {journal}
  {Phys.Rev.}\ }\textbf {\bibinfo {volume} {B74}},\ \bibinfo {pages}
  {064505}},\ \Eprint {http://arxiv.org/abs/cond-mat/0603603}
  {arXiv:cond-mat/0603603 [cond-mat]} \BibitemShut {NoStop}%
\bibitem [{\citenamefont {Gubankova}\ \emph {et~al.}(2003)\citenamefont
  {Gubankova}, \citenamefont {Vincent~Liu},\ and\ \citenamefont
  {Wilczek}}]{Gubankova:2003uj}%
  \BibitemOpen
  \bibfield  {author} {\bibinfo {author} {\bibnamefont {Gubankova},
  \bibfnamefont {E.}}, \bibinfo {author} {\bibfnamefont {W.}~\bibnamefont
  {Vincent~Liu}}, \ and\ \bibinfo {author} {\bibfnamefont {F.}~\bibnamefont
  {Wilczek}}} (\bibinfo {year} {2003}),\ \href {\doibase
  10.1103/PhysRevLett.91.032001} {\bibfield  {journal} {\bibinfo  {journal}
  {Phys.Rev.Lett.}\ }\textbf {\bibinfo {volume} {91}},\ \bibinfo {pages}
  {032001}},\ \Eprint {http://arxiv.org/abs/hep-ph/0304016}
  {arXiv:hep-ph/0304016 [hep-ph]} \BibitemShut {NoStop}%
\bibitem [{\citenamefont {Gubbels}\ and\ \citenamefont
  {Stoof}(2012)}]{Gubbels:2012je}%
  \BibitemOpen
  \bibfield  {author} {\bibinfo {author} {\bibnamefont {Gubbels}, \bibfnamefont
  {K.}}, \ and\ \bibinfo {author} {\bibfnamefont {H.}~\bibnamefont {Stoof}}}
  (\bibinfo {year} {2012}),\ \href@noop {} {\ }\Eprint
  {http://arxiv.org/abs/1205.0568} {arXiv:1205.0568 [cond-mat.quant-gas]}
  \BibitemShut {NoStop}%
\bibitem [{\citenamefont {{Gudmundsson}}\ \emph {et~al.}(1982)\citenamefont
  {{Gudmundsson}}, \citenamefont {{Pethick}},\ and\ \citenamefont
  {{Epstein}}}]{Gudmundsson:1982}%
  \BibitemOpen
  \bibfield  {author} {\bibinfo {author} {\bibnamefont {{Gudmundsson}},
  \bibfnamefont {E.~H.}}, \bibinfo {author} {\bibfnamefont {C.~J.}\
  \bibnamefont {{Pethick}}}, \ and\ \bibinfo {author} {\bibfnamefont {R.~I.}\
  \bibnamefont {{Epstein}}}} (\bibinfo {year} {1982}),\ \href {\doibase
  10.1086/183840} {\bibfield  {journal} {\bibinfo  {journal} {\apjl}\ }\textbf
  {\bibinfo {volume} {259}},\ \bibinfo {pages} {L19}}\BibitemShut {NoStop}%
\bibitem [{\citenamefont {{Haensel}}(2003)}]{Haensel2003}%
  \BibitemOpen
  \bibfield  {author} {\bibinfo {author} {\bibnamefont {{Haensel}},
  \bibfnamefont {P.}}} (\bibinfo {year} {2003}),\ in\ \href {\doibase
  10.1051/eas:2003043} {\emph {\bibinfo {booktitle} {EAS Publications
  Series}}},\ \bibinfo {series} {EAS Publications Series}, Vol.~\bibinfo
  {volume} {7},\ \bibinfo {editor} {edited by\ \bibinfo {editor} {\bibfnamefont
  {C.}~\bibnamefont {{Motch}}}\ and\ \bibinfo {editor} {\bibfnamefont {J.-M.}\
  \bibnamefont {{Hameury}}}},\ p.\ \bibinfo {pages} {249},\ \Eprint
  {http://arxiv.org/abs/arXiv:astro-ph/0301073} {arXiv:astro-ph/0301073}
  \BibitemShut {NoStop}%
\bibitem [{\citenamefont {Hanany}\ and\ \citenamefont
  {Tong}(2003)}]{Hanany:2003hp}%
  \BibitemOpen
  \bibfield  {author} {\bibinfo {author} {\bibnamefont {Hanany}, \bibfnamefont
  {A.}}, \ and\ \bibinfo {author} {\bibfnamefont {D.}~\bibnamefont {Tong}}}
  (\bibinfo {year} {2003}),\ \href@noop {} {\bibfield  {journal} {\bibinfo
  {journal} {JHEP}\ }\textbf {\bibinfo {volume} {0307}},\ \bibinfo {pages}
  {037}},\ \Eprint {http://arxiv.org/abs/hep-th/0306150} {arXiv:hep-th/0306150
  [hep-th]} \BibitemShut {NoStop}%
\bibitem [{\citenamefont {Haskell}\ \emph {et~al.}(2007)\citenamefont
  {Haskell}, \citenamefont {Andersson}, \citenamefont {Jones},\ and\
  \citenamefont {Samuelsson}}]{Haskell:2007zz}%
  \BibitemOpen
  \bibfield  {author} {\bibinfo {author} {\bibnamefont {Haskell}, \bibfnamefont
  {B.}}, \bibinfo {author} {\bibfnamefont {N.}~\bibnamefont {Andersson}},
  \bibinfo {author} {\bibfnamefont {D.~I.}\ \bibnamefont {Jones}}, \ and\
  \bibinfo {author} {\bibfnamefont {L.}~\bibnamefont {Samuelsson}}} (\bibinfo
  {year} {2007}),\ \href {\doibase 10.1103/PhysRevLett.99.231101} {\bibfield
  {journal} {\bibinfo  {journal} {Phys.Rev.Lett.}\ }\textbf {\bibinfo {volume}
  {99}},\ \bibinfo {pages} {231101}}\BibitemShut {NoStop}%
\bibitem [{\citenamefont {He}\ \emph {et~al.}(2007)\citenamefont {He},
  \citenamefont {Jin},\ and\ \citenamefont {Zhuang}}]{He:2006vr}%
  \BibitemOpen
  \bibfield  {author} {\bibinfo {author} {\bibnamefont {He}, \bibfnamefont
  {L.}}, \bibinfo {author} {\bibfnamefont {M.}~\bibnamefont {Jin}}, \ and\
  \bibinfo {author} {\bibfnamefont {P.}~\bibnamefont {Zhuang}}} (\bibinfo
  {year} {2007}),\ \href {\doibase 10.1103/PhysRevD.75.036003} {\bibfield
  {journal} {\bibinfo  {journal} {Phys.Rev.}\ }\textbf {\bibinfo {volume}
  {D75}},\ \bibinfo {pages} {036003}},\ \Eprint
  {http://arxiv.org/abs/hep-ph/0610121} {arXiv:hep-ph/0610121 [hep-ph]}
  \BibitemShut {NoStop}%
\bibitem [{\citenamefont {Heinke}\ and\ \citenamefont
  {Ho}(2010)}]{Heinke:2010cr}%
  \BibitemOpen
  \bibfield  {author} {\bibinfo {author} {\bibnamefont {Heinke}, \bibfnamefont
  {C.~O.}}, \ and\ \bibinfo {author} {\bibfnamefont {W.~C.~G.}\ \bibnamefont
  {Ho}}} (\bibinfo {year} {2010}),\ \href@noop {} {\bibfield  {journal}
  {\bibinfo  {journal} {Astrophys.J.}\ }\textbf {\bibinfo {volume} {719}},\
  \bibinfo {pages} {L167}},\ \Eprint {http://arxiv.org/abs/1007.4719}
  {arXiv:1007.4719 [astro-ph.HE]} \BibitemShut {NoStop}%
\bibitem [{\citenamefont {Hess}\ and\ \citenamefont
  {Sedrakian}(2011)}]{Hess:2011qw}%
  \BibitemOpen
  \bibfield  {author} {\bibinfo {author} {\bibnamefont {Hess}, \bibfnamefont
  {D.}}, \ and\ \bibinfo {author} {\bibfnamefont {A.}~\bibnamefont
  {Sedrakian}}} (\bibinfo {year} {2011}),\ \href {\doibase
  10.1103/PhysRevD.84.063015} {\bibfield  {journal} {\bibinfo  {journal}
  {Phys.Rev.}\ }\textbf {\bibinfo {volume} {D84}},\ \bibinfo {pages}
  {063015}},\ \Eprint {http://arxiv.org/abs/1104.1706} {arXiv:1104.1706
  [astro-ph.HE]} \BibitemShut {NoStop}%
\bibitem [{\citenamefont {Ho}\ \emph {et~al.}(2012)\citenamefont {Ho},
  \citenamefont {Glampedakis},\ and\ \citenamefont {Andersson}}]{Ho:2011aa}%
  \BibitemOpen
  \bibfield  {author} {\bibinfo {author} {\bibnamefont {Ho}, \bibfnamefont
  {W.~C.~G.}}, \bibinfo {author} {\bibfnamefont {K.}~\bibnamefont
  {Glampedakis}}, \ and\ \bibinfo {author} {\bibfnamefont {N.}~\bibnamefont
  {Andersson}}} (\bibinfo {year} {2012}),\ \href {\doibase
  10.1111/j.1365-2966.2012.20826.x} {\bibfield  {journal} {\bibinfo  {journal}
  {Mon.Not.Roy.Astron.Soc.}\ }\textbf {\bibinfo {volume} {422}},\ \bibinfo
  {pages} {2632}},\ \Eprint {http://arxiv.org/abs/1112.1415} {arXiv:1112.1415
  [astro-ph.HE]} \BibitemShut {NoStop}%
\bibitem [{\citenamefont {Hong}(2001)}]{Hong:2000ck}%
  \BibitemOpen
  \bibfield  {author} {\bibinfo {author} {\bibnamefont {Hong}, \bibfnamefont
  {D.~K.}}} (\bibinfo {year} {2001}),\ \href@noop {} {\bibfield  {journal}
  {\bibinfo  {journal} {Acta Phys.Polon.}\ }\textbf {\bibinfo {volume} {B32}},\
  \bibinfo {pages} {1253}},\ \Eprint {http://arxiv.org/abs/hep-ph/0101025}
  {arXiv:hep-ph/0101025 [hep-ph]} \BibitemShut {NoStop}%
\bibitem [{\citenamefont {Hong}(2000{\natexlab{a}})}]{Hong:1998tn}%
  \BibitemOpen
  \bibfield  {author} {\bibinfo {author} {\bibnamefont {Hong}, \bibfnamefont
  {D.~K.}}} (\bibinfo {year} {2000}{\natexlab{a}}),\ \href {\doibase
  10.1016/S0370-2693(99)01472-0} {\bibfield  {journal} {\bibinfo  {journal}
  {Phys.Lett.}\ }\textbf {\bibinfo {volume} {B473}},\ \bibinfo {pages} {118}},\
  \Eprint {http://arxiv.org/abs/hep-ph/9812510} {arXiv:hep-ph/9812510 [hep-ph]}
  \BibitemShut {NoStop}%
\bibitem [{\citenamefont {Hong}(2000{\natexlab{b}})}]{Hong:1999ru}%
  \BibitemOpen
  \bibfield  {author} {\bibinfo {author} {\bibnamefont {Hong}, \bibfnamefont
  {D.~K.}}} (\bibinfo {year} {2000}{\natexlab{b}}),\ \href {\doibase
  10.1016/S0550-3213(00)00330-8} {\bibfield  {journal} {\bibinfo  {journal}
  {Nucl.Phys.}\ }\textbf {\bibinfo {volume} {B582}},\ \bibinfo {pages} {451}},\
  \Eprint {http://arxiv.org/abs/hep-ph/9905523} {arXiv:hep-ph/9905523 [hep-ph]}
  \BibitemShut {NoStop}%
\bibitem [{\citenamefont {Hong}(2005)}]{Hong:2005jv}%
  \BibitemOpen
  \bibfield  {author} {\bibinfo {author} {\bibnamefont {Hong}, \bibfnamefont
  {D.~K.}}} (\bibinfo {year} {2005}),\ \href@noop {} {\ }\Eprint
  {http://arxiv.org/abs/hep-ph/0506097} {arXiv:hep-ph/0506097 [hep-ph]}
  \BibitemShut {NoStop}%
\bibitem [{\citenamefont {Hong}\ \emph {et~al.}(2000)\citenamefont {Hong},
  \citenamefont {Miransky}, \citenamefont {Shovkovy},\ and\ \citenamefont
  {Wijewardhana}}]{Hong:1999fh}%
  \BibitemOpen
  \bibfield  {author} {\bibinfo {author} {\bibnamefont {Hong}, \bibfnamefont
  {D.~K.}}, \bibinfo {author} {\bibfnamefont {V.~A.}\ \bibnamefont {Miransky}},
  \bibinfo {author} {\bibfnamefont {I.~A.}\ \bibnamefont {Shovkovy}}, \ and\
  \bibinfo {author} {\bibfnamefont {L.~C.~R.}\ \bibnamefont {Wijewardhana}}}
  (\bibinfo {year} {2000}),\ \href {\doibase 10.1103/PhysRevD.61.056001,
  10.1103/PhysRevD.62.059903} {\bibfield  {journal} {\bibinfo  {journal}
  {Phys.Rev.}\ }\textbf {\bibinfo {volume} {D61}},\ \bibinfo {pages}
  {056001}},\ \Eprint {http://arxiv.org/abs/hep-ph/9906478}
  {arXiv:hep-ph/9906478 [hep-ph]} \BibitemShut {NoStop}%
\bibitem [{\citenamefont {Hsu}(2000)}]{Hsu:2000sy}%
  \BibitemOpen
  \bibfield  {author} {\bibinfo {author} {\bibnamefont {Hsu}, \bibfnamefont
  {S.~D.}}} (\bibinfo {year} {2000}),\ \href@noop {} {\ }\Eprint
  {http://arxiv.org/abs/hep-ph/0003140} {arXiv:hep-ph/0003140 [hep-ph]}
  \BibitemShut {NoStop}%
\bibitem [{\citenamefont {Huang}\ and\ \citenamefont
  {Shovkovy}(2003)}]{Huang:2003xd}%
  \BibitemOpen
  \bibfield  {author} {\bibinfo {author} {\bibnamefont {Huang}, \bibfnamefont
  {M.}}, \ and\ \bibinfo {author} {\bibfnamefont {I.}~\bibnamefont {Shovkovy}}}
  (\bibinfo {year} {2003}),\ \href {\doibase 10.1016/j.nuclphysa.2003.10.005}
  {\bibfield  {journal} {\bibinfo  {journal} {Nucl.Phys.}\ }\textbf {\bibinfo
  {volume} {A729}},\ \bibinfo {pages} {835}},\ \Eprint
  {http://arxiv.org/abs/hep-ph/0307273} {arXiv:hep-ph/0307273 [hep-ph]}
  \BibitemShut {NoStop}%
\bibitem [{\citenamefont {Huang}\ and\ \citenamefont
  {Shovkovy}(2004{\natexlab{a}})}]{Huang:2004bg}%
  \BibitemOpen
  \bibfield  {author} {\bibinfo {author} {\bibnamefont {Huang}, \bibfnamefont
  {M.}}, \ and\ \bibinfo {author} {\bibfnamefont {I.~A.}\ \bibnamefont
  {Shovkovy}}} (\bibinfo {year} {2004}{\natexlab{a}}),\ \href {\doibase
  10.1103/PhysRevD.70.051501} {\bibfield  {journal} {\bibinfo  {journal}
  {Phys.Rev.}\ }\textbf {\bibinfo {volume} {D70}},\ \bibinfo {pages}
  {051501}},\ \Eprint {http://arxiv.org/abs/hep-ph/0407049}
  {arXiv:hep-ph/0407049 [hep-ph]} \BibitemShut {NoStop}%
\bibitem [{\citenamefont {Huang}\ and\ \citenamefont
  {Shovkovy}(2004{\natexlab{b}})}]{Huang:2004am}%
  \BibitemOpen
  \bibfield  {author} {\bibinfo {author} {\bibnamefont {Huang}, \bibfnamefont
  {M.}}, \ and\ \bibinfo {author} {\bibfnamefont {I.~A.}\ \bibnamefont
  {Shovkovy}}} (\bibinfo {year} {2004}{\natexlab{b}}),\ \href {\doibase
  10.1103/PhysRevD.70.094030} {\bibfield  {journal} {\bibinfo  {journal}
  {Phys.Rev.}\ }\textbf {\bibinfo {volume} {D70}},\ \bibinfo {pages}
  {094030}},\ \Eprint {http://arxiv.org/abs/hep-ph/0408268}
  {arXiv:hep-ph/0408268 [hep-ph]} \BibitemShut {NoStop}%
\bibitem [{\citenamefont {{Huxley}}\ \emph {et~al.}(1993)\citenamefont
  {{Huxley}}, \citenamefont {{Paulson}}, \citenamefont {{Laborde}},
  \citenamefont {{Tholence}}, \citenamefont {{Sanchez}}, \citenamefont
  {{Junod}},\ and\ \citenamefont {{Calemczuk}}}]{Huxley}%
  \BibitemOpen
  \bibfield  {author} {\bibinfo {author} {\bibnamefont {{Huxley}},
  \bibfnamefont {A.~D.}}, \bibinfo {author} {\bibfnamefont {C.}~\bibnamefont
  {{Paulson}}}, \bibinfo {author} {\bibfnamefont {O.}~\bibnamefont
  {{Laborde}}}, \bibinfo {author} {\bibfnamefont {J.~L.}\ \bibnamefont
  {{Tholence}}}, \bibinfo {author} {\bibfnamefont {D.}~\bibnamefont
  {{Sanchez}}}, \bibinfo {author} {\bibfnamefont {A.}~\bibnamefont {{Junod}}},
  \ and\ \bibinfo {author} {\bibfnamefont {R.}~\bibnamefont {{Calemczuk}}}}
  (\bibinfo {year} {1993}),\ \href {\doibase 10.1088/0953-8984/5/41/018}
  {\bibfield  {journal} {\bibinfo  {journal} {Journal of Physics Condensed
  Matter}\ }\textbf {\bibinfo {volume} {5}},\ \bibinfo {pages}
  {7709}}\BibitemShut {NoStop}%
\bibitem [{\citenamefont {Iida}\ and\ \citenamefont
  {Baym}(2002)}]{Iida:2002ev}%
  \BibitemOpen
  \bibfield  {author} {\bibinfo {author} {\bibnamefont {Iida}, \bibfnamefont
  {K.}}, \ and\ \bibinfo {author} {\bibfnamefont {G.}~\bibnamefont {Baym}}}
  (\bibinfo {year} {2002}),\ \href {\doibase 10.1103/PhysRevD.66.014015}
  {\bibfield  {journal} {\bibinfo  {journal} {Phys.Rev.}\ }\textbf {\bibinfo
  {volume} {D66}},\ \bibinfo {pages} {014015}},\ \Eprint
  {http://arxiv.org/abs/hep-ph/0204124} {arXiv:hep-ph/0204124 [hep-ph]}
  \BibitemShut {NoStop}%
\bibitem [{\citenamefont {Iida}\ and\ \citenamefont
  {Fukushima}(2006)}]{Iida:2006df}%
  \BibitemOpen
  \bibfield  {author} {\bibinfo {author} {\bibnamefont {Iida}, \bibfnamefont
  {K.}}, \ and\ \bibinfo {author} {\bibfnamefont {K.}~\bibnamefont
  {Fukushima}}} (\bibinfo {year} {2006}),\ \href {\doibase
  10.1103/PhysRevD.74.074020} {\bibfield  {journal} {\bibinfo  {journal}
  {Phys.Rev.}\ }\textbf {\bibinfo {volume} {D74}},\ \bibinfo {pages}
  {074020}},\ \Eprint {http://arxiv.org/abs/hep-ph/0603179}
  {arXiv:hep-ph/0603179 [hep-ph]} \BibitemShut {NoStop}%
\bibitem [{\citenamefont {Iida}\ \emph {et~al.}(2004)\citenamefont {Iida},
  \citenamefont {Matsuura}, \citenamefont {Tachibana},\ and\ \citenamefont
  {Hatsuda}}]{Iida:2003cc}%
  \BibitemOpen
  \bibfield  {author} {\bibinfo {author} {\bibnamefont {Iida}, \bibfnamefont
  {K.}}, \bibinfo {author} {\bibfnamefont {T.}~\bibnamefont {Matsuura}},
  \bibinfo {author} {\bibfnamefont {M.}~\bibnamefont {Tachibana}}, \ and\
  \bibinfo {author} {\bibfnamefont {T.}~\bibnamefont {Hatsuda}}} (\bibinfo
  {year} {2004}),\ \href {\doibase 10.1103/PhysRevLett.93.132001} {\bibfield
  {journal} {\bibinfo  {journal} {Phys.Rev.Lett.}\ }\textbf {\bibinfo {volume}
  {93}},\ \bibinfo {pages} {132001}},\ \Eprint
  {http://arxiv.org/abs/hep-ph/0312363} {arXiv:hep-ph/0312363 [hep-ph]}
  \BibitemShut {NoStop}%
\bibitem [{\citenamefont {Ippolito}\ \emph {et~al.}(2008)\citenamefont
  {Ippolito}, \citenamefont {Ruggieri}, \citenamefont {Rischke}, \citenamefont
  {Sedrakian},\ and\ \citenamefont {Weber}}]{Ippolito:2007hn}%
  \BibitemOpen
  \bibfield  {author} {\bibinfo {author} {\bibnamefont {Ippolito},
  \bibfnamefont {N.}}, \bibinfo {author} {\bibfnamefont {M.}~\bibnamefont
  {Ruggieri}}, \bibinfo {author} {\bibfnamefont {D.}~\bibnamefont {Rischke}},
  \bibinfo {author} {\bibfnamefont {A.}~\bibnamefont {Sedrakian}}, \ and\
  \bibinfo {author} {\bibfnamefont {F.}~\bibnamefont {Weber}}} (\bibinfo {year}
  {2008}),\ \href {\doibase 10.1103/PhysRevD.77.023004} {\bibfield  {journal}
  {\bibinfo  {journal} {Phys.Rev.}\ }\textbf {\bibinfo {volume} {D77}},\
  \bibinfo {pages} {023004}},\ \Eprint {http://arxiv.org/abs/0710.3874}
  {arXiv:0710.3874 [astro-ph]} \BibitemShut {NoStop}%
\bibitem [{\citenamefont {Ippolito}\ \emph {et~al.}(2007)\citenamefont
  {Ippolito}, \citenamefont {Nardulli},\ and\ \citenamefont
  {Ruggieri}}]{Ippolito:2007uz}%
  \BibitemOpen
  \bibfield  {author} {\bibinfo {author} {\bibnamefont {Ippolito},
  \bibfnamefont {N.~D.}}, \bibinfo {author} {\bibfnamefont {G.}~\bibnamefont
  {Nardulli}}, \ and\ \bibinfo {author} {\bibfnamefont {M.}~\bibnamefont
  {Ruggieri}}} (\bibinfo {year} {2007}),\ \href {\doibase
  10.1088/1126-6708/2007/04/036} {\bibfield  {journal} {\bibinfo  {journal}
  {JHEP}\ }\textbf {\bibinfo {volume} {0704}},\ \bibinfo {pages} {036}},\
  \Eprint {http://arxiv.org/abs/hep-ph/0701113} {arXiv:hep-ph/0701113 [hep-ph]}
  \BibitemShut {NoStop}%
\bibitem [{\citenamefont {{Ivanenko}}\ and\ \citenamefont
  {{Kurdgelaidze}}(1969)}]{Ivanenko1969}%
  \BibitemOpen
  \bibfield  {author} {\bibinfo {author} {\bibnamefont {{Ivanenko}},
  \bibfnamefont {D.~D.}}, \ and\ \bibinfo {author} {\bibfnamefont {D.~F.}\
  \bibnamefont {{Kurdgelaidze}}}} (\bibinfo {year} {1969}),\ \href@noop {}
  {\bibfield  {journal} {\bibinfo  {journal} {Nuovo Cimento Lettere}\ }\textbf
  {\bibinfo {volume} {1}},\ \bibinfo {pages} {13}}\BibitemShut {NoStop}%
\bibitem [{\citenamefont {{Ivanenko}}\ and\ \citenamefont
  {{Kurdgelaidze}}(1965)}]{Ivanenko1965}%
  \BibitemOpen
  \bibfield  {author} {\bibinfo {author} {\bibnamefont {{Ivanenko}},
  \bibfnamefont {D.~D.}}, \ and\ \bibinfo {author} {\bibfnamefont {D.~F.}\
  \bibnamefont {{Kurdgelaidze}}}} (\bibinfo {year} {1965}),\ \href {\doibase
  10.1007/BF01042830} {\bibfield  {journal} {\bibinfo  {journal}
  {Astrophysics}\ }\textbf {\bibinfo {volume} {1}},\ \bibinfo {pages}
  {251}}\BibitemShut {NoStop}%
\bibitem [{\citenamefont {{Iwamoto}}(1980)}]{Iwamoto1}%
  \BibitemOpen
  \bibfield  {author} {\bibinfo {author} {\bibnamefont {{Iwamoto}},
  \bibfnamefont {N.}}} (\bibinfo {year} {1980}),\ \href {\doibase
  10.1103/PhysRevLett.44.1637} {\bibfield  {journal} {\bibinfo  {journal}
  {Phys. Rev. Lett.}\ }\textbf {\bibinfo {volume} {44}},\ \bibinfo {pages}
  {1637}}\BibitemShut {NoStop}%
\bibitem [{\citenamefont {{Iwamoto}}(1981)}]{Iwamoto2}%
  \BibitemOpen
  \bibfield  {author} {\bibinfo {author} {\bibnamefont {{Iwamoto}},
  \bibfnamefont {N.}}} (\bibinfo {year} {1981}),\ \emph {\bibinfo {title}
  {{Neutrino processes in dense matter}}},\ \href@noop {} {Ph.D. thesis}\
  (\bibinfo  {school} {Illinois Univ., Urbana-Champaign.})\BibitemShut
  {NoStop}%
\bibitem [{\citenamefont {{Iwamoto}}(1982)}]{Iwamoto1982An}%
  \BibitemOpen
  \bibfield  {author} {\bibinfo {author} {\bibnamefont {{Iwamoto}},
  \bibfnamefont {N.}}} (\bibinfo {year} {1982}),\ \href {\doibase
  10.1016/0003-4916(82)90271-8} {\bibfield  {journal} {\bibinfo  {journal}
  {Annals Phys.}\ }\textbf {\bibinfo {volume} {141}},\ \bibinfo {pages}
  {1}}\BibitemShut {NoStop}%
\bibitem [{\citenamefont {Jaikumar}\ \emph {et~al.}(2006)\citenamefont
  {Jaikumar}, \citenamefont {Roberts},\ and\ \citenamefont
  {Sedrakian}}]{Jaikumar:2005hy}%
  \BibitemOpen
  \bibfield  {author} {\bibinfo {author} {\bibnamefont {Jaikumar},
  \bibfnamefont {P.}}, \bibinfo {author} {\bibfnamefont {C.~D.}\ \bibnamefont
  {Roberts}}, \ and\ \bibinfo {author} {\bibfnamefont {A.}~\bibnamefont
  {Sedrakian}}} (\bibinfo {year} {2006}),\ \href {\doibase
  10.1103/PhysRevC.73.042801} {\bibfield  {journal} {\bibinfo  {journal}
  {Phys.Rev.}\ }\textbf {\bibinfo {volume} {C73}},\ \bibinfo {pages}
  {042801}},\ \Eprint {http://arxiv.org/abs/nucl-th/0509093}
  {arXiv:nucl-th/0509093 [nucl-th]} \BibitemShut {NoStop}%
\bibitem [{\citenamefont {Kaplan}\ and\ \citenamefont
  {Reddy}(2002)}]{Kaplan:2001qk}%
  \BibitemOpen
  \bibfield  {author} {\bibinfo {author} {\bibnamefont {Kaplan}, \bibfnamefont
  {D.}}, \ and\ \bibinfo {author} {\bibfnamefont {S.}~\bibnamefont {Reddy}}}
  (\bibinfo {year} {2002}),\ \href {\doibase 10.1103/PhysRevD.65.054042}
  {\bibfield  {journal} {\bibinfo  {journal} {Phys.Rev.}\ }\textbf {\bibinfo
  {volume} {D65}},\ \bibinfo {pages} {054042}},\ \Eprint
  {http://arxiv.org/abs/hep-ph/0107265} {arXiv:hep-ph/0107265 [hep-ph]}
  \BibitemShut {NoStop}%
\bibitem [{\citenamefont {{Ketterle}}\ and\ \citenamefont
  {{Zwierlein}}(2008)}]{ketterle-review}%
  \BibitemOpen
  \bibfield  {author} {\bibinfo {author} {\bibnamefont {{Ketterle}},
  \bibfnamefont {W.}}, \ and\ \bibinfo {author} {\bibfnamefont {M.~W.}\
  \bibnamefont {{Zwierlein}}}} (\bibinfo {year} {2008}),\ \href {\doibase
  10.1393/ncr/i2008-10033-1} {\bibfield  {journal} {\bibinfo  {journal} {Nuovo
  Cimento Rivista Serie}\ }\textbf {\bibinfo {volume} {31}},\ \bibinfo {pages}
  {247}},\ \Eprint {http://arxiv.org/abs/0801.2500} {arXiv:0801.2500
  [cond-mat.other]} \BibitemShut {NoStop}%
\bibitem [{\citenamefont {Kiriyama}(2006)}]{Kiriyama:2006jp}%
  \BibitemOpen
  \bibfield  {author} {\bibinfo {author} {\bibnamefont {Kiriyama},
  \bibfnamefont {O.}}} (\bibinfo {year} {2006}),\ \href {\doibase
  10.1103/PhysRevD.74.114011} {\bibfield  {journal} {\bibinfo  {journal}
  {Phys.Rev.}\ }\textbf {\bibinfo {volume} {D74}},\ \bibinfo {pages}
  {114011}},\ \Eprint {http://arxiv.org/abs/hep-ph/0609185}
  {arXiv:hep-ph/0609185 [hep-ph]} \BibitemShut {NoStop}%
\bibitem [{\citenamefont {Kiriyama}\ \emph {et~al.}(2006)\citenamefont
  {Kiriyama}, \citenamefont {Rischke},\ and\ \citenamefont
  {Shovkovy}}]{Kiriyama:2006ui}%
  \BibitemOpen
  \bibfield  {author} {\bibinfo {author} {\bibnamefont {Kiriyama},
  \bibfnamefont {O.}}, \bibinfo {author} {\bibfnamefont {D.}~\bibnamefont
  {Rischke}}, \ and\ \bibinfo {author} {\bibfnamefont {I.}~\bibnamefont
  {Shovkovy}}} (\bibinfo {year} {2006}),\ \href {\doibase
  10.1016/j.physletb.2006.10.067} {\bibfield  {journal} {\bibinfo  {journal}
  {Phys.Lett.}\ }\textbf {\bibinfo {volume} {B643}},\ \bibinfo {pages} {331}},\
  \Eprint {http://arxiv.org/abs/hep-ph/0606030} {arXiv:hep-ph/0606030 [hep-ph]}
  \BibitemShut {NoStop}%
\bibitem [{\citenamefont {Klahn}\ \emph {et~al.}(2007)\citenamefont {Klahn},
  \citenamefont {Blaschke}, \citenamefont {Sandin}, \citenamefont {Fuchs},
  \citenamefont {Faessler} \emph {et~al.}}]{Klahn:2006iw}%
  \BibitemOpen
  \bibfield  {author} {\bibinfo {author} {\bibnamefont {Klahn}, \bibfnamefont
  {T.}}, \bibinfo {author} {\bibfnamefont {D.}~\bibnamefont {Blaschke}},
  \bibinfo {author} {\bibfnamefont {F.}~\bibnamefont {Sandin}}, \bibinfo
  {author} {\bibfnamefont {C.}~\bibnamefont {Fuchs}}, \bibinfo {author}
  {\bibfnamefont {A.}~\bibnamefont {Faessler}},  \emph {et~al.}} (\bibinfo
  {year} {2007}),\ \href {\doibase 10.1016/j.physletb.2007.08.048} {\bibfield
  {journal} {\bibinfo  {journal} {Phys.Lett.}\ }\textbf {\bibinfo {volume}
  {B654}},\ \bibinfo {pages} {170}},\ \Eprint
  {http://arxiv.org/abs/nucl-th/0609067} {arXiv:nucl-th/0609067 [nucl-th]}
  \BibitemShut {NoStop}%
\bibitem [{\citenamefont {Klahn}\ \emph {et~al.}(2012)\citenamefont {Klahn},
  \citenamefont {Blaschke},\ and\ \citenamefont {Weber}}]{Klahn:2012uq}%
  \BibitemOpen
  \bibfield  {author} {\bibinfo {author} {\bibnamefont {Klahn}, \bibfnamefont
  {T.}}, \bibinfo {author} {\bibfnamefont {D.}~\bibnamefont {Blaschke}}, \ and\
  \bibinfo {author} {\bibfnamefont {F.}~\bibnamefont {Weber}}} (\bibinfo {year}
  {2012}),\ \href {\doibase 10.1134/S1547477112060118} {\bibfield  {journal}
  {\bibinfo  {journal} {Phys.Part.Nucl.Lett.}\ }\textbf {\bibinfo {volume}
  {9}},\ \bibinfo {pages} {484}}\BibitemShut {NoStop}%
\bibitem [{\citenamefont {Knippel}\ and\ \citenamefont
  {Sedrakian}(2009)}]{Knippel:2009st}%
  \BibitemOpen
  \bibfield  {author} {\bibinfo {author} {\bibnamefont {Knippel}, \bibfnamefont
  {B.}}, \ and\ \bibinfo {author} {\bibfnamefont {A.}~\bibnamefont
  {Sedrakian}}} (\bibinfo {year} {2009}),\ \href {\doibase
  10.1103/PhysRevD.79.083007} {\bibfield  {journal} {\bibinfo  {journal}
  {Phys.Rev.}\ }\textbf {\bibinfo {volume} {D79}},\ \bibinfo {pages}
  {083007}},\ \Eprint {http://arxiv.org/abs/0901.4637} {arXiv:0901.4637
  [astro-ph.SR]} \BibitemShut {NoStop}%
\bibitem [{\citenamefont {Kojo}\ \emph {et~al.}(2010)\citenamefont {Kojo},
  \citenamefont {Pisarski},\ and\ \citenamefont {Tsvelik}}]{Kojo:2010fe}%
  \BibitemOpen
  \bibfield  {author} {\bibinfo {author} {\bibnamefont {Kojo}, \bibfnamefont
  {T.}}, \bibinfo {author} {\bibfnamefont {R.~D.}\ \bibnamefont {Pisarski}}, \
  and\ \bibinfo {author} {\bibfnamefont {A.}~\bibnamefont {Tsvelik}}} (\bibinfo
  {year} {2010}),\ \href {\doibase 10.1103/PhysRevD.82.074015} {\bibfield
  {journal} {\bibinfo  {journal} {Phys.Rev.}\ }\textbf {\bibinfo {volume}
  {D82}},\ \bibinfo {pages} {074015}},\ \Eprint
  {http://arxiv.org/abs/1007.0248} {arXiv:1007.0248 [hep-ph]} \BibitemShut
  {NoStop}%
\bibitem [{\citenamefont {Kryjevski}(2008)}]{Kryjevski:2005qq}%
  \BibitemOpen
  \bibfield  {author} {\bibinfo {author} {\bibnamefont {Kryjevski},
  \bibfnamefont {A.}}} (\bibinfo {year} {2008}),\ \href {\doibase
  10.1103/PhysRevD.77.014018} {\bibfield  {journal} {\bibinfo  {journal}
  {Phys.Rev.}\ }\textbf {\bibinfo {volume} {D77}},\ \bibinfo {pages}
  {014018}},\ \Eprint {http://arxiv.org/abs/hep-ph/0508180}
  {arXiv:hep-ph/0508180 [hep-ph]} \BibitemShut {NoStop}%
\bibitem [{\citenamefont {Kryjevski}\ and\ \citenamefont
  {SchŠfer}(2005)}]{Kryjevski:2004jw}%
  \BibitemOpen
  \bibfield  {author} {\bibinfo {author} {\bibnamefont {Kryjevski},
  \bibfnamefont {A.}}, \ and\ \bibinfo {author} {\bibfnamefont
  {T.}~\bibnamefont {SchŠfer}}} (\bibinfo {year} {2005}),\ \href {\doibase
  10.1016/j.physletb.2004.11.081} {\bibfield  {journal} {\bibinfo  {journal}
  {Phys.Lett.}\ }\textbf {\bibinfo {volume} {B606}},\ \bibinfo {pages} {52}},\
  \Eprint {http://arxiv.org/abs/hep-ph/0407329} {arXiv:hep-ph/0407329 [hep-ph]}
  \BibitemShut {NoStop}%
\bibitem [{\citenamefont {Kundu}\ and\ \citenamefont
  {Rajagopal}(2002)}]{Kundu:2001tt}%
  \BibitemOpen
  \bibfield  {author} {\bibinfo {author} {\bibnamefont {Kundu}, \bibfnamefont
  {J.}}, \ and\ \bibinfo {author} {\bibfnamefont {K.}~\bibnamefont
  {Rajagopal}}} (\bibinfo {year} {2002}),\ \href {\doibase
  10.1103/PhysRevD.65.094022} {\bibfield  {journal} {\bibinfo  {journal}
  {Phys.Rev.}\ }\textbf {\bibinfo {volume} {D65}},\ \bibinfo {pages}
  {094022}},\ \Eprint {http://arxiv.org/abs/hep-ph/0112206}
  {arXiv:hep-ph/0112206 [hep-ph]} \BibitemShut {NoStop}%
\bibitem [{\citenamefont {{Landau}}\ and\ \citenamefont
  {{Lifshit's}}(1959)}]{Landau:Elastic}%
  \BibitemOpen
  \bibfield  {author} {\bibinfo {author} {\bibnamefont {{Landau}},
  \bibfnamefont {L.~D.}}, \ and\ \bibinfo {author} {\bibfnamefont {E.~M.}\
  \bibnamefont {{Lifshit's}}}} (\bibinfo {year} {1959}),\ \href@noop {} {\emph
  {\bibinfo {title} {Theory of elasticity, by Landau, L.~D.; Lifshit's, E.~M.~
  London, Pergamon Press; Reading, Mass., Addison-Wesley Pub.~Co.,
  1959.~Addison-Wesley physics books}}}\BibitemShut {NoStop}%
\bibitem [{\citenamefont {Larkin}\ and\ \citenamefont
  {Ovchinnikov}(1964)}]{LO}%
  \BibitemOpen
  \bibfield  {author} {\bibinfo {author} {\bibnamefont {Larkin}, \bibfnamefont
  {A.~I.}}, \ and\ \bibinfo {author} {\bibfnamefont {Y.~N.}\ \bibnamefont
  {Ovchinnikov}}} (\bibinfo {year} {1964}),\ \href@noop {} {\bibfield
  {journal} {\bibinfo  {journal} {Zh. Eksp. Teor. Fiz.}\ }\textbf {\bibinfo
  {volume} {47(3)}},\ \bibinfo {pages} {1136}}\BibitemShut {NoStop}%
\bibitem [{\citenamefont {Lattimer}\ and\ \citenamefont
  {Prakash}(2001)}]{Lattimer:2000nx}%
  \BibitemOpen
  \bibfield  {author} {\bibinfo {author} {\bibnamefont {Lattimer},
  \bibfnamefont {J.~M.}}, \ and\ \bibinfo {author} {\bibfnamefont
  {M.}~\bibnamefont {Prakash}}} (\bibinfo {year} {2001}),\ \href {\doibase
  10.1086/319702} {\bibfield  {journal} {\bibinfo  {journal} {Astrophys.J.}\
  }\textbf {\bibinfo {volume} {550}},\ \bibinfo {pages} {426}},\ \Eprint
  {http://arxiv.org/abs/astro-ph/0002232} {arXiv:astro-ph/0002232 [astro-ph]}
  \BibitemShut {NoStop}%
\bibitem [{\citenamefont {Lattimer}\ and\ \citenamefont
  {Prakash}(2007)}]{Lattimer:2006xb}%
  \BibitemOpen
  \bibfield  {author} {\bibinfo {author} {\bibnamefont {Lattimer},
  \bibfnamefont {J.~M.}}, \ and\ \bibinfo {author} {\bibfnamefont
  {M.}~\bibnamefont {Prakash}}} (\bibinfo {year} {2007}),\ \href {\doibase
  10.1016/j.physrep.2007.02.003} {\bibfield  {journal} {\bibinfo  {journal}
  {Phys.Rept.}\ }\textbf {\bibinfo {volume} {442}},\ \bibinfo {pages} {109}},\
  \Eprint {http://arxiv.org/abs/astro-ph/0612440} {arXiv:astro-ph/0612440
  [astro-ph]} \BibitemShut {NoStop}%
\bibitem [{\citenamefont {Lattimer}\ \emph {et~al.}(1991)\citenamefont
  {Lattimer}, \citenamefont {Prakash}, \citenamefont {Pethick},\ and\
  \citenamefont {Haensel}}]{Lattimer:1991ib}%
  \BibitemOpen
  \bibfield  {author} {\bibinfo {author} {\bibnamefont {Lattimer},
  \bibfnamefont {J.~M.}}, \bibinfo {author} {\bibfnamefont {M.}~\bibnamefont
  {Prakash}}, \bibinfo {author} {\bibfnamefont {C.~J.}\ \bibnamefont
  {Pethick}}, \ and\ \bibinfo {author} {\bibfnamefont {P.}~\bibnamefont
  {Haensel}}} (\bibinfo {year} {1991}),\ \href {\doibase
  10.1103/PhysRevLett.66.2701} {\bibfield  {journal} {\bibinfo  {journal}
  {Phys.Rev.Lett.}\ }\textbf {\bibinfo {volume} {66}},\ \bibinfo {pages}
  {2701}}\BibitemShut {NoStop}%
\bibitem [{\citenamefont {{Lattimer}}\ \emph {et~al.}(1994)\citenamefont
  {{Lattimer}}, \citenamefont {{van Riper}}, \citenamefont {{Prakash}},\ and\
  \citenamefont {{Prakash}}}]{Lattimer1994}%
  \BibitemOpen
  \bibfield  {author} {\bibinfo {author} {\bibnamefont {{Lattimer}},
  \bibfnamefont {J.~M.}}, \bibinfo {author} {\bibfnamefont {K.~A.}\
  \bibnamefont {{van Riper}}}, \bibinfo {author} {\bibfnamefont
  {M.}~\bibnamefont {{Prakash}}}, \ and\ \bibinfo {author} {\bibfnamefont
  {M.}~\bibnamefont {{Prakash}}}} (\bibinfo {year} {1994}),\ \href {\doibase
  10.1086/174025} {\bibfield  {journal} {\bibinfo  {journal} {\apj}\ }\textbf
  {\bibinfo {volume} {425}},\ \bibinfo {pages} {802}}\BibitemShut {NoStop}%
\bibitem [{\citenamefont {Le~Bellac}(2000)}]{bellac2000thermal}%
  \BibitemOpen
  \bibfield  {author} {\bibinfo {author} {\bibnamefont {Le~Bellac},
  \bibfnamefont {M.}}} (\bibinfo {year} {2000}),\ \href
  {http://books.google.it/books?id=00\_x6GR8GXoC} {\emph {\bibinfo {title}
  {Thermal Field Theory}}},\ Cambridge Monographs on Mathematical Physics\
  (\bibinfo  {publisher} {Cambridge University Press})\BibitemShut {NoStop}%
\bibitem [{\citenamefont {Leibovich}\ \emph {et~al.}(2001)\citenamefont
  {Leibovich}, \citenamefont {Rajagopal},\ and\ \citenamefont
  {Shuster}}]{Leibovich:2001xr}%
  \BibitemOpen
  \bibfield  {author} {\bibinfo {author} {\bibnamefont {Leibovich},
  \bibfnamefont {A.~K.}}, \bibinfo {author} {\bibfnamefont {K.}~\bibnamefont
  {Rajagopal}}, \ and\ \bibinfo {author} {\bibfnamefont {E.}~\bibnamefont
  {Shuster}}} (\bibinfo {year} {2001}),\ \href {\doibase
  10.1103/PhysRevD.64.094005} {\bibfield  {journal} {\bibinfo  {journal}
  {Phys.Rev.}\ }\textbf {\bibinfo {volume} {D64}},\ \bibinfo {pages}
  {094005}},\ \Eprint {http://arxiv.org/abs/hep-ph/0104073}
  {arXiv:hep-ph/0104073 [hep-ph]} \BibitemShut {NoStop}%
\bibitem [{\citenamefont {Leutwyler}(1997)}]{Leutwyler:1996er}%
  \BibitemOpen
  \bibfield  {author} {\bibinfo {author} {\bibnamefont {Leutwyler},
  \bibfnamefont {H.}}} (\bibinfo {year} {1997}),\ \href@noop {} {\bibfield
  {journal} {\bibinfo  {journal} {Helv.Phys.Acta}\ }\textbf {\bibinfo {volume}
  {70}},\ \bibinfo {pages} {275}},\ \Eprint
  {http://arxiv.org/abs/hep-ph/9609466} {arXiv:hep-ph/9609466 [hep-ph]}
  \BibitemShut {NoStop}%
\bibitem [{\citenamefont {Lin}(2007)}]{Lin:2007rz}%
  \BibitemOpen
  \bibfield  {author} {\bibinfo {author} {\bibnamefont {Lin}, \bibfnamefont
  {L.-M.}}} (\bibinfo {year} {2007}),\ \href {\doibase
  10.1103/PhysRevD.76.081502} {\bibfield  {journal} {\bibinfo  {journal}
  {Phys.Rev.}\ }\textbf {\bibinfo {volume} {D76}},\ \bibinfo {pages}
  {081502}},\ \Eprint {http://arxiv.org/abs/0708.2965} {arXiv:0708.2965
  [astro-ph]} \BibitemShut {NoStop}%
\bibitem [{\citenamefont {Liu}\ and\ \citenamefont
  {Wilczek}(2003)}]{Liu:2002gi}%
  \BibitemOpen
  \bibfield  {author} {\bibinfo {author} {\bibnamefont {Liu}, \bibfnamefont
  {W.~V.}}, \ and\ \bibinfo {author} {\bibfnamefont {F.}~\bibnamefont
  {Wilczek}}} (\bibinfo {year} {2003}),\ \href {\doibase
  10.1103/PhysRevLett.90.047002} {\bibfield  {journal} {\bibinfo  {journal}
  {Phys.Rev.Lett.}\ }\textbf {\bibinfo {volume} {90}},\ \bibinfo {pages}
  {047002}},\ \Eprint {http://arxiv.org/abs/cond-mat/0208052}
  {arXiv:cond-mat/0208052 [cond-mat]} \BibitemShut {NoStop}%
\bibitem [{\citenamefont {Logoteta}\ \emph {et~al.}(2012)\citenamefont
  {Logoteta}, \citenamefont {Providencia}, \citenamefont {Vidana},\ and\
  \citenamefont {Bombaci}}]{Logoteta:2012ms}%
  \BibitemOpen
  \bibfield  {author} {\bibinfo {author} {\bibnamefont {Logoteta},
  \bibfnamefont {D.}}, \bibinfo {author} {\bibfnamefont {C.}~\bibnamefont
  {Providencia}}, \bibinfo {author} {\bibfnamefont {I.}~\bibnamefont {Vidana}},
  \ and\ \bibinfo {author} {\bibfnamefont {I.}~\bibnamefont {Bombaci}}}
  (\bibinfo {year} {2012}),\ \href {\doibase 10.1103/PhysRevC.85.055807}
  {\bibfield  {journal} {\bibinfo  {journal} {Phys.Rev.}\ }\textbf {\bibinfo
  {volume} {C85}},\ \bibinfo {pages} {055807}},\ \Eprint
  {http://arxiv.org/abs/1204.5909} {arXiv:1204.5909 [nucl-th]} \BibitemShut
  {NoStop}%
\bibitem [{\citenamefont {Lyne}\ \emph {et~al.}(2004)\citenamefont {Lyne},
  \citenamefont {Burgay}, \citenamefont {Kramer}, \citenamefont {Possenti},
  \citenamefont {Manchester} \emph {et~al.}}]{Lyne:2004cj}%
  \BibitemOpen
  \bibfield  {author} {\bibinfo {author} {\bibnamefont {Lyne}, \bibfnamefont
  {A.~G.}}, \bibinfo {author} {\bibfnamefont {M.}~\bibnamefont {Burgay}},
  \bibinfo {author} {\bibfnamefont {M.}~\bibnamefont {Kramer}}, \bibinfo
  {author} {\bibfnamefont {A.}~\bibnamefont {Possenti}}, \bibinfo {author}
  {\bibfnamefont {R.~N.}\ \bibnamefont {Manchester}},  \emph {et~al.}}
  (\bibinfo {year} {2004}),\ \href {\doibase 10.1126/science.1094645}
  {\bibfield  {journal} {\bibinfo  {journal} {Science}\ }\textbf {\bibinfo
  {volume} {303}},\ \bibinfo {pages} {1153}},\ \Eprint
  {http://arxiv.org/abs/astro-ph/0401086} {arXiv:astro-ph/0401086 [astro-ph]}
  \BibitemShut {NoStop}%
\bibitem [{\citenamefont {Maieron}\ \emph {et~al.}(2004)\citenamefont
  {Maieron}, \citenamefont {Baldo}, \citenamefont {Burgio},\ and\ \citenamefont
  {Schulze}}]{Maieron:2004af}%
  \BibitemOpen
  \bibfield  {author} {\bibinfo {author} {\bibnamefont {Maieron}, \bibfnamefont
  {C.}}, \bibinfo {author} {\bibfnamefont {M.}~\bibnamefont {Baldo}}, \bibinfo
  {author} {\bibfnamefont {G.~F.}\ \bibnamefont {Burgio}}, \ and\ \bibinfo
  {author} {\bibfnamefont {H.~J.}\ \bibnamefont {Schulze}}} (\bibinfo {year}
  {2004}),\ \href {\doibase 10.1103/PhysRevD.70.043010} {\bibfield  {journal}
  {\bibinfo  {journal} {Phys.Rev.}\ }\textbf {\bibinfo {volume} {D70}},\
  \bibinfo {pages} {043010}},\ \Eprint {http://arxiv.org/abs/nucl-th/0404089}
  {arXiv:nucl-th/0404089 [nucl-th]} \BibitemShut {NoStop}%
\bibitem [{\citenamefont {Mannarelli}\ \emph
  {et~al.}(2006{\natexlab{a}})\citenamefont {Mannarelli}, \citenamefont
  {Nardulli},\ and\ \citenamefont {Ruggieri}}]{Mannarelli:2006hr}%
  \BibitemOpen
  \bibfield  {author} {\bibinfo {author} {\bibnamefont {Mannarelli},
  \bibfnamefont {M.}}, \bibinfo {author} {\bibfnamefont {G.}~\bibnamefont
  {Nardulli}}, \ and\ \bibinfo {author} {\bibfnamefont {M.}~\bibnamefont
  {Ruggieri}}} (\bibinfo {year} {2006}{\natexlab{a}}),\ \href {\doibase
  10.1103/PhysRevA.74.033606} {\bibfield  {journal} {\bibinfo  {journal}
  {Phys.Rev.}\ }\textbf {\bibinfo {volume} {A74}},\ \bibinfo {pages}
  {033606}},\ \Eprint {http://arxiv.org/abs/cond-mat/0604579}
  {arXiv:cond-mat/0604579 [cond-mat]} \BibitemShut {NoStop}%
\bibitem [{\citenamefont {Mannarelli}\ \emph
  {et~al.}(2006{\natexlab{b}})\citenamefont {Mannarelli}, \citenamefont
  {Rajagopal},\ and\ \citenamefont {Sharma}}]{Mannarelli:2006fy}%
  \BibitemOpen
  \bibfield  {author} {\bibinfo {author} {\bibnamefont {Mannarelli},
  \bibfnamefont {M.}}, \bibinfo {author} {\bibfnamefont {K.}~\bibnamefont
  {Rajagopal}}, \ and\ \bibinfo {author} {\bibfnamefont {R.}~\bibnamefont
  {Sharma}}} (\bibinfo {year} {2006}{\natexlab{b}}),\ \href {\doibase
  10.1103/PhysRevD.73.114012} {\bibfield  {journal} {\bibinfo  {journal}
  {Phys.Rev.}\ }\textbf {\bibinfo {volume} {D73}},\ \bibinfo {pages}
  {114012}},\ \Eprint {http://arxiv.org/abs/hep-ph/0603076}
  {arXiv:hep-ph/0603076 [hep-ph]} \BibitemShut {NoStop}%
\bibitem [{\citenamefont {Mannarelli}\ \emph {et~al.}(2007)\citenamefont
  {Mannarelli}, \citenamefont {Rajagopal},\ and\ \citenamefont
  {Sharma}}]{Mannarelli:2007bs}%
  \BibitemOpen
  \bibfield  {author} {\bibinfo {author} {\bibnamefont {Mannarelli},
  \bibfnamefont {M.}}, \bibinfo {author} {\bibfnamefont {K.}~\bibnamefont
  {Rajagopal}}, \ and\ \bibinfo {author} {\bibfnamefont {R.}~\bibnamefont
  {Sharma}}} (\bibinfo {year} {2007}),\ \href {\doibase
  10.1103/PhysRevD.76.074026} {\bibfield  {journal} {\bibinfo  {journal}
  {Phys.Rev.}\ }\textbf {\bibinfo {volume} {D76}},\ \bibinfo {pages}
  {074026}},\ \Eprint {http://arxiv.org/abs/hep-ph/0702021}
  {arXiv:hep-ph/0702021 [hep-ph]} \BibitemShut {NoStop}%
\bibitem [{\citenamefont {{Matsuda}}\ and\ \citenamefont
  {{Shimahara}}(2007)}]{Matsuda:2007}%
  \BibitemOpen
  \bibfield  {author} {\bibinfo {author} {\bibnamefont {{Matsuda}},
  \bibfnamefont {Y.}}, \ and\ \bibinfo {author} {\bibfnamefont
  {H.}~\bibnamefont {{Shimahara}}}} (\bibinfo {year} {2007}),\ \href {\doibase
  10.1143/JPSJ.76.051005} {\bibfield  {journal} {\bibinfo  {journal} {Journal
  of the Physical Society of Japan}\ }\textbf {\bibinfo {volume}
  {76}}~(\bibinfo {number} {5}),\ \bibinfo {pages} {051005}},\ \Eprint
  {http://arxiv.org/abs/arXiv:cond-mat/0702481} {arXiv:cond-mat/0702481}
  \BibitemShut {NoStop}%
\bibitem [{\citenamefont {Maxwell}\ \emph {et~al.}(1977)\citenamefont
  {Maxwell}, \citenamefont {Brown}, \citenamefont {Campbell}, \citenamefont
  {Dashen},\ and\ \citenamefont {Manassah}}]{Maxwell:1977zz}%
  \BibitemOpen
  \bibfield  {author} {\bibinfo {author} {\bibnamefont {Maxwell}, \bibfnamefont
  {O.}}, \bibinfo {author} {\bibfnamefont {G.~E.}\ \bibnamefont {Brown}},
  \bibinfo {author} {\bibfnamefont {D.~K.}\ \bibnamefont {Campbell}}, \bibinfo
  {author} {\bibfnamefont {R.~F.}\ \bibnamefont {Dashen}}, \ and\ \bibinfo
  {author} {\bibfnamefont {J.~T.}\ \bibnamefont {Manassah}}} (\bibinfo {year}
  {1977}),\ \href {\doibase 10.1086/155447} {\bibfield  {journal} {\bibinfo
  {journal} {Astrophys.J.}\ }\textbf {\bibinfo {volume} {216}},\ \bibinfo
  {pages} {77}}\BibitemShut {NoStop}%
\bibitem [{\citenamefont {McLerran}\ and\ \citenamefont
  {Pisarski}(2007)}]{McLerran:2007qj}%
  \BibitemOpen
  \bibfield  {author} {\bibinfo {author} {\bibnamefont {McLerran},
  \bibfnamefont {L.}}, \ and\ \bibinfo {author} {\bibfnamefont {R.~D.}\
  \bibnamefont {Pisarski}}} (\bibinfo {year} {2007}),\ \href {\doibase
  10.1016/j.nuclphysa.2007.08.013} {\bibfield  {journal} {\bibinfo  {journal}
  {Nucl.Phys.}\ }\textbf {\bibinfo {volume} {A796}},\ \bibinfo {pages} {83}},\
  \Eprint {http://arxiv.org/abs/0706.2191} {arXiv:0706.2191 [hep-ph]}
  \BibitemShut {NoStop}%
\bibitem [{\citenamefont {Muther}\ and\ \citenamefont
  {Sedrakian}(2002)}]{Muther:2002mc}%
  \BibitemOpen
  \bibfield  {author} {\bibinfo {author} {\bibnamefont {Muther}, \bibfnamefont
  {H.}}, \ and\ \bibinfo {author} {\bibfnamefont {A.}~\bibnamefont
  {Sedrakian}}} (\bibinfo {year} {2002}),\ \href {\doibase
  10.1103/PhysRevLett.88.252503} {\bibfield  {journal} {\bibinfo  {journal}
  {Phys.Rev.Lett.}\ }\textbf {\bibinfo {volume} {88}},\ \bibinfo {pages}
  {252503}},\ \Eprint {http://arxiv.org/abs/cond-mat/0202409}
  {arXiv:cond-mat/0202409 [cond-mat]} \BibitemShut {NoStop}%
\bibitem [{\citenamefont {Muto}\ and\ \citenamefont
  {Tatsumi}(1988)}]{Muto:1988vw}%
  \BibitemOpen
  \bibfield  {author} {\bibinfo {author} {\bibnamefont {Muto}, \bibfnamefont
  {T.}}, \ and\ \bibinfo {author} {\bibfnamefont {T.}~\bibnamefont {Tatsumi}}}
  (\bibinfo {year} {1988}),\ \href {\doibase 10.1143/PTP.79.461} {\bibfield
  {journal} {\bibinfo  {journal} {Prog.Theor.Phys.}\ }\textbf {\bibinfo
  {volume} {79}},\ \bibinfo {pages} {461}}\BibitemShut {NoStop}%
\bibitem [{\citenamefont {Nardulli}(2002)}]{Nardulli:2002ma}%
  \BibitemOpen
  \bibfield  {author} {\bibinfo {author} {\bibnamefont {Nardulli},
  \bibfnamefont {G.}}} (\bibinfo {year} {2002}),\ \href@noop {} {\bibfield
  {journal} {\bibinfo  {journal} {Riv.Nuovo Cim.}\ }\textbf {\bibinfo {volume}
  {25N3}},\ \bibinfo {pages} {1}},\ \Eprint
  {http://arxiv.org/abs/hep-ph/0202037} {arXiv:hep-ph/0202037 [hep-ph]}
  \BibitemShut {NoStop}%
\bibitem [{\citenamefont {Nickel}(2009)}]{Nickel:2009wj}%
  \BibitemOpen
  \bibfield  {author} {\bibinfo {author} {\bibnamefont {Nickel}, \bibfnamefont
  {D.}}} (\bibinfo {year} {2009}),\ \href {\doibase 10.1103/PhysRevD.80.074025}
  {\bibfield  {journal} {\bibinfo  {journal} {Phys.Rev.}\ }\textbf {\bibinfo
  {volume} {D80}},\ \bibinfo {pages} {074025}},\ \Eprint
  {http://arxiv.org/abs/0906.5295} {arXiv:0906.5295 [hep-ph]} \BibitemShut
  {NoStop}%
\bibitem [{\citenamefont {Nickel}\ and\ \citenamefont
  {Buballa}(2009)}]{Nickel:2008ng}%
  \BibitemOpen
  \bibfield  {author} {\bibinfo {author} {\bibnamefont {Nickel}, \bibfnamefont
  {D.}}, \ and\ \bibinfo {author} {\bibfnamefont {M.}~\bibnamefont {Buballa}}}
  (\bibinfo {year} {2009}),\ \href {\doibase 10.1103/PhysRevD.79.054009}
  {\bibfield  {journal} {\bibinfo  {journal} {Phys.Rev.}\ }\textbf {\bibinfo
  {volume} {D79}},\ \bibinfo {pages} {054009}},\ \Eprint
  {http://arxiv.org/abs/0811.2400} {arXiv:0811.2400 [hep-ph]} \BibitemShut
  {NoStop}%
\bibitem [{\citenamefont {{Oppenheimer}}\ and\ \citenamefont
  {{Volkoff}}(1939)}]{Oppenheimer-Volkoff}%
  \BibitemOpen
  \bibfield  {author} {\bibinfo {author} {\bibnamefont {{Oppenheimer}},
  \bibfnamefont {J.~R.}}, \ and\ \bibinfo {author} {\bibfnamefont {G.~M.}\
  \bibnamefont {{Volkoff}}}} (\bibinfo {year} {1939}),\ \href {\doibase
  10.1103/PhysRev.55.374} {\bibfield  {journal} {\bibinfo  {journal} {Phys.
  Rev.}\ }\textbf {\bibinfo {volume} {55}},\ \bibinfo {pages}
  {374}}\BibitemShut {NoStop}%
\bibitem [{\citenamefont {Orsaria}\ \emph {et~al.}(2013)\citenamefont
  {Orsaria}, \citenamefont {Rodrigues}, \citenamefont {Weber},\ and\
  \citenamefont {Contrera}}]{Orsaria:2012je}%
  \BibitemOpen
  \bibfield  {author} {\bibinfo {author} {\bibnamefont {Orsaria}, \bibfnamefont
  {M.}}, \bibinfo {author} {\bibfnamefont {H.}~\bibnamefont {Rodrigues}},
  \bibinfo {author} {\bibfnamefont {F.}~\bibnamefont {Weber}}, \ and\ \bibinfo
  {author} {\bibfnamefont {G.}~\bibnamefont {Contrera}}} (\bibinfo {year}
  {2013}),\ \href {\doibase 10.1103/PhysRevD.87.023001} {\bibfield  {journal}
  {\bibinfo  {journal} {Phys.Rev.}\ }\textbf {\bibinfo {volume} {D87}},\
  \bibinfo {pages} {023001}},\ \Eprint {http://arxiv.org/abs/1212.4213}
  {arXiv:1212.4213 [astro-ph.SR]} \BibitemShut {NoStop}%
\bibitem [{\citenamefont {Owen}(2005)}]{Owen:2005fn}%
  \BibitemOpen
  \bibfield  {author} {\bibinfo {author} {\bibnamefont {Owen}, \bibfnamefont
  {B.~J.}}} (\bibinfo {year} {2005}),\ \href {\doibase
  10.1103/PhysRevLett.95.211101} {\bibfield  {journal} {\bibinfo  {journal}
  {Phys.Rev.Lett.}\ }\textbf {\bibinfo {volume} {95}},\ \bibinfo {pages}
  {211101}},\ \Eprint {http://arxiv.org/abs/astro-ph/0503399}
  {arXiv:astro-ph/0503399 [astro-ph]} \BibitemShut {NoStop}%
\bibitem [{\citenamefont {Page}\ \emph {et~al.}(2004)\citenamefont {Page},
  \citenamefont {Lattimer}, \citenamefont {Prakash},\ and\ \citenamefont
  {Steiner}}]{Page:2004fy}%
  \BibitemOpen
  \bibfield  {author} {\bibinfo {author} {\bibnamefont {Page}, \bibfnamefont
  {D.}}, \bibinfo {author} {\bibfnamefont {J.~M.}\ \bibnamefont {Lattimer}},
  \bibinfo {author} {\bibfnamefont {M.}~\bibnamefont {Prakash}}, \ and\
  \bibinfo {author} {\bibfnamefont {A.~W.}\ \bibnamefont {Steiner}}} (\bibinfo
  {year} {2004}),\ \href {\doibase 10.1086/424844} {\bibfield  {journal}
  {\bibinfo  {journal} {Astrophys.J.Suppl.}\ }\textbf {\bibinfo {volume}
  {155}},\ \bibinfo {pages} {623}},\ \Eprint
  {http://arxiv.org/abs/astro-ph/0403657} {arXiv:astro-ph/0403657 [astro-ph]}
  \BibitemShut {NoStop}%
\bibitem [{\citenamefont {Page}\ \emph {et~al.}(2011)\citenamefont {Page},
  \citenamefont {Prakash}, \citenamefont {Lattimer},\ and\ \citenamefont
  {Steiner}}]{Page:2010aw}%
  \BibitemOpen
  \bibfield  {author} {\bibinfo {author} {\bibnamefont {Page}, \bibfnamefont
  {D.}}, \bibinfo {author} {\bibfnamefont {M.}~\bibnamefont {Prakash}},
  \bibinfo {author} {\bibfnamefont {J.~M.}\ \bibnamefont {Lattimer}}, \ and\
  \bibinfo {author} {\bibfnamefont {A.~W.}\ \bibnamefont {Steiner}}} (\bibinfo
  {year} {2011}),\ \href {\doibase 10.1103/PhysRevLett.106.081101} {\bibfield
  {journal} {\bibinfo  {journal} {Phys.Rev.Lett.}\ }\textbf {\bibinfo {volume}
  {106}},\ \bibinfo {pages} {081101}},\ \Eprint
  {http://arxiv.org/abs/1011.6142} {arXiv:1011.6142 [astro-ph.HE]} \BibitemShut
  {NoStop}%
\bibitem [{\citenamefont {Palomba}(2012)}]{Palomba:2012wn}%
  \BibitemOpen
  \bibfield  {author} {\bibinfo {author} {\bibnamefont {Palomba}, \bibfnamefont
  {C.}} (\bibinfo {collaboration} {LIGO Scientific Collaboration, Virgo
  Collaboration})} (\bibinfo {year} {2012}),\ \href@noop {} {\ }\Eprint
  {http://arxiv.org/abs/1201.3176} {arXiv:1201.3176 [astro-ph.IM]} \BibitemShut
  {NoStop}%
\bibitem [{\citenamefont {Pao}\ \emph {et~al.}(2006)\citenamefont {Pao},
  \citenamefont {Wu},\ and\ \citenamefont {Yip}}]{Pao}%
  \BibitemOpen
  \bibfield  {author} {\bibinfo {author} {\bibnamefont {Pao}, \bibfnamefont
  {C.-H.}}, \bibinfo {author} {\bibfnamefont {S.-T.}\ \bibnamefont {Wu}}, \
  and\ \bibinfo {author} {\bibfnamefont {S.-K.}\ \bibnamefont {Yip}}} (\bibinfo
  {year} {2006}),\ \href {\doibase 10.1103/PhysRevB.73.132506} {\bibfield
  {journal} {\bibinfo  {journal} {Phys. Rev. B}\ }\textbf {\bibinfo {volume}
  {73}},\ \bibinfo {pages} {132506}}\BibitemShut {NoStop}%
\bibitem [{\citenamefont {{Partridge}}\ \emph {et~al.}(2006)\citenamefont
  {{Partridge}}, \citenamefont {{Li}}, \citenamefont {{Kamar}}, \citenamefont
  {{Liao}},\ and\ \citenamefont {{Hulet}}}]{Partridge}%
  \BibitemOpen
  \bibfield  {author} {\bibinfo {author} {\bibnamefont {{Partridge}},
  \bibfnamefont {G.~B.}}, \bibinfo {author} {\bibfnamefont {W.}~\bibnamefont
  {{Li}}}, \bibinfo {author} {\bibfnamefont {R.~I.}\ \bibnamefont {{Kamar}}},
  \bibinfo {author} {\bibfnamefont {Y.-a.}\ \bibnamefont {{Liao}}}, \ and\
  \bibinfo {author} {\bibfnamefont {R.~G.}\ \bibnamefont {{Hulet}}}} (\bibinfo
  {year} {2006}),\ \href {\doibase 10.1126/science.1122876} {\bibfield
  {journal} {\bibinfo  {journal} {Science}\ }\textbf {\bibinfo {volume}
  {311}},\ \bibinfo {pages} {503}},\ \Eprint
  {http://arxiv.org/abs/arXiv:cond-mat/0511752} {arXiv:cond-mat/0511752}
  \BibitemShut {NoStop}%
\bibitem [{\citenamefont {{Pethick}}(1992)}]{Pethick-review-Urca}%
  \BibitemOpen
  \bibfield  {author} {\bibinfo {author} {\bibnamefont {{Pethick}},
  \bibfnamefont {C.~J.}}} (\bibinfo {year} {1992}),\ \href {\doibase
  10.1103/RevModPhys.64.1133} {\bibfield  {journal} {\bibinfo  {journal} {Rev.
  Mod. Phys.}\ }\textbf {\bibinfo {volume} {64}},\ \bibinfo {pages}
  {1133}}\BibitemShut {NoStop}%
\bibitem [{\citenamefont {Pisarski}\ and\ \citenamefont
  {Rischke}(2000{\natexlab{a}})}]{Pisarski:1999tv}%
  \BibitemOpen
  \bibfield  {author} {\bibinfo {author} {\bibnamefont {Pisarski},
  \bibfnamefont {R.~D.}}, \ and\ \bibinfo {author} {\bibfnamefont {D.~H.}\
  \bibnamefont {Rischke}}} (\bibinfo {year} {2000}{\natexlab{a}}),\ \href
  {\doibase 10.1103/PhysRevD.61.074017} {\bibfield  {journal} {\bibinfo
  {journal} {Phys.Rev.}\ }\textbf {\bibinfo {volume} {D61}},\ \bibinfo {pages}
  {074017}},\ \Eprint {http://arxiv.org/abs/nucl-th/9910056}
  {arXiv:nucl-th/9910056 [nucl-th]} \BibitemShut {NoStop}%
\bibitem [{\citenamefont {Pisarski}\ and\ \citenamefont
  {Rischke}(2000{\natexlab{b}})}]{Pisarski:1999bf}%
  \BibitemOpen
  \bibfield  {author} {\bibinfo {author} {\bibnamefont {Pisarski},
  \bibfnamefont {R.~D.}}, \ and\ \bibinfo {author} {\bibfnamefont {D.~H.}\
  \bibnamefont {Rischke}}} (\bibinfo {year} {2000}{\natexlab{b}}),\ \href
  {\doibase 10.1103/PhysRevD.61.051501} {\bibfield  {journal} {\bibinfo
  {journal} {Phys.Rev.}\ }\textbf {\bibinfo {volume} {D61}},\ \bibinfo {pages}
  {051501}},\ \Eprint {http://arxiv.org/abs/nucl-th/9907041}
  {arXiv:nucl-th/9907041 [nucl-th]} \BibitemShut {NoStop}%
\bibitem [{\citenamefont {Polchinski}(1992)}]{Polchinski:1992ed}%
  \BibitemOpen
  \bibfield  {author} {\bibinfo {author} {\bibnamefont {Polchinski},
  \bibfnamefont {J.}}} (\bibinfo {year} {1992}),\ \href@noop {} {\ }\Eprint
  {http://arxiv.org/abs/hep-th/9210046} {arXiv:hep-th/9210046 [hep-th]}
  \BibitemShut {NoStop}%
\bibitem [{\citenamefont {Prakash}\ \emph {et~al.}(2001)\citenamefont
  {Prakash}, \citenamefont {Lattimer}, \citenamefont {Pons}, \citenamefont
  {Steiner},\ and\ \citenamefont {Reddy}}]{Prakash:2000jr}%
  \BibitemOpen
  \bibfield  {author} {\bibinfo {author} {\bibnamefont {Prakash}, \bibfnamefont
  {M.}}, \bibinfo {author} {\bibfnamefont {J.~M.}\ \bibnamefont {Lattimer}},
  \bibinfo {author} {\bibfnamefont {J.~A.}\ \bibnamefont {Pons}}, \bibinfo
  {author} {\bibfnamefont {A.~W.}\ \bibnamefont {Steiner}}, \ and\ \bibinfo
  {author} {\bibfnamefont {S.}~\bibnamefont {Reddy}}} (\bibinfo {year}
  {2001}),\ \href@noop {} {\bibfield  {journal} {\bibinfo  {journal}
  {Lect.Notes Phys.}\ }\textbf {\bibinfo {volume} {578}},\ \bibinfo {pages}
  {364}},\ \Eprint {http://arxiv.org/abs/astro-ph/0012136}
  {arXiv:astro-ph/0012136 [astro-ph]} \BibitemShut {NoStop}%
\bibitem [{\citenamefont {{Radhakrishnan}}\ and\ \citenamefont
  {{Manchester}}(1969)}]{Radhakrishnan1969}%
  \BibitemOpen
  \bibfield  {author} {\bibinfo {author} {\bibnamefont {{Radhakrishnan}},
  \bibfnamefont {V.}}, \ and\ \bibinfo {author} {\bibfnamefont {R.~N.}\
  \bibnamefont {{Manchester}}}} (\bibinfo {year} {1969}),\ \href {\doibase
  10.1038/222228a0} {\bibfield  {journal} {\bibinfo  {journal} {\nat}\ }\textbf
  {\bibinfo {volume} {222}},\ \bibinfo {pages} {228}}\BibitemShut {NoStop}%
\bibitem [{\citenamefont {Radzihovsky}\ and\ \citenamefont
  {Sheehy}(2010)}]{Radzihovsky}%
  \BibitemOpen
  \bibfield  {author} {\bibinfo {author} {\bibnamefont {Radzihovsky},
  \bibfnamefont {L.}}, \ and\ \bibinfo {author} {\bibfnamefont {D.~E.}\
  \bibnamefont {Sheehy}}} (\bibinfo {year} {2010}),\ \href {\doibase
  10.1088/0034-4885/73/7/076501} {\bibfield  {journal} {\bibinfo  {journal}
  {Reports on Progress in Physics}\ }\textbf {\bibinfo {volume} {73}}~(\bibinfo
  {number} {7}),\ \bibinfo {eid} {076501}},\ \Eprint
  {http://arxiv.org/abs/0911.1740} {arXiv:0911.1740 [cond-mat.quant-gas]}
  \BibitemShut {NoStop}%
\bibitem [{\citenamefont {Rajagopal}\ and\ \citenamefont
  {Sharma}(2006)}]{Rajagopal:2006ig}%
  \BibitemOpen
  \bibfield  {author} {\bibinfo {author} {\bibnamefont {Rajagopal},
  \bibfnamefont {K.}}, \ and\ \bibinfo {author} {\bibfnamefont
  {R.}~\bibnamefont {Sharma}}} (\bibinfo {year} {2006}),\ \href {\doibase
  10.1103/PhysRevD.74.094019} {\bibfield  {journal} {\bibinfo  {journal}
  {Phys.Rev.}\ }\textbf {\bibinfo {volume} {D74}},\ \bibinfo {pages}
  {094019}},\ \Eprint {http://arxiv.org/abs/hep-ph/0605316}
  {arXiv:hep-ph/0605316 [hep-ph]} \BibitemShut {NoStop}%
\bibitem [{\citenamefont {Rajagopal}\ and\ \citenamefont
  {Wilczek}(2000)}]{Rajagopal:2000wf}%
  \BibitemOpen
  \bibfield  {author} {\bibinfo {author} {\bibnamefont {Rajagopal},
  \bibfnamefont {K.}}, \ and\ \bibinfo {author} {\bibfnamefont
  {F.}~\bibnamefont {Wilczek}}} (\bibinfo {year} {2000}),\ \href@noop {} {\
  }\Eprint {http://arxiv.org/abs/hep-ph/0011333} {arXiv:hep-ph/0011333
  [hep-ph]} \BibitemShut {NoStop}%
\bibitem [{\citenamefont {Rapp}\ \emph {et~al.}(1998)\citenamefont {Rapp},
  \citenamefont {Schafer}, \citenamefont {Shuryak},\ and\ \citenamefont
  {Velkovsky}}]{Rapp:1997zu}%
  \BibitemOpen
  \bibfield  {author} {\bibinfo {author} {\bibnamefont {Rapp}, \bibfnamefont
  {R.}}, \bibinfo {author} {\bibfnamefont {T.}~\bibnamefont {Schafer}},
  \bibinfo {author} {\bibfnamefont {E.~V.}\ \bibnamefont {Shuryak}}, \ and\
  \bibinfo {author} {\bibfnamefont {M.}~\bibnamefont {Velkovsky}}} (\bibinfo
  {year} {1998}),\ \href {\doibase 10.1103/PhysRevLett.81.53} {\bibfield
  {journal} {\bibinfo  {journal} {Phys.Rev.Lett.}\ }\textbf {\bibinfo {volume}
  {81}},\ \bibinfo {pages} {53}},\ \Eprint
  {http://arxiv.org/abs/hep-ph/9711396} {arXiv:hep-ph/9711396 [hep-ph]}
  \BibitemShut {NoStop}%
\bibitem [{\citenamefont {Rapp}\ \emph {et~al.}(2000)\citenamefont {Rapp},
  \citenamefont {Schafer}, \citenamefont {Shuryak},\ and\ \citenamefont
  {Velkovsky}}]{Rapp:1999qa}%
  \BibitemOpen
  \bibfield  {author} {\bibinfo {author} {\bibnamefont {Rapp}, \bibfnamefont
  {R.}}, \bibinfo {author} {\bibfnamefont {T.}~\bibnamefont {Schafer}},
  \bibinfo {author} {\bibfnamefont {E.~V.}\ \bibnamefont {Shuryak}}, \ and\
  \bibinfo {author} {\bibfnamefont {M.}~\bibnamefont {Velkovsky}}} (\bibinfo
  {year} {2000}),\ \href {\doibase 10.1006/aphy.1999.5991} {\bibfield
  {journal} {\bibinfo  {journal} {Annals Phys.}\ }\textbf {\bibinfo {volume}
  {280}},\ \bibinfo {pages} {35}},\ \Eprint
  {http://arxiv.org/abs/hep-ph/9904353} {arXiv:hep-ph/9904353 [hep-ph]}
  \BibitemShut {NoStop}%
\bibitem [{\citenamefont {Rapp}\ \emph {et~al.}(2001)\citenamefont {Rapp},
  \citenamefont {Shuryak},\ and\ \citenamefont {Zahed}}]{Rapp:2000zd}%
  \BibitemOpen
  \bibfield  {author} {\bibinfo {author} {\bibnamefont {Rapp}, \bibfnamefont
  {R.}}, \bibinfo {author} {\bibfnamefont {E.~V.}\ \bibnamefont {Shuryak}}, \
  and\ \bibinfo {author} {\bibfnamefont {I.}~\bibnamefont {Zahed}}} (\bibinfo
  {year} {2001}),\ \href {\doibase 10.1103/PhysRevD.63.034008} {\bibfield
  {journal} {\bibinfo  {journal} {Phys.Rev.}\ }\textbf {\bibinfo {volume}
  {D63}},\ \bibinfo {pages} {034008}},\ \Eprint
  {http://arxiv.org/abs/hep-ph/0008207} {arXiv:hep-ph/0008207 [hep-ph]}
  \BibitemShut {NoStop}%
\bibitem [{\citenamefont {Reddy}\ and\ \citenamefont
  {Rupak}(2005)}]{Reddy:2004my}%
  \BibitemOpen
  \bibfield  {author} {\bibinfo {author} {\bibnamefont {Reddy}, \bibfnamefont
  {S.}}, \ and\ \bibinfo {author} {\bibfnamefont {G.}~\bibnamefont {Rupak}}}
  (\bibinfo {year} {2005}),\ \href {\doibase 10.1103/PhysRevC.71.025201}
  {\bibfield  {journal} {\bibinfo  {journal} {Phys.Rev.}\ }\textbf {\bibinfo
  {volume} {C71}},\ \bibinfo {pages} {025201}},\ \Eprint
  {http://arxiv.org/abs/nucl-th/0405054} {arXiv:nucl-th/0405054 [nucl-th]}
  \BibitemShut {NoStop}%
\bibitem [{\citenamefont {{Reichley}}\ and\ \citenamefont
  {{Downs}}(1969)}]{Reichley1969}%
  \BibitemOpen
  \bibfield  {author} {\bibinfo {author} {\bibnamefont {{Reichley}},
  \bibfnamefont {P.~E.}}, \ and\ \bibinfo {author} {\bibfnamefont {G.~S.}\
  \bibnamefont {{Downs}}}} (\bibinfo {year} {1969}),\ \href {\doibase
  10.1038/222229a0} {\bibfield  {journal} {\bibinfo  {journal} {\nat}\ }\textbf
  {\bibinfo {volume} {222}},\ \bibinfo {pages} {229}}\BibitemShut {NoStop}%
\bibitem [{\citenamefont {{Richards}}\ \emph {et~al.}(1969)\citenamefont
  {{Richards}}, \citenamefont {{Pettengill}}, \citenamefont {{Roberts}},
  \citenamefont {{Counselman}},\ and\ \citenamefont {{Rankin}}}]{Richards1969}%
  \BibitemOpen
  \bibfield  {author} {\bibinfo {author} {\bibnamefont {{Richards}},
  \bibfnamefont {D.~W.}}, \bibinfo {author} {\bibfnamefont {G.~H.}\
  \bibnamefont {{Pettengill}}}, \bibinfo {author} {\bibfnamefont {J.~A.}\
  \bibnamefont {{Roberts}}}, \bibinfo {author} {\bibfnamefont {C.~C.}\
  \bibnamefont {{Counselman}}}, \ and\ \bibinfo {author} {\bibfnamefont
  {J.}~\bibnamefont {{Rankin}}}} (\bibinfo {year} {1969}),\ \href@noop {}
  {\bibfield  {journal} {\bibinfo  {journal} {\iaucirc}\ }\textbf {\bibinfo
  {volume} {2181}},\ \bibinfo {pages} {1}}\BibitemShut {NoStop}%
\bibitem [{\citenamefont {Rischke}(2000{\natexlab{a}})}]{Rischke:2000ra}%
  \BibitemOpen
  \bibfield  {author} {\bibinfo {author} {\bibnamefont {Rischke}, \bibfnamefont
  {D.~H.}}} (\bibinfo {year} {2000}{\natexlab{a}}),\ \href {\doibase
  10.1103/PhysRevD.62.054017} {\bibfield  {journal} {\bibinfo  {journal}
  {Phys.Rev.}\ }\textbf {\bibinfo {volume} {D62}},\ \bibinfo {pages}
  {054017}},\ \Eprint {http://arxiv.org/abs/nucl-th/0003063}
  {arXiv:nucl-th/0003063 [nucl-th]} \BibitemShut {NoStop}%
\bibitem [{\citenamefont {Rischke}(2000{\natexlab{b}})}]{Rischke:2000qz}%
  \BibitemOpen
  \bibfield  {author} {\bibinfo {author} {\bibnamefont {Rischke}, \bibfnamefont
  {D.~H.}}} (\bibinfo {year} {2000}{\natexlab{b}}),\ \href {\doibase
  10.1103/PhysRevD.62.034007} {\bibfield  {journal} {\bibinfo  {journal}
  {Phys.Rev.}\ }\textbf {\bibinfo {volume} {D62}},\ \bibinfo {pages}
  {034007}},\ \Eprint {http://arxiv.org/abs/nucl-th/0001040}
  {arXiv:nucl-th/0001040 [nucl-th]} \BibitemShut {NoStop}%
\bibitem [{\citenamefont {Rischke}(2004)}]{Rischke:2003mt}%
  \BibitemOpen
  \bibfield  {author} {\bibinfo {author} {\bibnamefont {Rischke}, \bibfnamefont
  {D.~H.}}} (\bibinfo {year} {2004}),\ \href {\doibase
  10.1016/j.ppnp.2003.09.002} {\bibfield  {journal} {\bibinfo  {journal}
  {Prog.Part.Nucl.Phys.}\ }\textbf {\bibinfo {volume} {52}},\ \bibinfo {pages}
  {197}},\ \Eprint {http://arxiv.org/abs/nucl-th/0305030}
  {arXiv:nucl-th/0305030 [nucl-th]} \BibitemShut {NoStop}%
\bibitem [{\citenamefont {Rischke}\ and\ \citenamefont
  {Shovkovy}(2002)}]{Rischke:2002rz}%
  \BibitemOpen
  \bibfield  {author} {\bibinfo {author} {\bibnamefont {Rischke}, \bibfnamefont
  {D.~H.}}, \ and\ \bibinfo {author} {\bibfnamefont {I.~A.}\ \bibnamefont
  {Shovkovy}}} (\bibinfo {year} {2002}),\ \href {\doibase
  10.1103/PhysRevD.66.054019} {\bibfield  {journal} {\bibinfo  {journal}
  {Phys.Rev.}\ }\textbf {\bibinfo {volume} {D66}},\ \bibinfo {pages}
  {054019}},\ \Eprint {http://arxiv.org/abs/nucl-th/0205080}
  {arXiv:nucl-th/0205080 [nucl-th]} \BibitemShut {NoStop}%
\bibitem [{\citenamefont {Rischke}\ \emph {et~al.}(2001)\citenamefont
  {Rischke}, \citenamefont {Son},\ and\ \citenamefont
  {Stephanov}}]{Rischke:2000cn}%
  \BibitemOpen
  \bibfield  {author} {\bibinfo {author} {\bibnamefont {Rischke}, \bibfnamefont
  {D.~H.}}, \bibinfo {author} {\bibfnamefont {D.~T.}\ \bibnamefont {Son}}, \
  and\ \bibinfo {author} {\bibfnamefont {M.~A.}\ \bibnamefont {Stephanov}}}
  (\bibinfo {year} {2001}),\ \href {\doibase 10.1103/PhysRevLett.87.062001}
  {\bibfield  {journal} {\bibinfo  {journal} {Phys.Rev.Lett.}\ }\textbf
  {\bibinfo {volume} {87}},\ \bibinfo {pages} {062001}},\ \Eprint
  {http://arxiv.org/abs/hep-ph/0011379} {arXiv:hep-ph/0011379 [hep-ph]}
  \BibitemShut {NoStop}%
\bibitem [{\citenamefont {{Rizzi}}\ \emph {et~al.}(2008)\citenamefont
  {{Rizzi}}, \citenamefont {{Polini}}, \citenamefont {{Cazalilla}},
  \citenamefont {{Bakhtiari}}, \citenamefont {{Tosi}},\ and\ \citenamefont
  {{Fazio}}}]{Rizzi}%
  \BibitemOpen
  \bibfield  {author} {\bibinfo {author} {\bibnamefont {{Rizzi}}, \bibfnamefont
  {M.}}, \bibinfo {author} {\bibfnamefont {M.}~\bibnamefont {{Polini}}},
  \bibinfo {author} {\bibfnamefont {M.~A.}\ \bibnamefont {{Cazalilla}}},
  \bibinfo {author} {\bibfnamefont {M.~R.}\ \bibnamefont {{Bakhtiari}}},
  \bibinfo {author} {\bibfnamefont {M.~P.}\ \bibnamefont {{Tosi}}}, \ and\
  \bibinfo {author} {\bibfnamefont {R.}~\bibnamefont {{Fazio}}}} (\bibinfo
  {year} {2008}),\ \href {\doibase 10.1103/PhysRevB.77.245105} {\bibfield
  {journal} {\bibinfo  {journal} {Phys. Rev. B}\ }\textbf {\bibinfo {volume}
  {77}}~(\bibinfo {number} {24}),\ \bibinfo {eid} {245105}},\ \Eprint
  {http://arxiv.org/abs/0712.3364} {arXiv:0712.3364 [cond-mat.str-el]}
  \BibitemShut {NoStop}%
\bibitem [{\citenamefont {Roberge}\ and\ \citenamefont
  {Weiss}(1986)}]{Roberge:1986mm}%
  \BibitemOpen
  \bibfield  {author} {\bibinfo {author} {\bibnamefont {Roberge}, \bibfnamefont
  {A.}}, \ and\ \bibinfo {author} {\bibfnamefont {N.}~\bibnamefont {Weiss}}}
  (\bibinfo {year} {1986}),\ \href {\doibase 10.1016/0550-3213(86)90582-1}
  {\bibfield  {journal} {\bibinfo  {journal} {Nucl.Phys.}\ }\textbf {\bibinfo
  {volume} {B275}},\ \bibinfo {pages} {734}}\BibitemShut {NoStop}%
\bibitem [{\citenamefont {{Ruderman}}(1969)}]{Ruderman1969}%
  \BibitemOpen
  \bibfield  {author} {\bibinfo {author} {\bibnamefont {{Ruderman}},
  \bibfnamefont {M.}}} (\bibinfo {year} {1969}),\ \href {\doibase
  10.1038/223597b0} {\bibfield  {journal} {\bibinfo  {journal} {\nat}\ }\textbf
  {\bibinfo {volume} {223}},\ \bibinfo {pages} {597}}\BibitemShut {NoStop}%
\bibitem [{\citenamefont {Ruderman}(1972)}]{Ruderman1972}%
  \BibitemOpen
  \bibfield  {author} {\bibinfo {author} {\bibnamefont {Ruderman},
  \bibfnamefont {M.}}} (\bibinfo {year} {1972}),\ \href {\doibase
  10.1146/annurev.aa.10.090172.002235} {\bibfield  {journal} {\bibinfo
  {journal} {Annu. Rev. Astron. Astrophys.}\ }\textbf {\bibinfo {volume}
  {10}},\ \bibinfo {pages} {427}}\BibitemShut {NoStop}%
\bibitem [{\citenamefont {Ruester}\ \emph {et~al.}(2005)\citenamefont
  {Ruester}, \citenamefont {Werth}, \citenamefont {Buballa}, \citenamefont
  {Shovkovy},\ and\ \citenamefont {Rischke}}]{Ruester:2005jc}%
  \BibitemOpen
  \bibfield  {author} {\bibinfo {author} {\bibnamefont {Ruester}, \bibfnamefont
  {S.~B.}}, \bibinfo {author} {\bibfnamefont {V.}~\bibnamefont {Werth}},
  \bibinfo {author} {\bibfnamefont {M.}~\bibnamefont {Buballa}}, \bibinfo
  {author} {\bibfnamefont {I.~A.}\ \bibnamefont {Shovkovy}}, \ and\ \bibinfo
  {author} {\bibfnamefont {D.~H.}\ \bibnamefont {Rischke}}} (\bibinfo {year}
  {2005}),\ \href {\doibase 10.1103/PhysRevD.72.034004} {\bibfield  {journal}
  {\bibinfo  {journal} {Phys.Rev.}\ }\textbf {\bibinfo {volume} {D72}},\
  \bibinfo {pages} {034004}},\ \Eprint {http://arxiv.org/abs/hep-ph/0503184}
  {arXiv:hep-ph/0503184 [hep-ph]} \BibitemShut {NoStop}%
\bibitem [{\citenamefont {Ruester}\ \emph
  {et~al.}(2006{\natexlab{a}})\citenamefont {Ruester}, \citenamefont {Werth},
  \citenamefont {Buballa}, \citenamefont {Shovkovy},\ and\ \citenamefont
  {Rischke}}]{Ruester:2006aj}%
  \BibitemOpen
  \bibfield  {author} {\bibinfo {author} {\bibnamefont {Ruester}, \bibfnamefont
  {S.~B.}}, \bibinfo {author} {\bibfnamefont {V.}~\bibnamefont {Werth}},
  \bibinfo {author} {\bibfnamefont {M.}~\bibnamefont {Buballa}}, \bibinfo
  {author} {\bibfnamefont {I.~A.}\ \bibnamefont {Shovkovy}}, \ and\ \bibinfo
  {author} {\bibfnamefont {D.~H.}\ \bibnamefont {Rischke}}} (\bibinfo {year}
  {2006}{\natexlab{a}}),\ \href@noop {} {\ }\Eprint
  {http://arxiv.org/abs/nucl-th/0602018} {arXiv:nucl-th/0602018 [nucl-th]}
  \BibitemShut {NoStop}%
\bibitem [{\citenamefont {Ruester}\ \emph
  {et~al.}(2006{\natexlab{b}})\citenamefont {Ruester}, \citenamefont {Werth},
  \citenamefont {Buballa}, \citenamefont {Shovkovy},\ and\ \citenamefont
  {Rischke}}]{Ruester:2005ib}%
  \BibitemOpen
  \bibfield  {author} {\bibinfo {author} {\bibnamefont {Ruester}, \bibfnamefont
  {S.~B.}}, \bibinfo {author} {\bibfnamefont {V.}~\bibnamefont {Werth}},
  \bibinfo {author} {\bibfnamefont {M.}~\bibnamefont {Buballa}}, \bibinfo
  {author} {\bibfnamefont {I.~A.}\ \bibnamefont {Shovkovy}}, \ and\ \bibinfo
  {author} {\bibfnamefont {D.~H.}\ \bibnamefont {Rischke}}} (\bibinfo {year}
  {2006}{\natexlab{b}}),\ \href {\doibase 10.1103/PhysRevD.73.034025}
  {\bibfield  {journal} {\bibinfo  {journal} {Phys.Rev.}\ }\textbf {\bibinfo
  {volume} {D73}},\ \bibinfo {pages} {034025}},\ \Eprint
  {http://arxiv.org/abs/hep-ph/0509073} {arXiv:hep-ph/0509073 [hep-ph]}
  \BibitemShut {NoStop}%
\bibitem [{\citenamefont {Rupak}\ and\ \citenamefont
  {Jaikumar}(2012)}]{Rupak:2012wk}%
  \BibitemOpen
  \bibfield  {author} {\bibinfo {author} {\bibnamefont {Rupak}, \bibfnamefont
  {G.}}, \ and\ \bibinfo {author} {\bibfnamefont {P.}~\bibnamefont {Jaikumar}}}
  (\bibinfo {year} {2012}),\ \href@noop {} {\ }\Eprint
  {http://arxiv.org/abs/1209.4343} {arXiv:1209.4343 [nucl-th]} \BibitemShut
  {NoStop}%
\bibitem [{\citenamefont {Saint-James}\ \emph {et~al.}(1969)\citenamefont
  {Saint-James}, \citenamefont {Sarma},\ and\ \citenamefont
  {Thomas}}]{Sarma-book}%
  \BibitemOpen
  \bibfield  {author} {\bibinfo {author} {\bibnamefont {Saint-James},
  \bibfnamefont {D.}}, \bibinfo {author} {\bibfnamefont {G.}~\bibnamefont
  {Sarma}}, \ and\ \bibinfo {author} {\bibfnamefont {E.~J.}\ \bibnamefont
  {Thomas}}} (\bibinfo {year} {1969}),\ \href@noop {} {\emph {\bibinfo {title}
  {Type II superconductivity}}}\ (\bibinfo  {publisher} {Pergamon Press Oxford,
  New York})\BibitemShut {NoStop}%
\bibitem [{\citenamefont {Sandin}\ and\ \citenamefont
  {Blaschke}(2007)}]{Sandin:2007zr}%
  \BibitemOpen
  \bibfield  {author} {\bibinfo {author} {\bibnamefont {Sandin}, \bibfnamefont
  {F.}}, \ and\ \bibinfo {author} {\bibfnamefont {D.}~\bibnamefont {Blaschke}}}
  (\bibinfo {year} {2007}),\ \href {\doibase 10.1103/PhysRevD.75.125013}
  {\bibfield  {journal} {\bibinfo  {journal} {Phys.Rev.}\ }\textbf {\bibinfo
  {volume} {D75}},\ \bibinfo {pages} {125013}},\ \Eprint
  {http://arxiv.org/abs/astro-ph/0701772} {arXiv:astro-ph/0701772 [astro-ph]}
  \BibitemShut {NoStop}%
\bibitem [{\citenamefont {Sarma}(1963)}]{Sarma19631029}%
  \BibitemOpen
  \bibfield  {author} {\bibinfo {author} {\bibnamefont {Sarma}, \bibfnamefont
  {G.}}} (\bibinfo {year} {1963}),\ \href {\doibase
  10.1016/0022-3697(63)90007-6} {\bibfield  {journal} {\bibinfo  {journal}
  {Journal of Physics and Chemistry of Solids}\ }\textbf {\bibinfo {volume}
  {24}}~(\bibinfo {number} {8}),\ \bibinfo {pages} {1029 }}\BibitemShut
  {NoStop}%
\bibitem [{\citenamefont {Schafer}(2000{\natexlab{a}})}]{Schafer:1999fe}%
  \BibitemOpen
  \bibfield  {author} {\bibinfo {author} {\bibnamefont {Schafer}, \bibfnamefont
  {T.}}} (\bibinfo {year} {2000}{\natexlab{a}}),\ \href {\doibase
  10.1016/S0550-3213(00)00063-8} {\bibfield  {journal} {\bibinfo  {journal}
  {Nucl.Phys.}\ }\textbf {\bibinfo {volume} {B575}},\ \bibinfo {pages} {269}},\
  \Eprint {http://arxiv.org/abs/hep-ph/9909574} {arXiv:hep-ph/9909574 [hep-ph]}
  \BibitemShut {NoStop}%
\bibitem [{\citenamefont {Schafer}(2000{\natexlab{b}})}]{Schafer:2000tw}%
  \BibitemOpen
  \bibfield  {author} {\bibinfo {author} {\bibnamefont {Schafer}, \bibfnamefont
  {T.}}} (\bibinfo {year} {2000}{\natexlab{b}}),\ \href {\doibase
  10.1103/PhysRevD.62.094007} {\bibfield  {journal} {\bibinfo  {journal}
  {Phys.Rev.}\ }\textbf {\bibinfo {volume} {D62}},\ \bibinfo {pages}
  {094007}},\ \Eprint {http://arxiv.org/abs/hep-ph/0006034}
  {arXiv:hep-ph/0006034 [hep-ph]} \BibitemShut {NoStop}%
\bibitem [{\citenamefont {Schafer}(2003{\natexlab{a}})}]{Schafer:2003jn}%
  \BibitemOpen
  \bibfield  {author} {\bibinfo {author} {\bibnamefont {Schafer}, \bibfnamefont
  {T.}}} (\bibinfo {year} {2003}{\natexlab{a}}),\ \href {\doibase
  10.1016/j.nuclphysa.2003.08.028} {\bibfield  {journal} {\bibinfo  {journal}
  {Nucl.Phys.}\ }\textbf {\bibinfo {volume} {A728}},\ \bibinfo {pages} {251}},\
  \Eprint {http://arxiv.org/abs/hep-ph/0307074} {arXiv:hep-ph/0307074 [hep-ph]}
  \BibitemShut {NoStop}%
\bibitem [{\citenamefont {Schafer}(2003{\natexlab{b}})}]{Schafer:2003vz}%
  \BibitemOpen
  \bibfield  {author} {\bibinfo {author} {\bibnamefont {Schafer}, \bibfnamefont
  {T.}}} (\bibinfo {year} {2003}{\natexlab{b}}),\ \href@noop {} {\ ,\ \bibinfo
  {pages} {185}}\Eprint {http://arxiv.org/abs/hep-ph/0304281}
  {arXiv:hep-ph/0304281 [hep-ph]} \BibitemShut {NoStop}%
\bibitem [{\citenamefont {Schafer}\ and\ \citenamefont
  {Shuryak}(1998)}]{Schafer:1996wv}%
  \BibitemOpen
  \bibfield  {author} {\bibinfo {author} {\bibnamefont {Schafer}, \bibfnamefont
  {T.}}, \ and\ \bibinfo {author} {\bibfnamefont {E.~V.}\ \bibnamefont
  {Shuryak}}} (\bibinfo {year} {1998}),\ \href {\doibase
  10.1103/RevModPhys.70.323} {\bibfield  {journal} {\bibinfo  {journal}
  {Rev.Mod.Phys.}\ }\textbf {\bibinfo {volume} {70}},\ \bibinfo {pages}
  {323}},\ \Eprint {http://arxiv.org/abs/hep-ph/9610451} {arXiv:hep-ph/9610451
  [hep-ph]} \BibitemShut {NoStop}%
\bibitem [{\citenamefont {Schafer}\ and\ \citenamefont
  {Wilczek}(1999{\natexlab{a}})}]{Schafer:1998ef}%
  \BibitemOpen
  \bibfield  {author} {\bibinfo {author} {\bibnamefont {Schafer}, \bibfnamefont
  {T.}}, \ and\ \bibinfo {author} {\bibfnamefont {F.}~\bibnamefont {Wilczek}}}
  (\bibinfo {year} {1999}{\natexlab{a}}),\ \href {\doibase
  10.1103/PhysRevLett.82.3956} {\bibfield  {journal} {\bibinfo  {journal}
  {Phys.Rev.Lett.}\ }\textbf {\bibinfo {volume} {82}},\ \bibinfo {pages}
  {3956}},\ \Eprint {http://arxiv.org/abs/hep-ph/9811473} {arXiv:hep-ph/9811473
  [hep-ph]} \BibitemShut {NoStop}%
\bibitem [{\citenamefont {Schafer}\ and\ \citenamefont
  {Wilczek}(1999{\natexlab{b}})}]{Schafer:1998na}%
  \BibitemOpen
  \bibfield  {author} {\bibinfo {author} {\bibnamefont {Schafer}, \bibfnamefont
  {T.}}, \ and\ \bibinfo {author} {\bibfnamefont {F.}~\bibnamefont {Wilczek}}}
  (\bibinfo {year} {1999}{\natexlab{b}}),\ \href {\doibase
  10.1016/S0370-2693(99)00162-8} {\bibfield  {journal} {\bibinfo  {journal}
  {Phys.Lett.}\ }\textbf {\bibinfo {volume} {B450}},\ \bibinfo {pages} {325}},\
  \Eprint {http://arxiv.org/abs/hep-ph/9810509} {arXiv:hep-ph/9810509 [hep-ph]}
  \BibitemShut {NoStop}%
\bibitem [{\citenamefont {Schafer}\ and\ \citenamefont
  {Wilczek}(1999{\natexlab{c}})}]{Schafer:1999jg}%
  \BibitemOpen
  \bibfield  {author} {\bibinfo {author} {\bibnamefont {Schafer}, \bibfnamefont
  {T.}}, \ and\ \bibinfo {author} {\bibfnamefont {F.}~\bibnamefont {Wilczek}}}
  (\bibinfo {year} {1999}{\natexlab{c}}),\ \href {\doibase
  10.1103/PhysRevD.60.114033} {\bibfield  {journal} {\bibinfo  {journal}
  {Phys.Rev.}\ }\textbf {\bibinfo {volume} {D60}},\ \bibinfo {pages}
  {114033}},\ \Eprint {http://arxiv.org/abs/hep-ph/9906512}
  {arXiv:hep-ph/9906512 [hep-ph]} \BibitemShut {NoStop}%
\bibitem [{\citenamefont {Schertler}\ \emph {et~al.}(2000)\citenamefont
  {Schertler}, \citenamefont {Greiner}, \citenamefont {Schaffner-Bielich},\
  and\ \citenamefont {Thoma}}]{Schertler:2000xq}%
  \BibitemOpen
  \bibfield  {author} {\bibinfo {author} {\bibnamefont {Schertler},
  \bibfnamefont {K.}}, \bibinfo {author} {\bibfnamefont {C.}~\bibnamefont
  {Greiner}}, \bibinfo {author} {\bibfnamefont {J.}~\bibnamefont
  {Schaffner-Bielich}}, \ and\ \bibinfo {author} {\bibfnamefont
  {M.}~\bibnamefont {Thoma}}} (\bibinfo {year} {2000}),\ \href {\doibase
  10.1016/S0375-9474(00)00305-5} {\bibfield  {journal} {\bibinfo  {journal}
  {Nucl.Phys.}\ }\textbf {\bibinfo {volume} {A677}},\ \bibinfo {pages} {463}},\
  \Eprint {http://arxiv.org/abs/astro-ph/0001467} {arXiv:astro-ph/0001467
  [astro-ph]} \BibitemShut {NoStop}%
\bibitem [{\citenamefont {Schmitt}(2005)}]{Schmitt:2004et}%
  \BibitemOpen
  \bibfield  {author} {\bibinfo {author} {\bibnamefont {Schmitt}, \bibfnamefont
  {A.}}} (\bibinfo {year} {2005}),\ \href {\doibase 10.1103/PhysRevD.71.054016}
  {\bibfield  {journal} {\bibinfo  {journal} {Phys.Rev.}\ }\textbf {\bibinfo
  {volume} {D71}},\ \bibinfo {pages} {054016}},\ \Eprint
  {http://arxiv.org/abs/nucl-th/0412033} {arXiv:nucl-th/0412033 [nucl-th]}
  \BibitemShut {NoStop}%
\bibitem [{\citenamefont {Schmitt}\ \emph {et~al.}(2002)\citenamefont
  {Schmitt}, \citenamefont {Wang},\ and\ \citenamefont
  {Rischke}}]{Schmitt:2002sc}%
  \BibitemOpen
  \bibfield  {author} {\bibinfo {author} {\bibnamefont {Schmitt}, \bibfnamefont
  {A.}}, \bibinfo {author} {\bibfnamefont {Q.}~\bibnamefont {Wang}}, \ and\
  \bibinfo {author} {\bibfnamefont {D.~H.}\ \bibnamefont {Rischke}}} (\bibinfo
  {year} {2002}),\ \href {\doibase 10.1103/PhysRevD.66.114010} {\bibfield
  {journal} {\bibinfo  {journal} {Phys.Rev.}\ }\textbf {\bibinfo {volume}
  {D66}},\ \bibinfo {pages} {114010}},\ \Eprint
  {http://arxiv.org/abs/nucl-th/0209050} {arXiv:nucl-th/0209050 [nucl-th]}
  \BibitemShut {NoStop}%
\bibitem [{\citenamefont {Schmitt}\ \emph {et~al.}(2004)\citenamefont
  {Schmitt}, \citenamefont {Wang},\ and\ \citenamefont
  {Rischke}}]{Schmitt:2003aa}%
  \BibitemOpen
  \bibfield  {author} {\bibinfo {author} {\bibnamefont {Schmitt}, \bibfnamefont
  {A.}}, \bibinfo {author} {\bibfnamefont {Q.}~\bibnamefont {Wang}}, \ and\
  \bibinfo {author} {\bibfnamefont {D.~H.}\ \bibnamefont {Rischke}}} (\bibinfo
  {year} {2004}),\ \href {\doibase 10.1103/PhysRevD.69.094017} {\bibfield
  {journal} {\bibinfo  {journal} {Phys.Rev.}\ }\textbf {\bibinfo {volume}
  {D69}},\ \bibinfo {pages} {094017}},\ \Eprint
  {http://arxiv.org/abs/nucl-th/0311006} {arXiv:nucl-th/0311006 [nucl-th]}
  \BibitemShut {NoStop}%
\bibitem [{\citenamefont {Sedrakian}(2007)}]{Sedrakian:2006mq}%
  \BibitemOpen
  \bibfield  {author} {\bibinfo {author} {\bibnamefont {Sedrakian},
  \bibfnamefont {A.}}} (\bibinfo {year} {2007}),\ \href {\doibase
  10.1016/j.ppnp.2006.02.002} {\bibfield  {journal} {\bibinfo  {journal}
  {Prog.Part.Nucl.Phys.}\ }\textbf {\bibinfo {volume} {58}},\ \bibinfo {pages}
  {168}},\ \Eprint {http://arxiv.org/abs/nucl-th/0601086}
  {arXiv:nucl-th/0601086 [nucl-th]} \BibitemShut {NoStop}%
\bibitem [{\citenamefont {Sedrakian}(2013)}]{Sedrakian:2013pva}%
  \BibitemOpen
  \bibfield  {author} {\bibinfo {author} {\bibnamefont {Sedrakian},
  \bibfnamefont {A.}}} (\bibinfo {year} {2013}),\ \href@noop {} {\bibfield
  {journal} {\bibinfo  {journal} {Astron. and Astrophys. 555,}\ }\textbf
  {\bibinfo {volume} {L10}}},\ \Eprint {http://arxiv.org/abs/1303.5380}
  {arXiv:1303.5380 [astro-ph.HE]} \BibitemShut {NoStop}%
\bibitem [{\citenamefont {{Shapiro}}\ and\ \citenamefont
  {{Teukolsky}}(1983)}]{Shapiro-Teukolsky}%
  \BibitemOpen
  \bibfield  {author} {\bibinfo {author} {\bibnamefont {{Shapiro}},
  \bibfnamefont {S.~L.}}, \ and\ \bibinfo {author} {\bibfnamefont {S.~A.}\
  \bibnamefont {{Teukolsky}}}} (\bibinfo {year} {1983}),\ \href@noop {} {\emph
  {\bibinfo {title} {Research supported by the National Science Foundation.~New
  York, Wiley-Interscience, 1983, 663 p.}}}\BibitemShut {Stop}%
\bibitem [{\citenamefont {Sharma}\ and\ \citenamefont
  {Reddy}(2008)}]{Sharma:2008rc}%
  \BibitemOpen
  \bibfield  {author} {\bibinfo {author} {\bibnamefont {Sharma}, \bibfnamefont
  {R.}}, \ and\ \bibinfo {author} {\bibfnamefont {S.}~\bibnamefont {Reddy}}}
  (\bibinfo {year} {2008}),\ \href {\doibase 10.1103/PhysRevA.78.063609}
  {\bibfield  {journal} {\bibinfo  {journal} {Phys.Rev.}\ }\textbf {\bibinfo
  {volume} {A78}},\ \bibinfo {pages} {063609}},\ \Eprint
  {http://arxiv.org/abs/0804.2280} {arXiv:0804.2280 [cond-mat.supr-con]}
  \BibitemShut {NoStop}%
\bibitem [{\citenamefont {Sheehy}\ and\ \citenamefont
  {Radzihovsky}(2006)}]{Sheehy:2006qc}%
  \BibitemOpen
  \bibfield  {author} {\bibinfo {author} {\bibnamefont {Sheehy}, \bibfnamefont
  {D.~E.}}, \ and\ \bibinfo {author} {\bibfnamefont {L.}~\bibnamefont
  {Radzihovsky}}} (\bibinfo {year} {2006}),\ \href {\doibase
  10.1103/PhysRevLett.96.060401} {\bibfield  {journal} {\bibinfo  {journal}
  {Phys. Rev. Lett.}\ }\textbf {\bibinfo {volume} {96}},\ \bibinfo {pages}
  {060401}}\BibitemShut {NoStop}%
\bibitem [{\citenamefont {{Shin}}\ \emph {et~al.}(2008)\citenamefont {{Shin}},
  \citenamefont {{Schunck}}, \citenamefont {{Schirotzek}},\ and\ \citenamefont
  {{Ketterle}}}]{ketterle3}%
  \BibitemOpen
  \bibfield  {author} {\bibinfo {author} {\bibnamefont {{Shin}}, \bibfnamefont
  {Y.-I.}}, \bibinfo {author} {\bibfnamefont {C.~H.}\ \bibnamefont
  {{Schunck}}}, \bibinfo {author} {\bibfnamefont {A.}~\bibnamefont
  {{Schirotzek}}}, \ and\ \bibinfo {author} {\bibfnamefont {W.}~\bibnamefont
  {{Ketterle}}}} (\bibinfo {year} {2008}),\ \href {\doibase
  10.1038/nature06473} {\bibfield  {journal} {\bibinfo  {journal} {\nat}\
  }\textbf {\bibinfo {volume} {451}},\ \bibinfo {pages} {689}},\ \Eprint
  {http://arxiv.org/abs/0709.3027} {arXiv:0709.3027 [cond-mat.soft]}
  \BibitemShut {NoStop}%
\bibitem [{\citenamefont {Shovkovy}\ \emph {et~al.}(2003)\citenamefont
  {Shovkovy}, \citenamefont {Hanauske},\ and\ \citenamefont
  {Huang}}]{Shovkovy:2003ps}%
  \BibitemOpen
  \bibfield  {author} {\bibinfo {author} {\bibnamefont {Shovkovy},
  \bibfnamefont {I.}}, \bibinfo {author} {\bibfnamefont {M.}~\bibnamefont
  {Hanauske}}, \ and\ \bibinfo {author} {\bibfnamefont {M.}~\bibnamefont
  {Huang}}} (\bibinfo {year} {2003}),\ \href@noop {} {\bibfield  {journal}
  {\bibinfo  {journal} {eConf}\ }\textbf {\bibinfo {volume} {C030614}},\
  \bibinfo {pages} {039}},\ \Eprint {http://arxiv.org/abs/hep-ph/0310286}
  {arXiv:hep-ph/0310286 [hep-ph]} \BibitemShut {NoStop}%
\bibitem [{\citenamefont {Shovkovy}\ and\ \citenamefont
  {Huang}(2003)}]{Shovkovy:2003uu}%
  \BibitemOpen
  \bibfield  {author} {\bibinfo {author} {\bibnamefont {Shovkovy},
  \bibfnamefont {I.}}, \ and\ \bibinfo {author} {\bibfnamefont
  {M.}~\bibnamefont {Huang}}} (\bibinfo {year} {2003}),\ \href {\doibase
  10.1016/S0370-2693(03)00748-2} {\bibfield  {journal} {\bibinfo  {journal}
  {Phys.Lett.}\ }\textbf {\bibinfo {volume} {B564}},\ \bibinfo {pages} {205}},\
  \Eprint {http://arxiv.org/abs/hep-ph/0302142} {arXiv:hep-ph/0302142 [hep-ph]}
  \BibitemShut {NoStop}%
\bibitem [{\citenamefont {Shovkovy}\ and\ \citenamefont
  {Wijewardhana}(1999)}]{Shovkovy:1999mr}%
  \BibitemOpen
  \bibfield  {author} {\bibinfo {author} {\bibnamefont {Shovkovy},
  \bibfnamefont {I.}}, \ and\ \bibinfo {author} {\bibfnamefont
  {L.}~\bibnamefont {Wijewardhana}}} (\bibinfo {year} {1999}),\ \href {\doibase
  10.1016/S0370-2693(99)01297-6} {\bibfield  {journal} {\bibinfo  {journal}
  {Phys.Lett.}\ }\textbf {\bibinfo {volume} {B470}},\ \bibinfo {pages} {189}},\
  \Eprint {http://arxiv.org/abs/hep-ph/9910225} {arXiv:hep-ph/9910225 [hep-ph]}
  \BibitemShut {NoStop}%
\bibitem [{\citenamefont {Shternin}\ \emph {et~al.}(2011)\citenamefont
  {Shternin}, \citenamefont {Yakovlev}, \citenamefont {Heinke}, \citenamefont
  {Ho},\ and\ \citenamefont {Patnaude}}]{Shternin:2010qi}%
  \BibitemOpen
  \bibfield  {author} {\bibinfo {author} {\bibnamefont {Shternin},
  \bibfnamefont {P.~S.}}, \bibinfo {author} {\bibfnamefont {D.~G.}\
  \bibnamefont {Yakovlev}}, \bibinfo {author} {\bibfnamefont {C.~O.}\
  \bibnamefont {Heinke}}, \bibinfo {author} {\bibfnamefont {W.~C.~G.}\
  \bibnamefont {Ho}}, \ and\ \bibinfo {author} {\bibfnamefont {D.~J.}\
  \bibnamefont {Patnaude}}} (\bibinfo {year} {2011}),\ \href@noop {} {\bibfield
   {journal} {\bibinfo  {journal} {Mon.Not.Roy.Astron.Soc.}\ }\textbf {\bibinfo
  {volume} {412}},\ \bibinfo {pages} {L108}},\ \Eprint
  {http://arxiv.org/abs/1012.0045} {arXiv:1012.0045 [astro-ph.SR]} \BibitemShut
  {NoStop}%
\bibitem [{\citenamefont {Son}(2005)}]{Son:2005ak}%
  \BibitemOpen
  \bibfield  {author} {\bibinfo {author} {\bibnamefont {Son}, \bibfnamefont
  {D.}}} (\bibinfo {year} {2005}),\ \href {\doibase
  10.1103/PhysRevLett.94.175301} {\bibfield  {journal} {\bibinfo  {journal}
  {Phys.Rev.Lett.}\ }\textbf {\bibinfo {volume} {94}},\ \bibinfo {pages}
  {175301}},\ \Eprint {http://arxiv.org/abs/cond-mat/0501658}
  {arXiv:cond-mat/0501658 [cond-mat]} \BibitemShut {NoStop}%
\bibitem [{\citenamefont {Son}(1999)}]{Son:1998uk}%
  \BibitemOpen
  \bibfield  {author} {\bibinfo {author} {\bibnamefont {Son}, \bibfnamefont
  {D.~T.}}} (\bibinfo {year} {1999}),\ \href {\doibase
  10.1103/PhysRevD.59.094019} {\bibfield  {journal} {\bibinfo  {journal}
  {Phys.Rev.}\ }\textbf {\bibinfo {volume} {D59}},\ \bibinfo {pages}
  {094019}},\ \Eprint {http://arxiv.org/abs/hep-ph/9812287}
  {arXiv:hep-ph/9812287 [hep-ph]} \BibitemShut {NoStop}%
\bibitem [{\citenamefont {Son}(2002)}]{Son:2002zn}%
  \BibitemOpen
  \bibfield  {author} {\bibinfo {author} {\bibnamefont {Son}, \bibfnamefont
  {D.~T.}}} (\bibinfo {year} {2002}),\ \href@noop {} {\ }\Eprint
  {http://arxiv.org/abs/hep-ph/0204199} {arXiv:hep-ph/0204199 [hep-ph]}
  \BibitemShut {NoStop}%
\bibitem [{\citenamefont {Son}\ and\ \citenamefont
  {Stephanov}(2000{\natexlab{a}})}]{Son:1999cm}%
  \BibitemOpen
  \bibfield  {author} {\bibinfo {author} {\bibnamefont {Son}, \bibfnamefont
  {D.~T.}}, \ and\ \bibinfo {author} {\bibfnamefont {M.~A.}\ \bibnamefont
  {Stephanov}}} (\bibinfo {year} {2000}{\natexlab{a}}),\ \href {\doibase
  10.1103/PhysRevD.61.074012} {\bibfield  {journal} {\bibinfo  {journal}
  {Phys.Rev.}\ }\textbf {\bibinfo {volume} {D61}},\ \bibinfo {pages}
  {074012}},\ \Eprint {http://arxiv.org/abs/hep-ph/9910491}
  {arXiv:hep-ph/9910491 [hep-ph]} \BibitemShut {NoStop}%
\bibitem [{\citenamefont {Son}\ and\ \citenamefont
  {Stephanov}(2000{\natexlab{b}})}]{Son:2000tu}%
  \BibitemOpen
  \bibfield  {author} {\bibinfo {author} {\bibnamefont {Son}, \bibfnamefont
  {D.~T.}}, \ and\ \bibinfo {author} {\bibfnamefont {M.~A.}\ \bibnamefont
  {Stephanov}}} (\bibinfo {year} {2000}{\natexlab{b}}),\ \href {\doibase
  10.1103/PhysRevD.62.059902} {\bibfield  {journal} {\bibinfo  {journal}
  {Phys.Rev.}\ }\textbf {\bibinfo {volume} {D62}},\ \bibinfo {pages}
  {059902}},\ \Eprint {http://arxiv.org/abs/hep-ph/0004095}
  {arXiv:hep-ph/0004095 [hep-ph]} \BibitemShut {NoStop}%
\bibitem [{\citenamefont {{Son}}\ and\ \citenamefont
  {{Stephanov}}(2006)}]{Son:2005qx}%
  \BibitemOpen
  \bibfield  {author} {\bibinfo {author} {\bibnamefont {{Son}}, \bibfnamefont
  {D.~T.}}, \ and\ \bibinfo {author} {\bibfnamefont {M.~A.}\ \bibnamefont
  {{Stephanov}}}} (\bibinfo {year} {2006}),\ \href@noop {} {\bibfield
  {journal} {\bibinfo  {journal} {Phys. Rev. A}\ }\textbf {\bibinfo {volume}
  {74}}},\ \Eprint {http://arxiv.org/abs/arXiv:cond-mat/0507586}
  {arXiv:cond-mat/0507586} \BibitemShut {NoStop}%
\bibitem [{\citenamefont {Steiner}\ \emph {et~al.}(2002)\citenamefont
  {Steiner}, \citenamefont {Reddy},\ and\ \citenamefont
  {Prakash}}]{Steiner:2002gx}%
  \BibitemOpen
  \bibfield  {author} {\bibinfo {author} {\bibnamefont {Steiner}, \bibfnamefont
  {A.~W.}}, \bibinfo {author} {\bibfnamefont {S.}~\bibnamefont {Reddy}}, \ and\
  \bibinfo {author} {\bibfnamefont {M.}~\bibnamefont {Prakash}}} (\bibinfo
  {year} {2002}),\ \href {\doibase 10.1103/PhysRevD.66.094007} {\bibfield
  {journal} {\bibinfo  {journal} {Phys.Rev.}\ }\textbf {\bibinfo {volume}
  {D66}},\ \bibinfo {pages} {094007}},\ \Eprint
  {http://arxiv.org/abs/hep-ph/0205201} {arXiv:hep-ph/0205201 [hep-ph]}
  \BibitemShut {NoStop}%
\bibitem [{\citenamefont {{Strohmayer}}\ \emph {et~al.}(1991)\citenamefont
  {{Strohmayer}}, \citenamefont {{van Horn}}, \citenamefont {{Ogata}},
  \citenamefont {{Iyetomi}},\ and\ \citenamefont {{Ichimaru}}}]{Strohmayer}%
  \BibitemOpen
  \bibfield  {author} {\bibinfo {author} {\bibnamefont {{Strohmayer}},
  \bibfnamefont {T.}}, \bibinfo {author} {\bibfnamefont {H.~M.}\ \bibnamefont
  {{van Horn}}}, \bibinfo {author} {\bibfnamefont {S.}~\bibnamefont {{Ogata}}},
  \bibinfo {author} {\bibfnamefont {H.}~\bibnamefont {{Iyetomi}}}, \ and\
  \bibinfo {author} {\bibfnamefont {S.}~\bibnamefont {{Ichimaru}}}} (\bibinfo
  {year} {1991}),\ \href {\doibase 10.1086/170231} {\bibfield  {journal}
  {\bibinfo  {journal} {\apj}\ }\textbf {\bibinfo {volume} {375}},\ \bibinfo
  {pages} {679}}\BibitemShut {NoStop}%
\bibitem [{\citenamefont {{Tolman}}(1939)}]{Tolman}%
  \BibitemOpen
  \bibfield  {author} {\bibinfo {author} {\bibnamefont {{Tolman}},
  \bibfnamefont {R.~C.}}} (\bibinfo {year} {1939}),\ \href {\doibase
  10.1103/PhysRev.55.364} {\bibfield  {journal} {\bibinfo  {journal} {Phys.
  Rev.}\ }\textbf {\bibinfo {volume} {55}},\ \bibinfo {pages}
  {364}}\BibitemShut {NoStop}%
\bibitem [{\citenamefont {Uji}\ \emph {et~al.}(2006)\citenamefont {Uji},
  \citenamefont {Terashima}, \citenamefont {Nishimura}, \citenamefont
  {Takahide}, \citenamefont {Konoike}, \citenamefont {Enomoto}, \citenamefont
  {Cui}, \citenamefont {Kobayashi}, \citenamefont {Kobayashi}, \citenamefont
  {Tanaka}, \citenamefont {Tokumoto}, \citenamefont {Choi}, \citenamefont
  {Tokumoto}, \citenamefont {Graf},\ and\ \citenamefont {Brooks}}]{Uji:2006}%
  \BibitemOpen
  \bibfield  {author} {\bibinfo {author} {\bibnamefont {Uji}, \bibfnamefont
  {S.}}, \bibinfo {author} {\bibfnamefont {T.}~\bibnamefont {Terashima}},
  \bibinfo {author} {\bibfnamefont {M.}~\bibnamefont {Nishimura}}, \bibinfo
  {author} {\bibfnamefont {Y.}~\bibnamefont {Takahide}}, \bibinfo {author}
  {\bibfnamefont {T.}~\bibnamefont {Konoike}}, \bibinfo {author} {\bibfnamefont
  {K.}~\bibnamefont {Enomoto}}, \bibinfo {author} {\bibfnamefont
  {H.}~\bibnamefont {Cui}}, \bibinfo {author} {\bibfnamefont {H.}~\bibnamefont
  {Kobayashi}}, \bibinfo {author} {\bibfnamefont {A.}~\bibnamefont
  {Kobayashi}}, \bibinfo {author} {\bibfnamefont {H.}~\bibnamefont {Tanaka}},
  \bibinfo {author} {\bibfnamefont {M.}~\bibnamefont {Tokumoto}}, \bibinfo
  {author} {\bibfnamefont {E.~S.}\ \bibnamefont {Choi}}, \bibinfo {author}
  {\bibfnamefont {T.}~\bibnamefont {Tokumoto}}, \bibinfo {author}
  {\bibfnamefont {D.}~\bibnamefont {Graf}}, \ and\ \bibinfo {author}
  {\bibfnamefont {J.~S.}\ \bibnamefont {Brooks}}} (\bibinfo {year} {2006}),\
  \href {\doibase 10.1103/PhysRevLett.97.157001} {\bibfield  {journal}
  {\bibinfo  {journal} {Phys. Rev. Lett.}\ }\textbf {\bibinfo {volume} {97}},\
  \bibinfo {pages} {157001}}\BibitemShut {NoStop}%
\bibitem [{\citenamefont {Vanderheyden}\ and\ \citenamefont
  {Jackson}(2000)}]{Vanderheyden:2000ti}%
  \BibitemOpen
  \bibfield  {author} {\bibinfo {author} {\bibnamefont {Vanderheyden},
  \bibfnamefont {B.}}, \ and\ \bibinfo {author} {\bibfnamefont {A.~D.}\
  \bibnamefont {Jackson}}} (\bibinfo {year} {2000}),\ \href {\doibase
  10.1103/PhysRevD.62.094010} {\bibfield  {journal} {\bibinfo  {journal}
  {Phys.Rev.}\ }\textbf {\bibinfo {volume} {D62}},\ \bibinfo {pages}
  {094010}},\ \Eprint {http://arxiv.org/abs/hep-ph/0003150}
  {arXiv:hep-ph/0003150 [hep-ph]} \BibitemShut {NoStop}%
\bibitem [{\citenamefont {{Weber}}(1999)}]{Weber-book}%
  \BibitemOpen
  \bibinfo {editor} {\bibnamefont {{Weber}}, \bibfnamefont {F.}},\ Ed.
  (\bibinfo {year} {1999}),\ \href@noop {} {\emph {\bibinfo {title} {Pulsars as
  astrophysical laboratories for nuclear and particle physics /F. Weber.
  Bristol, U.K. : Institute of Physics, c1999. QB 464 W42 1999.
  DA}}}\BibitemShut {NoStop}%
\bibitem [{\citenamefont {Weber}\ \emph {et~al.}(2007)\citenamefont {Weber},
  \citenamefont {Negreiros},\ and\ \citenamefont {Rosenfield}}]{Weber:2007ch}%
  \BibitemOpen
  \bibfield  {author} {\bibinfo {author} {\bibnamefont {Weber}, \bibfnamefont
  {F.}}, \bibinfo {author} {\bibfnamefont {R.}~\bibnamefont {Negreiros}}, \
  and\ \bibinfo {author} {\bibfnamefont {P.}~\bibnamefont {Rosenfield}}}
  (\bibinfo {year} {2007}),\ \href@noop {} {\ }\Eprint
  {http://arxiv.org/abs/0705.2708} {arXiv:0705.2708 [astro-ph]} \BibitemShut
  {NoStop}%
\bibitem [{\citenamefont {Wu}\ and\ \citenamefont {Yip}(2003)}]{Wu:2003zzh}%
  \BibitemOpen
  \bibfield  {author} {\bibinfo {author} {\bibnamefont {Wu}, \bibfnamefont
  {S.-T.}}, \ and\ \bibinfo {author} {\bibfnamefont {S.}~\bibnamefont {Yip}}}
  (\bibinfo {year} {2003}),\ \href {\doibase 10.1103/PhysRevA.67.053603}
  {\bibfield  {journal} {\bibinfo  {journal} {Phys.Rev.}\ }\textbf {\bibinfo
  {volume} {A67}},\ \bibinfo {pages} {053603}}\BibitemShut {NoStop}%
\bibitem [{\citenamefont {Xu}(2003)}]{Xu:2003xe}%
  \BibitemOpen
  \bibfield  {author} {\bibinfo {author} {\bibnamefont {Xu}, \bibfnamefont
  {R.-X.}}} (\bibinfo {year} {2003}),\ \href {\doibase 10.1086/379209}
  {\bibfield  {journal} {\bibinfo  {journal} {Astrophys.J.}\ }\textbf {\bibinfo
  {volume} {596}},\ \bibinfo {pages} {L59}},\ \Eprint
  {http://arxiv.org/abs/astro-ph/0302165} {arXiv:astro-ph/0302165 [astro-ph]}
  \BibitemShut {NoStop}%
\bibitem [{\citenamefont {Yang}(2005)}]{Yang:2005}%
  \BibitemOpen
  \bibfield  {author} {\bibinfo {author} {\bibnamefont {Yang}, \bibfnamefont
  {K.}}} (\bibinfo {year} {2005}),\ \href@noop {} {\bibfield  {journal}
  {\bibinfo  {journal} {Phys.Rev.Lett.}\ }}\Eprint
  {http://arxiv.org/abs/cond-mat/0508484} {arXiv:cond-mat/0508484 [cond-mat]}
  \BibitemShut {NoStop}%
\bibitem [{\citenamefont {Yang}(2006)}]{Yang:2006ez}%
  \BibitemOpen
  \bibfield  {author} {\bibinfo {author} {\bibnamefont {Yang}, \bibfnamefont
  {K.}}} (\bibinfo {year} {2006}),\ \href@noop {} {\ }\Eprint
  {http://arxiv.org/abs/cond-mat/0603190} {arXiv:cond-mat/0603190 [cond-mat]}
  \BibitemShut {NoStop}%
\bibitem [{\citenamefont {{Zwierlein}}\ and\ \citenamefont
  {{Ketterle}}(2006)}]{ketterle2}%
  \BibitemOpen
  \bibfield  {author} {\bibinfo {author} {\bibnamefont {{Zwierlein}},
  \bibfnamefont {M.~W.}}, \ and\ \bibinfo {author} {\bibfnamefont
  {W.}~\bibnamefont {{Ketterle}}}} (\bibinfo {year} {2006}),\ \href@noop {} {\
  }\Eprint {http://arxiv.org/abs/cond-mat/0603489} {arXiv:cond-mat/0603489
  [cond-mat]} \BibitemShut {NoStop}%
\bibitem [{\citenamefont {{Zwierlein}}\ \emph {et~al.}(2006)\citenamefont
  {{Zwierlein}}, \citenamefont {{Schirotzek}}, \citenamefont {{Schunck}},\ and\
  \citenamefont {{Ketterle}}}]{ketterle1}%
  \BibitemOpen
  \bibfield  {author} {\bibinfo {author} {\bibnamefont {{Zwierlein}},
  \bibfnamefont {M.~W.}}, \bibinfo {author} {\bibfnamefont {A.}~\bibnamefont
  {{Schirotzek}}}, \bibinfo {author} {\bibfnamefont {C.~H.}\ \bibnamefont
  {{Schunck}}}, \ and\ \bibinfo {author} {\bibfnamefont {W.}~\bibnamefont
  {{Ketterle}}}} (\bibinfo {year} {2006}),\ \href {\doibase
  10.1126/science.1122318} {\bibfield  {journal} {\bibinfo  {journal}
  {Science}\ }\textbf {\bibinfo {volume} {311}},\ \bibinfo {pages} {492}},\
  \Eprint {http://arxiv.org/abs/arXiv:cond-mat/0511197}
  {arXiv:arXiv:cond-mat/0511197 [cond-mat]} \BibitemShut {NoStop}%
\end{thebibliography}
\end{document}